\documentclass[aps,prl,twocolumn,floats,superscriptaddress]{revtex4-2}
\usepackage{graphicx}
\usepackage{dcolumn}
\usepackage{bm}
\usepackage{amsmath}
\usepackage{color}
\usepackage{amssymb}
\usepackage{ulem}
\usepackage{xcolor}
\usepackage{subfigure}
\usepackage{amssymb}
\usepackage{latexsym}
\usepackage{epsfig}
\usepackage{psfrag}
\usepackage{amsmath,amsfonts,amssymb,bm,ulem}
\usepackage{color}
\usepackage{float}

\graphicspath{{figs/}}

\newcommand{\be}{\begin{equation}}
\newcommand{\ee}{\end{equation}}
\newcommand{\bea}{\begin{eqnarray}}
\newcommand{\eea}{\end{eqnarray}}
\newcommand{\beq}{\begin{equation}}
\newcommand{\eeq}{\end{equation}}

\begin{document}
%%%%%%%%%%%%%%%%%%%%%%%%%%%%%%%%%%%%%% AUTHORS %%%%%%%%%%%%%%%%%%%%%%%%%

\author{Andrey Chubukov}
\affiliation{Department of Physics, University of Minnesota, Minneapolis, MN 55455, USA}

\author{Ilya Esterlis}
\affiliation{Department of Physics, University of Wisconsin-Madison, Madison, WI 53706-1390, USA}

\author{Artem Abanov}
\affiliation{Department of Physics,  Texas A\&M University, College Station,  TX 77843, USA USA}

\author{Nikolay Prokof'ev}
\affiliation{Department of Physics, University of Massachusetts, Amherst, MA 01003, USA}

%%%%%%%%%%%%%%%%%%%%%%%%%%%%%%%%%%%%%%%%%%%%%%%%%%%%%%%%%%%%%%%%%%%%%%%%%%%%%%
\title{Breakdown of the Migdal-Eliashberg theory
for electron-phonon systems. Role of  polarons/bipolarons }
%%%%%%%%%%%%%%%%%%%%%%%%%%%%%%%%%%%%%%%%%%%%%%%%%%%%%%%%%%%%%%%%%%%%%%%%%%%%%%
\begin{abstract}
The Migdal-Eliashberg theory (MET) describes electrons interacting with phonons in the adiabatic limit when the phonon Debye frequency is much smaller than the Fermi energy. A conventional belief is that MET holds even at strong coupling, when electron self-energy is large, and for an interaction with an optical phonon breaks down only near the point where the dressed phonon spectrum softens to near zero. We analyze numerically and analytically a different option for this case—a collapse to a polaronic/bipolaronic ground state. The last scenario has never been analyzed in precise quantitative terms for a generic electron density. Using variational considerations, we establish rigorous upper bounds on the coupling $\lambda$, at which a Fermi-liquid state transforms into the bipolaron/polaron state. We show that at small and near-maximum densities, this happens well before a dressed phonon softens. This is true both in two- and three-dimensional systems; in the latter, the upper bound on $\lambda$ tends to zero in the limit of small or near-full density. We present analytical reasoning for this behavior based on hints extracted from exact diagrammatic treatment of the on-site Holstein model for the spin polarized case and argue that polarons are produced by fermions with energies comparable to the bandwidth; i.e., polaron formation is outside the realm of MET. Closer to half-filling, the leading instability upon increasing $\lambda$ is toward a charge-density-wave state (CDW), and there exists a strong coupling regime of MET near this instability, while the polaron/bipolaron state develops at larger $\lambda$ out of a CDW-ordered state and inherits a CDW order over some range of coupling.
\end{abstract}

\maketitle
%\tableofcontents
%%%%%%%%%%%%%%%%%%%%%%%%%%%%%%%%%%%
\section{Introduction}
\label{sec:1}

An incredibly rich and fascinating physics displayed by a system of fermions interacting
with lattice vibrations (phonons) continues to attract significant interest among researchers
interested in various aspects of strong correlations and superconductivity (see, e.g., the review articles  \cite{Devreese_2005,Aleksandrov_2010,Franchini_2021,Dai_2025}).
Models of coupled electron-phonon (e-ph) systems, particularly that of fermions interacting with a single Einstein phonon, have long been considered the best examples
of the applicability of Migdal-Eliashberg theory (MET), both in the normal and superconducting states; see, e.g., Refs.~\cite{Migdal,Eliashberg,AGD}.
At weak coupling, MET is synonymous with perturbation theory/BCS theory of superconductivity,
with the advantage that no artificial high-energy cutoff is required; e.g., superconducting $T_c$
can be computed exactly, including the prefactor~\cite{Mazin_2005,*Wang_Ch_2013,Andrey_review,Mirabi_2020,Kiessling2025,*Kiessling2025_1,*Gnezdilov_2025}
     \footnote{At weak coupling, $\lambda \ll 1$, Eliashberg $T_c = 1.13 \sqrt{e} \omega_0 e^{-1/\lambda} = 0.69 \omega e^{-1/\lambda}$, where $\omega_0$ is the Debye frequency (see \cite{Mazin_2005,*Wang_Ch_2013} and references therein), 
differs by $\sqrt{e}$ from the often cited  value 1.13, which holds  for the case when the phonon susceptibility is set to be constant at frequencies smaller than $\omega_0$ and zero otherwise.}.
At strong coupling, MET states that in the adiabatic limit,  when the Debye frequency, $\omega_0$, is much smaller than the Fermi energy, $E_F$,  one can proceed beyond perturbation theory and
rigorously analyze the properties of both  normal and superconducting states,
despite a large enhancement of the
fermionic mass and a non-FL behavior in a wide range of frequencies.
Specifically, MET states that at strong coupling, one can still use a self-consistent one-loop approximation because vertex corrections remain small in
$\gamma=\omega_0/E_F \ll 1$. The physical argument is that
in the scattering processes, accounting for vertex corrections, fermions with typical momenta of the order $k_F$ vibrate at frequencies comparable to $\omega_0$, far away from their own resonance at $E_F$.

More recent studies, which have intensified over the last few years with the advent of more powerful analytical and numerical methods, 
questioned the existence of the strong-coupling regime 
%described by 
MET for the interaction with an Einstein phonon. 
One argument
here is technical - in many cases, the effective Debye frequency dressed by interaction with fermions acquires rather strong momentum dependence, which qualitatively  changes
the structure of the one-loop self-consistent theory and the rationale for neglecting vertex corrections.
Another argument, outlined in  Ref. \cite{Yuzbashyan,*Yuzbashyan_1,*Yuzbashyan_2,*Yuzbashyan_3},
points to a possibility that
in the strong coupling regime, the MET theory  undergoes a kinetic instability at which the phonon and electron subsystems develop a temperature difference.
And the most drastic argument
is that even before the system enters a strong coupling regime, it may undergo
a radical reconstruction of low-energy excitations due to the emergence of
polarons and bipolarons  as relevant degrees of freedom~\cite{Alexandrov, Mott,*Mott_1}.

In this communication, we analyze the interplay between MET and polaron/bipolaron formation
at $T=0$ for several systems (in two- and three-dimensions) with Holstein-type e-ph interaction, including interaction with an Einstein phonon with frequency $\omega_0$.
We combine a variational numerical analysis and an analytical (diagrammatic) approach and analyze the evolution of the system's behavior with increasing 
e-ph coupling at various fermionic filling factors $n$.
We keep the ratio of the e-ph interaction and the Debye frequency
fixed at a large value and vary the bandwidth $W$.  We analyze how the fermionic
density of state (DOS) evolves as $W$ decreases and eventually vanishes.
We introduce fermions that behave as free quasiparticles
in the limit $W \to \infty$  and express the DOS of the FL, described by MET,
and of a  polaron state  in terms of the same fermions.
In other words, we  do not introduce new fermionic operators for electron-phonon bound states, as is often done in the analysis of a polaron state (see, e.g., \cite{Mahan00}).

 \begin{figure}[]
 \centering
    \noindent
    \includegraphics[]{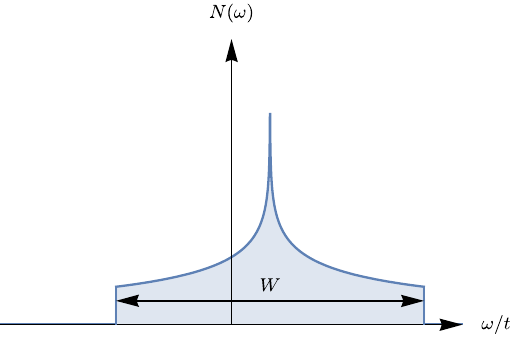}
 \includegraphics[]{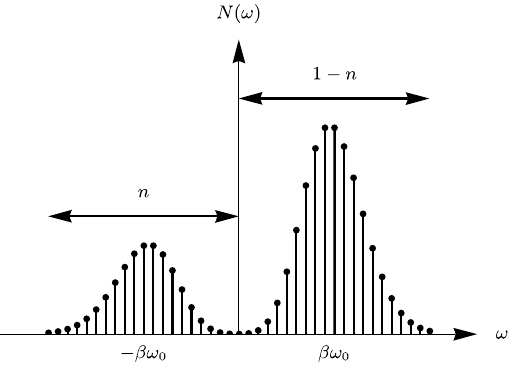}
  \caption{Density of states $N(\omega)$ for 2D tight-binding model with nearest-neighbor hopping $t$ and bandwidth $W = 8t$. The parameter $\beta \gg 1$ is controlling the ratio between the e-ph interaction and Debye frequency $\omega_0$.  (a) The weak coupling limit of large $W \gg \beta \omega_0$. The density of states is a continuous function of frequency. 
  (b)  The limit  $t=W=0$ (the single site Holstein model).  The density of states consists  of a set of delta function peaks marked by vertical lines, separated by the phonon energy $\omega_0$. The length of each vertical line is proportional to the residue of the corresponding $\delta$-function.}
 \label{fig:dos_ss_Holstein}
 \end{figure}
 
 For an Einstein phonon with frequency $\omega_0$, the limit $W=0$ is described by the exactly solvable Holstein model. In the latter, the DOS of spinless (spin-polarized) fermions consists of a set of $\delta$-functional peaks separated by $\omega_0$ (Fig.~\ref{fig:dos_ss_Holstein}b).
These peaks correspond to polarons, which can be interpreted as 
``bound states" of an electron and strong lattice deformation 
(see more on this interpretation below).  
The peak with the smallest energy is the ground state, whereas other peaks are states
featuring multiple excited phonons relative to the ground state.
For spinful fermions, phonon-mediated pairing interaction between polarons made of fermions with opposite spins binds polarons into bipolarons. This leads to vanishing of the DOS around $\omega=0$.
For both polarons and bipolarons, the DOS is quite different from that at small e-ph coupling $\lambda$, which is  continuous at $\omega =0$ and has a width $W$ (Fig. ~\ref{fig:dos_ss_Holstein}a).

\begin{figure*}[htbp]
	\includegraphics[width=0.4\linewidth]{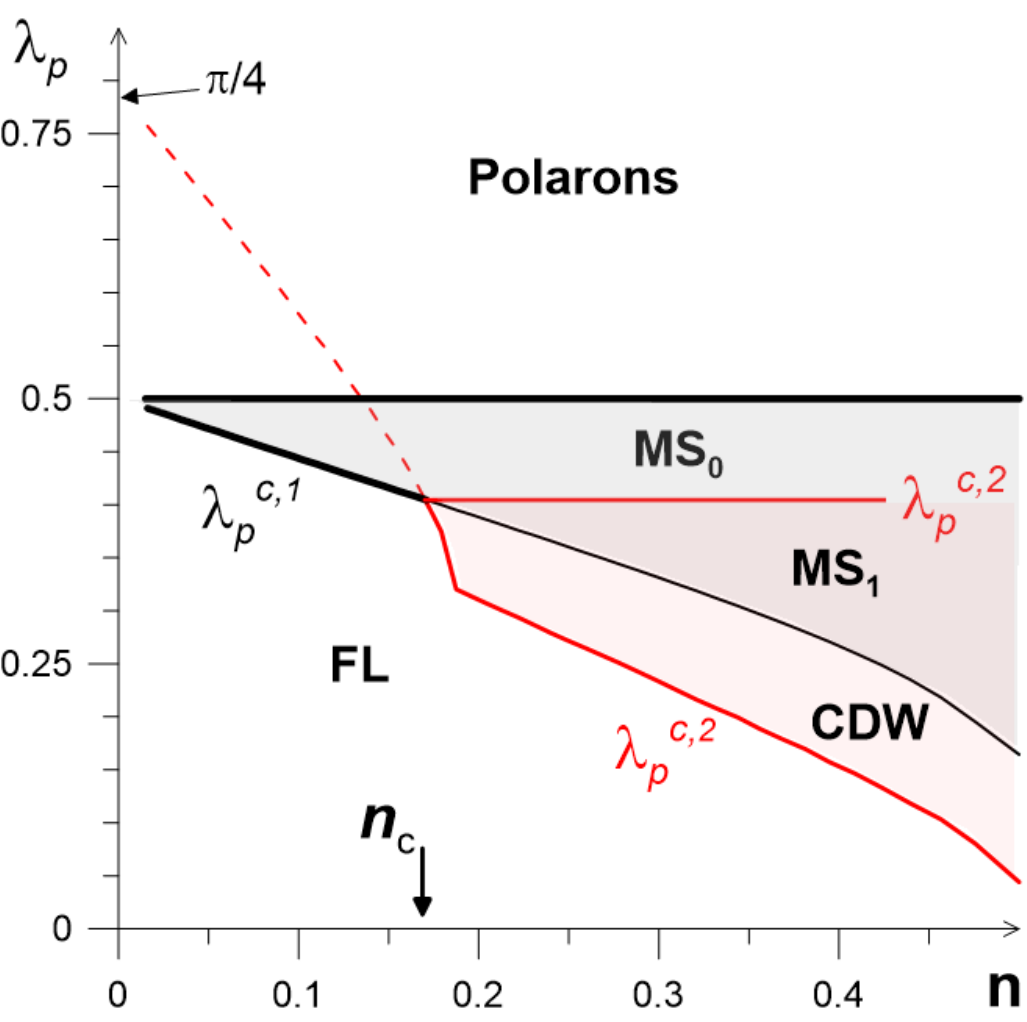}
    ~~~~~~~~~
    \includegraphics[width=0.44\linewidth]{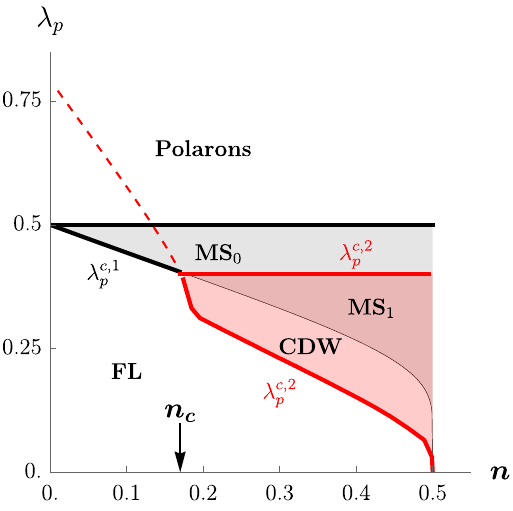}
\caption{Phase diagram obtained in variational (left) and analytical (right) analysis for a homogeneous polaron order on the $(\lambda_p, n)$ plane, where $\lambda_p$ is the ratio of polaron binding energy and fermionic bandwidth and $n$ is the density of fermions.
 The variational phase diagram has been obtained for spinful electrons and bipolarons. It is adjusted here to spinless fermions and polarons for comparison with the analytical phase diagram.  We see that the two phase diagrams are identical. We explain in the text why this is so.  
The phase diagram contains the regions of a Fermi liquid (FL), described by MET, a homogeneous polaron state, a charge-ordered 
state of electrons (CDW) and  mixed states, in which both polaron and electronic  components are present.
The notations MS$_0$ and MS$_1$ 
denote, respectively, the mixed states in which  the  electronic component is homogeneous and where it is CDW-ordered. }
\label{fig:phase_diag_ms1}
\end{figure*}

\begin{figure}[htbp]
    \includegraphics[width=\linewidth]{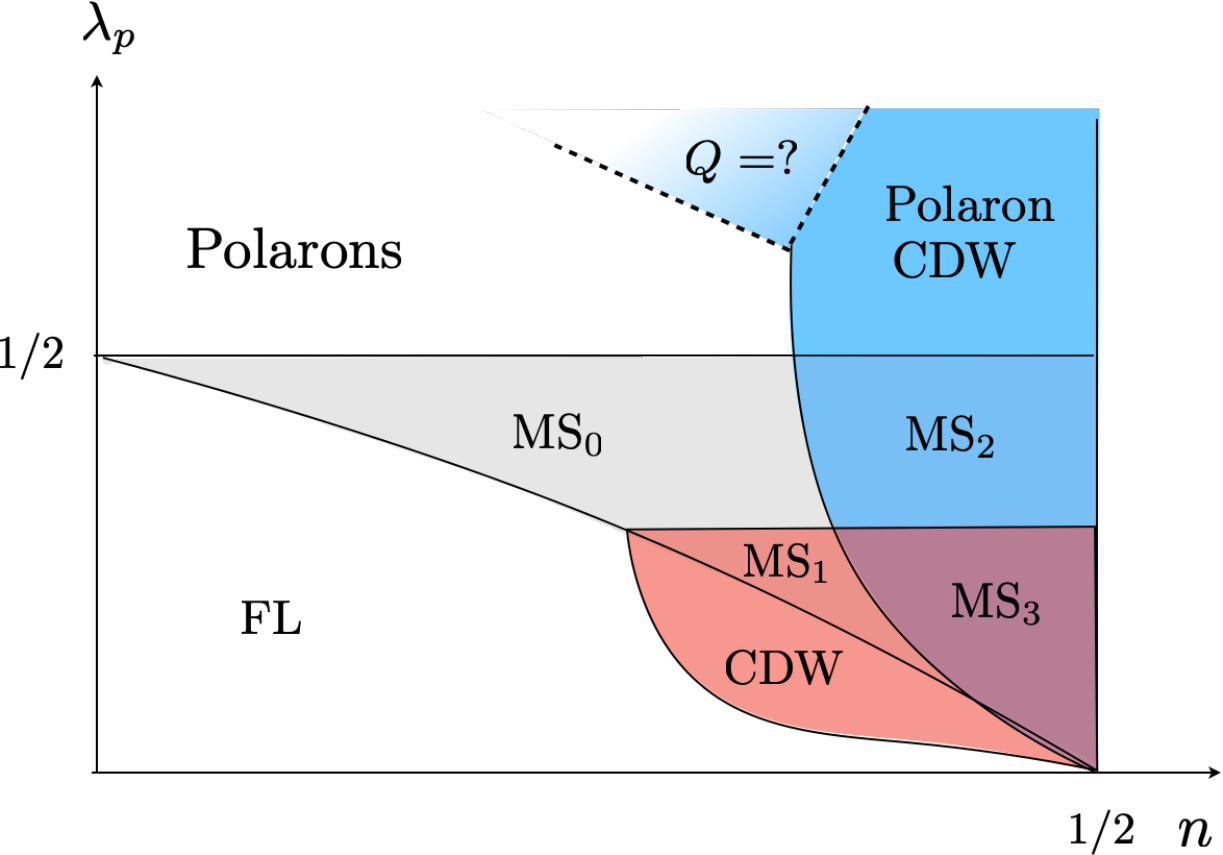}
\caption{Schematic phase diagram for an electron-phonon system as a function density $n$ and dimensionless coupling strength $\lambda_p$, displaying the full set of phases which we considered in this study.  The phase diagram is for spinless fermions and polarons.  The phase diagram for spinful fermions and bipolarons is quite similar.   The diagram contains  pure polaron  states with homogeneous (P) and checkerboard (P-CDW) order  
 and, potentially, the states with a polaron order with an incommensurate momentum $Q$, an FL state, a CDW-ordered  state of fermions and a set of mixed states in which density $n$ is split between polaron and  FL/CDW components. The mixed state  MS$_0$ has P and FL components, 
MS$_1$ has P and CDW components, MS$_2$ has P-CDW and FL components, and MS$_3$ has P-CDW and CDW components.
 The lines approaching $n=1/2$ should be viewed as describing system behavior at densities arbitrary close  but not exactly equal to $1/2$.  For  $n  \equiv 1/2$, the behavior is special because of degeneracy between an electronic and polaronic CDW. }
\label{fig:phase_diag_full}
\end{figure}

Using numerical and analytical studies, we present the theoretical scenario for the evolution of the system at different $W$ and fermionic density $n$  (see Figs.~\ref{fig:phase_diag_ms1} and \ref{fig:phase_diag_full}).
In the numerical study, we consider $S = 1/2$ fermions, for which the polaron state is
that of heavy bipolarons (bound states of two electrons dressed by lattice distortion).
We  compare ground state energies and chemical potentials of the FL state, described by MET, and the variational state consisting of localized bipolarons, identify the ranges of couplings at which either the pure bipolaron state or the mixed state with bipolaron and FL components 
is energetically favorable, and compare this coupling with the one at which the dressed phonon spectrum softens to zero. 
 
 In the analytical study, we consider spinless
fermions (a spin-polarized system), in which case there are polarons but no bipolarons.
We  first develop a diagrammatic description of the polaron state for an Einstein phonon in the limit $W=0$ (a dispersionless fermion),
identify the key features of the diagrammatic series, and then
show how these features evolve when $W$ becomes finite and identify 
the range of $W$ (i.e., the range of couplings), where
the polaron state and the mixed state with both polaron and FL components present are  energetically favorable compared to a pure 
FL state described by MET. We also analyze how a polaron state emerges  if one departs from the FL and increases the coupling. 

In both numerical and analytical study we  consider a particle-hole symmetric model for which the behavior at small fermionic densities is the same as near the full filling.  For spinful fermions, the phase diagram is symmetric with respect to $n=1$; for spinless fermions, it is symmetric around $n =1/2$.
%AC 
A short summary of the results is presented in the companion paper~\cite{chubukov_prl_2026}.

\subsection{Notations and summary of the results}

For convenience of the reader, we present below the list of notation and a short summary of the results of our
numerical and analytical studies, focusing on
$n \leq 1$ for spinful fermions and $n \leq 1/2$ for spinless fermions.   

\subsubsection{\bf {Notations}}
\label{subsec_Notations}

We define the electron-phonon coupling as $g(q)$ and the bare phonon frequency as $\omega_0 (q)$. For numerical studies, we define the dimensionless coupling constant $\lambda$ as $\lambda_0 = N_F \langle g^2 (q) /\omega^2_0 (q)\rangle_{FS}$, where $N_F$ is the density of states at the Fermi level, the averaging is over the Fermi surface with $q$ connecting points on the Fermi surface.  We define $\lambda^{c,1}_0$  as the critical coupling for the instability of an FL state toward the development of polarons/bipolarons and $\lambda^{c,2}_0$ as the critical coupling for the instability of an FL state toward a charge-density-wave (CDW) order.  
 To account for the softening of the phonon mode in MET due to interaction with electrons, from $\omega_0 (q)$ to $\omega_r (q)$,  we also introduce
the renormalized coupling 
$\lambda_r = N_F \langle g^2 (q) /\omega^2_r (q)\rangle_{FS}$.  
     
For analytical studies, we focus on 2D,  set $g$ and $\omega_0$ to be momentum independent, introduce the bandwidth $W$  and define
an alternative 
dimensionless coupling $\lambda_p = g^2/(2 \omega^2_0 W) = \beta \omega_0/W$, where $\beta = g^2/(2 \omega^3_0)$. 
We set $\beta$ as a large number and vary $\lambda_p$ by varying $W$. 
     
We obtain the same two critical couplings in terms of $\lambda_p$: $\lambda^{c,1}_p$ and $\lambda^{c,2}_p$. We argue that at small density $\lambda_p$ is more adequate for the analysis of polaron formation than  $\lambda_0 = (g^2/\omega_0^2) N_F$ because we
 will argue in the following that polarons are produced by fermions with energies comparable to the bandwidth $W$ rather than the Fermi energy $E_F$.          

 \subsubsection{\bf {Numerical results}}
 
We  considered $S=1/2$ fermions with a tight-binding dispersion relation on
the simple cubic lattice in three dimensions (3D) and the square lattice in 2D.
We introduced a local density-displacement-type coupling to lattice vibrations and probed  several typical phonon dispersions. For the FL state, we compute 
the critical value $\lambda_0^{c,2}$  for an electronic CDW instability, 
 obtain the momenta at which the instability takes place, and evaluate the dressed coupling $\lambda_r$. We then compared the 
energy and chemical potential of the FL  state at various fermionic densities with those of the variational bipolaron (BP) state composed of 
localized fermion pairs.  For the cases where the bottom of the dispersion of bare phonons is close to $q=0$, we considered a homogeneous state with the highest density $n_i =2$ at $N n/2$ sites ($N$ is the total number of sites). For the cases where the bottom of the bare dispersion is near the antiferromagnetic momentum ${\bf Q}$, we considered the state with the checkerboard arrangement of pairs on $Nn$ sites. We label this state BP-CDW.    
We  determined the critical coupling $\lambda^{c,1}_0$ for the transformation of the FL state into the mixed FL + BP state (which is likely to be  phase separated) by analyzing the coupling at which the chemical
potentials of the FL and BP states become equal. 
We emphasize that once bipolarons  appear, the description based on MET 
becomes invalid, although the theory itself remains internally stable against low-energy fluctuations. 
We then identified the larger critical coupling, corresponding to the upper end of the mixed state, as the one at which the chemical potential of the BP state reaches the bottom of the electron dispersion. 
At even larger coupling, the ground state is a BP state with localized fermionic pairs.  

 We show the results in Figs.~\ref{Upin3D} and \ref{all2D} in units of $\lambda_0$ and reexpress some of these results 
 in units of $\lambda_p$  in Figs. \ref{fig:phase_diag_ms1}  and \ref{fig21}. 
In 3D (Fig.~\ref{Upin3D}), we found that in the low density limit $\lambda_0^{c,1}$ is proportional to the Fermi momentum, i.e. it scales as  $n^{1/3}$, while $\lambda_0^{c,2}= O(1)$ ($\lambda_0^{c,2} = 1/2$ for 
fermions with a parabolic dispersion). In this situation,  
bipolarons appear, and the description within the MET becomes invalid 
already at weak coupling, when the MET is well inside its stability region.
In other words, the low-density (continuous) limit of the MET in 3D is ill-defined.
 As density increases, $\lambda_0^{c,1}$ also increases, while $\lambda_0^{c,2}$ decreases. We showed that for some phonon dispersions,  there exists a range of fermion densities around half-filling $n_c <n < 1$ where $\lambda_0^{c,2} < \lambda_0^{c,1}$. In this range, the internal instability of the MET toward a CDW order develops before the polaron formation.
 
 We also analyzed the structure of the mixed  state in this range and found 
 the line at which the FL component in the mixed state loses a CDW order as  $\lambda_0$ increases.   
This line can be viewed as an extension of the 
$\lambda_0^{c,2}$ line to the mixed state. We found that 
$\lambda_0^{c,2} (n)$  displays a reentrant behavior at $n>n_c$. 
       
As  $\lambda$ approaches $\lambda_0^{c,2}$ from below, the system 
should nominally go into the strong coupling regime of MET, where the dressed $\lambda_r$ becomes large. However, we found that in most cases that we studied, $\lambda_r$ remains $O(1)$ even in the vicinity of $\lambda_0^{c,2}$.
We associate this with the fact that for finite $n$,  the phonon 
spectrum softens only at isolated momentum points, and the phase space around the softening is not large enough to bring the system into the strong coupling regime
in a detectable range of $\lambda_0 \leq \lambda_0^{c,2}$. 

We found very similar results for 2D systems (Fig. \ref{all2D})  with one notable exception:
in the low density limit $\lambda_0^{c,1}$ approaches a finite value. This is directly related to the fact that the density of states of a 2D Fermi gas is independent of $k_F$ at small $k_F$. Nevertheless, we found numerically  that 
at low fermion densities $\lambda_0^{c,1} < \lambda_0^{c,2}$ , i.e. the prime reason for breakdown of the MET at $n <<1$ is an instability toward the formation of bipolarons. 

 \subsubsection{\bf {Analytical results}}

We considered 2D spinless fermions with tight-binding dispersion coupled to a dispersionless phonon with frequency $\omega_0$.  We  obtained three sets of analytical results: one for an infinitesimally small density $n =0+$, another for an infinitesimally small deviation for full filling $n = 1-$, and the third for arbitrary $n$. 
We present the analytical phase diagrams for the homogeneous and checkerboard polaron configurations in the right panel of 
 Fig. \ref{fig:phase_diag_ms1} and of Fig. \ref{fig:phase_diag_ms2+ms3}, respectively.
 For cases $n=0$ and $n=1$, we first demonstrate that the exact DOS of the Holstein model in the  limit $W=0$ is reproduced using the eikonal diagrammatic approach. 
We then extend  the diagrammatic analysis to finite $W$  and argue that  the DOS     
 can be approximated by a
 set of narrow, nearly $\delta-$functional peaks at low energies, $\omega = m \omega_0$, $m =0, 1, 2$,  etc. for $n=0$ and  $m =0, -1, -2$ for $n=1$, 
  which describe heavy polarons
with  exponentially large mass  $e^{\beta}$
and  a free-particle-like continua of  width $W$, centered at much larger 
  $ \omega = \beta \omega_0$ at $n=0$ and at $\omega = -\beta \omega_0$ at $n=1$.    As $W$ increases, the lower end of the continuum moves toward $\omega =0$, absorbing  low-energy polarons  one by one. 
 The last polaron at $\omega =0$ is absorbed at $\lambda^{c,1}_p =1/2$. At smaller $\lambda_p$,  the system is in an FL state described by MET.  There is no mixed state for $n =0$ and $n=1$.  For these $n$, we found $\lambda^{c,2}_p = \pi/4$, which is larger than $\lambda^{c,1}_p =1/2$. In this situation, an FL state becomes energetically unfavorable compared  to the polaron state when MET is internally state.  We also 
 approached  $\lambda^{c,1}_p =1/2$ from smaller $\lambda_p$  and argued that the onset of the  polaron instability is the development of a  peak in the DOS at the lower end of the continuum, which can be interpreted as a polaron bound state.  We argued that the bound state is {\it not} a low-energy phenomenon despite that it appears at $\omega$ equal to the chemical potential, because the corresponding contribution to the fermionic self-energy comes from 
 high-order, $O(\beta)$,  in the diagrammatic series and involves fermions with energies comparable to $W$.
  Fluctuations of these  high-energy fermions are disconnected from  fluctuations of low-energy  fermions  with energies well within $W$, which  determine the stability of MET. 
 
At a finite $n$, we found that one has to do a calculation in three steps to reproduce the exact DOS of the Holstein model at $W=0$. First, one has to add an ancilla fermion ${\tilde c}$ -- a hole-like excitation with the opposite value of the chemical potential compared to $\mu_n = (2n-1) \beta \omega_0$  for the original electron $c$ (for each, $\mu_n$ is the chemical potential without the Hartree contribution, the  total 
 chemical potential $\mu_P = - \beta \omega_0$)  
   Second, one has to   
 introduce a condensate  order  parameter $\Delta = <c^\dagger c>$ and use it to decouple the phonon-induced interaction between the densities of the original and ancilla fermions.  We obtained a nonzero value of $\Delta$ from the minimization of the ground-state energy  and verified that the energy is smaller at a finite $\Delta = 2\beta_0 (n(1-n)^{1/2}$.   Diagonalizing the  effective quadratic Hamiltonian we found two sets of fermions with energies $\pm \beta \omega_0$, which can be identified as describing the filled and empty states.  The electron-phonon interaction decouples between these new fermions, and the effective Hamiltonian becomes the sum of the  two terms: one is the same as for $n=0$, and the other is the same as for $n=1$.  Third, one has to do eikonal calculations for $n=0$ and $n=1$ components. Using this procedure, we reproduced the exact DOS for the physical fermion consisting of two sets of $\delta$-functions at $\omega \geq 0$ and $\omega \leq 0$,  the first with the overall factor $1-n$ and the second with the overall factor $n$. 
 
We then extended the calculations at $0<n<1/2$ to  finite $W$.  Now the condensate order parameter $\Delta$ has to be selected with a particular momentum ${\bf q}$.  We analyzed two polaron states: a homogeneous one with 
$\Delta_{q=0} =\Delta_0$ and a checkerboard one with $\Delta_{q=Q} =\Delta_Q$, where $Q= (\pi,\pi)$.
We verified that the $\Delta_0$ state  is  localized with no density fluctuations between occupied and non-occupied states,
We showed that the DOS of each polaron state  can be approximated by a
 set of narrow, nearly $\delta-$functional peaks at low energies, $\omega = n \omega_0$, $n =0, \pm1, \pm 2$,  etc.,
  which describe heavy polarons
with  exponentially large mass  $e^{\beta}$ and  two free continua at higher energies, centered at
  $ \omega = \pm \beta \omega_0$, each with width $W$.  
 As $W$ increases and $\lambda_p$ decreases,  the lower edges of the two continua extend to smaller
$|\omega|$
 and absorb low-energy polarons  one by one. 
 
 For the homogeneous polaron state, the edges of the two continua merge at $\omega =0$ at $\lambda_p =1/2$. We found that at smaller 
 $\lambda_p$,  the ground state is a mixed state, in which a portion 
  of a system with density $n-\delta$ is in the homogeneous polaron state and the other portion, with density $\delta$ is in an FL state. 
    The parameter $\delta$ increases with decreasing $\lambda_p$ and becomes equal to $n$ at 
   the critical $\lambda^{c,1}_p (n) < 1/2$. 
   At $\lambda_p < \lambda^{c,1}_p (n)$ the ground state is an FL described by MET. 
 For  larger $n >n_c$, the phase diagram is modified because 
  the critical $\lambda^{c,2}_p$, at which the phonon spectrum in a MET softens at some momentum $q$, becomes smaller than $\lambda^{c,1}_p$ (the two lines cross at $n=n_c$).   In this situation, the leading instability upon    
    increasing $\lambda_p$ is in a state with a CDW electronic order. We argued that  near the onset of CDW order
     there is a range of strong coupling behavior within MET. 
     \footnote{We emphasize that vertex corrections are still small in this regime, as $\omega_r (q)$ is smaller than $E_F$ for all $q$.  In this respect strong coupling MET for electron-phonon interaction
  is different from an effective MET for electrons interacting with soft fluctuations in spin or charge channel
 (see Refs. \cite{Zhang_1,*Zhang_2} for more details).}.
  For the same $n >n_c$,  the $\lambda^{c,2}_p (n)$ line  displays  a reentrant behavior inside the mixed state, splitting it into states with and without CDW order of the FL component.
  
 For the checkerboard polaron order with $q =Q$, the DOS at large but finite $\lambda_p$  again consists of the two continua and polaron patches at smaller frequencies, and the lower edges of the continua extend to smaller $\omega$ as $\lambda_p$ decreases. 
 However,  they do not reach $\omega=0$ at the onset of the mixed phase, which leads to a somewhat different form of the DOS in the mixed phase compared to the $q=0$ case. The phase diagram  for $q =Q$ is  similar to that for $q=0$ (see Fig.  \ref{fig:phase_diag_ms2+ms3} ).
 However,  the $q=Q$ state is not localized beyond 
 the mean-field approximation, and to obtain the phase diagram, we treated the effects due to non-localization phenomenologically.  

We  combined the phase diagrams for $q=0$ and $q=Q$ polaron orders,  Figs. \ref{fig:phase_diag_ms1} and \ref{fig:phase_diag_ms2+ms3} , 
 and  obtained the complete phase diagram, Fig.~\ref{fig:phase_diag_full}, 
 assuming that the homogeneous polaron state develops in smaller $n$ and the checkerboard polaron state develops near $n =1/2$.

\subsubsection{\bf {Structure of the paper}} 

 In Sec. \ref{sec:Summary} we introduce the model, set up the notation, and briefly review MET.  In Section~\ref{sec:Numerics} we examine the physics of spin-$1/2$ fermions numerically using the variational approach.
In Section~\ref{sec:Analytics} we present our analytical results for spinless fermions.
We first consider the limiting cases of zero and filled band densities and then
present results for arbitrary $n$.
We compare our analytical and numerical results in Sec. \ref{sec:comp} in 2D and 3D and
discuss a comparison with previous numerical studies of the system evolution with increasing coupling.
We present our conclusions in Sec.\ref{sec:concl}.  Some technical details are presented in the Appendices 
In particular, in Appendix~\ref{app_A} we rationalize the need to include vertex corrections into diagrammatic treatment by 
comparing the eikonal technique, which treats self-energy and vertex corrections on equal footing and reproduces the  exact result for the Holstein model,  with rainbow and self-consistent
one-loop approximations, both of which neglect vertex corrections.

 Our conclusions regarding bipolaron/polaron instability as 
the leading reason for MET breakdown at small enough density $n$ 
are generic. Due to vanishing DOS at
small 
 $n$ 
 in 3D,  continuous space models of e-ph interaction 
have zero 
radius of convergence and become unstable toward bipolarons/polarons  already 
at vanishingly small $\lambda_0$, while 
lattice 
 models 
 become 
  unstable 
   at
$\lambda_0 \ll 1$. When it comes to the phase diagrams and 
the  locations of the transition lines,
 the results 
are model dependent.  
 In particular, we  demonstrated in our variational study 
 that 
 phonon dispersion 
 strongly affects the phase diagram
  already 
 for the simplest Holstein model.
  Modifications to  the phase diagram are also expected if one changes the coupling, 
the lattice type, the number of 
 phonon branches, or 
adds Coulomb interaction to the model.
 In this respect, 
different Hamiltonians should be investigated on the case-by-case basis, 
and our work points to the key steps that need to be taken in the analysis. 

\section{Theoretical setup and MET}
\label{sec:Summary}

\subsection{Theoretical setup}

We consider a lattice model of fermionic density coupled to lattice vibrations. In its simplest form, the
coupling is between the electron density and a single phonon mode.
The corresponding interaction Hamiltonian is (in notation closely following the textbook [\onlinecite{AGD}])
\begin{equation}
H_{\rm int} = \sum_{{\mathbf q} } \, \frac{g({\mathbf q})}{\sqrt{2\omega_0({\mathbf q})}} \,
\left[ n_{\mathbf q} b_{\mathbf q} + h.c. \right] ,
\label{Hint}
\end{equation}
where $n_{\mathbf q}= (1/N) \sum_k c^\dagger_{k+q} c_k$ 
is the Fourier transform of the total electron density 
($N$  is the total number of sites),
and $b_{\mathbf q}$ is the annihilation operator of a phonon with 
momentum ${\mathbf q}$ and frequency $\omega_0({\mathbf q})$.
We assume a tight binding model for electrons on a square/qubic lattice with dispersion
$\epsilon_{{\mathbf k}}=-2t\sum_{\alpha =1}^{d} \cos (k_{\alpha} a)$ and set the  lattice spacing $a$ as a unit
of length (i.e., we set $a=1$ in the rest of the text). In these notations
${\bar g} ({\mathbf q}) = g({\mathbf q})/\sqrt{2 \omega_0 ({\mathbf q})}$ has the dimension of energy.
The bare fermion propagator in Matsubara frequencies is
$G_0({\mathbf k}, \omega_m) = 1/(i \omega_m - \epsilon_{\mathbf k}+\mu)$.
Similarly, the bare phonon propagator is $\chi ({\mathbf q}, \Omega_m) = 2\omega_0 ({\mathbf q})/(\Omega^2_m + \omega^2_0 ({\mathbf q}))$.

Equation (\ref{Hint}) is often simplified by considering  momentum-independent parameters
$g$, $\omega_0$, and ${\bar g} = g/\sqrt{2\omega_0}$. In this work, we also approximate
$g({\mathbf q})$ by $g$ but  keep the momentum dependence of $\omega_0 ({\mathbf q})$ (see Eq. (\ref{oo}))
in  our numerical analysis. In the analytical study, we approximate $\omega_0 (q)$ by $\omega_0$.

In general, there are four energy scales in the problem: electron-phonon interaction, Debye frequency,
Fermi energy, and the fermionic bandwidth, $W$.
For densities around half-filling, $E_F$ and $W$ are comparable, but at a small density, $E_F$ is much smaller than $W$, and for near-full filling $W-E_F$ is much smaller than $W$.

\subsection{Migdal-Eliashberg theory}
\label{MET_th}

MET is a theoretical tool to extract dressed fermionic and bosonic propagators from Eq. (\ref{Hint}) both in the normal and superconducting states. The theory uses as input the adiabatic condition $\omega_0 (q) \ll E_F$
and {\it assumes} that the interactions are relevant only for
fermions in the near vicinity of the Fermi surface and operates with two dimensionless  parameters.
The first is the dimensionless electron-phonon coupling
\begin{equation}
\lambda_0 = N_F \left\langle \frac{g^2}{\omega^2_0({\mathbf k}-{\mathbf k}')} \right\rangle _{FS} ,
\label{l0}
\end{equation}
where $\langle \dots \rangle$ stands for average over the Fermi surface (FS),
and the second is
\begin{equation}
\lambda^E_0 = \frac{N_F}{E_F}
\left\langle \frac{g^2}{\omega_0({\mathbf k}-{\mathbf k}')} \right\rangle _{FS} .
\label{l0_1}
\end{equation}
The latter is often called the Eliashberg parameter;
for constants $g$ and $\omega_0$ one has $\lambda^E_0 = \lambda_0 \omega_0/E_F$.
In perturbation theory,
$\lambda_0$ accounts for the renormalization of the fermionic residue via the self-energy
$\Sigma (\omega_m) = -i\lambda_0 \omega_m$ at the smallest $\omega$ (we use sign convention $G^{-1} = G^{-1}_0 - \Sigma$),  while a much smaller Eliashberg parameter
$\lambda^E_{0}$ controls the strength of vertex corrections and  the Landau damping term in the bosonic propagator.
Sometimes, the effective fermion-phonon coupling with dimension of energy is defined as
$u^2 = N_F g^2$
(see, e.g., \cite{combescot,paper_5}. In these notations,
$\lambda_0 = u^2/\omega^2_0$ and $\lambda^E_0 \sim u^2/(E_F \omega_0)$.
We emphasize that of four parameters, only three - $g$, $\omega_0$ and $E_F$ - appear in MET. The bandwidth scale $W$ does not appear there
as MET assumes that there are no interesting effects coming from fermions with energies comparable to $W$.
Moreover, the scale $E_F$  is important only in the adiabatic condition that
sets the validity of MET.

The distinction between MET and conventional perturbation theory follows from
the argument~\cite{Migdal,Eliashberg} that MET remains under control even
outside the applicability of a perturbative expansion when the renormalized dimensionless coupling
$\lambda_r$,  Eq.~(\ref{lr}), becomes larger than one. This holds
as long as the Eliashberg parameter remains small and vertex corrections can be neglected.
From a mathematical perspective, MET in the strong coupling regime is a self-consistent one-loop theory.
It was understood  early on~\cite{Migdal,Eliashberg,AGD} that the applicability condition of MET must
 be formulated not in terms of the bare couplings $(\lambda_0,\lambda^E_0)$
but in terms of the renormalized ones
$(\lambda_r, \lambda^E_r)$, based on the phonon spectrum softened by electron polarization.
In the adiabatic regime,  the dynamic renormalization of the phonon propagator due to Landau damping is small
and can be neglected for the same reason as vertex corrections, while the renormalization by the real part of the static polarization changes $\omega_0 (q)$ to dressed $\omega_r (q)$. Accordingly, the renormalized
e-ph couplings in MET theory are defined by
 \begin{equation}
 \lambda_r = N_F \left\langle \frac{g^2}{\omega^2_r({\mathbf k}-{\mathbf k}')} \right\rangle _{FS} ,
\label{lr}
\end{equation}
 and
\begin{equation}
\lambda^E_r = \frac{N_F}{E_F}
\left\langle \frac{g^2}{\omega_r({\mathbf k}-{\mathbf k}')} \right\rangle _{FS} ,
\label{lr_1}
\end{equation}
where the dressed phonon frequency is
\begin{equation}
\omega^2_r({\mathbf q}) = \omega^2_0({\mathbf q}) -g^2
 \Pi_{st}({\mathbf q}) .
\label{or}
\end{equation}
In the absence of vertex corrections, the static polarization is given by
the convolution of two Green's functions.
\begin{equation}
\Pi_{st}({\mathbf q}) = -2 T \sum_{\omega_m} \int \frac{d^dk}{(2\pi)^d} G({\mathbf k}, \omega_m) G({\mathbf k}+{\mathbf q}, \omega_m)
\label{k_2}
\end{equation}
where
$G({\mathbf k}, \omega_m) = 1/(i\omega_m - \Sigma ({\mathbf k}, \omega_m) - \epsilon_{\mathbf k})$.
In principle, this equation and the one for
 the fermionic self-energy
 \begin{equation}
 \Sigma ({\mathbf k}, \omega_n) = - g^2 T \sum_{\Omega_m} \int \frac{d^d q}{(2\pi)^d} \frac{G({\mathbf k}+{\mathbf q}, \omega_n + \Omega_m)}{\omega^2_r ({\mathbf q})+ \Omega^2_m}
 \label{k_1}
 \end{equation}
have to be solved self-consistently. However, for a generic $q \sim k_F$, self-consistency 
is not needed, as typical internal $\omega_m$ in (\ref{k_2}) are of order $E_F$, and for such high frequencies, $\Sigma ({\mathbf k}, \omega_m) \approx \Sigma (\omega_m)$ is smaller than  $\omega_m$ by the same $\lambda_r$ that governs the strength of vertex corrections \cite{AGD}.
This allows one to evaluate  $\Pi_{st} ({\mathbf q})$ using free-fermion propagators
(this is in line with the generic argument that within MET interactions are relevant
only for fermions with energies much smaller than $E_F$).
Summing up over frequency in (\ref{k_2}), we obtain
\beq
\Pi_{st} ({\mathbf q}) = \int \frac{d^d k}{(2\pi)^d}
 \frac{f_{{\mathbf k}}-f_{{\mathbf k}+{\mathbf q}}}
{\epsilon_{{\mathbf k}+{\mathbf q}} -\epsilon_{{\mathbf k}} } .
\label{Po}
\end{equation}
Here $f_{{\mathbf k}}$ is the Fermi distribution function.
Because $\Pi_{st} ({\mathbf q})$  is positive, coupling to electrons reduces the phonon frequency.
For a generic dispersion, $\omega_r ({\mathbf q})$ softens to zero at $q=q_c$
at some $\lambda_0=\lambda_{c} \sim  O(1)$.
Thus, MET cannot be extended beyond this $\lambda_{c}$. However, it has been argued that a controlled MET may still hold at $\lambda_0 \leq \lambda_{c}$ provided renormalized couplings obey the inequality $ \lambda^E_r \ll 1$, i.e. there still exists a range of $\lambda_0$ where $\lambda_r > 1$ but $\lambda^E_r \ll 1$.

The situation is somewhat special for an Einstein phonon and parabolic fermionic dispersion in 2D. Here, the free-fermion  polarization operator at $T=0$ does not depend on $q$ for  $|{\bf q}| <2k_F$, i.e., for any momentum transfer between fermions on the Fermi surface. As a result, $\omega_r = \omega_0 (1-2 \lambda_0)^{1/2}$ remains momentum independent, leading to $\lambda_r = \lambda_0/(1-2\lambda_0)$ and $\lambda^E_r = \lambda^E_0 /(1-2\lambda_0)^{1/2}$, with a stronger divergence of $\lambda_r$ as  $\lambda_{0,c} \to 1/2$.
The self-energy in this case is given by
\beq
\Sigma (\omega_m) =   - i \lambda_r \omega_r \arctan{\frac{\omega_m}{\omega_r}} .
\label{1_c_1}
\eeq
At small $\omega_m$, $\Sigma (\omega_m) =  - i\lambda_r \omega_m$.
The strong coupling condition $\lambda_r \gg 1$, $\lambda^E_r \ll 1$ is satisfied at $1 \gg 1-2 \lambda_0 \gg \omega_0^2/(2E_F)^2$.   Within this regime (if it holds) the mass renormalization is strong but vertex corrections are still  small.
However, even in this regime, the self-energy is relevant only for fermions with  frequencies
 $\omega_m < E_F \lambda^E_r$, which remain smaller than $E_F$ as long as the dressed Eliashberg parameter $\lambda^E_r$ remains small.

In our analysis, we explore the relevance of the third dimensionless parameter, $E_F/W$.
It is of order one at densities around half-filling, but is small at low fermionic density $n$.
We show that at small $E_F/W$, the polaron state wins energetically over the FL state already at small $\lambda_0$ (parametrically small  in 3D and numerically small in 2D), i.e. the development of a polaron state
is not related to the softening of the phonon spectrum.
We further argue that polaron formation is not a low-energy phenomenon.
Rather, it involves electronic states with energies exceeding the bandwidth,
i.e. when the conduction band is nearly empty,  the FL state becomes unstable
because of fluctuations outside the low-frequency range considered in MET.
To put it differently, fermions with energies comparable to the bandwidth
rather than the ones near the Fermi surface  control which state is the true vacuum
(by particle-hole symmetry, the same holds in the opposite limit of nearly completely occupied band,
when $W-E_F << W$).

For fermionic densities around half-filling, the leading instability upon increasing $\lambda_0$ is phonon softening at some $q_c$, while a polaron state develops at a larger coupling.
\begin{figure*}[htbp]
	\centering
    \noindent
	\includegraphics[width=0.325\linewidth]{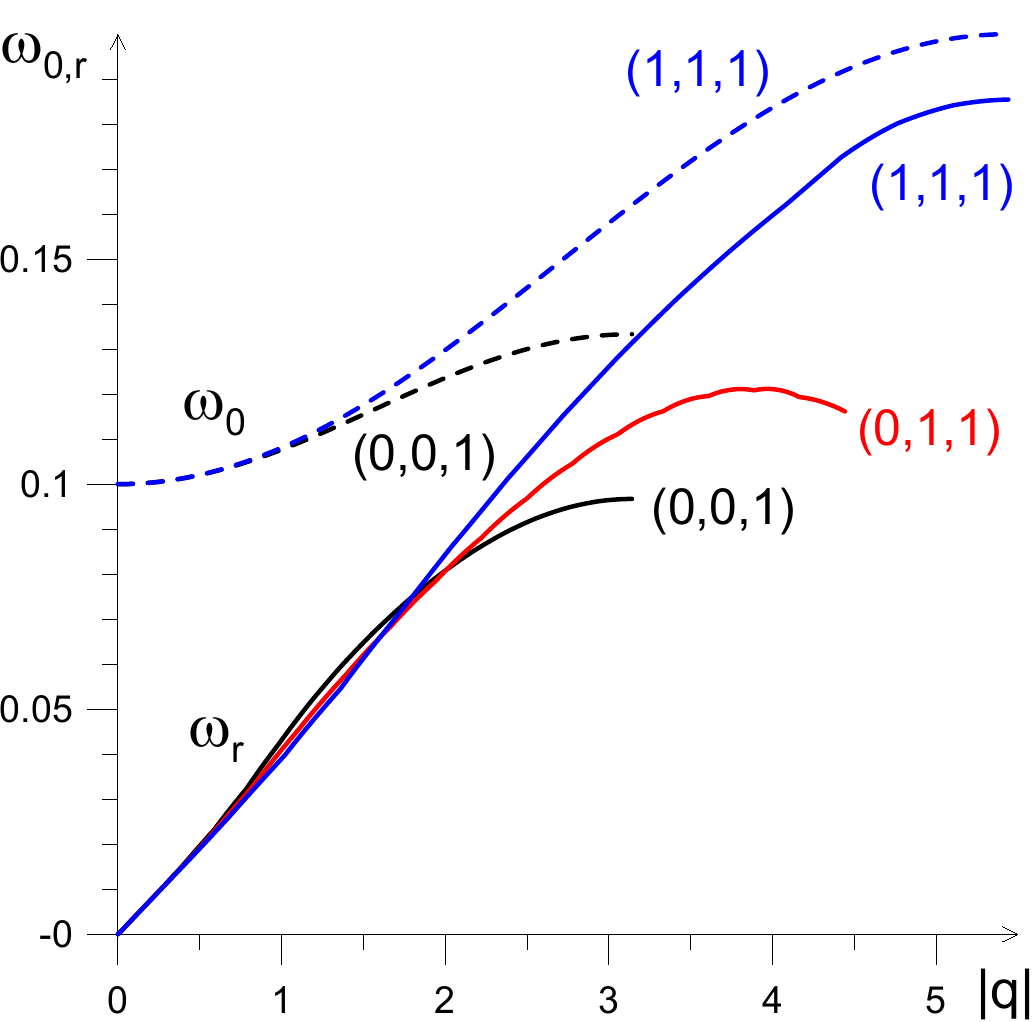} % {OmRUP2in3D.pdf}
	\includegraphics[width=0.325\linewidth]{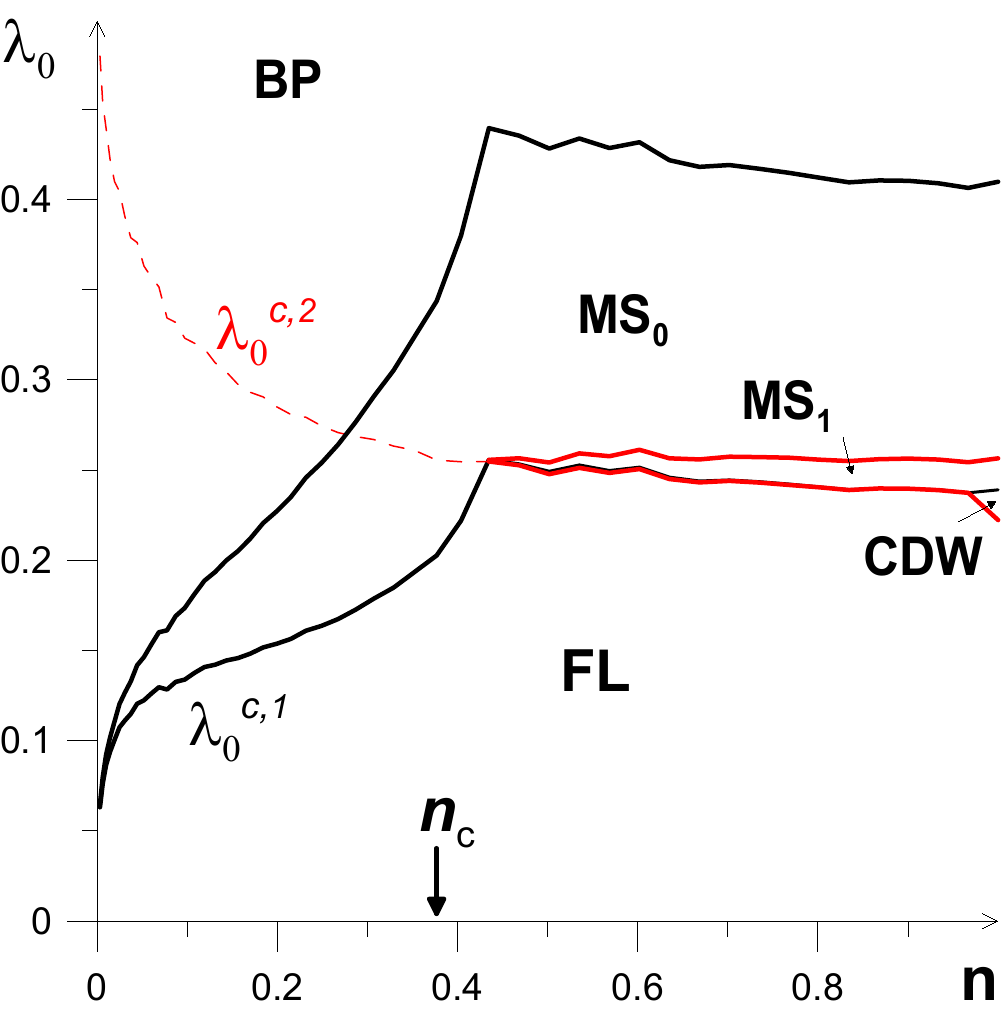} %  {BUP23D.pdf}
    \includegraphics[width=0.325\linewidth]{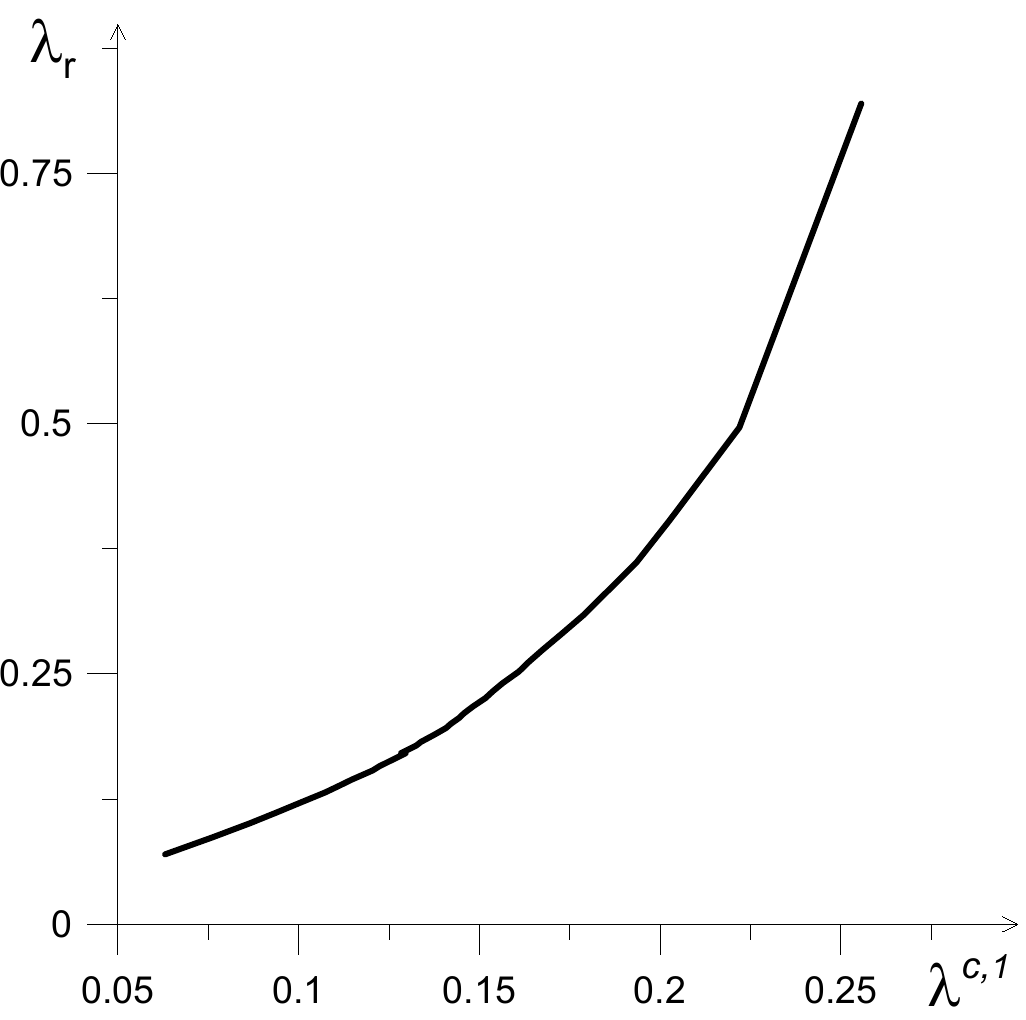} %  {LLCUP23D.pdf}
\\
	\includegraphics[width=0.325\linewidth]{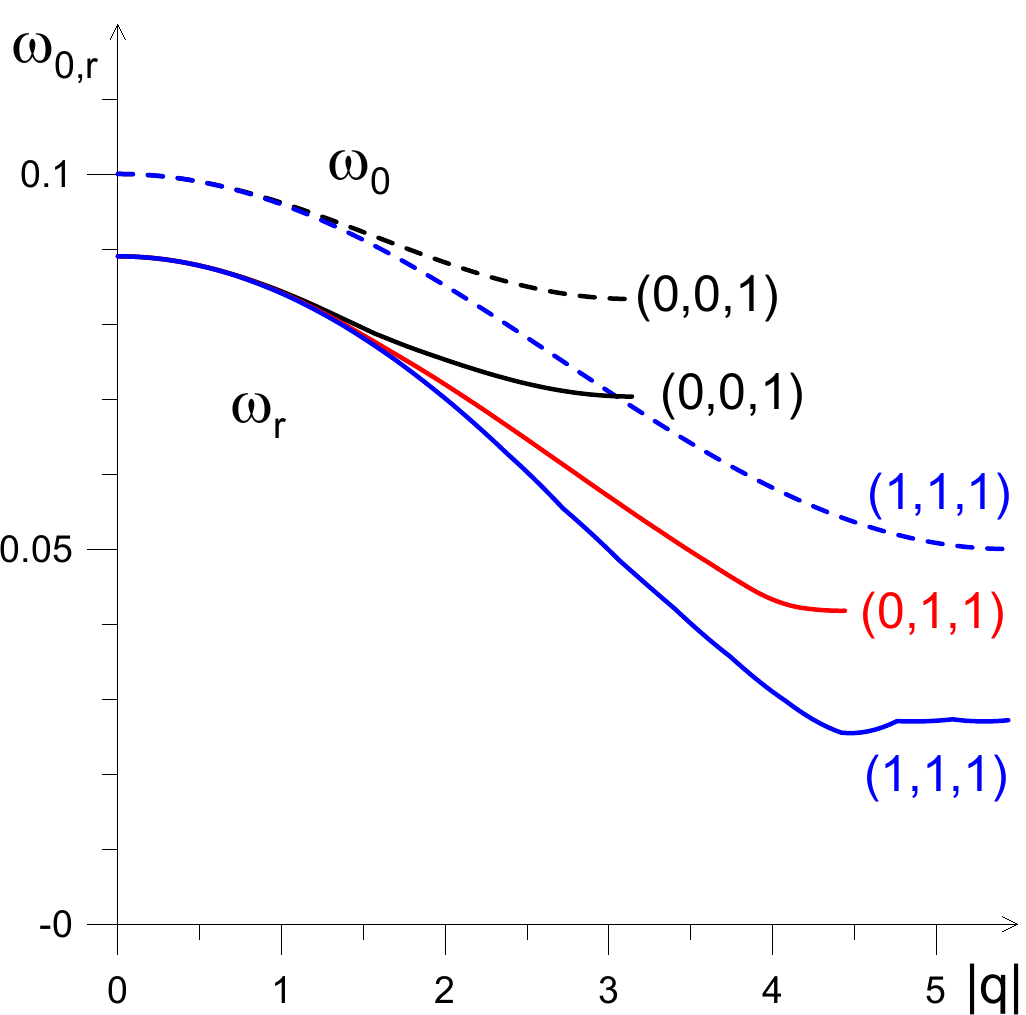} % {OmRDOWNin3D.pdf}
	\includegraphics[width=0.325\linewidth]{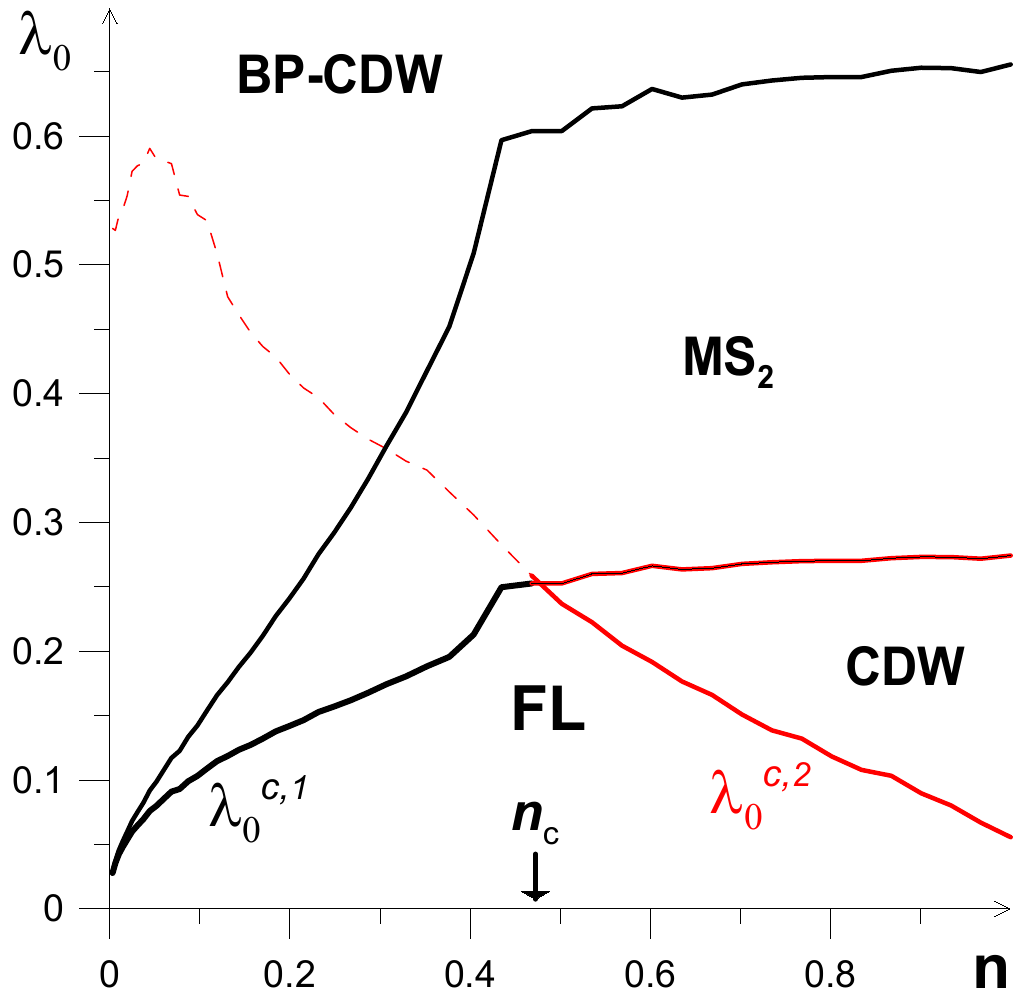} % {BDOWN3D.pdf}
    \includegraphics[width=0.325\linewidth]{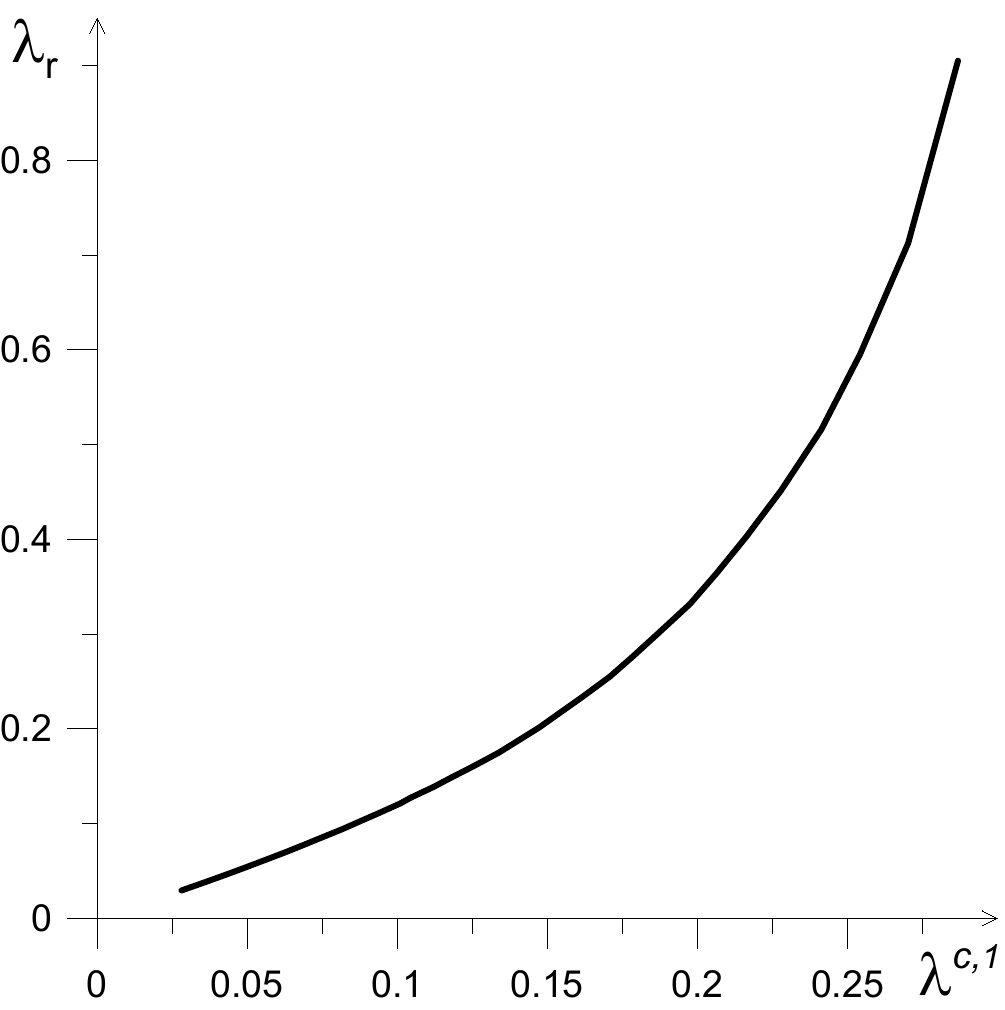} % {LLCDOWN3D.pdf}
\\
	\includegraphics[width=0.325\linewidth]{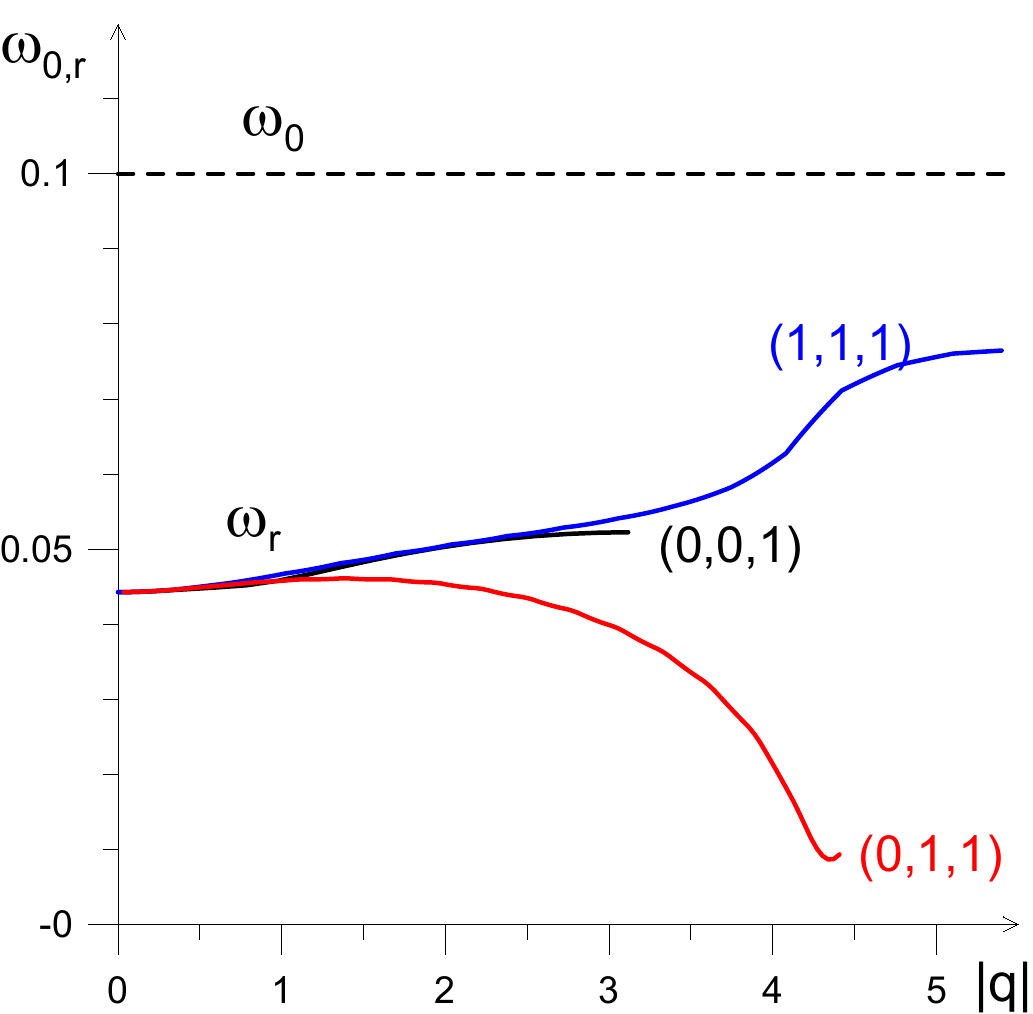} % {OmRFLATin3D.pdf}
	\includegraphics[width=0.325\linewidth]{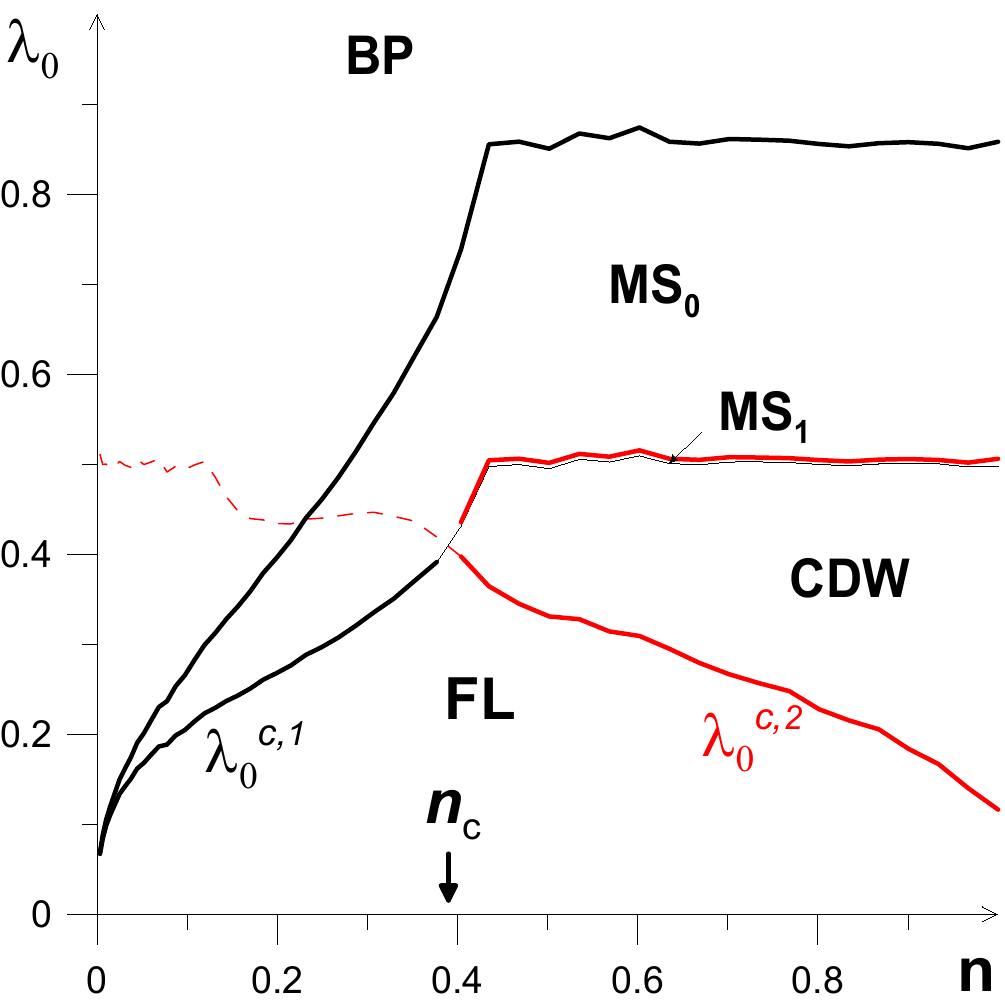} % {BFLAT3D.pdf}
    \includegraphics[width=0.325\linewidth]{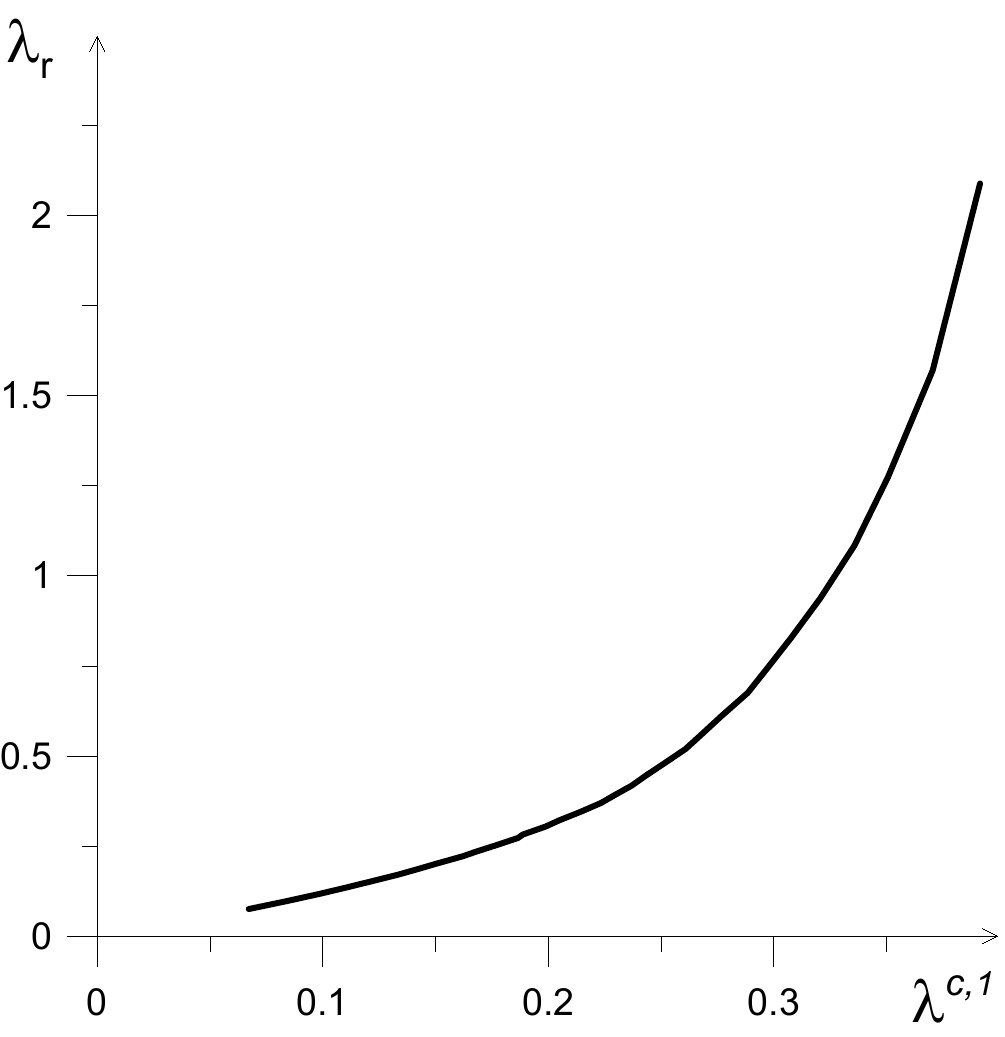} % {LLCFLAT3D.pdf}
\caption{
Numerical results for 3D fermions with tight-binding dispersion, coupled to a phonon with dispersion given by Eq. (\ref{oo}), parametrized by $\omega_0$ and $\tau$.
Left panel: Bare (dashed) and renormalized (solid) phonon spectra
along different directions in the Brillouin zone for 
$\omega_0/t =0.1$, $\tau / t=1/120$ (upper row),
$\omega_0/t =0.1$, $\tau / t=-1/240$ (middle row),
and for the dispersionless bare phonon frequency 
$\omega_0/t =0.1$, $\tau =0 $ (lower row).
The renormalized phonon spectrum is shown for the density marked 
by $n_c$ in the middle panel and the coupling $\lambda_0$  ($= \lambda_0^{c,1} = \lambda_0^{c,2}$ in the middle panel),  where it  first touches zero at some momentum $q_c$. 
Middle panel: The phase diagram. For the upper and lower panels, we consider a homogeneous BP state with the largest possible $n_i =2$ on $N n /2$ sites ($N$ is the total number of sites). 
For the middle panel, we consider a checkerboard arrangement of $Nn/2$ pairs (labeled as BP-CDW). The mixed states contain both FL and BP components, with equal chemical potentials but different volume fractions, which vary in opposite directions as $\lambda_0$ increases.  
It is located in between the upper solid black line and the line $\lambda_0^{c,1}(n)$.
At its lower  end, the volume fraction of a BP vanishes, at the upper end the volume fraction of an FL vanishes. 
Note that the end points of the MS  tend to zero in the low density limit, i.e., in this limit MET becomes unstable toward BP state already at infinitesimally small $\lambda_0$.
The red line marked as $\lambda_0^{c,2}$ is where the phonon spectrum in the FL state would  soften at some  ${\mathbf q}_c$ and symmetry related momenta at arbitrary $n$.
This line is  meaningful only when $\lambda_0^{c,2} < \lambda_0^{c,1}$. 
  This holds  at densities $n > n_c$, when
the system first develops a CDW electronic order in between $\lambda_0^{c,2}$ and $\lambda_0^{c,1}$  and the MS emerges at $\lambda_0^{c,1}$ out of a CDW state. In this situation, there exists another line in the MS (upper solid red line) at which CDW electron order disappears. 
Right panel: Renormalized coupling
$\lambda_r$ at the critical line $\lambda_0^{c,1}$ for $n$ at which MS emerges before the phonon spectrum softens.
The compositions of the mixed states is:  
MS$_0$ is a mixture of the BP and FL states; 
MS$_1$  of the BP and CDW states;
MS$_2$  of the BP-CDW and FL states.
}
\label{Upin3D}
\end{figure*}

\section{Numerical analysis}
\label{sec:Numerics}

We considered several systems with electron-phonon interaction (\ref{Hint})
in cubic and square lattices to determine the critical values of $\lambda_0$ and $\lambda_r$
as functions of the electron density, $n$, or $E_F/W$ ratio. 
 We increased the coupling and checked at which $\lambda_0$ 
(i) the chemical potential of a BP state, $\mu_{BP}$  becomes the same as the chemical potential of an FL state, $\mu_{FL}$,
at vanishing FL density ( this  will set the  upper end of the mixed state),
(ii) $\mu_{FL}$ at the full density $n$  becomes the same or higher than $\mu_{BP}$,
 provided the BP state has lower energy than an FL state at the  same density $n_{bp}$ ($E_{BP} (n_{bp} ) \le E_{FL} (n_{bp} )$)
( this will set the   lower end of the mixed state),
and (iii) the dressed phonon spectrum softens to zero at some momentum.

The bare phonon spectrum was parameterized as
\begin{equation}
\omega_0({\mathbf q}) = \omega_0 + 2 \tau \sum_{\alpha =1}^{d} \left[1- \cos (k_{\alpha} a) \right] .
\label{oo}
\end{equation}
For $\tau = 0$ the spectrum is dispersionless, while for positive/negative $\tau$ its minimum
is located at the origin or momentum $Q=(\pi,\pi,\pi)$, respectively.

The dominant contributions to the ground-state energy of an FL  with density $n$ are the kinetic energy of itinerant fermions 
$E_{\text{kin},FL}$ and the Hartree term $U = -N[g/\omega_0(0)]^2 n^2/2$,
i.e. $E_{FL}=E_{\text{kin},FL}+U$.  We approximate $E_{\text{kin},FL}$, and the fermion 
density by the corresponding expressions of a Fermi gas 
\beq
E_{\text{kin},FG}
 =2\sum_{\epsilon_{{\mathbf k}}<E_F} \epsilon_{{\mathbf k}}, \qquad
n=2\sum_{\epsilon_{{\mathbf k}}<E_F}
\eeq
In the adiabatic limit the difference between 
$E_{\text{kin},FL}$ and $E_{\text{kin},FG}$ is small in the Eliashberg (adiabatic) parameter $\lambda^E$. 
Within the same approximation, the chemical potential of an FL is 
$\mu_{FL}=E_F-[g/\omega_0(0)]^2 n$.

\begin{figure*}[htbp]
	\centering
	\includegraphics[width=0.325\linewidth]{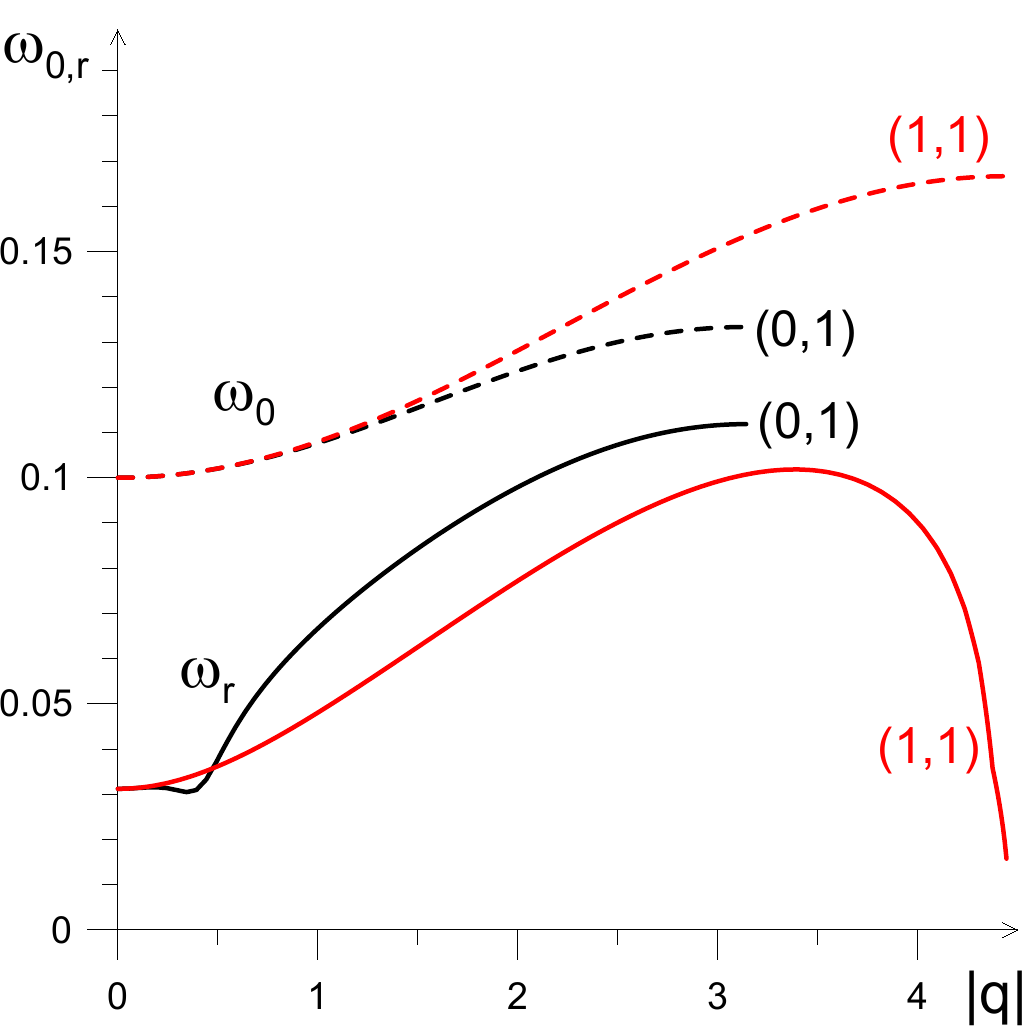}
	\includegraphics[width=0.325\linewidth]{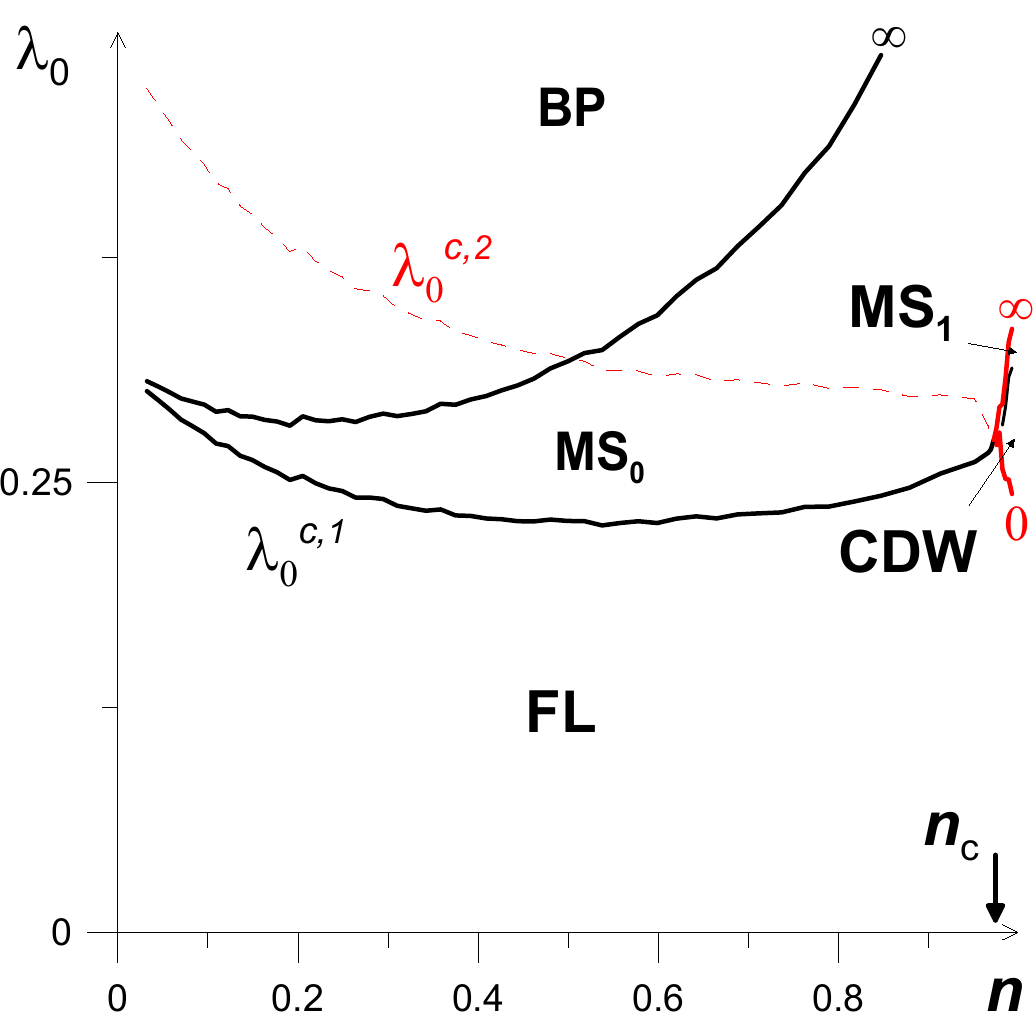}
    \includegraphics[width=0.325\linewidth]{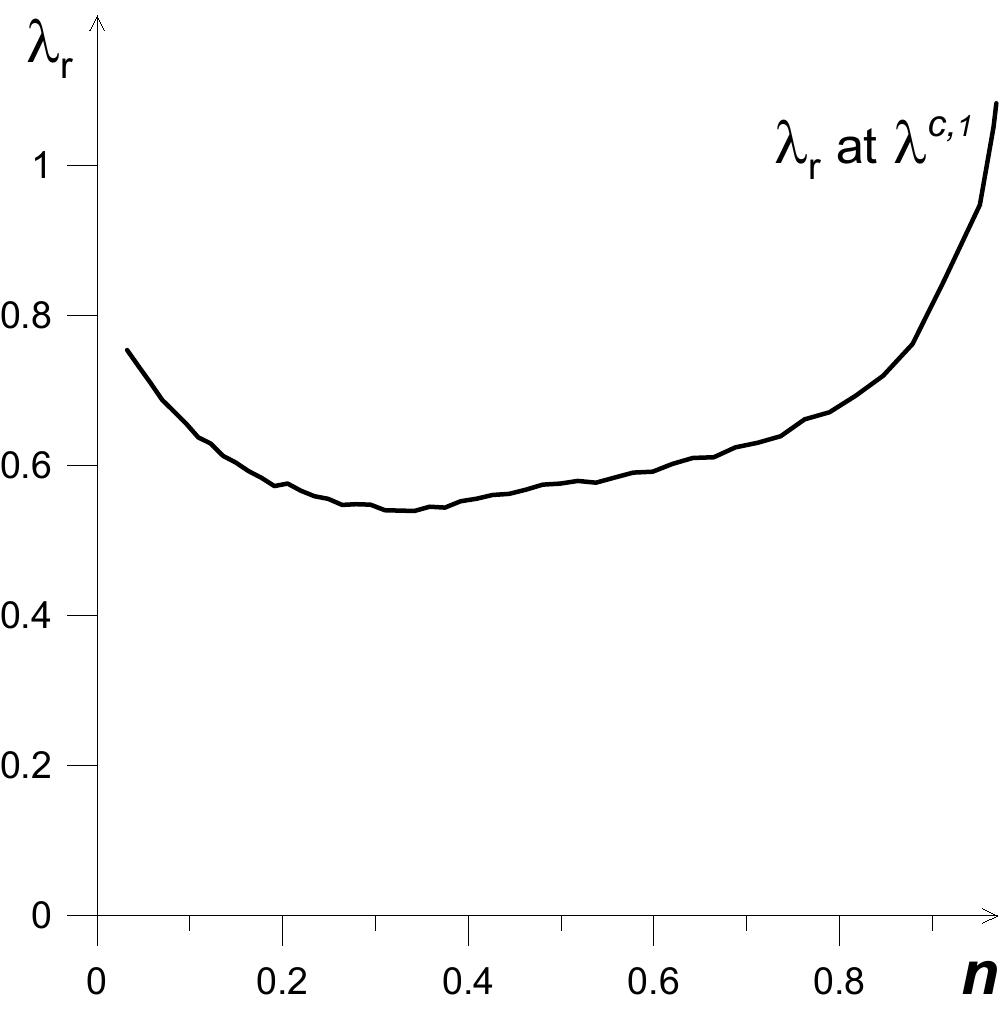}
\\
	\includegraphics[width=0.325\linewidth]{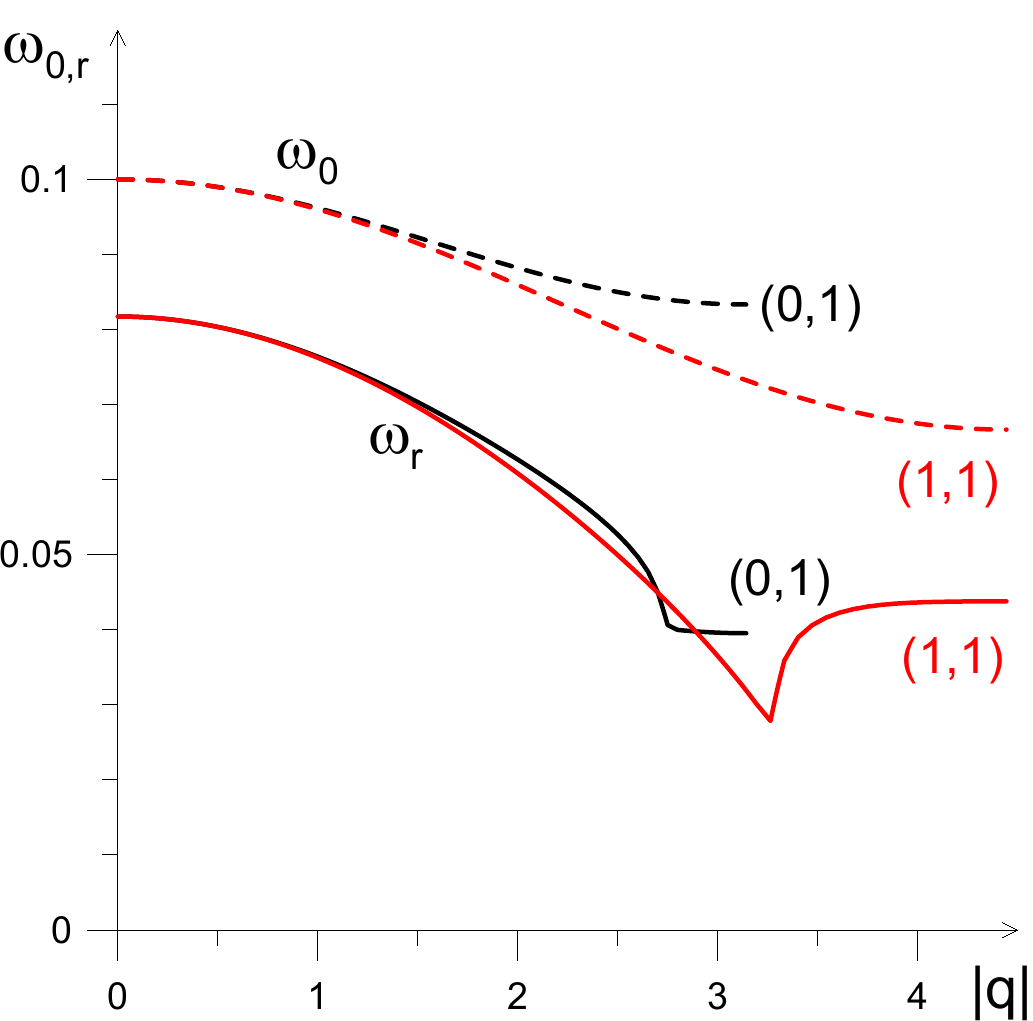}
	\includegraphics[width=0.325\linewidth]{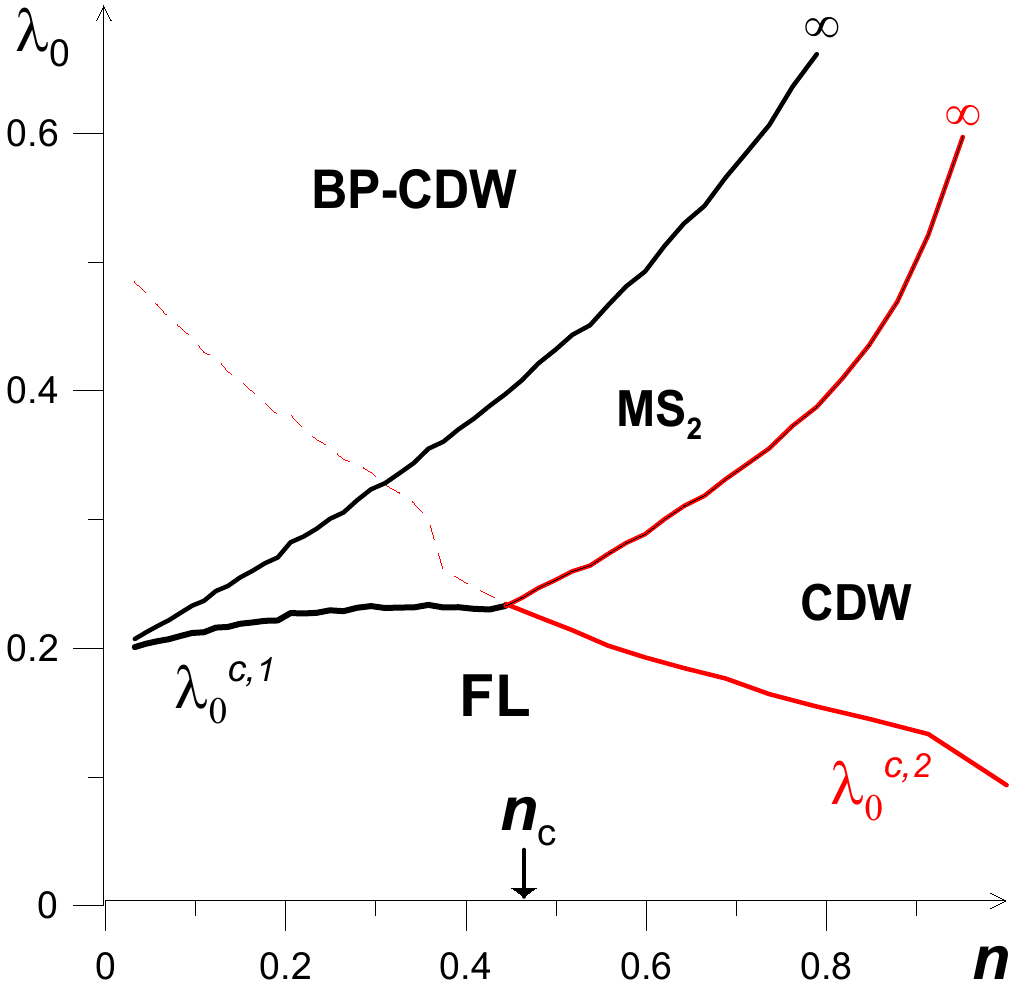}
    \includegraphics[width=0.325\linewidth]{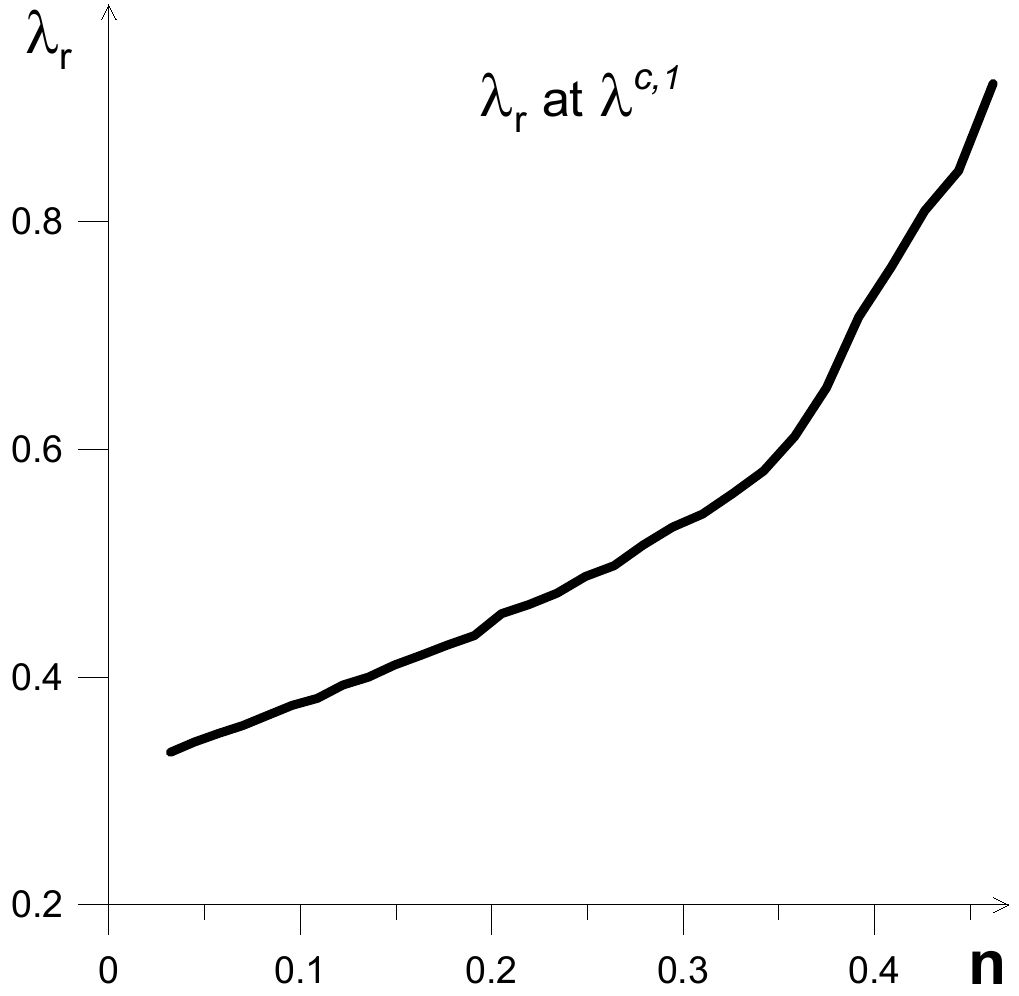}
\\
	\includegraphics[width=0.325\linewidth]{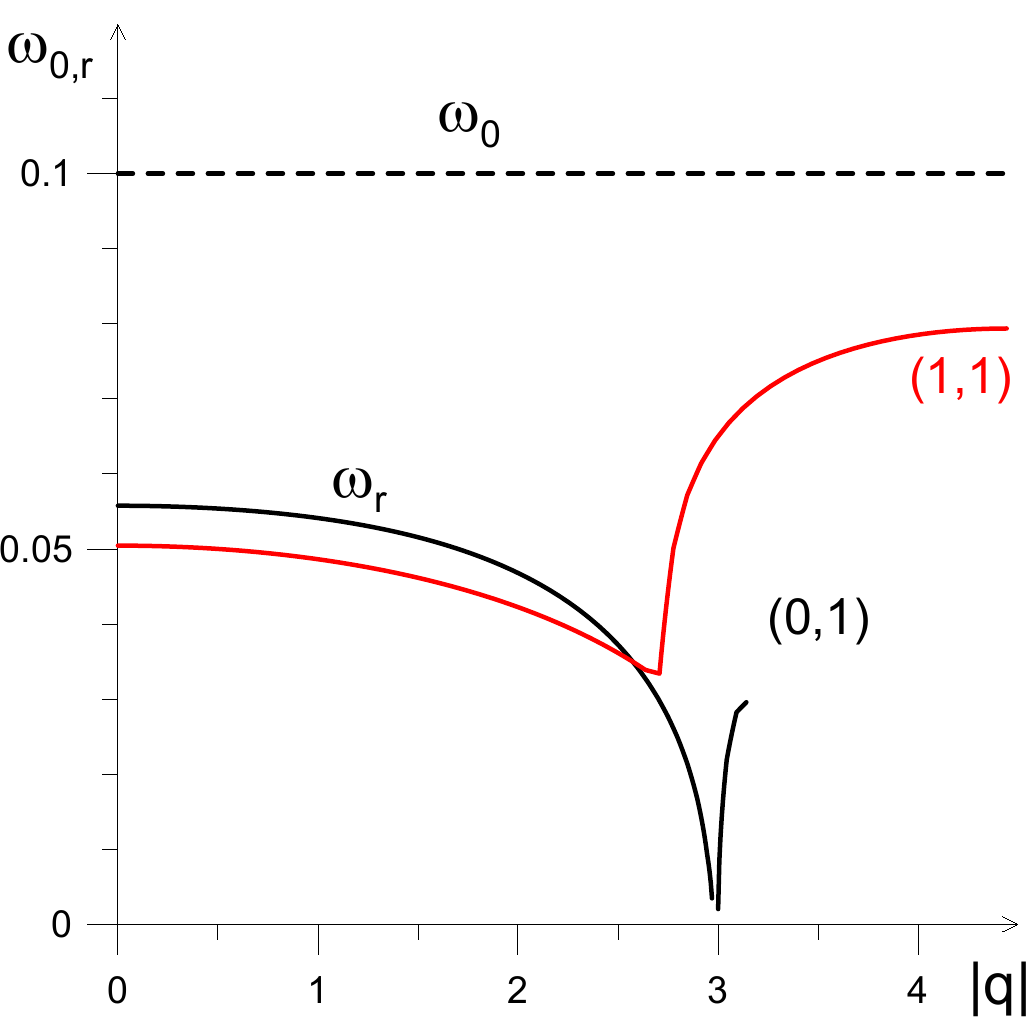}
	\includegraphics[width=0.325\linewidth]{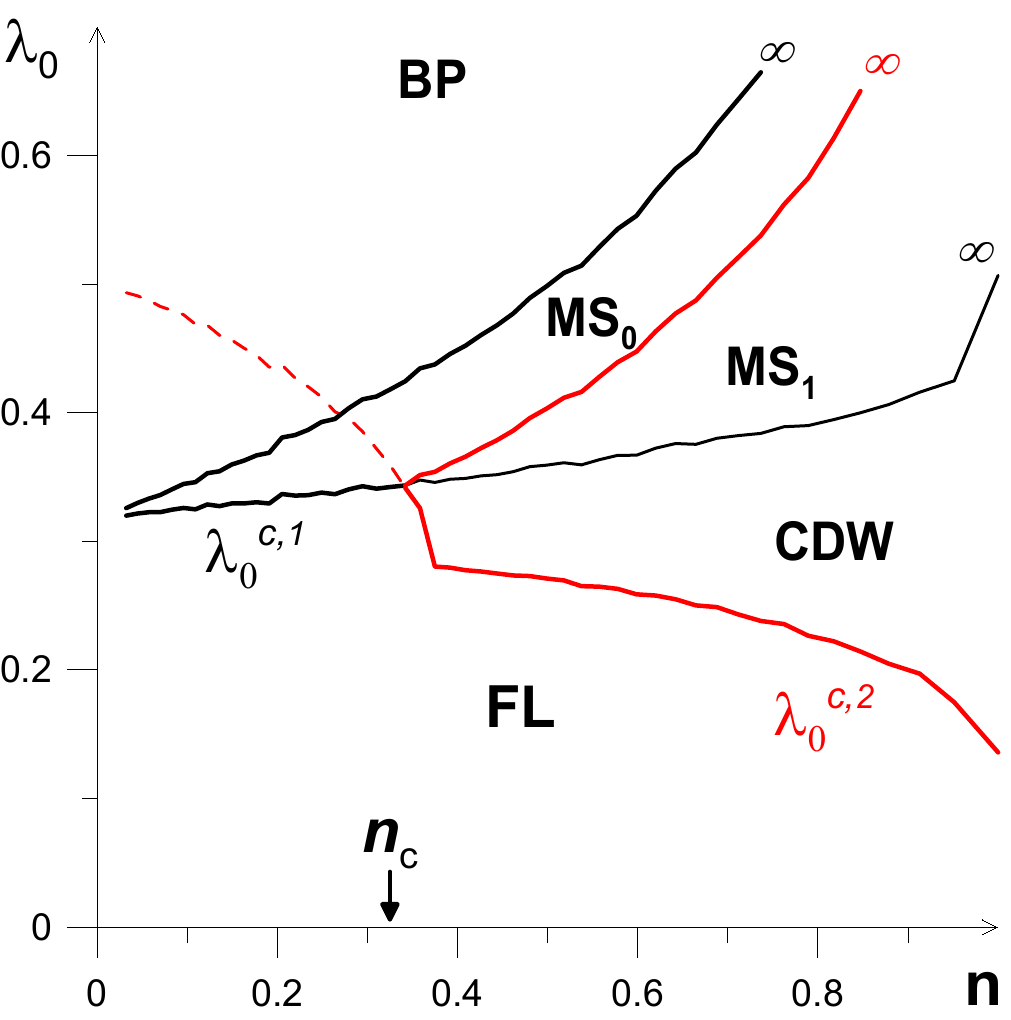}
    \includegraphics[width=0.325\linewidth]{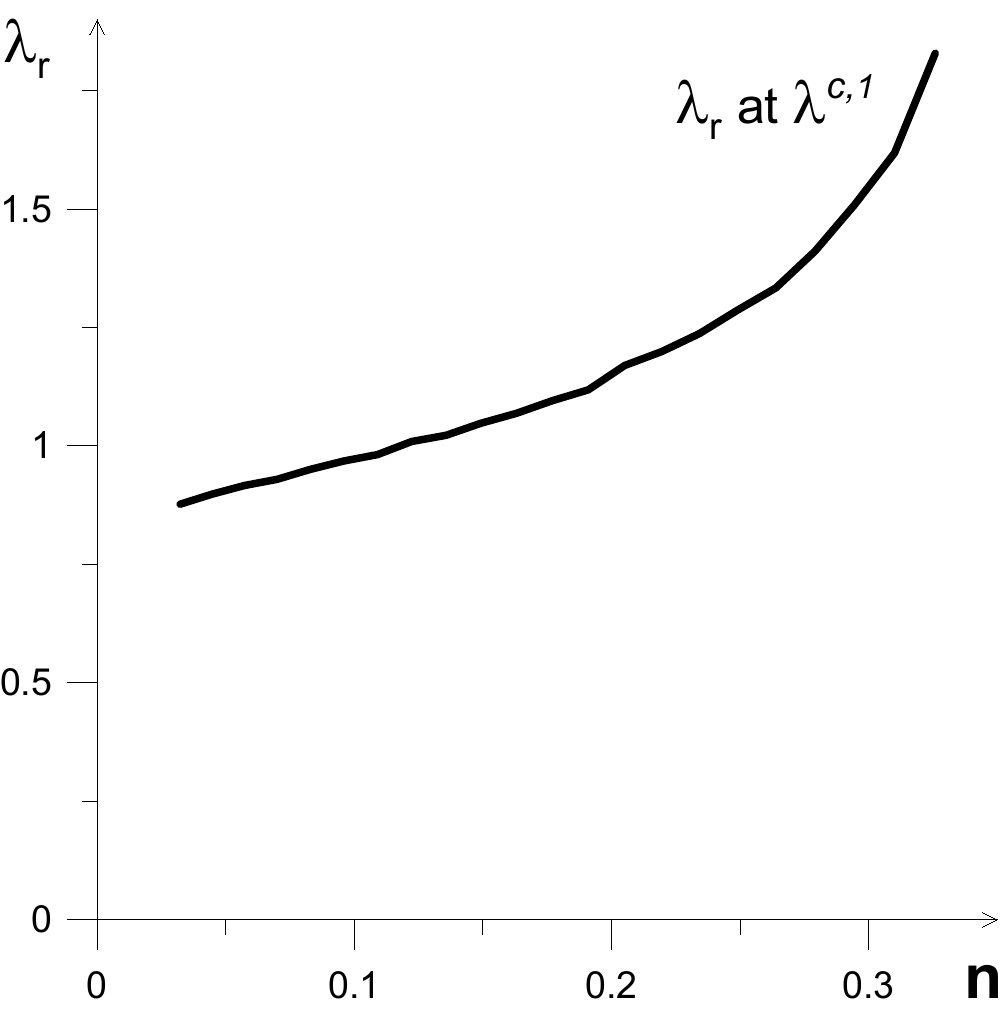}
\caption{
Same plots as in Fig.~\ref{Upin3D}, but now for a 2D system with
$\omega_0/t =0.1$, $\tau / t=1/120$ (upper row,  bare phonon dispersing is upwards),
$\omega_0/t =0.1$, $\tau / t=-1/240$ (middle row, bare phonon dispersing is downwards), and
$\omega_0/t =0.1$,  $\tau / t=0$ (lower row, bare phonon is dispersionless).
Left panel: Bare (dashed lines) and renormalized (solid lines) phonon spectra.  
The renormalized phonon
spectrum is shown for the density marked by $n_c$ in the middle panel and the coupling $\lambda_0$  ($= \lambda_0^{c,1} = \lambda_0^{c,2}$ in the middle panel),  where 
it  first touches zero at some momentum $q_c$. 
Middle panel: The phase diagram.
For the upper and lower panels, we consider a homogeneous BP state with the largest possible $n_i =2$ on $N n /2$ sites ($N$ is the total number of sites). 
For the middle panel, we consider a checkerboard arrangement of $Nn/2$ pairs. 
The MS state contains both FL and BP components, with equal chemical potentials but different volume fractions, which 
vary in opposite directions as $\lambda_0$ increases.   This state is located in between the upper  the solid black line and the line $\lambda_0^{c,1}(n)$.
At its lower  end, the volume fraction of a BP vanishes, at the upper end the volume fraction of an FL vanishes. 
The red line marked as $\lambda_0^{c,2}$ is where the phonon spectrum in the FL state would  soften at some  ${\mathbf q}_c$ and symmetry related momenta at arbitrary $n$. This line is  meaningful only when $\lambda_0^{c,2} < \lambda_0^{c,1}$. 
 At small $n$,  $\lambda_0^{c,1}$ and the upper solid line tend to a constant.  We emphasize that this constant is smaller than the putative $\lambda_0^{c,2}$ at $n \to 0$, i.e., at small $n$, an FL  state described by MET becomes unstable toward BP before the phonon spectrum softens. 
 Note also that $\lambda_0^{c,1}$  does not vary much  with $n$ 
except in the vicinity of $n=1$ where the density of states diverges.
It goes to zero at half-filling for the bare phonon dispersion with minimum at vector ${\mathbf Q}$. At densities $n > n_c$ for the upper and lower rows the system first develops a CDW electronic order in between $\lambda_0^{c,2}$ and $\lambda_0^{c,1}$,  and the MS emerges at $\lambda_0^{c,1}$ out of a CDW state. In this situation, there exists another line in the MS (upper solid red line) at which CDW electron order disappears upon increasing $\lambda_0$.   
Right panel: Renormalized coupling
$\lambda_r$ at the critical line $\lambda_0^{c,1}$ for $n$ at which MS emerges before the phonon spectrum softens.
The notations $MS_0$ - $MS_2$ are the same as in Fig. \ref{Upin3D}. }
\label{all2D}
\end{figure*}

\begin{figure*}[htbp]
 \includegraphics[width=0.4\linewidth]{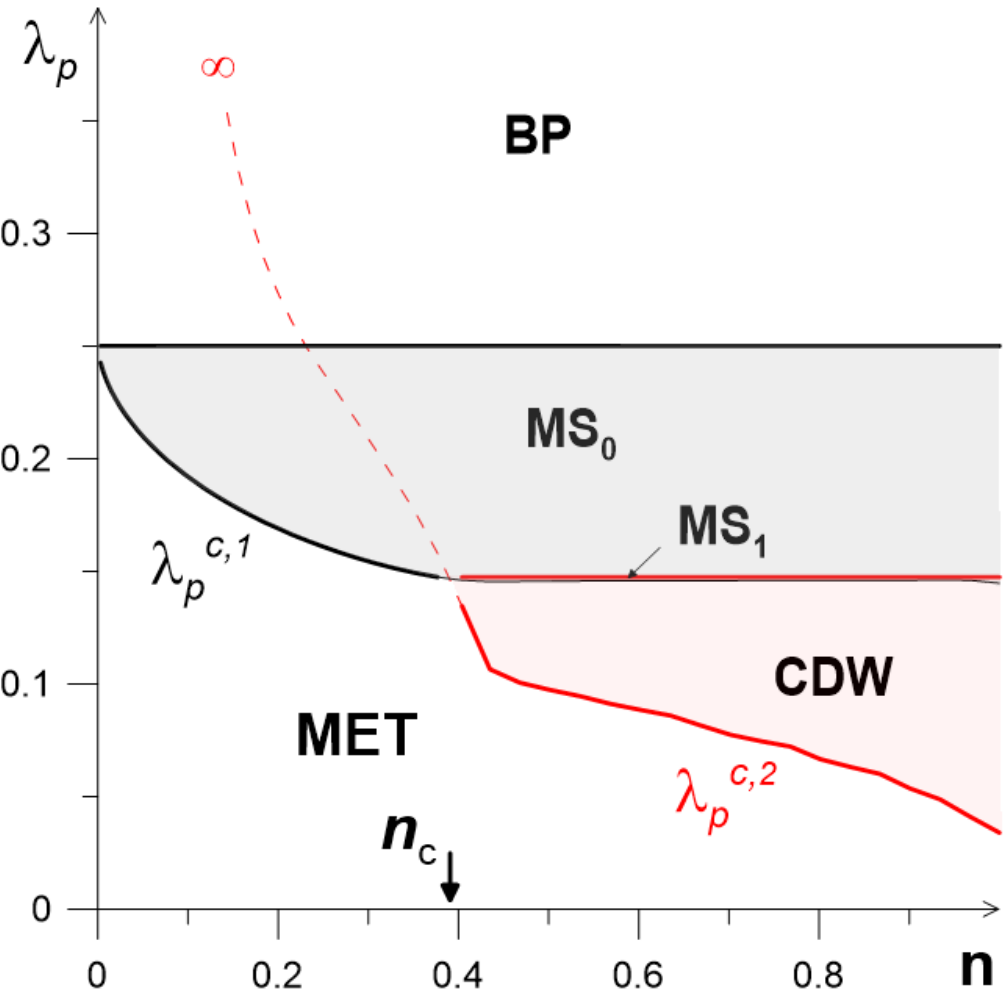}
 ~~
	\includegraphics[width=0.4\linewidth]{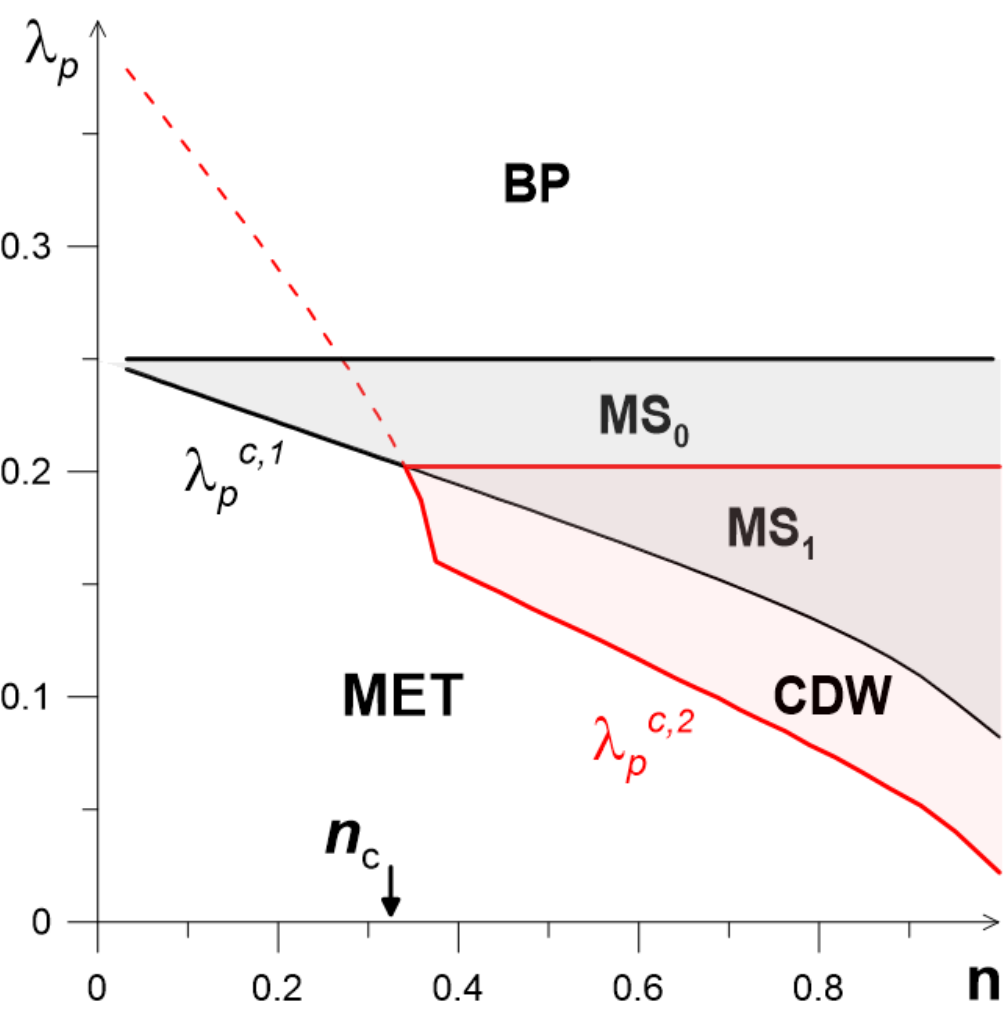}
    \caption{The  phase diagram for 3D and 2D systems (left and right panels, respectively) 
for a dispersionless bare phonon, in terms of the coupling $\lambda_p$, which we use in the analytical study. 
 The phase diagrams are the ones from the lower middle panels in Figs.  \ref{Upin3D} and \ref{all2D}), reexpressed in terms of $\lambda_p$} 
\label{fig:additional}
\end{figure*}

For convenience of a reader, in Fig. \ref{fig:additional} we also present the variational phase diagrams in 3D and 2D  for a dispersionless bare 
 phonon (the lower middle panels in Figs \ref{Upin3D} and \ref{all2D}) in terms of $\lambda_p$, which we use in the analytical study. 

As a competing variational state, we considered a set of \underline{localized} 
bipolarons with a density profile $n_i$. Its exact energy 
$E_{BP}$ is obtained by analyzing the shifts of the harmonic modes
\begin{equation}
E_{BP} = -\frac{g^2}{2} \sum_{{\mathbf q} } \frac{|n_{{\mathbf q}}|^2}{\omega^2_0({\mathbf q})}
\equiv \sum_{ij} n_i n_j F({\mathbf r}_{ij}),
\label{Ebi}
\end{equation}
where the interaction potential is
\begin{equation}
F({\mathbf r}) = - \frac{g^2}{2} \sum_{{\mathbf q} }
\frac{e^{i{\mathbf q}\cdot } {\mathbf r} } {\omega^2_0({\mathbf q})} .
\label{Fij}
\end{equation}
In our numerical analysis, we considered two bipolaron states. One is a homogeneous state with the highest possible density $n_i=2$ at $N_e/2$ sites ($N_e = N n$) and zero density at other sites.
This self-bound state is favored when the minimum of the bare phonon dispersion is 
at or near  zero momentum.  
According to Eq.~(\ref{Ebi}), the  energy of a homogeneous BP state is
$E_{BP}^{(a)} = (N_e/2) 4 \sum_i F({\mathbf r}_i)$.
As expected, when $n=2$ (filled band) $E_{BP}^{(a)}=U=E_{FL}$.
The chemical potential of this polaron state is $\mu_{BP} = E_{BP}^{(a)}/N_e$.
This state takes full advantage of the energy gain from the soft phonon modes at small $q=0$. 

For a dispersionless  Einstein phonons, $F({\mathbf r}) \propto \delta_{{\mathbf r},0}$,
and the energy and chemical potential of a homogeneous BP state are
\beq
E_{BP}^{(q=0)}= -N_e \left(\frac{g}{\omega_0}\right)^2, 
~~\mu^{(q=0)}_{BP} = -\left(\frac{g}{\omega_0}\right)^2.
\eeq
Within this approximation,  the energy of localized bipolarons
is the same regardless of their distribution in space. The degeneracy is removed by treating bipolarons as an interacting Bose liquid of hard-core bosons, however, the lifting of the degeneracy in the dilute regime is controlled by the inverse of the exponentially large bipolaron effective mass, $ m_{bi} \propto \exp \{ g^2/\omega_0^3 \}$, see Refs.~\cite{Chakraverty1,Chakraverty2}, and is very weak.
With exponential accuracy, this state can still be treated as that of localized randomly places pairs.

When the bottom of the bare phonon dispersion  is near ${\mathbf q} = {\mathbf Q}=(\pi,\pi,\pi)$, we considered the state with a checkerboard arrangement of sites with localized pairs, which in this case has a lower energy $E_{BP}^{{b}}$ than a homogeneous state. In the localized BP-CDW state,  if  a  given  site  has a bipolaron, then the neighboring site is empty and the next-nearest one again has a bipolaron. 
For the checkerboard arrangement of the $n <1$, $N_e\le N$ sites, other sites are just empty (for the $n>1$, $2N-N_e$ sites would have the checkerboard arrangement of the bipolarons, while the remaining $N_e-N$ sites would be doubly occupied by electrons).
The chemical potential of the checkerboard bipolaron state is $\mu_{BP} = E_{BP}^{(b)}/N_e$. 
This treatment is used throughout the manuscript with the exception of Sec. \ref{CDW_pol}), where we consider 
the checkerboard polaron order near half-filling.   The  numerical 
  computations  reduce to  evaluations of momentum sums for large system sizes 
($L=1024$ in 2D and $L=512$ in 3D) and can be easily done
on a laptop.

We show the results in Fig. \ref{Upin3D} for a 3D system and in Fig. \ref{all2D} for a 2D system. In the left panels of both figures, we show  the bare and renormalized  phonon spectra at the critical density/coupling when the spectrum first touches zero at some momentum $q_c$ for three types of a bare phonon dispersion, i.e. the plots are for a particular density $n =n_c$  marked in the middle panels.
In the middle panels, we show the phase diagrams in the $(\lambda_0,n)$ plane for $0<n<1$. 
  
We considered a variational  homogeneous  BP state  for the cases when the bare phonon spectrum has a minimum at momentum $q=0$ 
(the upper and lower rows) and a variational checkerboard BP state for the cases when the renormalized phonon spectrum has a minimum at momentum ${\mathbf Q}$ 
(the middle rows in both figures). In the right panels, we show the renormalized coupling 
$\lambda_r$ for couplings at which the mixed state emerges before the phonon spectrum softens. 

There are two key critical lines in the phase diagrams: $\lambda_0^{c,1}$, at which bipolarons emerge, and $\lambda_0^{c,2}$, at which the phonon spectrum softens at $q_c$. At densities $n < n_c$ we have $\lambda_0^{c,1} < \lambda_0^{c,2}$, i.e. a pure FL state becomes unstable toward bipolarons, 
while MET is still internally stable.  
For a 3D system, we found that $\lambda_0^{c,1}$ is smaller than $0.5$ for all three types of bare phonon dispersion in Fig.~\ref{Upin3D}. Furthermore, at small densities, corresponding to a near-spherical Fermi surface, 
$\lambda_0^{c,1}$  tends to zero as  $n^{1/3}$ due to vanishing $N_F \sim p_F \sim n^{1/3}$ while a putative $\lambda_0^{c,2}$  remains finite.
As a result, at $n \to 0$,  an FL  becomes unstable against bipolarons already at infinitesimally small $\lambda_0$.  This contradicts earlier claims that in the continuous limit, corresponding to small $n$,
an electron-phonon system is well described by the MET because the 
dressed  $\lambda_r$ is also small as $n^{1/3}$.
 
In 2D, $\lambda_0^{c,1}$  tends to a finite value $1/\pi$  at $n \to 0$ (see Fig. \ref{all2D}) because in this limit $N_F$ does not depend on the Fermi momentum. 
Still, we found that this value is 
smaller than the putative $\lambda_0^{c,2}$ at $n \to 0$, i.e., at small $n$  an FL again  becomes unstable against  bipolarons  before the phonon spectrum softens.  

In actual calculations, we  identified $\lambda_0^{c,1}$ as the line at which the chemical potentials of an FL and a  variational BP state become equal, $\mu_{BP} = \mu_{FL}$. We explicitly verified that  a  further increase of $\lambda_0$ leads to
a mixed (or phase separated) state consisting of heavy bipolarons at density $n_{bi}$, which increases in volume with increasing $\lambda_0$,  and FL fermions at density $n-2n_{bi}$, which decreases with increasing $\lambda_0$. The same holds if we increase $n$ keeping $\lambda_0$ fixed at a value above $\lambda_0^{c,1}$ and increase $n$:  extra electrons form bipolarons, while the number of itinerant electrons remains intact.   The chemical potentials $\mu_{BP}$ and  $\mu_{FL}$ remain equal within the mixed phase, as required by the Maxwell construction.
The mixed phases (with or without CDW order in its components) 
are located in the middle panels between $\lambda_0^{c,1}$ and 
the upper  solid black line, at which $2n_{bi} =n$.  At larger $\lambda_0$,
the  entire system  consists of bipolarons.  In 2D, the upper boundary of the mixed state diverges at half-filling as for the tight-binding dispersion the 2D DOS diverges at $n=1$.
   
At densities $n > n_c$,  we have $\lambda_0^{c,2} < \lambda_0^{c,1}$.  
 In this situation, the system first develops an electronic CDW order at $\lambda_0^{c,2}$,  and the mixed state emerges at $\lambda_0^{c,1}$ out of a CDW state. We found in our variational analysis that 
for these $n$, there exists another line inside the mixed state (solid red lines in the middle sections of Figs. \ref{Upin3D} and \ref{all2D}) at which the order of CDW electrons disappears as the number of $\lambda_0$.  For a 3D system with a dispersionless bare phonon, the range of $\lambda_0$, in which an FL component of the mixed phase is CDW-ordered, is tiny (the lower middle panel in Fig.~\ref{Upin3D}), but for other bare dispersions in 3D and for all bare dispersions in 2D it is sizable. We also note that a large value  of $n_c \leq 1$ in the upper middle panel of Fig.  \ref{all2D} is the consequence of our choice of a bare 2D dispersion with a minimum at $q=0$. For a more flat bare dispersion, $n_c$ is far from half-filling (see  the lower middle panel in  Fig.  \ref{all2D}). 
For the checkerboard bipolaron state, the variational analysis finds that on approach to $n_c$,  the $\lambda_0^{c,1}$ line is determined by the energy balance between the BP and FL states at the same density $n=1$. 
This leads to the discontinuous jump of the FL state density 
across the $\lambda_0^{c,1}$ line and a direct transition between the CDW and MS$_{2}$ states.        

Next, it is generally expected that near the onset of a CDW instability, a Fermi-liquid enters a strong-coupling regime where  the dressed $\lambda_r \gg 1$  (Refs. \cite{Zhang_1,*Zhang_2,paper_1},
 or becomes unstable  by some other mechanism~\cite{Yuzbashyan,*Yuzbashyan_1,*Yuzbashyan_2,*Yuzbashyan_3}.
However, because phonon softens only at discrete  points in momentum space and 
the dressed coupling, $\lambda_r \propto \langle \omega^{-2}_r (\mathbf{q} )\rangle_{FS}$, involves averaging over $\mathbf{q}$ 
connecting points on the Fermi surface,  the increase of $\lambda_r$ near the instability is far less drastic~\cite{marsiglio2020eliashberg,Andrey_review} than in the special case when  $\omega_r$ is treated as momentum independent and softens simultaneously at  all momenta (see~\cite{paper_5} and references therein).  
The dimensionality of space matters here:
the corresponding momentum integral over the Fermi surface is more singular in 2D. For
$\omega_r^2({\mathbf q}\to {\mathbf q}_c) \to \delta^2 + c({\mathbf q}-{\mathbf q}_c)^2$ with $\delta \ll \Omega $, we have 
$\lambda_r \propto |\ln[\delta ]|$ in a 3D system and 
 $\lambda_r \propto 1/\delta $ in 2D.
Thus, despite the dramatic softening of $\omega_r$ at some momentum, the 
increase of $\lambda_r$ is quite moderate, except for the very close vicinity of a CDW transition, which is difficult to reach numerically. 
We did  simulations with $N=256^3$ sites for 3D systems with various bare phonon spectra and found that $\lambda_r$ barely exceeds unity at $n = n_c$, 
see the right panels in Fig.~\ref{Upin3D}.   We found similar results in the 2D system with $N=1024^2$ sites (right panels in Fig. \ref{all2D}).  
Moreover, in real materials, a structural transition 
is often weakly first-order, i.e.,  $\delta $ is reduced near the transition but does not vanish. 

Equation (\ref{or}) also makes it clear that neglecting the momentum dependence of $\omega_r$ is highly unphysical because it implies that the bare phonon spectrum has to be adjusted to compensate for the momentum dependence of $\omega_r$ coming from a particle-hole polarization bubble.

Finally, we emphasize that  our variational analysis establishes the \underline{\textit{upper bounds}}  on $\lambda^{c,1}$ as we only analyzed two variational states with localized bipolarons.
Other potential variational bipolaron states include  extended bound states of pairs or electron delocalization in a homogeneous bipolaron state.  Also, even if  the polaron order emerges with zero or near-zero momentum at $n =n_c$, it may evolve as $n$ approaches half-filling. This is particularly true in 2D as the very near half-filling of the polarization bubble at ${\bf Q} = (\pi,\pi)$ is logarithmically singular, hence the polaron state is likely a checkerboard even if it was a homogeneous one at $n =n_c$.

\section{Analytic consideration}
\label{sec:Analytics}

For simplicity of the analytic treatment, we assume that the system is spin polarized, and fermions with only one spin component undergo a transformation from the FL to
a polaron state.
For spinless fermions, there is no phonon-mediated pairing into singlet pairs, hence no bipolarons.
Also, for a parabolic dispersion, $\omega_r = \omega_0 (1- \lambda_0)^{1/2}$ with no factor of 2 in front of $\lambda_0$.
Extension to the spinful fermions is straightforward, but requires a separate analysis.
We further assume that the bare $\omega_0 (q) = \omega_0$ is momentum independent
(an Einstein phonon).
Anticipating that the polaron physics involves energies of order $W$ rather than of order
$E_F$, we introduce another coupling constant
\beq
\lambda_p = \frac{g^2}{2\omega^2_0 W}
\label{k_8}
\eeq
For an Einstein phonon, $\lambda_0 = 2 N_F W \lambda_p$.   The difference
between $\lambda_p$ and $\lambda_0$ is most noticeable in small $n$.
In 3D,
$\lambda_0 \propto N_F$  scales as $n^{1/3}$, while $\lambda_p$ is independent of $n$.
In 2D, $\lambda_0$  remains finite at $n \to 0$, but still depends 
on the structure of fermionic excitations near the bottom of the band,
while $\lambda_p$ depends on the bandwidth $W$.
For comparison with
the numerical results from Sec. \ref{sec:Numerics} we
use the same tight-binding dispersion
$\epsilon_{{\mathbf k}}=-2t\sum_{\alpha =1}^{
2} \cos (k_{\alpha} a)$, with
bandwidth $W = 8t$

In the following, we present the
 analytical results for a 2D system. We discuss the 3D case in Sec. \ref{sec:comp}.
For practical purposes, we choose to fix $g$ and $\omega_0$ and vary
$\lambda_p$ by changing the bandwidth $W$. We introduce
\beq
\beta = \frac{g^2}{2\omega^3_0} \equiv  \frac{{\bar g}^2}{\omega^2_0}
\label{1}
\eeq
and express the coupling $\lambda_p$ as  $\lambda_p = \beta \omega_0/W$.
 We  assume
that  ${\bar g}  \gg \omega_0$, i.e., that $\beta \gg 1$.
We  discuss the case of smaller $\beta$ in Appendix ~\ref{app_B}.

\subsection{Input}

We used as input the exact results for the fermionic DOS $N(\omega)$ in the two limits. One is the limit of large $W$, where 
$\lambda_p \to 0$ and the system behaves as a weakly interacting  Fermi gas 
The DOS in this limit,
\bea
 &&N (\omega) =  N_0 (\omega) = -\frac{1}{\pi}  \int \frac{d^2k}{4\pi^2}~ {\text{Im}} G^{\text{ret}}_0 (k, \omega)  \nonumber \\
 &&=  \int \frac{d^2k}{4\pi^2}  \delta (\omega + \mu_{FG} - \epsilon_k)
 \eea
  where $G^{\text{ret}}_0  (\omega, k) = 1/(\omega + \mu_{FG} -\epsilon_k + i \delta)$ is a retarded Green's function of a free fermion.
  The DOS is non-zero for $-W/2 -\mu_{FG}< \omega < W/2-\mu_{FG}$.
  Within this range, it is a continuous function of $\omega$.
 The relation between the fermionic  density $n$ and the chemical potential $\mu_{FG}$ is
\beq
 n= \int_{-\infty}^0 N_0 (\omega) d \omega; ~~ 1-n = \int_{0}^{\infty} N_0(\omega) d \omega
\label{rr_a}
\eeq
For tight-binding dispersion in 2D,
\beq
N_0 (\omega) = \frac{4}{\pi^2 W} K \left(1 - \left(2 (\omega+ \mu_{FG}) /W\right)^2  \right),
\label{tt_3}
\eeq
and
\beq
 n = \frac{2}{\pi^2} \int_{-1}^{\hat{\mu}_{FG}} K\left(1-x^2\right) dx, ~~
1-n =  \frac{2}{\pi^2} \int_{\hat{\mu}_{FG}}^1 K\left(1-x^2\right) dx
\label{k_3}
\eeq
 where $ \hat{\mu}_{FG}= 2\mu_{FG}/W$.  We show $N (\omega)$  in Fig.~\ref{fig:dos_ss_Holstein}a.

In the small density limit, $\mu_{FG} \approx - W/2$ and 
 $N(\omega =0) = N_F \approx 2/(\pi W)$. 
For an Einstein phonon
we then have $\lambda_0 =(4/\pi) \lambda_p$.
This relation can be modified by changing the dispersion near the bottom of the band without
changing the bandwidth. In particular, one can engineer  $\lambda_0$ to remain small when $\lambda_p =1$.

In the calculations below  we will be using  
\beq
\frac{1}{N} \sum_k f(\epsilon_k) = \int \frac{d^2 k}{4\pi^2} f(\epsilon_k) = \int d \epsilon N_0 (\epsilon) f(\epsilon)
\eeq 
 with 
 \beq
 N_0 (\epsilon) = \frac{4}{\pi^2 W} K\left(1 - \left(\frac{2\epsilon}{W}\right)^2\right)
 \label{wed_2}
 \eeq

In the opposite limit $W =0$, i.e., $\lambda_p = \infty$,
the model of Eq. (\ref{Hint}) reduces to
the  single-site  Holstein model. The latter can be solved exactly by the canonical Lang-Firsov transformation~ (see Refs \cite{lang_firsov,Holstein1959,Ranninger_1993,Mahan00} and Appendix \ref{app_E}).
The exact retarded fermionic Green's function at density $n$, $G^{\text{ret}} (\omega, n) \equiv G^{H}(\omega, n)$
("H'' stands for Holstein)  is the sum of contributions from $n=0$ (empty sites) and $n=1$ (filled sites)
      \beq
      G^{H} (\omega,n) =  \left[(1-n) G^{H} (\omega, 0) + n G^{H} (\omega, 1)\right],
      \label{n_1}
      \eeq
      The Green's functions $G^{H} (\omega, 0)$ and $G^{H} (\omega, 1)$ are given by
      \beq
      G^{H} (\omega, 0) = e^{-\beta} \sum_{m=0}^\infty \frac{\beta^m}{m!} \frac{1}{\omega + i \delta -m \omega_0},
     \label{n_2}
      \eeq
      and
    \beq
      G^{H} (\omega, 1) = e^{-\beta} \sum_{m=0}^\infty \frac{\beta^m}{m!} \frac{1}{\omega + i \delta + m \omega_0}.
      \label{n_3}
      \eeq
The DOS of the Holstein model, $N^H(\omega ,n) = - (1/\pi) \text{Im}  G^{H} (\omega, n)$, consists of a set of $\delta-$functional peaks separated by $\omega_0$:
\bea
&&N^H (\omega, n) = \nonumber \\
&& e^{-\beta} \sum_{m=0}^\infty \frac{\beta^m}{m!} \left[(1-n)
 \delta (\omega -m \omega_0) + n \delta (\omega + m \omega_0) \right]. \nonumber \\
 &&
\label{d_25_8}
\eea
We plot $N^H(\omega,n)$ in Fig.~\ref{fig:dos_ss_Holstein}b.
For $n=0$ ($n=1$) the peaks are located only at $\omega \geq 0$ ($\omega \leq 0$) while for
$0<n<1$,
the peaks are located at both positive and negative frequencies.
The conditions in Eq.~(\ref{rr_a}) are obviously satisfied.

The $\delta-$functional peaks correspond to
polarons --
``bound states" of a fermions and a phonon cloud around it ( more on this below). 
The polaron peak at $\omega =0$  is the most
relevant one as it determines the properties of the system at finite $T \ll \omega_0$.
For large $\beta$, the residue of this peak, $Z_0 = e^{-\beta}$, is exponentially small.
The residues $Z_n$ of the peaks at $\omega = n \omega_0$ with $n \ll \beta$ are also exponentially small,
while  $Z$-factors of the peaks at $\omega$ near $\beta \omega_0$ are much larger and scale as $1/\sqrt{\beta}$

Comparing the two limits, we see that their DOS are fundamentally different - a continuous function at small
$\lambda_p$ and a set of polaron $\delta$-functional peaks at
$\lambda_p\to \infty$.
Our goal is to understand how DOS evolves from one form to another and at which
$\lambda_p$  polaron
  peaks  develop.
For this purpose, it is instructive to first consider the limits
$n \to 0$ and $n \to 1$ when only one
 set of $\delta$-functions is present, either at $\omega >0$ or $\omega <0$,
and then analyze an arbitrary $0<n<1$ when the DOS in the Holstein model has
$\delta-$functional peaks at both positive and negative $\omega$.

 \subsection{Limit of vanishing density }
\label{sec:van}
Before we proceed,  a comment is in order. The small density limit should be understood either as the
 small, but still finite $n$, in which the Fermi energy $E_F \sim n W$
 is much lower than $W$, but still higher than $\omega_0$, i.e. $1 \gg n \gg \omega_0/W$ (range A), or the range of the smallest $n$, where $E_F \ll \omega_0$ (range B). In range A, the Eliashberg parameter $\lambda^E_0$ is still parametrically smaller than $\lambda_0$, the dressed $\omega_r (q=0)$ scales as $(1-2\lambda_0)$,
    and MET is rigorously justified as long as $ 1-2\lambda_0 > (\omega_0/E_F)^2$
    In range B, which includes $n=0$,
the adiabatic condition does not hold. We explicitly verified 
 (see Appendix \ref{app:Q})
 that in this range
vertex corrections for low-energy fermions are controlled by the same $\lambda_0$
as the mass renormalization, i.e. $\lambda^E_0 \sim \lambda_0$ 
 (In other words, the Eliashberg parameter is (roughly)  $\omega_0/(E_F +\omega_0)$.)
 MET for low-energy excitations then holds
up to $\lambda_0 = O(1)$,
 but there is no region where a strong coupling MET is under theoretical control.
 We emphasize that in both region A and region B,  the ratio $\omega_0/W$ is small. In this respect, our regime B 
  is still different from a truly anti-adiabatic regime $\omega_0 > W$. 
   
   The numerical results for the smallest $n$ have been obtained in range A, and in our analysis of the interplay between polaron formation and the internal stability of MET, we will focus on $1 \gg n \gg \omega_0/W$
. Restricting to this range would be problematic if relevant changes in the spectral function at a finite $W$ occurred at $W \sim \omega_0$. We show, however, that the changes in the spectral function begin at much larger $W \sim \sqrt{\beta} \omega_0$, when $\omega_0/W \sim 1/\sqrt{\beta}$ and $\omega_0/E_F \sim 1/(n\sqrt{\beta})$.  We can then set minimal $n$ to be somewhat larger than $1/\sqrt{\beta}$ and treat it effectively as the limit $n \to 0$, as
 the corrections due to $n$ are $O(n)$, and for minimal $n$ they are small in $1/\sqrt{\beta}$.

\subsubsection{\bf {Zero bandwidth, Holstein model}}
\label{n_0_W_0}

\begin{figure}[]
 \includegraphics[]{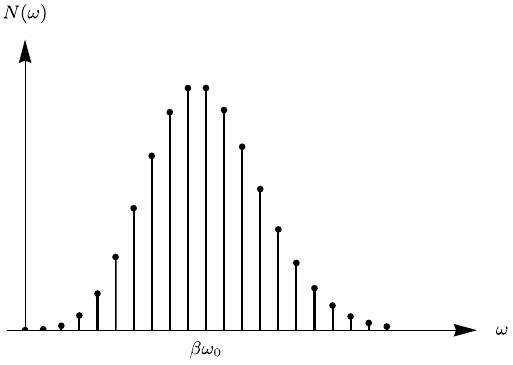}
 \caption{DOS $N(\omega)$  for vanishing density $n \to 0$. The DOS is non-zero only for $\omega \geq 0$.  The $\delta-$functional peaks are at $\omega = n \omega_0, n =0, 1, 2...$.   The residues of the peaks
  $Z_n$  are the largest at $n \approx \beta$, where $Z_n \sim 1/\sqrt{\beta}$.  At smaller $n = O(1)$, $Z_n \sim e^{-\beta}$ and at $n >> \beta$,  $Z_n \sim e^{-\beta} (e \beta/n)^n$. }
 \label{fig:dos_W0_n0}
 \end{figure}

At $W=0$, the limit $n \to 0$ can be taken by  setting $n=0$ in the exact solution (\ref{n_1})-(\ref{n_3}).
The poles of $G^H (\omega,0)$ are located at positive frequencies, and the
 smallest peak frequency is  $\omega =0$ (Fig.~\ref{fig:dos_W0_n0}).

To describe how this DOS evolves at a finite $W$,  we need an analytic
computational approach (a compliment to  the diagrammatic Monte Carlo technique
\cite{polarons1998,polarons2000})
 which, on the one hand, captures polaron bound states at $W=0$ and, on the other hand,
can be extended to a finite $W$. We argue, following earlier works, e.g. 
Refs. \cite{SadovskiiBook,Kuchinskiy_2024},  that polaron bound states at $\omega = n \omega_0$, $n =0,1,2,\ldots$  can be captured using diagrammatic expansion, which treats the renormalizations of the internal fermionic lines and  vertex corrections on equal footing. The approach is often termed "eikonal", where the  notation has been borrowed from scattering theory, where the diagrammatic treatment is similar~\cite{Levy1969}. 
More specifically, in  the following, we call "eikonal" the computational method which includes all diagrams linked to a  single line and neglects diagrams with fermionic loops.  We argue that the latter vanish identically in the limits of zero and maximal fermionic density and also vanish at a finite density for a homogeneous arrangement  of polarons.  
In Appendix \ref{app_A} we compare the eikonal approach with two approximate calculations, both neglecting vertex corrections: the rainbow approximation and the self-consistent one-loop approximation.  We show that the rainbow approximation  fails to reproduce the salient features of the exact solution of the Holstein model. The self-consistent one-loop approximation reproduces 
the discrete spectrum of the DOS but fails to reproduce the structure of the residues  of the peaks, particularly 
at large $\beta$.

The eikonal computational approach has been used to treat the effects of scattering by static impurities
 ~\cite{Lee1973,Efros1971,Sadovskii1974, Posazhennikova2003,Kiselev2009,*Efremov2022}, by charge fluctuations in one-dimensional systems
 ~\cite{Sadovskii1974SS,Tchernyshyov1999} 
  and near a nematic transition in 2D~\cite{Yamase_2012}, 
and  by spin
fluctuations~\cite{Vilk1997,Schmalian1998,Schmalian1999,
Sadovskii_review,*Sadovskii_extra,*Sadovskii_extra_1,Rohe_2005,Sedrakyan2010,Yamase_2016,Ye2023,*Ye2023_1}.

The point of departure for the eikonal diagrammatic treatment
is the expression for the propagator of a dispersionless fermion  $G^e_0 (\omega) = 1/(\omega + \mu)$  (the upper index $e$ stands for eikonal), where $\mu$ is the chemical potential that must be obtained from condition $n \to 0$.
 We assume and then verify that  $\mu = -\beta \omega_0$ and use this value of $\mu$ in the formulas below, i.e., set
 \beq
 G^e_0 (\omega) = \frac{1}{(\omega -\beta \omega_0)}
 \label{d_23_1}
 \eeq

 \begin{figure}[]
 \includegraphics[width=\linewidth]{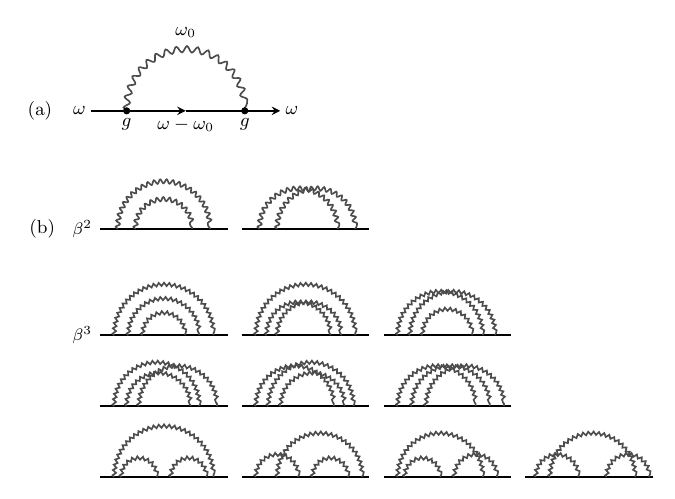}
 \caption{(a) One-loop electron self-energy. (b) Higher-order loop diagrams included in the eikonal expansion, shown 
   up to three-loop order.}
 \label{fig:eikonal_diags}
 \end{figure}
The diagram for the one-loop self-energy $\Sigma^{e,1}$
is shown in Fig.~\ref{fig:eikonal_diags}a. In analytical form
\beq
\Sigma^{e,1} (\omega) =   \frac{\beta \omega^2_0}{\omega  -(\beta+1) \omega_0}.
\label{xxx}
\eeq
A straightforward analysis shows that at higher orders
 one has to include series of  diagrams  stringed to a single line, see (Fig.~\ref{fig:eikonal_diags}b).
 These diagrams contain vertex corrections and the corrections to internal fermionic propagators.
There is no Migdal theorem for dispersionless fermions, hence both corrections have to be treated on equal footing.
Diagrams, which are not  stringed to a single line, all vanish because they contain
 bubbles made of fermionic propagators with the poles in the same half-plane of frequency.
For the same reason, the polarization operator that could potentially lead to renormalization of $\omega_0$,  also vanishes.

At the two-loop order,  eikonal series for the fermionic propagator are (see Fig.~\ref{fig:eikonal_diags})
  \begin{widetext}
    \beq
    G^{e}(\omega) =  \left(\frac{1}{(\omega -\beta \omega_0)} +  \frac{\beta \omega^2_0}{(\omega -\beta \omega_0)^2 ( \omega  -(\beta+1) \omega_0)} \left(1 + \frac{\beta \omega^2_0}{(\omega  -(\beta+1) \omega_0)} \left(\frac{1}{(\omega  -\beta \omega_0)} + \frac{2}{(\omega  -(\beta+2) \omega_0)}\right)\right) \right).
\label{2}
  \eeq
  \end{widetext}
The corresponding series for self-energy $\Sigma^e = 1/G^{e}_0 - 1/G^e$ are
\begin{eqnarray}
&&\Sigma^{e} (\omega) = \frac{\beta \omega^2_0}{\omega  -(\beta+1) \omega_0} \nonumber \\
&& +   \frac{2\beta^2 \omega^4_0}{(\omega -(\beta+1)\omega_0)^2 (\omega -(\beta+2)\omega_0)}.
\label{7}
\end{eqnarray}
Examining perturbation theory at higher-order, we find that
 it generates a set of higher-order poles in  $G^e (\omega)$
 at frequencies  $\omega =\omega_0 (\beta + m)$, $m \geq 0$, but does not generate  poles
 at smaller frequencies. This is clearly inconsistent with the exact solution of the Holstein model, which shows
  that $N^H (\omega)$ has $\delta-$functional peaks at $\omega = n \omega_0$, $n \geq 0$.
 Furthermore, near $\omega=0$, where the exact $G^H (\omega) \approx 1/(\omega e^{\beta})$,  the series in (\ref{2}) holds in powers of a small parameter $1/\beta$. Indeed,
  \beq
  G^{e}(0) = - \frac{1}{\beta \omega_0} \left(1 + \frac{1}{\beta} + \frac{2}{\beta^2} + ....\right),
  \label{2_1}
  \eeq
At first glance, these series should not significantly affect the bare value $G^{e}_0 (0) = - \frac{1}{\beta \omega_0}$.  However, this is a false impression.
We show in the following that the complete diagrammatic series gives rise to $ G^{e}(0) = \infty$ and $G^{e}(\omega)= e^{-\beta}/\omega$ in small $\omega$.  In more general terms, the full $G^e (\omega)$ contains polaron peaks at $\omega  = n \omega_0$ with $n < \beta$, which  are invisible in the order-by-order diagrammatics, yet are present because the series do not converge no matter how small the expansion parameter  $1/\beta$ is
 \footnote{The series  diverge for {\it any} $\omega$, including the largest  $\omega \gg \beta \omega_0$.}.
Indeed, inspection of the series of perturbations in (\ref{2_1}) shows that on the m-th order, the prefactor for $1/\beta^m$ scales as $m!$.  At large $m$, $m! \approx (1/2\pi m)^{1/2}  (m/e)^m$, and hence the expansion holds in $(m/e\beta)^m$. This shows that the expansion parameter remains small only up to $m_c = e\beta$. At larger $m$, the series starts to diverge no matter how small $1/\beta$ is.

  The complete series for $G (\omega)$ can be explicitly summed. We first show how to reproduce
   the exact $G^{H} (\omega) \propto 1/(\omega e^{\beta})$ at small frequencies, as this will be relevant to our analysis at a finite $W$ in the next subsection,  and then present the diagrammatic result for the  full Green's function at arbitrary $\omega$.

\begin{figure}[]
 \includegraphics[]{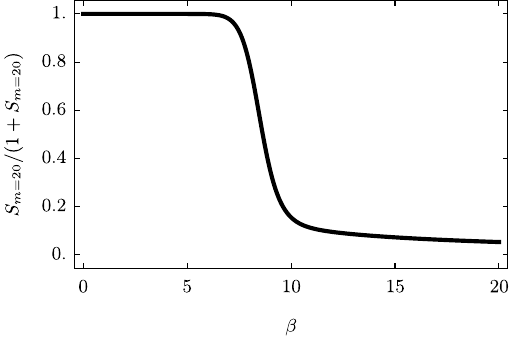}
 \caption{Function $S_m(\beta)/[1+ S_m (\beta)]$ with $S_m (\beta)$ from Eq.~\eqref{ww_5} for $m=20$ terms in the continued fractions expansion. Observe that $S_{20}(\beta)/[1+ S_{20} (\beta)] \approx 1$ for $\beta \lesssim 7.34 \approx 20/e$ 
 (i.e.,  $S_{20}(\beta)$ is large) and rapidly drops  at larger $\beta$.}
 \label{fig:sm}
 \end{figure}

  To reproduce the exact $G^H (\omega)$ we need to prove that (i) the series for $(G^e (0))^{-1}$ cancel out $(G^{e}_0 (0))^{-1}  = - \beta \omega_0$  (this will also justify using $\mu = -\beta \omega_0$)  and (ii) the  $\omega$ term in $(G^{e}_0 (\omega))^{-1}$ is renormalized to
 $(G^e (\omega))^{-1} =  e^\beta \omega$. To prove this, we use the fact~\cite{Ciuchi_1997,
 Pairault1998,*Pairault2000,Goodvin2006,Berciu_2006} that
     perturbation series for $G^{e} (\omega)$ in (\ref{2}) form continued fractions:
     \beq
     G^{e} (\omega) = \frac{1}{\omega - \beta \omega_0 - \beta \omega^2_0 G_1 (\omega)},
     \label{ww_1a}
     \eeq
     where
     \beq
      G_1 (\omega) =  \frac{1}{\omega - (\beta+1) \omega_0 - 2\beta \omega^2_0 G_2 (\omega)}
      \label{ww_1b}
     \eeq
      and
      \bea
      G_2 (\omega) &=&  \frac{1}{\omega - (\beta+2) \omega_0 - 3\beta \omega^2_0 G_3 (\omega)} \nonumber \\
      && .... \nonumber \\
      G_n (\omega) &=&  \frac{1}{\omega - (\beta+n) \omega_0 - (n+1)\beta \omega^2_0 G_{n+1} (\omega)}.
      \label{ww_1_c}
      \eea
 An examination of these series shows that one can obtain an analytic expression for $G_1 (0)$ even if we
  keep $m$ terms in the continued fractions. We find
    \beq
    \omega_0 G_{1,m} (0) = -\frac{1}{1+ \frac{1}{S_m(\beta)}} = -\frac{S_m (\beta)}{1 + S_m (\beta)},
     \label{ww_2}
    \eeq
    where
    \beq
    S_m(\beta) = \sum_{n=1}^m \frac{n!}{\beta^n}.
    \label{ww_5}
    \eeq
    Substituting into (\ref{ww_1}), we obtain
    \beq
    G^{e}_m (0) = - \frac{1 + S_m (\beta)}{\beta \omega_0}.
    \eeq
       If we keep the number of terms in the continued fractions $m < \beta e$,  we find $S_{m+1}  (\beta) < S_m (\beta)$ and   $(G^{e} (0)) = - 1/(\beta \omega_0)  (1 + 1/\beta +2/\beta^2 + ...) \approx - 1/(\beta \omega_0) $, in line with (\ref{2_1}).
      However, once $m$ exceeds  $\beta e$, a regular expansion in $1/\beta$ breaks down and $S_m (\beta)$ becomes large as $(m/\beta e)^m$.
      If we keep all terms in the continued fractions, we find $S_\infty (\beta) = \sum_{n=1}^\infty \frac{n!}{\beta^n} = \infty$ for any value of $\beta$.  Substituting into (\ref{ww_2}) we then obtain
     \beq
      G^{e} (0) = - \frac{1 + S_{\infty}}{\beta \omega_0} =\infty,
      \label{ww_6}
      \eeq
       in agreement with the exact solution.  As a side remark, we note  that to  detect a strong enhancement of $ G^{e} (0)$,
      one does not need to sum up an infinite number of terms in the continued fractions, taking
       $m$ terms with $m \geq \beta e$ is enough.
        In Fig.~\ref{fig:sm} we 
        plot $S_m(\beta)/(1+ S_m (\beta)) = -\omega_0 G_1$ for $m = 20$. We clearly see a sharp crossover
         from small to large $S(\beta)$  at $\beta \geq 7$, which is close to $20/e = 7.36$.
 
\begin{figure}[]
 \includegraphics[]{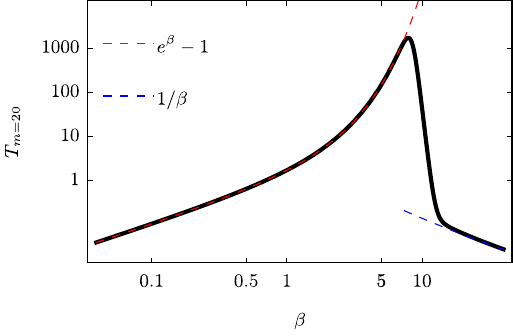}
 \caption{Function $T_m(\beta)$ from Eq.~\eqref{ww_8} for $m=20$ terms in the continued fractions expansion. Dashed line show the limiting behaviors for $m > \beta e$ (red)  and $m < \beta e $ (blue); see text.}
 \label{fig:tm}
 \end{figure}

      We next expand to first order in $\omega$ and obtain the prefactor for the linear in $\omega$ term in $(G^{e} (\omega))^{-1} = \omega(1 + T_\infty (\beta))$, where
      $T_\infty (\beta) = -\beta \omega^2_0 d G_1 (\omega)/d\omega$.  Analyzing the contributions from the $m$ terms in the continued fractions, we obtain a partial $T_m (\beta)$ in the  form
      \beq
    T_m (\beta) = \beta T^2_{m,0}   + 2 \beta^2   T^2_{m,0} T^2_{m,1}  + 6 \beta^3 T^2_{m,0} T^2_{m,1} T^2_{m,2} + ...
   \label{ww_4}
    \eeq
     and
     \beq
     T_{m,l} = T_{m,l} (\beta) =  \frac{1}{1+l + \frac{(1+l)!}{\sum_{n=1}^m \frac{(n+l)!}{\beta^n}}}
     \label{ww_7}
    \eeq
    The first term in these series $T_{m,0} (\beta)= 1/(1+ 1/S_m (\beta))$.
    If we set $m <  \beta e$, we find that $T_m (\beta) = 1/\beta + 4/\beta^2 +...$ is small and
     $(G^{e} (\omega))^{-1} \approx  \omega$.
     However, if we increase $m$ to $m > \beta e$, we find $T_{m,l} \approx 1/(1+l)$, independent of $m$.  Substituting into (\ref{ww_4}), we then obtain
     \bea
    T_m (\beta) &=& \beta  + \frac{2 \beta^2}{2^2} +  \frac{6 \beta^3}{6^2} + ...\nonumber\\
    &=& \beta  + \frac{\beta^2}{2} +  \frac{\beta^3}{6} + ... = e^\beta -1.
   \label{ww_8}
    \eea
    Hence,  $G^e (\omega) = \left(\omega \left(1 + T_m (\omega)\right)\right)^{-1} =
    e^{-\beta}/\omega$. This coincides with the exact Green's function of the Holstein model $G^H (\omega, 0)$ at $\omega \to 0$.   Extending $\omega$ to $\omega + i\delta$, we find that the DOS has a $\delta-$functional peak at $\omega =0$. We verify numerically that the continued fractions in (\ref{ww_1}) yield $G^e (\omega) = e^{-\beta}/\omega$ at the smallest $\omega$ once we take the $m$ terms in  (\ref{ww_1}) and set $m > \beta e$.  We
     show $T_m (\beta)$ as a function of $\beta$ at a given $m$ in Fig.~\ref{fig:tm}
     We again clearly see a sharp crossover from $T_m (\beta) \approx 1/\beta$ to $T_m (\beta) \approx e^\beta -1$ at $\beta \sim m/e$.
     
     This approach can be extended to  $\omega \approx k \omega_0$ where $k >0$ is an integer. We did not attempt to
 extract an analytical formula for $G_1 (\omega)$, but numerical analysis is quite straightforward since again only $m \geq \beta e$ terms are actually  needed.  At $\omega = k \omega_0$  we clearly see that $\omega_0 G_1 (0) = -(\beta -k)/\beta$, hence $G^e (k \omega_0) = - (1/\omega_0) 1/ \left((\beta - k) + \beta \omega_0  G_1 (0) \right) = \infty$, and at $\omega = k \omega_0 + {\tilde \omega}$,  $\omega^2_0 \beta (G_1 (\omega)- G_1 (k \omega_0)  =
        - {\tilde \omega} (e^{\beta} k!/(\beta^k) -1)$, hence $G^e (k \omega_0 + {\tilde \omega}) = (e^{-\beta} \beta^k/k!)/{\tilde \omega}$, the same as $G^H (k \omega_0 + {\tilde \omega}, 0)$
\footnote{Numerical calculations at a finite ${\tilde \omega}$ require some care as one has to select the contribution from a particular pole at $\omega = k \omega_0$, which can be done by either taking a very small ${\tilde \omega}$ or keeping it small but finite and comparing $G^e (k \omega_0 + {\tilde \omega})$ with the full form of $G^H (k \omega_0 + {\tilde \omega}, 0)$ from (\ref{n_2}).    For negative $k$, we verified that $(G^e (k \omega_0))^{-1}$ does not vanish, i.e., there are no poles at negative $\omega$. This is what we expect at the density $n =0+$.} 

This calculation also clarifies how we should view a polaron within this approach. Like we said in the introduction, a polaron can generally be viewed as a "bound state" between a fermion and lattice distortion, i.e., between a fermion and a phonon cloud. The 
 issue is how many phonons are in the cloud. 
  We found above that one needs $m \approx \beta e$
 terms in continued fractions to reproduce polaron peaks.  Each subsequent term in fractions adds an additional interaction with a phonon, so the $m-$th term can be viewed as the result of subsequent interaction with $m$ phonons.
 By this reasoning,  the  relevant number of phonons, which contribute to the formation of polarons, is  
 $\approx \beta e$.
 We note in passing that this is 
an  alternative way of looking at 
 polaron formation  compared to a traditional 
  Lang-Firsov approach~\cite{lang_firsov,Holstein1959,Mahan00}. In Lang-Firsov  approach,  a polaron is viewed as the result of a shift in the equilibrium position of a local oscillator in the presence of an electron.  The Hamiltonian of a displaced oscillator reduces to the sum of energies of a free dressed phonon and free electron, the latter with the energy shifted by 
  $-\beta \omega_0$. In these variables, a polaron is a free fermion  without a 
  cloud of dressed phonons.   The equivalence between Lang-Firsov and our approach is established when one computes the fermionic Green's function in Lang-Firsov approach by evaluating the overlap between the original and displaced phonon oscillator. 
        
        \begin{figure}[htbp]
        \centering
    \noindent
    \includegraphics[]{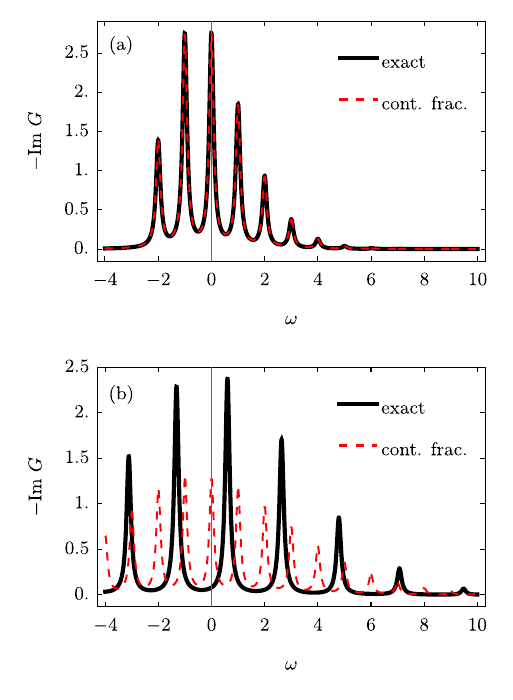} 
\caption{Comparison of the exact DOS with the one obtained by using the continued fractions representation for 
the Green's function with 20 terms. (a) $\beta =2$.  
The two DOS's are indistinguishable.  (b) $\beta =10$.
 The two DOS are clearly different. For better presentation we added the damping $\gamma =0.05$.}
\label{fig_21}
\end{figure}
In Fig. \ref{fig_21} we show the DOS obtained by taking 20 terms in continued fractions compared to the exact DOS $-1/\pi$ Im $G^H (\omega+ i0, 0)$ where $G^H$ is given by (\ref{n_2}).  We see that for $\beta \leq 1$ the agreement is essentially perfect. For larger $\beta > 20/e \approx 7.4$, the agreement is poor, implying that  more terms are needed in the continued fractions.

We can go further and obtain the analytical  expression for the eikonal series for  $G^e (\omega)$ at any $\omega$. This is most straightforwardly done in the imaginary time representation. To get the full $G^e (\tau)$, we rotate the one-loop self-energy (\ref{xxx}) to the Matsubara axis ($(\omega \to i \omega_m)$,  convert it to imaginary time and then exponentiate.   
This is an exact procedure, which correctly accounts 
for all higher-order diagrams.
Namely, 
 if we expand the exponent in Taylor series and convert each term back to momentum space, we reproduce order-by-order perturbation theory.     
The result For $G^{e} (\tau)$ is:
\begin{eqnarray}
G^{e}
(\tau)
  & = & \exp \left\{ \mu \tau +  \beta \omega_0^2 \int_0^{\tau} d\tau_1 \int_{\tau_1}^{\tau} d\tau_2 e^{-\omega_0 (\tau_2-\tau_1)}  \right\} \nonumber \\
               & = & \exp \left\{ -\beta  +\beta e^{-\omega_0 \tau }  \right\} .
\label{Gtau}
\end{eqnarray}
Upon Taylor series expansion and conversion back to real frequencies one immediately recovers Eq.~(\ref{n_2}) because each term in the series is a simple exponential function of $\tau$.   The conversion of  $G^{e}
(\tau)$ to retarded  $G^{e} (\omega + i \delta)$ immediately above the real frequencies can be done by first taking a Fourier transform to Matsubara $\omega_m$ and then replacing  $i\omega_m = \omega + i\delta$ to get the retarded Green's function.  The result is 
      \beq
     G^{e}(\omega) = \frac{1}{\omega + i\delta} {_1}F{_1} (1, 1 - \frac{\omega + i\delta}{\omega_0}, - \beta),
     \label{k_4}
     \eeq
where  ${_1}F{_1} (a,b,c)$ is the Kummer confluent hypergeometric function. 
The latter can be expressed in terms of  ordinary and upper generalized $\Gamma$-functions as
\bea
G^e (\omega)=& -& \frac{e^{-\beta}}{\omega_0} \frac{(\beta)^{\omega/\omega_0}}{\cos{(\pi \omega/\omega_0)}} \times \nonumber \\
&&\left[\Gamma(-\frac{\omega + i\delta}{\omega_0}) -{\text {Re}} \Gamma(-\frac{\omega}{\omega_0},-\beta)\right] . 
\label{k_4_a}
\eea
The  Green  function $G^e (\omega)$  has poles at $\omega = n \omega_0, n =0, 1, 2...$, which come from the poles in
the ordinary $\Gamma-$function.  We explicitly verified that this $G^e (\omega)$
coincides with $G^{H} (\omega, 0)$ in (\ref{n_2}) (see Appendix \ref{app_D}) for details).

Eq. (\ref{k_4_a})  has been previously obtained in Ref.~\cite{Kuchinskiy_2024} by expressing the full $G^e (\omega)$ in terms of full $G^e$ for intermediate fermions and the full vertex, the vertex in terms of $G^c$ using the generalized Ward identity, and
  solving recurrence relations for the Green's function.

\begin{figure}[]
 \includegraphics[]{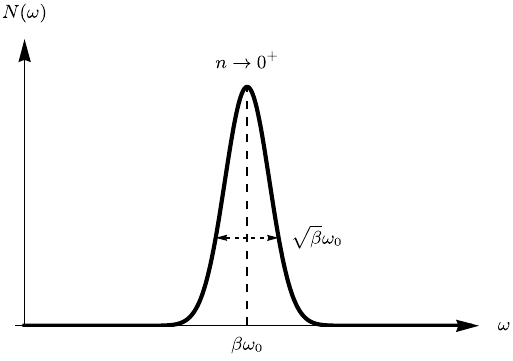}
 \caption{Envelope of the DOS, Eq.~\eqref{rr_14},  in the limit of vanishing density $n\to 0^+$.  Observe that at  large $\beta$,  the width of the envelope is parametrically smaller than the frequency at which it is  centered. }
 \label{fig:dos_ss_holstein_envelope_n0}
 \end{figure}

We emphasize that  the summation of eikonal diagrammatic series, which reproduces the exact $G^H (\omega, 0)$ with poles at the integer $\omega/\omega_0$, is
 different from the Borel summation, which expresses the divergent series via the integral over the axillary variable~\cite{Schmalian1998,*Schmalian1999,Sedrakyan2010,Ye2023,*Ye2023_1}.
 The Borel summation produces a continuous  Gaussian (Franck-Condon)  form of 
 Im $G^H (\omega, 0)$ and is applicable for large $\omega \sim \beta \omega_0$, when the distance $\omega_0$  between the peaks is much smaller than the peak position, and the DOS consisting of $\delta-$functions is well approximated by its continuous envelope.    To get the envelope in our case we approximate the eikonal series in (\ref{2}) by neglecting constants compared to $\beta$, i.e., replacing $\beta+1$, $\beta + 2$, etc as just $\beta$.  Doing this, we obtain $G^e (\omega) \approx  Q (X) /(\omega - \beta \omega_0)$, where
   \beq
    X =  \frac{\beta \omega^2_0}{(\omega - \beta \omega_0)^2}
   \label{rr_10}
   \eeq
   and
   \beq
  Q(X) = 1 + X + 3X^2 +15 X^3 + .. = \sum_{n=0}^\infty (2n-1)!! X^n.
  \label{rr_10a}
   \eeq
 The Borel summation of a formally divergent sum in (\ref{rr_10a}) uses the fact that $(2n-1)!! = \Gamma (n+1/2) 2^n/\sqrt{\pi}$ and invokes an integral representation of the $\Gamma$-function:  $\Gamma (n+1/2) = \int_0^{\infty} e^{-t} t^{n-1/2} dt$. One can verify that the order-by-order expansion in $X$ in the l.h.s of (\ref{rr_10a}) is reproduced if we take
  \beq
   Q(X) = \frac{1}{\sqrt{\pi}} \int_0^\infty \frac{dt e^{-t}}{\sqrt{t}} \frac{1}{1-2 tX}.
   \label{rr_11}
   \eeq
    The integral in (\ref{rr_11}) can be explicitly evaluated, and the result is
  \beq
  Q(X) = \frac{\sqrt{2\pi} e^{-\frac{1}{2X}}}{\sqrt{-X}} \left(1+ Erf \left( -\frac{1}{\sqrt{-2X}} \right)\right),
   \label{rr_12}
  \eeq
  where Erf (...) is an error function.  Separating real and imaginary parts in (\ref{rr_12}), we obtain a continuum DOS
  \beq
  N^{e}_{env} (\omega) \approx \sqrt{\frac{1}{2\pi \beta \omega^2_0}}  e^{-\frac{(\omega - \beta \omega_0)^2}{2 \beta \omega^2_0}}
    \label{rr_14}
  \eeq
 This DOS coincides with the envelope of the exact $ N^H (\omega)$, which can be instantly verified by
 expressing $N^H (\omega)$ as
   \beq
N^H (\omega, 0) = -\frac{1}{\pi} {\text Im} G^H (\omega,0) = e^{\beta} \sum_{m=0}^\infty \frac{\beta^m}{m!} \delta (\omega - m \omega_0), 
\eeq
 replacing $\sum_m$ by  $\int dm$,  and
   $m!$ by $\sqrt{2\pi m} (m/e)^m$.  The integrand is the largest at  $m \sim \beta$, and the integration over $m$ reproduces (\ref{rr_14}).
   The  DOS envelope  is centered at $\omega = \beta \omega_0$ and its width is $\sqrt{\beta} \omega_0$. We plot the envelope $N^{H}_{env} (\omega, 0) = N^e_{env} (\omega)$ in Fig.~\ref{fig:dos_ss_holstein_envelope_n0}.

We further note that (i) the total DOS of the envelope $\int d \omega N^e_{env} (\omega) \approx 1$, up to corrections of order $e^{-\beta}$,  and (ii)
  the width of the envelope is parametrically smaller than
its  peak position.
In this situation, the envelope can be further approximated by the DOS of a free fermion with energy $\beta \omega_0$, i.e., Im $G^e_0 (\omega)$. The  DOS  at $W=0$  and $n = 0+$ can then be viewed as consisting of a free fermion peak at $\omega = \beta \omega_0$  and polaron peaks at $\omega = n \omega_0$, $n \ll \beta$, with exponentially small residues $e^{-\beta}$.  This DOS will be the starting  point for our analysis of the DOS at finite $W$.

Before we continue, we elaborate on the remark that we made earlier on the relation between our DOS in terms of the original fermions ($c$ fermions in the notation of Eq. (\ref{Hint})) and the one in terms of $p$ fermions, which emerge after the canonical transformation ~\cite{lang_firsov} and describe composite objects -- fermions dressed by lattice distortions.   
In terms of $c$ fermions, the DOS as a function of $E = \omega + \mu$  consists of a free-fermion peak at $E =0$, a polaron peak at
$E = - \beta \omega_0$ with residue $Z = e^{-\beta}$ and other polaron peaks  further at negative
energies. In terms of p-fermions, the free-fermion peak and the  polaron peak are interchanged:
the polaron peak is at $E= -\beta \omega_0$ and has residue $Z=1$, and the "free-fermion" peak is at $E = 0$ and
has an exponentially small residue.

\subsubsection{\bf {Finite bandwidth.}}
\label{n_0_W_finite}

At a finite bandwidth $W$, we and re-analyze eikonal series by adding the dispersion $\epsilon_k$ to the bare fermionic propagator:
\beq
G^e_{0} ({\mathbf k}, \omega) =\frac{1}{\omega - (\epsilon_{\mathbf k} - \mu)}.
\label{tt_1}
\eeq
We assume and then verify that $\mu$ remains equal to $-\beta \omega_0$
up  to corrections of order $e^{-\beta}$, and that
the polaron state remains stable as long as $\beta \omega_0 > W/2$.  In this situation, the pole of
the time-ordered  $G^e_{0} ({\mathbf k}, \omega + i \delta {\text {sign}} \omega)$ remains in the
  lower half-plane of frequency for all momenta ${\bf k}$.  Then the
   diagrams, which are not linked to a single line,  still vanish and also $\omega_0$ is not renormalized.
 However, self-energy is generally affected by $\epsilon_{\mathbf k}$, and the polaron peaks at $\omega = n \omega_0$ acquire some momentum dependence~\cite{Berciu_2006}. 

We first analyze the Green's function $G^e (k, \omega)$  at the smallest $\omega$.  It still can be represented as the continued fractions (\ref{ww_1}), i.e.
   $(G^e ({\mathbf k}, \omega))^{-1} = \omega - \beta \omega_0 - \epsilon_{\mathbf k} - \beta \omega^2_0 G_1 ({\mathbf k}, \omega)$, but now $G_1 ({\mathbf k}, \omega)$ and all other $G_n ({\mathbf k}, \omega)$ are expressed via integrals over the momenta of internal fermions (Ref. \cite{Berciu_2006}).
We recall that at $W=0$, $-\beta \omega^2_0 G_1 (k, \omega) = \beta \omega_0 + \omega(e^{\beta} -1)$ and 
$G^e (\omega) = e^{-\beta}/\omega$,
 and the exponential behavior comes from the continued fractions at order $m = O(\beta)$, reflecting the fact that a polaron is a bound state of an electron and $O(\beta)$ phonons.  Relevant phonon energy at this order of continued fractions is of order $\beta \omega_0$. 
An inspection of the continued fractions in the presence of $\epsilon_{\mathbf k}$
shows that, as long as $W \ll \beta \omega_0$,  we have at $\omega =0$, 
     $-\beta \omega^2_0 G_1 ( {\mathbf k},0)= \beta \omega_0 + O(W)$ and hence
     $(G^e ({\mathbf k}, 0))^{-1} = O(W)$.
The same analysis at finite $\omega$ yields
$-\beta \omega^2_0 (G_1 ({\mathbf k}, \omega) - G_1 ({\mathbf k}, 0)) = \omega (e^{\beta} -1 + O(W/\beta \omega_0)$, i.e. $(G^e ({\mathbf k}, \omega))^{-1} = \omega (e^\beta + O(W/\beta \omega_0))$. The first term is exponentially large, the second is at most $O(1)$ and can be neglected for all $W < \beta \omega_0 e^{\beta}$.
  Combining  $\omega =0$ and linear in $\omega$ terms, we obtain at the smallest $\omega$
\beq
G^e({\mathbf k}, \omega) = \frac{e^{-\beta}}{\omega   - e^{-\beta}  f({\mathbf k})}.
\label{ww_1}
\eeq
We see that the mass of the polaron at $\omega =0$ is exponentially large, i.e.,  a polaron is   an exponentially heavy quasipraticle.
We also emphasize that the condition $n=0$ implies that $f(k) \geq 0$ for all $k$.   If the momentum dependence of the full Green's function retains the same form as for free fermions, $f(k) = \epsilon_k + W/2$.   

Integrating $G^e({\mathbf k}, \omega) $ in (\ref{ww_1}) over momentum, we find that the DOS of this polaron
is non-zero at   $\omega >0$  in an exponentially small patch $0< \omega < W e^{-\beta}$.   However, within this patch,  $N^e (\omega) \sim 1/W$.
This agrees with the earlier results~\cite{Ranninger_1992,Ranninger_1993,Berciu_2006}.  

The same holds for other polaron peaks at $\omega = m \omega_0$ with $m \ll \beta$  -- they all
transform into patches with $N^e (\omega) \sim 1/W$ and exponentially small width $W e^{-\beta} \beta^n/n!$. 
 The only difference with the patch next to $\omega =0$ is that  for $m>0$,  the momentum-dependent piece $f_m (k)$ can change sign as a function of $k$. 
This leads to the DOS at small $\omega$, consisting of narrow patches.
 Fermionic states with $\omega$ in between patches are localized, while those inside the patches are extended.
 The patches  {\it remain well separated
even at $W > \omega_0$}, when naive reasoning would suggest overlapping of the DOS of neighboring excited polaron states. This also justifies a'posteriori our conjecture that  at a non-zero $W$, $\mu = -\beta \omega_0$ up to $e^{-\beta}$ corrections.
Indeed,  as we just argued, for $\mu = -\beta \omega_0$, the prefactor for the term $\omega$ in $G^{-1} ({\mathbf k}, \omega)$ also remains
 $e^{\beta}$ at a finite $W$,   which leads to the $e^{-\beta}$ width of the polaron patch around $\omega =0$.
The maximal change in total density from this patch is then also $e^{-\beta}$.
 This change is compensated for by an exponentially small change of $\mu$ compared to $-\beta \omega_0$.
We recall that we neglected numerically small shifts of the chemical potential due to
virtual delocalization of polaron states to the nearest-neighbor sites (they are quantified in Sect. 
\ref{CDW_pol}).

Fermions at frequencies $\omega \sim \beta \omega_0$ display different behavior.
Like we said, at $W=0$, the DOS of these fermions 
can be well approximated  by the Gaussian centered at $\omega = \beta \omega_0$  
 with the width $\sqrt{\beta} \omega_0$  and can be
further  approximated by a single
$\delta-$function at $\omega = \beta \omega_0$ for frequencies $|\omega - \beta \omega_0|$ larger than the 
  width of the Gaussian,  $\sqrt{\beta} \omega_0$. 
We emphasize that this holds for the large $\beta$ we consider here. For $\beta = O(1)$  there is no clear separation between 
  polaron peaks at different frequencies (see Appendix \ref{app_B}).    
   We emphasize that the transformation of the DOS from its form in the atomic limit begins at $W \sim \sqrt{\beta} \omega_0 \gg \omega_0$.  There are no notable changes in DOS at $W \sim \omega_0$.

The approximation of states at $\omega \sim \beta \omega_0$ by a single free-fermion peak 
becomes even better at a non-zero $W$. The residues of individual $\delta$-functions  with $m \approx \beta$ 
 scale as $Z_m \sim 1/\sqrt{\beta}$, i.e., individual peaks begin to overlap at $W \sim \sqrt{\beta} \omega_0$. 
  At larger $W$, individual peaks are no longer distinguishable, and also the bandwidth  exceeds the width  
  of the Gaussian made out of these states. As a consequence,  the DOS near $\omega = \beta \omega_0$  can be interpreted as that of a free dispersing fermion, described by a free-fermion Green's function $G^e_{0} ({\mathbf k}, \omega)$, Eq. (\ref{tt_1}). This DOS is a continuum of width $W$. 
The full DOS then consists of  this  continuum and a set of narrow patches near $\omega =0$,
which are the DOS's of  weakly dispersing heavy polarons. We show this DOS  in Fig.~\ref{fig:schematic_dos_finiteW_n0p}a.
  In Appendix \ref{app_E} we discuss in more detail how individual peaks near $\omega = \beta \omega_0$ transform into a continuum 
  when $W \sim \sqrt{\beta} \omega_0$.

\begin{figure}[t]
 \includegraphics[]{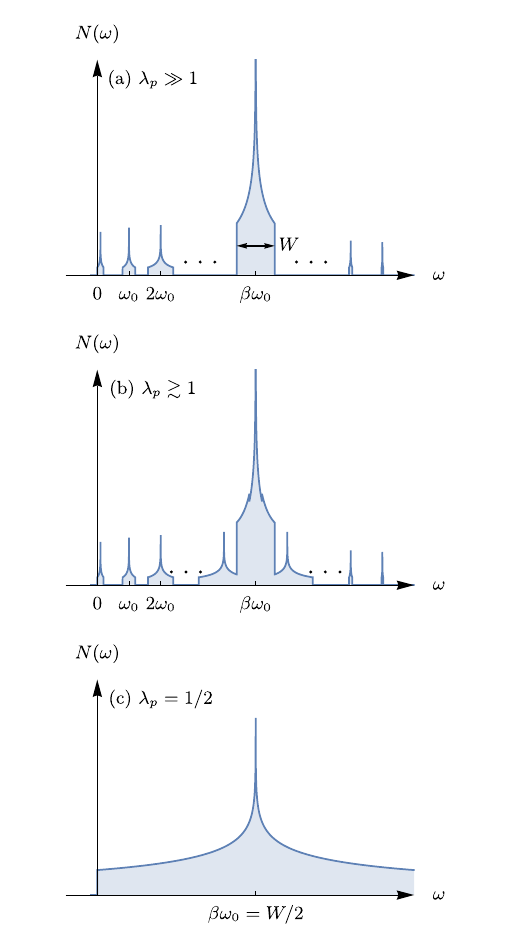}
 \caption{Schematic electron DOS in the regime $W > \sqrt \beta \omega_0$, for density $n=0$. The width of the peak at $\omega = \beta \omega_0$ is set by $W$; the width of the peaks nears $\omega = 0$ is exponentially small $\sim We^{-\beta}$. Panels (a)-(c) show evolution with increasing $W$, as polaron ``patches" are absorbed by the continuum. 
 The height of each low-energy patch is $O(1)$,  for its structure we assumed a tight-binding dispersion.}
 \label{fig:schematic_dos_finiteW_n0p}
 \end{figure}

As $W$ increases, the lower end of the continuum extends to smaller $\omega$.
In this process, it absorbs polaron patches one by one, starting from one ones closest to the continuum. We show this in  Fig.~\ref{fig:schematic_dos_finiteW_n0p}b.
At $\beta \omega_0 =W/2$, i.e., at $\lambda_p =1/2$, the continuum absorbs the last polaron at $\omega =0$ (Fig.~\ref{fig:schematic_dos_finiteW_n0p}c). This is the end point of the polaron state. At larger $W$,  the system is in the FL phase and the DOS $N(\omega)$ is
continuous  between $\omega =0$ and $\omega = W$.
That critical $\lambda_p =1/2$ is corroborated by the observation that for this $\lambda_p$,  the chemical potential of the polaron state $\mu =-\beta \omega_0$ coincides with the
chemical potential of a Fermi gas $\mu_{FG} = -W/2$ at $n = 0^+$. 
Furthermore,  we will show below that at $n = 0+$, the
energies of the polaron state $E_P = - \beta \omega_0 n$ and of the Fermi gas $E_{FG} = - W n/2 = -\beta \omega_0 (2 \lambda_p)$ also coincide at $\lambda_p =1/2$.

A few comments are in order here. First, strictly speaking, we need to compare $\mu$ and $E_P$ with the chemical potential $\mu_{FL}$ and the ground-state energy $E_{FL}$ of an FL.  However, as we said in Sec. \ref{sec:Numerics}, the difference between the kinetic energies $E_{{\text {kin}}, FL}$ and $E_{{\text {kin}}, FG}$  and between $\mu_{FG}$ and $\mu_{FL}$) is small in the Eliashberg parameter, which we treat as infinitesimally small. 
For this reason, in the following we approximate $\mu_{FL}$ by $\mu_{FG}$ and $E_{{\text {kin}}, FL}$  by $E_{{\text {kin}}, FG}$. 
Second, in principle, there exist corrections to the chemical potential of the polaron state $\mu =-\beta \omega_0$,  arising from virtual delocalization of the polaron states (fluctuations of the order parameter $\Delta$ with finite $q$). These corrections shift the critical value of $\lambda_p$ down from $1/2$. 
In our current analysis, we assume that $\mu$ remains equal to $-\beta \omega_0$ at a finite $W$. We will return to this issue in Sec. \ref{CDW_pol}.
The absorption of  polarons with energies of order $\omega_0$ the growing continuum can be equivalently described by the growth of the spectral weight in between polaron patches.  We do not have rigorous description for this, but we emphasize that the logics behind exponentially small  ($e^{\beta}$) width of the polaron peak  is that the factor
 $e^{\beta}$ comes from continued fractions of order $m \sim \beta$, when the relevant phonon frequency is comparable to $\beta \omega_0$.   At $W$ parametrically smaller than $\beta \omega$,  the corrections from the  fermionic dispersion to continued fraction result at this order are small and can be neglcted.    However, at $W \sim \beta \omega_0$, they become  relevant at eventually destroy  polaron patches.

 It is also instructive to discuss the evolution of the system behavior starting from the opposite limit of small $\lambda_p$ (large $W$), where MET is valid. We  recall that  MET
is assumed to remain internally stable as long as the dressed phonon Debye frequency $\omega_r$ is positive~\cite{Migdal,Eliashberg,AGD,Sadovskii:2022_ufn} (barring potential dynamical
  instabilities at smaller $\lambda_p$~\cite{Yuzbashyan,*Yuzbashyan_1,*Yuzbashyan_2,*Yuzbashyan_3}).
As we said above,
in our model $\omega_r$ remains positive up to $\lambda_0 =1$. In terms of
$\lambda_p  = \pi \lambda_0/4$, the stability holds up to
$\lambda_p = \pi/4$. This coupling is larger than $1/2$, meaning
that there is a range of $\lambda_p$ where the FL state is a local minimum with respect to low-energy perturbations within MET, yet the polaron state has a lower energy.

This last statement may sound contradictory because Fig. \ref{fig:schematic_dos_finiteW_n0p},
viewed from right to left,  
shows that the polaron bound state develops at $\omega =0$, which at a face value looks like a low-energy instability. However, we argue that polaron formation is a high-energy phenomenon. This
becomes clear from the analysis of the fermionic self-energy. It has two contributions.
One comes from low loop-orders in perturbation theory ($m = O(1)$, where $m$ is a loop order) and involves
internal fermions with  frequencies well within the bandwidth.  This contribution is captured within MET.  It becomes singular when the dressed phonon softens, but is  regular at smaller $\lambda_p$, including $\lambda_p =1/2$.    Another  comes from high-loop orders $m \geq W/\omega_0$,  from the processes in which
multiple phonon scattering moves an internal fermion from small $\omega$  to outside of the bandwidth.
The contribution to self-energy  from these processes is of the order
$(\lambda_p)^{W/\omega_0}$.  When $\lambda_p$ is small,  the corresponding self-energy is irrelevant.
However, when $\lambda_p >1 $, it  becomes exponentially  large and causes a dip in $N(\omega)$ at $\omega =0$, which is the beginning of the formation of the polaron  state.  A more sophisticated analysis is needed to  verify that this occurs at $\lambda_p  =1/2$.

We also note that the Debye frequency changes discontinuously at $\lambda_p =1/2$ from $\omega_r (q)$ in MET to $\omega_0$ in the polaron state.  However, at $n=0^+$, this discontinuity affects only infinitesimally small $q$  as $\omega_r$ in a 2D FL gets reduced with increasing $\lambda_p$ only for $q< 2k_F$, where $k_F \propto n^{1/2}$, while  at larger $q$, 
$\omega_r (q) \approx \omega_0$.  \\

\subsection{Case of full filling,  $n  = 1$}
\label{n_1a}

The analysis for full filling is similar to that for zero filling. Like there, we focus on the range where $n$ is very near $1$, yet $\omega_D/E_F \ll 1$, where $E_F$  is now counted from the top of the band.\\

\subsubsection{\bf {Zero bandwidth, Holstein model}}

\begin{figure}[]
 \includegraphics[]{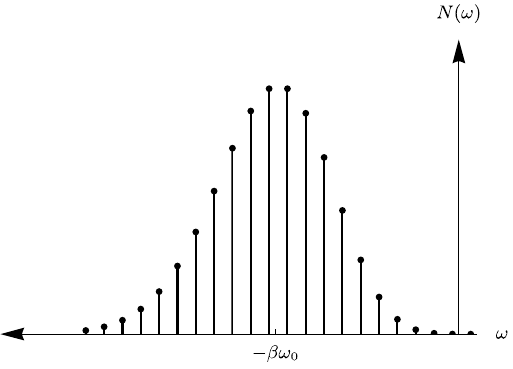}
 \caption{DOS $N(\omega)$  for the case of near-full filling $n \to 1$. The DOS is non-zero only for $\omega \leq 0$.  The $\delta-$functional peaks are at $\omega = -n \omega_0, n =0, 1, 2...$.   The residues of the peaks
  $Z_n$  are the largest at $n \approx \beta$, where $Z_n \sim 1/\sqrt{\beta}$.  At smaller $n = O(1)$, $Z_n \sim e^{-\beta}$ and at $n \gg \beta$,  $Z_n \sim e^{-\beta} (e \beta/n)^n$. }
 \label{fig:dos_W0_n1}
 \end{figure}

The diagrammatic treatment at $W=0$ proceeds in the same way as at $n=0+$ with one distinction:
the  self-energy contains an additional off-shot (Hartree) term $\Sigma^{off-shot} = -2 \beta \omega_0$.
Incorporating it into the Green's function of a free dispersionless fermion, we find that the off-shell self-energy changes the bare chemical potential $\mu= -\beta \omega_0$ to $\mu - \Sigma^{off-shot} = +\beta \omega_0$.
With this modification, the bare Green's function for eikonal calculations
 is
\beq
G^e_0 (\omega) = \frac{1}{\omega + \beta \omega_0}.
\eeq
The eikonal series then become
\begin{widetext}
    \beq
    G^e(\omega) = \frac{1}{\omega +\beta \omega_0} +  \frac{\beta \omega^2_0}{(\omega_m +\beta \omega_0)^2 (\omega +(\beta+1) \omega_0)} \left(1 + \frac{\beta \omega^2_0}{(\omega +(\beta+1) \omega_0)}
    \left(\frac{1}{(\omega +\beta \omega_0)} + \frac{2}{(\omega +(\beta+2) \omega_0)}\right)\right) + ... 
\label{2_2}
  \eeq
  \end{widetext}
The poles in the order-by-order expansion of $G(\omega)$ are now at negative frequencies $\omega < -\beta \omega_0$.  Like for the case $n =0+$, the series do not converge for any frequency and have to be evaluated in the same way as in the previous Section. The  fully dressed retarded Green's function is
   \beq
   G^e(\omega +i\delta ) =\frac{1}{\omega + i\delta} {_1}F{_1} (1, 1 + \frac{\omega + i\delta}{\omega_0}, - \beta),
     \label{4_1}
     \eeq
where, we recall, ${_1}F{_1} (a,b,c)$ is the Kummer confluent hypergeometric function. It has
 poles
at $\omega = -n \omega_0, n =1, 2...$, the additional pole
in  $G^e(\omega)$  at  $\omega =0$ comes from the $1/(\omega+ i\delta)$ prefactor in (\ref{4_1}).
The DOS, $N^e (\omega) =N^H (\omega, 1)$, then consists of the set of $\delta-$functional peaks at $\omega = -m \omega_0$, $m =0,1, 2..$ (Fig.~\ref{fig:dos_W0_n1}).

\subsubsection{\bf {Finite bandwidth}}

The analysis at a finite $W$ also parallels that at $n \to 0$.
  The bare Green's function is
 \beq
 G^{e}_0 ({\mathbf k}, \omega) =\frac{1}{\omega - (\epsilon_{\mathbf k} - \beta \omega_0 )}.
\label{tt_1_1}
\eeq
  The full $ G^{e} ({\mathbf k}, \omega)$ at $n = 1^{-}$ consists of patches of heavy polarons, with exponentially small  
   width $e^{-\beta}$  and a continuum with width $W$, centered at $\omega = - \beta \omega_0$.
    For the polaron next to $\omega =0$,
    \beq
G^e({\mathbf k}, \omega) = \frac{e^{-\beta}}{\omega   - e^{-\beta}  {\tilde f}({\mathbf k})}, 
\label{ww_1_1}
\eeq
   where now ${\tilde f} (k) \leq 0$ for all $k$.    If the momentum dependence of the full Green's function retains the same form as for free fermions, $f(k) = \epsilon_k -W/2$.   
 
\begin{figure}[t]
 \includegraphics[]{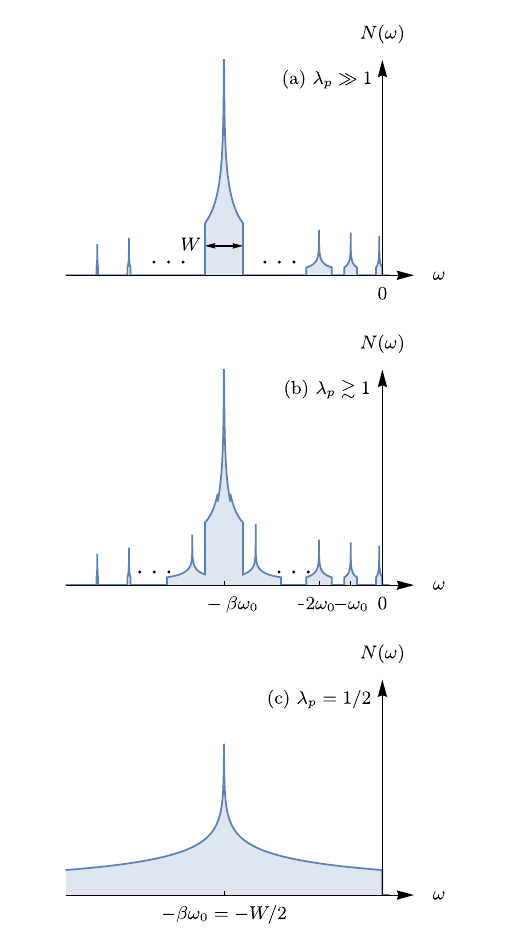}
 \caption{Schematic electron DOS in the regime $W > \sqrt \beta \omega_0$, for density $n=1^-$. The width of the peak at $\omega = -\beta \omega_0$ is set by $W$; the width of the peaks nears $\omega = 0$ is exponentially small $\sim We^{-\beta}$. Panels (a)-(c) show evolution with increasing $W$, a polaron ``patches" are absorbed by the continuum.}
 \label{fig:schematic_dos_finiteW_n1m}
 \end{figure}

 As $W$ increases ($\lambda_p$ decreases), the lower end of the continuum extends to smaller negative $\omega$ and absorbs heavy polarons one by one.  The last polaron at $\omega =0$ is absorbed at $\lambda_p =1/2$, and at smaller $\lambda_p$ the ground state is the FL described by MET. We show the evolution of DOS with increasing $W$ in Fig.~\ref{fig:schematic_dos_finiteW_n1m}.
 Like at $n = 0^+$,  the MET remains internally stable up to $\lambda_p = \pi/4 >1/2$,  and  the polaron formation at  $\lambda_p =1/2$, coming from smaller $\lambda_p$, involves high-loop-order self-energy diagrams and processes involving fermions with energies outside of the bandwidth.

\subsection{ Arbitrary filling, $0<n <1$}
\label{arb_n}

\subsubsection{\bf {Zero bandwidth, Holstein model}}

The exact spectral function  of the Holstein model $A^{H} (\omega, n)$
 at $0<n<1$ is presented in Eq. (\ref{d_25_8}).
It consists of a set of $\delta-$functions at both positive and negative $\omega_n = n \omega_0$, $n = 0, \pm 1, \pm 2$, etc.
The total spectral weight at positive frequencies is $1-n$ and that at negative frequencies is $n$.
We argued earlier that at large $|\omega| \approx \beta \omega_0$, where
the distance between the peaks is much smaller than their positions,  the spectral function
 can be approximated by a  Gaussian envelope. For $0<n<1$, there are two envelopes, one 
centered at $\omega = \beta \omega_0$ and the other  at $\omega = - \beta \omega_0$.
Each envelope has a small width $\sqrt{\beta}\omega_0 \ll \beta \omega_0$
and can be further approximated  by a $\delta$-function at $\omega = \pm \beta \omega_0$
with residue $Z=1-n$ for negative $\omega$ and $Z=n$ for positive $\omega$:
\beq
 A (\omega, n) =  n \delta(\omega + \beta \omega_0) + (1-n) \delta(\omega - \beta \omega_0) \,.
\label{qq_1}
 \eeq
The corresponding "free-fermion" retarded Green's function is
  \beq
  G^{ret} (\omega, n) = \frac{n}{\omega + \beta \omega_0 + i \delta} + \frac{1-n}{\omega - \beta \omega_0  + i \delta}.
  \label{k_9}
\eeq
 The two-peak structure in (\ref{qq_1}) is qualitatively different from the one-peak structure of $A (\omega)$ at $n =0$ and $n =1$. There, the free-fermion part of the Green's function
coincides with the bare $G^e_0 (\omega)$ in the eikonal  approach.
 This is clearly not the case at a finite $n$ as starting from a Green's function with a single pole one has to include the self-energy to reproduce the two-pole structure of $G (\omega, n)$ in (\ref{k_9}).

To get an idea of how to modify the eikonal calculations at $0<n<1$, we
introduce the bare Green's function for the eikonal calculations
\beq
G^e_0 (\omega,n) = \frac{1}{\omega + \mu_n},
\label{9}
\eeq
where $\mu_n = -\beta \omega_0 (1-2n)$  incorporates the off-shot (Hartree) self-energy $\Sigma^{\text {off-shot}} = -2\beta \omega_0 n$,  express  $G (\omega,n)$  from (\ref{k_9})
in terms of $G^e_0 (\omega, n)$,  and extract the self-energy
$\Sigma (\omega, n) = ((G^e_0 (\omega, n))^{-1} -(G (\omega, n))^{-1})^{-1}$.   We find
\beq
    \Sigma  (\omega, n) = \frac{|\Delta|^2}{\omega - {\mu_n}},
\eeq
 where we introduced
 \beq
  |\Delta|^2 = 4n (1-n) (\beta \omega_0)^2
 \label{rr_15}
 \eeq
This self-energy has a pole at $\omega = {\mu_n}$ whose frequency is opposite to that of the pole of
$ G^e_0 (\omega,n)$ in Eq. (\ref{9}).

A self-energy of this form emerges if we assume that there exists another branch of excitations with  opposite sign of ${\mu_n}$ (an ancilla fermion) and that there is a bilinear coupling between the physical and ancilla fermions.
A somewhat similar approach has been used  in the analysis of pseudogap behavior for hole doped cuprates (see  Ref. \cite{Bonetti_2025} for a recent review). In our case, the physical reasoning for two-fermion description is that at $0<n<1$ each lattice site is either occupied or empty with a certain probability.
  In this respect, a physical fermion describes states occupied with probability $n$ and an  ancilla fermion  describes states occupied with probability $1-n$. The chemical potential for the ancilla fermion is $-\beta \omega_0 (1-2 (1-n))= -{\mu_n}$ and its bare Green's function is ${\bar G}^e_0 (\omega, n) = \frac{1}{\omega - {\mu_n}}$.
Because the ancilla fermion is a mirror image of the physical fermion, it couples to a phonon by the same Eq. (\ref{Hint}).
 The full Hamiltonian  is the same of the Hamiltonians for the physical and ancilla fermions:
\beq
   H' =  -{\mu_n} \left(c^\dagger c - {\tilde c}^\dagger {\tilde c} \right)  + \frac{g}{\sqrt{2\omega_0}} \left(c^\dagger c + {\tilde c}^\dagger {\tilde c}\right) (b + b^\dagger),
   \label{10_d}
   \eeq
where operators $c$ and ${\tilde c}$ describe physical and ancillary fermions, respectively
 \footnote{We follow~\cite{Lifshitz2006} and call $H'$ the Hamiltonian.}.

Now, although $H'$ in (\ref{10_d}) splits between  $c$ and ${\tilde c}$ fermions,  the two couple to the same phonon. This implies that if we  kept  $H'$ as is, there  would appear  an  effective phonon-mediated interaction between $c$ and ${\tilde c}$  fermions.  In the static limit, this effective interaction is
\beq
U_{eff} = - 2 \beta \omega_0 c^\dagger   c{\tilde c}^\dagger {\tilde c}.
\label{10_a}
\eeq
 To compensate for this interaction, we have to introduce the counter-term
 \beq
U_{c,{\tilde c}} =  2 \beta \omega_0 c^\dagger   c{\tilde c}^\dagger {\tilde c} = -2 c^\dagger   {\tilde c}{\tilde c}^\dagger  c.
\label{10_a_a}
\eeq
 The total Hamiltonian for the physical  fermion and its ancilla mirror image  is then
 \beq
 H = H' + U_{c,{\tilde c}}.
 \label{10_a_a_a}
\eeq
Taking the Hamiltonian in this form guarantees that there is no Hartree renormalization for either $c$  or ${\tilde c}$ fermion by a mixed  off-shot diagram (we recall that Hartree renormalization involving only $c$ or only ${\tilde c}$ fermions is already incorporated into $\mu$).  

The  next step is central to our analysis. We assume and then verify that we can lower the ground-state energy  by introducing a  composite $U(1)$
order parameter  (a condensate) 
\beq
\Delta = - 2 \beta \omega_0  \langle c^\dagger {\tilde c} \rangle
\label{10_b}
\eeq
 Using  it to decouple $U_{c,{\tilde c}}$ in (\ref{10_a_a}), we  obtain the effective $H'$ in the form
\bea
   H^{'}_{eff}  &=& \frac{|\Delta|^2}{2\beta \omega_0} -{\mu_n} \left(c^\dagger c - {\tilde c}^\dagger {\tilde c} \right) + \Delta c^\dagger {\tilde c} + \Delta^* {\tilde c}^\dagger c \nonumber \\
   && + \frac{g}{\sqrt{2\omega_0}} \left(c^\dagger c + {\tilde c}^\dagger {\tilde c}\right) (b + b^\dagger).
   \label{10}
   \eea
 Note that we flipped the sign of the $|\Delta|^2$ term in (\ref{10}) compared to $\langle U_{c,{\tilde c}}\rangle = - \frac{|\Delta|^2}{2\beta \omega_0}$.
    This is done to avoid triple counting in the calculation of $\langle H^{'}_{eff}\rangle $ 
   as $\langle U_{c,{\tilde c}}\rangle $ appears there three  times: directly and twice after averaging in the quadratic form. The same procedure has been used in the derivation of the effective Hamiltonian  for superconductivity (see e.g.,  ~\cite{Lifshitz2006,AGD} ). 

  It is convenient to spilt  the effective Hamiltonian into the quadratic and interaction parts:  $  H^{'}_{eff} =   H^{'}_{eff,2} +   H^{'}_{eff,e-ph}$,  where
 \bea
   H^{'}_{eff,2}  &=& \frac{|\Delta|^2}{2\beta \omega_0} -{\mu_n} \left(c^\dagger c - {\tilde c}^\dagger {\tilde c} \right) + \Delta c^\dagger {\tilde c} + \Delta^* {\tilde c}^\dagger c, \nonumber \\
  H^{'}_{eff,e-ph} &=&  \frac{g}{\sqrt{2\omega_0}} \left(c^\dagger c + {\tilde c}^\dagger {\tilde c}\right) (b + b^\dagger).
   \label{10_bb}
   \eea  
 The quadratic part 
  can be diagonalized by a Bogolyubov transformation.
 For $\Delta = |\Delta| e^{\phi}$ we take
   \bea
   c &=& \left(\cos{\theta}~ \alpha + \sin{\theta}~ {\tilde \alpha}\right) e^{i\phi/2} \nonumber \\
   {\tilde c} &=& \left(\cos{\theta}~ {\tilde \alpha} - \sin{\theta}~ \alpha \right) e^{-i\phi/2}.
   \eea
The Hamiltonian in terms of $\alpha$ and ${\tilde \alpha}$ becomes diagonal once we choose $\tan(2\theta) = |\Delta|/{\mu_n}$.  Using the sign convention, in which ${\text {sign}} \cos{2\theta} = - {\text {sign}}{\mu_n}$, we obtain
\beq
\cos{\theta} = \frac{1}{\sqrt{2}} \left(1 - \frac{\mu_n}{E}\right)^{1/2}, ~~\sin{\theta} = -\frac{1}{\sqrt{2}} \left(1 + \frac{\mu_n}{E}\right)^{1/2},
\eeq
and
\beq
   H^{'}_{eff,2} =   \frac{|\Delta|^2}{2\beta \omega_0} + E \left( \alpha^\dagger \alpha - {\tilde \alpha}^\dagger {\tilde \alpha} \right)
   \label{11_b}
   \eeq
   where $E =({\mu_n}^2_0 + |\Delta|^2)^{1/2} >0$.
    Applying this transformation to $H^{'}_{eff,e-ph}$, we  find that it has the same form in terms of $\alpha$ and ${\tilde \alpha}$ as in terms of $c$ and ${\tilde c}$:
   \beq
    H^{'}_{eff,e-ph} =  \frac{g}{\sqrt{2\omega_0}} \left(\alpha^\dagger \alpha + {\tilde \alpha}^\dagger {\tilde \alpha} \right) (b + b^\dagger). 
       \label{11_bb} 
    \eeq   
We see that  the {\it {full}}  
\beq
H^{'}_{eff} =  H^{'}_{\alpha} + H^{'}_{\tilde \alpha},
      \label{11_bbb} 
    \eeq   
i.e,  
the system decouples into two independent subsystems: in one, described by ${\tilde \alpha}$, all states are empty, and in the other, described by ${\alpha}$, all states are occupied.
In this respect,  ${\tilde \alpha}$ and $\alpha$ describe  realizations in which a given lattice site is filled or empty.
We emphasize that both $\alpha$ and ${\tilde \alpha}$ are linear combinations of the original and ancilla fermions. 

Let us momentarily neglect 
  $ H^{'}_{eff,e-ph}$ and    
  treat $\alpha$ and ${\tilde \alpha}$  as free fermions,  described by $H^{'}_{eff,2}$. 
  The value of $|\Delta|$ is determined from the self-consistent condition,  Eq. (\ref{10_b})
    \beq
     |\Delta| = - \beta \omega_0 \frac{|\Delta|}{E}  \left(\langle\alpha^\dagger \alpha\rangle - \langle{\tilde \alpha}^\dagger {\tilde \alpha} \rangle\right).
     \label{d_24_3}
     \eeq
      Canceling $|\Delta|$ on both sides and using $\langle\alpha^\dagger \alpha\rangle =0$,  $\langle{\tilde \alpha}^\dagger {\tilde \alpha} \rangle =1$, we obtain
     \beq
     E = \sqrt{{\mu_n}^2 + |\Delta|^2} = \beta \omega_0.
    \label{d_24_4}
     \eeq
 This yields
\beq
|\Delta| = 2 \beta \omega_0 (n(1-n))^{1/2},
\label{d_24_6}
     \eeq
the same as in (\ref{rr_15}).
The condition on the fermion density
\beq
\frac{1}{2} \left(1 + \frac{{\mu_n}}{E}\right)  =n
\label{d_24_5}
     \eeq
is satisfied, as it should.
The retarded Green's function of the physical $c-$fermion is
\bea
   G^{c} (\omega) &= &
    \frac{1}{2}\, \frac{1-\mu_n /E}{\omega - E + i\delta} +
    \frac{1}{2}\, \frac{1+\mu_n /E}{\omega + E + i\delta} \label{qq_7_2_5} \nonumber \\
    &=& \frac{1-n}{\omega -\beta \omega_0 + i \delta} + \frac{n}{\omega  + \beta \omega_0 +  i \delta}.
   \label{qq_7_2}
   \eea
It coincides with (\ref{k_9}).

We can also  use  the first line in Eq. (\ref{qq_7_2_5})  for the approximate $G^c (\omega)$ with $E = \sqrt{{\mu_n}^2 + |\Delta|^2}$
and
 obtain the  corresponding 
 grand potential  $\Omega ({\mu_n}, |\Delta|)$ with   ${\mu_n}$ and $|\Delta|$ as parameters. 
  The equilibrium values of ${\mu_n}$ and $|\Delta|$ are then obtained from
   $\partial\Omega/\partial {\mu_n} =-n $ and
  $\partial\Omega/\partial {|\Delta|} =0$.
   The potential  $\Omega ({\mu_n}, |\Delta|)$  is expressed via the time-ordered $G^c$ (the one with $i\delta {\text sign} \omega$ instead of $i\delta$)  as
   ~\cite{Lifshitz2006,AGD}
  \bea
\Omega ({\mu_n}, |\Delta|)&=& i \int^{\mu_n} d \mu^* \int \frac{d\omega}{2\pi}  \lim_{t \to -0} G^c (\omega, \mu^*, |\Delta|) e^{-i\omega t} \nonumber \\
 &&+ \Delta^2/(4\beta \omega_0).
  \label{qq_7_2_2}
   \eea
   The last term is half of the term $|\Delta|^2/(2\beta \omega_0)$ in (\ref{10}), which has to be split between the physical and the ancillary fermions.  The contribution from the lower end of the integral over ${\mu_n}$ has to be chosen to satisfy
   the boundary condition $\Omega ({\mu_n}, |\Delta|) =0$ at $n=0$.
 The factor $e^{-i\omega t}$ in (\ref{qq_7_2_2}) requires that the integral over $\omega$ be taken over the upper half-plane of frequency and selects the pole in $G^c(\omega, \mu^*, |\Delta|)$ at $\omega = -E= - \sqrt{(\mu^*)^2 + |\Delta|^2}$.  Evaluating the frequency integral explicitly, we  obtain
     \beq
\Omega ({\mu_n}, |\Delta|)= -\frac{1}{2} \left ({\mu_n} + \sqrt{{\mu_n}^2 + |\Delta|^2}\right) + \Delta^2/(4\beta \omega_0).
    \label{qq_7_3}
   \eeq

Evaluating the derivatives over ${\mu_n}$ and ${|\Delta|}$ we obtain the set of two coupled equations
 \bea
 &&1 + \frac{{\mu_n}}{\sqrt{{\mu_n}^2 + |\Delta|^2}} =2n \nonumber \\
 && \sqrt{{\mu_n}^2 + |\Delta|^2} = \beta \omega_0.
  \label{qq_7_4}
 \eea
Solving them,  we obtain ${\mu_n} = (2n-1) \beta \omega_0$ and $|\Delta| = 2 \beta \omega_0 (n (1-n))^{1/2}$,  the same as we obtained before.  Substituting these ${\mu_n}$ and $|\Delta|$ into (\ref{qq_7_3}), we find the equilibrium value $\Omega^{eq} (n) = - \beta \omega_0 n^2$. 
Evaluating then $E_n= \Omega^{eq} (n)  + {\mu_n} n$,
  we obtain
\beq
E_n = -\beta \omega_0 n (1-n) = -\beta \omega_0 n (1-n).
 \label{qq_7_5}
\eeq
The relation $d E _n /dn = {\mu_n} (n) = -\beta \omega_0 (1-2n)$ is satisfied, as it should be.

 Eq. (\ref{qq_7_5}) could also be obtained by  directly  evaluating $\langle H^{'}_{eff} \rangle + \mu_n (n - (1-n))$  with equilibrium $\mu_n $ and $\Delta$ and taking half of the result as the energy of $c$ fermions.   Computing it this way, we see that the two contributions  containing $\mu_n$  ( the direct one and the one in $\langle H^{'}_{eff} \rangle$) cancel out, as they indeed should.  This cancellation also eliminates the contribution to the energy from  the Hartree term, as we treated it as the self-energy correction to $\mu$.  We then have to add    $E_{\text {Hartree}} =-\beta \omega_0 n^2$.  to the ground-state energy.  With this addition, the full 
 energy of the polaron state becomes
\beq
E_P (n)  = -\beta \omega_0 n. 
 \label{qq_7_6}
\eeq  

 \begin{figure}[]
 \includegraphics[width=0.4\textwidth]{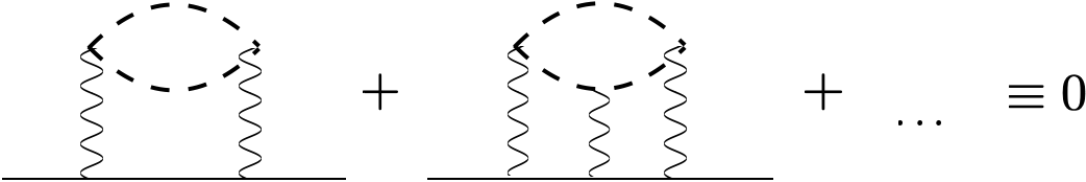}
    ~\\
    ~\\
    ~\\
 \includegraphics[width=0.4\textwidth]{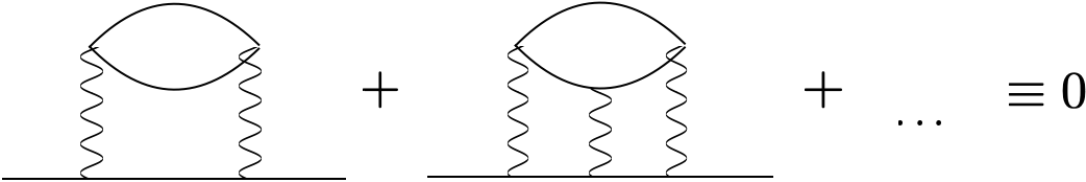}
 \caption{Potential contributions to $\alpha$ fermion self-energy from $\tilde\alpha$ fermions (top) and $\alpha$ fermions (bottom). Solid (dashed) lines correspond to $\alpha$ ($\tilde\alpha$) fermions. In both cases, the corrections vanish. }
 \label{fig:aab_diag}
 \end{figure}
We now include 
 $H^{'}_{eff,e-ph}$ in (\ref{11_bb}). Because  the fermionic energy $E = \beta \omega_0$,   the full  $H^{'}_{\alpha}$  and $H^{'}_{{\tilde \alpha}}$  in (\ref{11_bbb}) 
are  the same as  the electron-phonon Hamiltonian at $n=0$ and  $n=1$, respectively.
   We verified that
 the self-energy for the $\alpha$ fermion is fully determined by $H^{'}_{\alpha}$ without an input from $H^{'}_{{\tilde \alpha}}$ and vise versa (the only potential  contributions to the self-energy for an $\alpha$  fermion from
  the processes involving  ${\tilde \alpha}$ fermions  come from the diagrams shown in 
   Fig.~\ref{fig:aab_diag}.
  These diagrams contain loops made out of  ${\tilde \alpha}$ fermions and vanish after integrating over the running frequency in the loop because  the poles of $G^{{\tilde \alpha}} (\omega)$ are in upper half-plane of frequency and the integration contour can be closed in the lower half-plane (for the same reason the diagrams with a loop made of 
   $\alpha$-fermions also vanish).  Using the results from the previous two sections, we then immediately find that
 the full spectral functions $ A^{\alpha} (\omega)$ and $A^{{\tilde \alpha}} (\omega)$
 are the same as the eikonal spectral functions at $n=0$ and $n=1$, respectively, i.e., each contains the set of $\delta$-functional peaks at positive (negative) $\omega = p \omega_0$.  The full Green's functions are
  \bea
   G^{\alpha}_{full} (\omega) &=& G^{e} (\omega, n=0) = \nonumber \\
   \frac{1}{\omega + i\delta}~ {_1}F{_1} (1, 1 - \frac{\omega + i\delta}{\omega_0}, - \beta), \nonumber \\
   G^{{\tilde \alpha}}_{full} (\omega) &=& G^{e} (\omega, n=1) = \nonumber \\
    \frac{1}{\omega + i\delta}~ {_1}F{_1} (1, 1 + \frac{\omega + i\delta}{\omega_0}, - \beta), \nonumber
     \eea
      where  ${_1}F{_1} (a,b,c)$ is the Kummer confluent hypergeometric function.
The full retarded Green's function for the original (physical) $c$ fermion is 
 \beq
   G^{c}_{full}  (\omega) =  (1-n) G^{\alpha} (\omega) + n G^{{\tilde \alpha}} (\omega).
  \label{qq_7}
   \eeq
this Green's function is  identical to
the retarded $G^{H} (\omega, n)$ given by Eq. (\ref{n_1}).   
It contains  $\delta$-functional peaks at $\omega = l \omega_0$, $l = 0, \pm 1, \pm 2...$, each with additional residue $Z_l <1$ ($Z_l \sim e^{-l}$ at small $l$)  instead  
 of two free-fermion peaks  $\omega = \pm \beta \omega_0$  in $G^c (\omega)$ in (\ref{qq_7_2}).   
 We see therefore  that the exact Green's function  of the Holstein model
 for $0<n <1$ is fully reproduced in the 
  diagrammatic analysis,  however, one has to introduce an
ancilla fermion and a $U(1)$ particle-hole condensate order parameter $\Delta$  consisting of a physical and an ancilla fermion.
made couple it  bi-linearly to the original fermion by introducing the $U(1)$ order parameter $\Delta$.  

We conclude this section with a brief discussion of the accuracy of  approximating the full energy of the polaron state by 
the free-fermion $E_P (n)$  from (\ref{qq_7_6}).  As we  found in the previous two sections,  $\delta-$functional peaks at frequencies  $\omega \approx \pm \beta \omega_0$ are dense in the sense that the distance between the neighboring peaks  $\delta \omega = \omega_0$ is   small  in $1/\beta$  compared to 
 the envelope 
 peak positions at $\omega \approx \pm \beta \omega_0$.    These peaks are then 
    well  approximated by the  two gaussians (one at $\omega >0$, another at $\omega <0$), each with the width $\sqrt{\beta} \omega_0$ and can be 
     further approximated  by free-fermion peaks at $\omega = \pm \beta \omega_0$ because the width of a gaussian is parameterically smaller than the position of its center.  For these peaks, free-fermion approximation of $G^c_{full} (\omega)$ by free-fermion $G^c (\omega)$ and the approximation of the ground-state energy  by the free-fermion $E_P (n)$   is valid up to corrections, which are small in $1/\beta$.   The gaussian approximation misses  discrete peaks  at  small $\omega = l \omega_0$.  However,  the residues of these peaks $Z_l\sim e^{-\beta} \beta^|l|$  are exponentially small, and we expect that these peaks will only give exponentially small corrections to the free-fermion $E_P (n)$. 
   
 \subsubsection{\bf {Finite bandwidth.  A homogeneous polaron order.}}
\label{arb_n_arb_W}

At a finite $W$, we add the dispersion $\epsilon_{\mathbf k}$
to the original and ancillary fermions. The dispersion  comes with the same sign for both fermions.
The Hamiltonian $H'$ in terms of physical and ancillary fermions is $H' = H'_2 + H'_{e-ph}$, where
\beq
   H'_2 =  \sum_{\mathbf k}  \left(\epsilon_{{\mathbf k}} -{\mu_n}\right) c^\dagger_{\mathbf k} c_{\mathbf k} + \left(\epsilon_{{\mathbf k}} +{\mu_n}\right)  {\tilde c}^\dagger_{\mathbf k} {\tilde c}_{\mathbf k}
\eeq
     and
   \beq
   H'_{e-ph} =  \frac{g}{\sqrt{2N\omega_0}}
   \sum_{{\mathbf k},{\mathbf q}} \left(c_{\mathbf k}^\dagger c_{{\mathbf k}+{\mathbf q}} + {\tilde c}^\dagger_{\mathbf k} {\tilde c}_{{\mathbf k}+{\mathbf q}} \right) (b^\dagger_{\mathbf q} + b_{-{\mathbf q}}).
   \label{qq_9_b}
   \eeq
 We keep the notation $\mu_n$ for the chemical potential, but keep in mind that $\mu_n$ has to be obtained from the condition on the fermionic density and may by itself become the function of $\lambda_p$.

The counter term at finite $W$ has the form
\beq
U_{c,{\tilde c}} = -2 \beta \omega_0 \frac{1}{N} \sum_{{\mathbf k},{\mathbf p}, {\mathbf q}}
c_{\mathbf k}^\dagger {\tilde c}_{{\mathbf k}+{\mathbf q}} {\tilde c}^\dagger_{{\mathbf p}} c_{{\mathbf p}-
{\mathbf q}},
\label{zzz_30}
\eeq
  We again decouple this term  by introducing the $U(1)$  condensate $\Delta_{\mathbf q} \propto \sum_{\mathbf k} \langle c^\dagger_{\mathbf k} {\tilde c}_{{\mathbf k}+{\mathbf q}}\rangle$  with some momentum ${\mathbf q}$.
In this section, we assume that  $\Delta$ is uniform, i.e., the polaron state is spatially homogeneous. We have 
\beq
\Delta_{q=0} = \Delta_0 = -2\beta \omega_0 \frac{1}{N} \sum_{{\mathbf k}} \langle c_{\mathbf k}^\dagger {\tilde c}_{{\mathbf k}}\rangle.
\eeq
Decoupling $U_{eff}$ as we did at $W=0$, we obtain the effective Hamiltonian
 $H'_{eff} = H'_{eff,2} + H'_{eff, e-ph}$,  where
  \bea
   H'_{eff,2} &=&  \sum_{\mathbf k}  \left(\epsilon_{{\mathbf k}} -{\mu_n}\right) c^\dagger_{\mathbf k} c_{\mathbf k} + \left(\epsilon_{{\mathbf k}} +{\mu_n}\right) {\tilde c}^\dagger_{\mathbf k} {\tilde c}_{\mathbf k}  \nonumber \\
    &+& \sum_{\mathbf k} \left[\Delta_0 c^\dagger_{\mathbf k} {\tilde c}_{\mathbf k} + \Delta_0^* {\tilde c}^\dagger_{\mathbf k} c_{\mathbf k} + \frac{|\Delta_0|^2}{2\beta \omega_0}\right],
    \label{qq_9_a}
    \eea
     and
   \beq
   H'_{eff, e-ph} =  \frac{g}{\sqrt{2N\omega_0}}
   \sum_{{\mathbf k},{\mathbf q}} \left(c_{\mathbf k}^\dagger c_{{\mathbf k}+{\mathbf q}} + {\tilde c}^\dagger_{\mathbf k} {\tilde c}_{{\mathbf k}+{\mathbf q}} \right) (b^\dagger_{\mathbf q} + b_{-{\mathbf q}}).
   \eeq
  The last term is the same as  the original $H'_{e-ph}$.
   Diagonalizing the quadratic part of the Hamiltonian
    we obtain
     \beq
   H'_{eff,2} = \sum_{\mathbf k} \left[\left(\epsilon_{\mathbf k} + E\right) \alpha^\dagger_{\mathbf k} \alpha_{\mathbf k}  + \left(\epsilon_{\mathbf k} - E\right)
    {\tilde \alpha}^\dagger_{\mathbf k} {\tilde \alpha}_{\mathbf k} + \frac{|\Delta_0|^2}{2\beta \omega_0}\right],
 \label{qq_10_a}
 \eeq
  where
  \beq
  E =  \sqrt{{\mu_n}^2 + |\Delta_0|^2}
 \label{qq_10_b}
  \eeq
  and
 \beq
  H'_{eff,e-ph} =  \frac{g}{\sqrt{2N\omega_0}}
   \sum_{{\mathbf k},{\mathbf q}} \left(\alpha_{\mathbf k}^\dagger \alpha_{{\mathbf k}+{\mathbf q}} + {\tilde \alpha}^\dagger_{\mathbf k} {\tilde \alpha}_{{\mathbf k}+{\mathbf q}} \right) (b^\dagger_{\mathbf q} + b_{-{\mathbf q}}).
   \label{qq_9_b_1}
   \eeq
   Note that $E$ is independent of the momentum.
  The Green's function of the physical $c$ fermion is a weighted sum of the Green's functions of the $\alpha$ and ${\tilde \alpha}$ fermions:
   \beq
   G^{c} (\omega, \epsilon_{\mathbf k}) =    \frac{E-{\mu_n}}{2E}\, G^{\alpha} (\omega, \epsilon_{\mathbf k}) +
   \frac{E+{\mu_n}}{2E}\,
    G^{{\tilde \alpha}} (\omega,\epsilon_{\mathbf k}).
  \label{qq_7_1_a_1}
   \eeq
 Keeping only the contributions from $H^{'}_{eff,2}$, we obtain
\beq
   G^{c}  (\omega, \epsilon_{\mathbf k}) = \frac 12 \frac{1 - \mu_n/E}{\omega - (\epsilon_{\mathbf k} + E )} + \frac 12  \frac{1 + \mu_n /E}{\omega - (\epsilon_{\mathbf k} - E  )}.
  \label{qq_7_1_a_2}
   \eeq
The condition that the fermionic density is $n$ is 
\beq
\frac{{\mu_n}}{E} \left[\frac{1}{N} \sum_{\mathbf k} \left( \langle\alpha^\dagger_{\mathbf k} \alpha_{\mathbf k}\rangle- \langle{\tilde \alpha}^\dagger_{\mathbf k} {\tilde \alpha}_{\mathbf k}\rangle \right)\right]  =1-2n,
\label{d_24_5_1}
     \eeq
and the self-consistent equation on $|\Delta_0| \neq 0$ is
   \beq
      1= - \frac{\beta \omega_0}{E} \frac{1}{N} \left[ \sum_{\mathbf k} \left(\langle\alpha^\dagger_{\mathbf k} \alpha_{\mathbf k}\rangle -
      \langle{\tilde \alpha}^\dagger_{\mathbf k} {\tilde \alpha}_{\mathbf k} \rangle\right)\right].
     \label{d_24_3_1}
     \eeq
    The solution of these two equations  is different for $\lambda_p >1/2$ and $\lambda_p <1/2$. 
     We consider these two cases separately.\\
     
     \begin{centering}
     {\it {The case $\lambda_p >1/2$.}} \\
     \end{centering}
     \vspace{\baselineskip}
     We assume and then verify that for $\lambda_p >1/2$, $E > W/2$. Then $\langle\alpha^\dagger_{\mathbf k}\alpha_{\mathbf k}\rangle =0$ and $\langle{\tilde \alpha}^\dagger_{\mathbf k}{\tilde \alpha}_{\mathbf k}\rangle=1$.  Using this in  (\ref{d_24_5_1}) and  (\ref{d_24_3_1}),
     we find that $\mu_n$ retains its bare value  ${\mu_n} = \beta \omega_0 (2n-1)$,  and $E = \beta \omega_0$ is also the same as at $W=0$.
     We see that $E > W/2$, as we assumed.  Extracting  $|\Delta_0|$ from $E$,  we find
  \beq
|\Delta_0| = 2 (\beta \omega_0) (n(1-n))^{1/2}, ~~
\label{d_24_6_1}
     \eeq
 the same  at $W=0$.

 \begin{figure}[t]
 \includegraphics[]{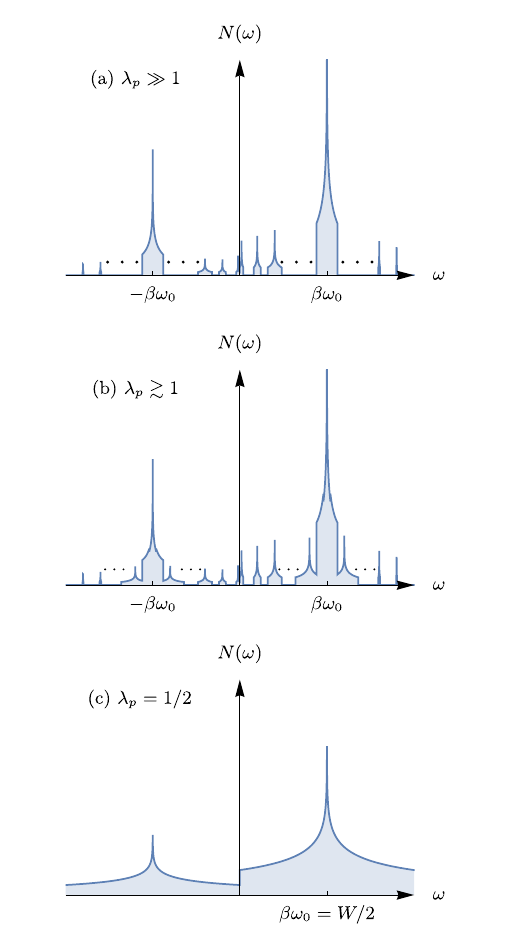}
 \caption{Schematic electron DOS in 2D in the regime $W > \sqrt \beta \omega_0$, for a generic density $n<1/2$. The width of the peak at $\omega = \beta \omega_0$ is set by $W$; the width of the peaks nears $\omega = 0$ is exponentially small $\sim We^{-\beta}$. Panels (a)-(c) show evolution with increasing $W$,
   toward $\beta \omega_0/2$ (i.e., toward $\lambda_p =1/2$) . As $W$ increases,  the lower  ends of the two continua move to smaller $\omega$, absorbing polaron patches one by one.}
 \label{fig:schematic_dos_finiteW}
 \end{figure}
 
\begin{figure}[t]
 \includegraphics[]{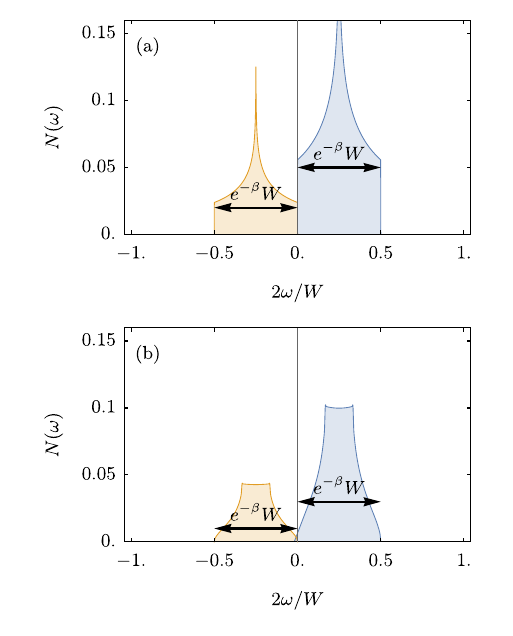}
 \caption{Schematic form of the DOS   inside the patch centered at $\omega =0$ in 2D (upper panel) and 3D (lower panel), for a homogeneous condensate $\Delta_0$.   In both cases, the DOS consists of two  components, which do not couple to each other.  At $\omega >0$, the DOS  is the same as for $n=0$,  and at $\omega <0$ is the same as for $n=1$.   Each component of  the  DOS 
  vanishes strictly at  $\omega =0$, hence the DOS at zero frequency is equal to zero.   This implies that for a  
homogeneous $\Delta_0$, there is no Fermi surface in the polaron state.    in 2D, the DOS vanishes at $\omega =0$, but discontinuously jumps to a finite value  at $\omega \neq 0$. In 3D, the DOS gradually  approaches zero at $\omega \to 0$.   Features of both DOS at finite $\omega$ do not play any role in our analysis. }
 \label{fig:new_patches}
 \end{figure}

   We now consider the full $H^{'}_{eff}$. Because the cubic term $H^{'}_{eff,e-ph}$ does not contain the mixed terms in the form $\alpha^\dagger {\tilde \alpha}$ and  ${\tilde \alpha}^\dagger \alpha$,  the full $H^{'}_{eff}$  is decoupled into  $H'_{eff}  = H'_{\alpha} + H'_{\tilde \alpha}$,
     where $H'_{\alpha}$ is the same as at $n=0$ and
  $H'_{{\tilde \alpha}}$ is the same as  at $n=1$. Using the results of the previous two Sections, we then find the DOS consisting  of two independent pieces, one at positive and one at negative $\omega$.  For $\omega_0 \sqrt{\beta} < W < \omega_0 \beta$, the  DOS is a combination of a free-particle continuum  with width $W$, centered at $\omega = \pm \beta \omega_0$, and exponentially narrow patches of polaron DOS,  centered at $\omega = \pm n \omega_0$ (Fig. \ref{fig:schematic_dos_finiteW} a).  
   The patches centered at $\omega = m \omega_0 $ and $m >0$ are,  to first approximation,   miniature  versions of the DOS for free fermions, but for a much smaller width.    The patch centered at $\omega=0$ consists of two parts: the part at $\omega >0$ comes solely from $\alpha$ fermions and the part at $\omega <0$ comes solely from ${\tilde \alpha}$ fermions.  There is no mixture between the two because the full Hamiltonian    completely decouples between $\alpha$ and ${\tilde \alpha}$ fermions.   DOS  of $\alpha$ fermions is the  same as at $n=0$  and the DOS of ${\tilde \alpha}$ fermions is the same as at $n=1$, both vanish at  $\omega =0$.  As a result, the DOS in this  patch   vanishes at $\omega =0$.  This last result implies that there  is no Fermi surface in the polaron state. 
 We illustrate this in Fig. \ref{fig:new_patches}.

As $W$ increases ($\lambda_p$ decreases), the two continua get wider,  and their lower ends move to smaller $\omega$, absorbing polaron patches one by one (Fig. \ref{fig:schematic_dos_finiteW} b).
  The last polaron gets absorbed at $\lambda_p =1/2$, independent on the value of  $n$.   At this $\lambda_p$, the DOS consists of two continua touching each other at $\omega =0$ (Fig. \ref{fig:schematic_dos_finiteW} c).
  We emphasize that this DOS is still very different from that  of the FL.  
  Namely, the DOS has two peaks (two van Hove singularities), and  the total width of the range where it is non-zero is $2W$ as opposed to $W$ in an FL.  

The grand  potential and the ground-state energy for
  for $\lambda_p >1/2$ can be  obtained  in the same way as before -- as the contribution from the free-fermion  continua.  Contributions from the patches of heavy polarons at low $\omega$   are relatively small in $e^{-\beta}$.

  To obtain the grand potential  $\Omega ({\mu_n}, \Delta_0)$ for $c-$ fermions, 
  we again use the relation between   $\Omega$ and the time-ordered Green's function $G^c (\omega, \epsilon_k)$,  Eq. (\ref{qq_7_2_2}). For non-zero $W$, the grand potential per unit area is 
  \begin{widetext}
   \beq
 \Omega ({\mu_n}, |\Delta_0|)= i \int^{\mu_n} d \mu^* \int_{-W/2}^{W/2} d\epsilon N_0 (\epsilon)  \int  \frac{d\omega}{2\pi}
 \lim_{t \to -0} G^c  (\omega, \epsilon, \mu^*, |\Delta_0|) e^{-i\omega t} + |\Delta_0|^2/(4\beta \omega_0),
  \label{qq_7_2_2_1}
   \eeq
  \end{widetext}
   where $\epsilon = \epsilon_{\mathbf k}$ and $N(\epsilon)$, subject to  $\int_{-W/2}^{W/2} d\epsilon N(\epsilon) =1$,  is the free-fermion DOS for the tight-binding dispersion, see Eq.  (\ref{tt_3}). 
   Using (\ref{qq_7_1_a_2}) for $G^c$ and noticing that the location of the poles in the complex plane of frequency is not affected by $\epsilon$ as long as $\lambda_p >1/2$, we find that $\Omega ({\mu_n}, |\Delta_0|)$ retains the same form,
   Eq. (\ref{qq_7_3}), as at $W=0$:
      \beq
 \Omega({\mu_n}, |\Delta_0|)= -\frac{1}{2} \left ({\mu_n} + \sqrt{{\mu_n}^2 + |\Delta_0|^2}\right) + \frac{\Delta_0^2}{4\beta \omega_0}.
    \label{qq_7_3_1}
   \eeq
 Conditions $\partial \Omega/\partial {\mu_n} = -n$ and
$\partial \Omega/\partial {|\Delta_0|} = 0$ then produce the same equations as
(\ref{d_24_5_1}) and (\ref{d_24_3_1}), whose solutions are $\mu_n = \beta \omega_0 (2n-1)$ and $|\Delta_0| = 2 \beta \omega_0 \sqrt{n(1-n)}$.  Substituting these ${\mu_n}$ and $|\Delta_0|$ into (\ref{qq_7_3_1}), we find the equilibrium value $\Omega (n) = - \beta \omega_0 n^2$.    
 The corresponding energy is 
\beq
E (n) = -\beta \omega_0 n (1-n) = -\beta \omega_0 n  + \beta \omega_0 n^2.
 \label{qq_7_5_1}
\eeq
Adding the Hartree  term $\beta \omega_0 n^2$,  we obtain the full  ground-state energy of the homogeneous polaron state, which we label $E_{P,0} (n)$ (the subindex $0$ stands for homogeneous).  We have
\beq
E_{P,0} (n) = -\beta \omega_0 n.
 \label{qq_7_5_2}
\eeq
 We emphasize that this energy does not depend on $W$
  and from this perspective the full energy  of the homogeneous polaron state is the potential energy.
 The independence on $W$ is in line with the general argument that we presented in Sec. \ref{sec:Numerics}, where we considered a variational state with localized polarons and argued that its energy is exactly $-\beta \omega_0 n$ because the absence of fluctuations of fermionic density at a given site  implies that there are no contributions to the energy from fermionic hopping.  For ${\bf q} =0$, we showed above that after diagonalization,  the
 effective Hamiltonian at a finite $W$ completely decouples between the filled and empty states.  This implies that there are no fluctuations of fermionic density at a given site, which explains Eq. (\ref{qq_7_5_2}).

Next, we show that at $\lambda_p =1/2$ and $n >0$, $E_{P,0}$  is
  smaller than the energy  of an FL state described by MET. The latter is 
      $E_{FL} = E_{{\text{kin}},FL} - n^2 \beta \omega_0$, where
      the first term is the kinetic energy of a Fermi gas with density $n$ and the second is the Hartree term. 
      For tight-binding dispersion,
      \beq
    E_{ {\text{kin}},FL} = - \frac{W}{\pi^2} \left(E(1-{\hat \mu}^2_{FL}) - {\hat \mu}^2_{FL} K(1-{\hat \mu}^2_{FG})\right).
     \label{zz_1}
     \eeq
       Here $E (...)$ is an elliptic integral of the second kind and ${\hat \mu}_{FL} = 2 \mu_{FL} /W$. The relation between ${\hat \mu}_{FL}$ and  density $n$ of a Fermi gas is 
      \beq
      n = \frac{2}{\pi^2} \int_{-1}^{{\hat \mu}_{FL}} dy K(1-y^2).
      \label{zz_2}
      \eeq
      We compare $E_{FL}$ and $E_{P,0}$ at small $n$,  where we can get the analytical expression for $E_{FL}$.  Indeed, at  small $n$, ${\hat \mu}_{FL} \approx -1 + \pi n
      - \pi^2 n^2/4 + O(n^3)$. Substituting into (\ref{zz_1}) and expanding in $n$,  we find 
       $E_{{\text{kin}},FL} = - Wn/2 (1 - \pi n/2)$.
      Adding the  Hartree term and
       expressing  $E_{FL}$ in units of $\beta \omega_0$, we obtain
       \beq
       E_{FL} = - \beta \omega_0 \left(n^2 + \frac{n}{2 \lambda_p} \left(1- \frac{\pi n}{2}\right) \right).
       \label{eq:e_fl}
      \eeq
      At $\lambda_p =1/2$, $E_{FL} = E_{P,0} (n) + \beta \omega_0 n^2 (\pi-2)/\pi$. 
      At $n = 0$, the two energies coincide, as we found earlier, but at a finite $n$, $E_{FL} > E_{P,0} (n)$, i.e., 
the polaron state at $\lambda_p =1/2$ is energetically favorable  than a  FL one.\\

	 \begin{figure}[]
	 \includegraphics[]{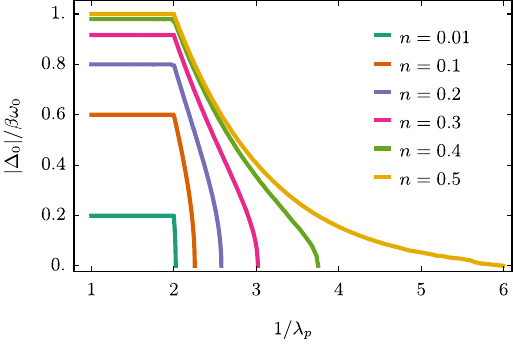}
	 \caption{The condensate order parameter $|\Delta_0|$ vs $1/\lambda_p$.  For $1/\lambda_p <2$ ($\lambda_p >1/2$), $|\Delta_0| = 2\beta \omega_0 (n (1-n))^{1/2}$ is independent on $\lambda_p$. At smaller $\lambda_p$, $|\Delta_0|$ decreases and eventually vanishes at $\lambda_p = \lambda^{c,1}_p$ (see text). }
	 \label{fig:delta_vs_lambda}
	 \end{figure}

      \begin{centering}
     {\it {The case $\lambda_p <1/2$.}} \\
     \end{centering}
\vspace{\baselineskip}

      We now move to $\lambda_p <1/2$.   The self-consistency equation on $\Delta$ is still given by
      (\ref{d_24_3_1}), but now  $(1/N) \sum_{\mathbf k}\langle\alpha^\dagger_{\mathbf k}\alpha_{\mathbf k}\rangle$  and $(1/N) \sum_{\mathbf k}\langle{\tilde \alpha}^\dagger_{\mathbf k}{\tilde \alpha}_{\mathbf k}\rangle$ are both non-zero.  Evaluating the averages, we obtain
      \beq
      1 = \frac{\beta \omega_0}{E}
      \left(\int_{-W/2}^{E}
      N_0(\epsilon) d \epsilon - \int_{E}
      ^{W/2} N(\epsilon) d \epsilon \right).
      \label{d_25_1}
      \eeq
      The condition on the  fermionic density becomes
      \beq
       n= \frac{1}{2} \left (1 + \frac{{\mu_n}}{E}
       \left(\int_{-W/2}^{E}
       N_0(\epsilon) d \epsilon - \int_{E}
       ^{W/2} N_0(\epsilon) d \epsilon \right)\right).
      \label{d_25_2}
      \eeq
      Solving (\ref{d_25_1}) and (\ref{d_25_2}), we find that ${\mu_n} = \beta \omega_0 (2n-1)$   does not change, but $|\Delta_0|$ and $E = \sqrt{{\mu_n}^2 + |\Delta_0|^2}$ differ from their expressions 
 for $\lambda_p >1/2$. For the tight-binding dispersion with $N(\epsilon)$ given by
      (\ref{tt_3}), we define ${\hat E} = 2E/W$, express $|\Delta_0|$ in terms of ${\hat E}$  as 
      \beq
      |\Delta_0|^2 = \frac{W^2}{4} \left({\hat E}^2 - 4 \lambda^2_p (2n-1)^2\right),
      \label{d_25_3}
      \eeq
      and obtain the self-consistent  equation on ${\hat E}$ in the form
      \beq
      1 -\frac{\hat E}{2\lambda_p}  = \frac{4}{\pi^2} \int_{\hat E}^1 K(1-x^2) dx
      \label{d_25_4}.
      \eeq
      Solving this equation and substituting the solution  into (\ref{d_25_3}), we obtain $|\Delta_0|$ as a function of $\lambda_p$. We plot this function in Fig.~\ref{fig:delta_vs_lambda}.
	 
      We see that $|\Delta_0|$ decreases with decreasing $\lambda_p$ and vanishes at a density-dependent  value $\lambda^{c,1}_p$.  For small $n$, $\lambda^{c,1}_p =0.5 (1 - (\pi -2) n)$, for  $n$ near $1/2$, $\lambda^{c,1}_p \approx (8/\pi^2) 1/|\log{(1-2n)}|)$.

	\begin{figure*}[]
	 \includegraphics[]{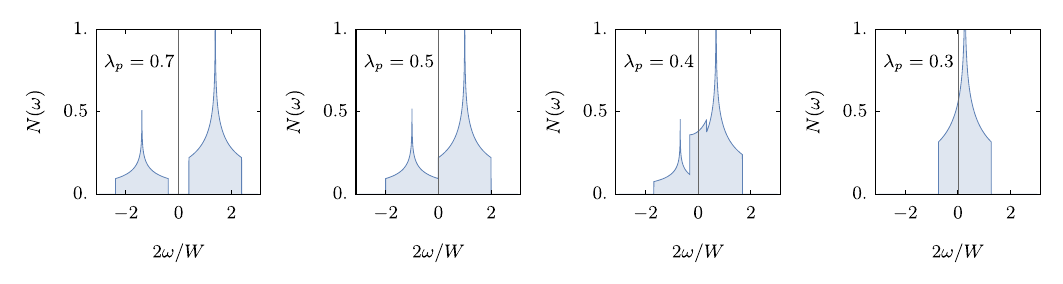}
	 \caption{Evolution of the DOS (in units of $2/W$) between $\lambda_p >1/2$ and $\lambda_p <1/2$ for density $n =0.15$,
  for which $\lambda^{c,1}_p \approx 0.3$.  
In the mixed phase at $0.3 < \lambda_p <1/2$,   the DOS displays  a pseudogap behavior (two peaks at a finite $\omega$ and a non-zero DOS between the peaks). }
	 \label{fig:dos_vs_lambda}
	 \end{figure*}
	 
      The DOS also evolves with $\lambda_p$.
       At $\lambda_p \leq 1/2$, it  displays  two peaks 
        at $\omega = \pm E$ and is non-zero between the peaks (see Fig.~\ref{fig:dos_vs_lambda}).
        The profile of the DOS evolves due to the evolution of $E$ and of the coherence factors $(1/2) (1 \pm {\mu_n}/\sqrt{{\mu_n}^2+ |\Delta_0|^2})$, 
         but as long as $\Delta_0$ is non-zero, $N(\omega)$ has two peaks (van Hove singularities) at finite frequencies 
         $\omega = \pm E$ and is non-zero between the peaks.      
         The structure of this DOS is similar to that in the pseudogap (PG) phase in the cuprates, and  by analogy
        we  label  this intermediate state as the PG state.  It does not possess a set of polaron patches, which are absorbed by the two continua, but is still qualitatively different from the DOS in an FL.  
       Finally, at $\lambda_p = \lambda^{c,1}_p$, the residue of  the one of the two peaks in $G^c (\omega, \epsilon_k)$ in (\ref{qq_7_1_a_2}) 
       vanishes,  the Green's function and the DOS  become the same as in an FL, and   the chemical potential ${\mu_n}$ coincides with
       $\mu_{FL}$, introduced in (\ref{zz_1}), (\ref{zz_2}).
        At smaller $\lambda_p$, $|\Delta_0| =0$, the ancilla fermion does not play a role, and the system is in the FL state, described by MET.

       We next compute the grand potential and the ground-state energy for the PG state.
        We use the same relation between $\Omega (\mu_n, |\Delta_0|) $ and the integral of $G^{c} (\omega, \epsilon_{\mathbf k})$,  as in Eq. (\ref{qq_7_2_2_1}), but now set the lower limit of integration over $\mu^*$   to match the grand potential in an FL in the limit  $|\Delta_0| \to 0$.  Using (\ref{qq_7_1_a_2}) for
       $G^{c} (\omega)$,  we obtain the grand potential in the pseudogap state,  $\Omega ({\mu_n}, |\Delta_0|) = \Omega_{PG} $, in the form 
        \begin{widetext}
       \bea
        && \Omega_{PG}  = \frac{|\Delta_0|^2}{4\beta \omega_0} - \frac{{\mu_n}}{2} \label{d_25_5}  \\
         &&- \frac{\sqrt{{\mu_n}^2 + |\Delta_0|^2}}{2} \left(\int_{-W/2}^{\sqrt{{\mu_n}^2 + |\Delta_0|^2}}
       N_0 (\epsilon) d \epsilon - \int_{\sqrt{{\mu_n}^2 + |\Delta_0|^2}}
       ^{W/2} N_0 (\epsilon) d \epsilon \right) + \frac{1}{2} \left(\int_{-W/2}^{\sqrt{{\mu_n}^2 + |\Delta_0|^2}}
       \epsilon N_0 (\epsilon) d \epsilon - \int_{\sqrt{{\mu_n}^2 + |\Delta_0|^2}}
       ^{W/2} \epsilon N_0 (\epsilon) d \epsilon \right) \nonumber
      \eea
      \end{widetext}
     We explicitly  verified that conditions $\partial \Omega_{PG}/\partial {\mu_n} =-n$
       and $\partial \Omega_{PG}/\partial {|\Delta_0|} =0$
       yield the same set of equations for ${\mu_n}$ and $|\Delta_0|$ as (\ref{d_24_5_1}) and (\ref{d_24_3_1}).
          Substituting the equilibrium values of ${\mu_n}$ and $|\Delta_0|$ into (\ref{d_25_5}) and adding
         ${\mu_n} n$ and the Hartree term  $- n^2 \beta \omega_0$, we obtain the full $E_{PG} (n, \lambda_p) $. 
         We verified that $\partial E_{PG}/\partial n = -\beta \omega_0$, is as it should be.  

         For $\lambda_{p} \leq 1/2$, we obtained after some algebra
         \beq
         E_{PG} = -\beta \omega_0 \left ( n + \frac{(1-2\lambda_p)^2}{2(\pi-2)}\right)
         \label{d_31_1}
         \eeq
          For a generic $\lambda_p$ we obtained $E_{PG}$  numerically and verified that it can be cast into
          $E_{PG} = -\beta \omega_0 n  - \beta \omega_0 \Psi (\lambda_p)$, where the last term does not depend on $n$ and does not affect the form of ${\mu_n}$.
         We also verified that at $\lambda_p = \lambda^{c,1}_p$,  the grand potential (\ref{d_25_5})  coincides with its value in an FL: 
        \beq
      \Omega_{PG} ({\mu_n},0) = -\mu_{FL} n + \int_{-W/2}^{{\mu_n}_{FL}}  \epsilon N_0 (\epsilon) d \epsilon 
       \label{d_25_6}
      \eeq
      
      	\begin{figure}[]
	 \includegraphics[]{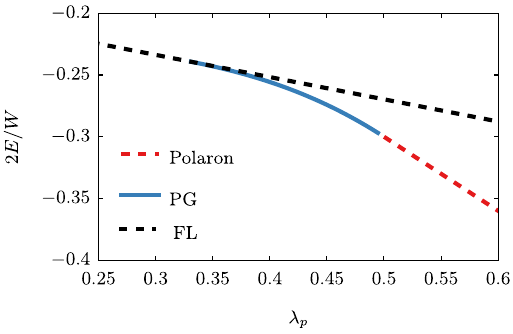}
	 \caption{ground-state energy as a function of $\lambda_p$ for density $n=0.3$. We show energies of the three different states: the polaron state, the pseudogap (PG) state, and the Fermi-liquid (FL).}
	 \label{fig:e_vs_lambda}
	 \end{figure}
      
     We plot $E_{PG} (n, \lambda_p)$ as a function of $\lambda_p$ for a given $n$  in Fig. \ref{fig:e_vs_lambda}
     along with the full energy of a  FL, $E_{FL} (n)$.   We see that  $E_{PG}
     (n, \lambda_p)$ is smaller than $E_{FL} (n)$ as long as $|\Delta_0|$  is non-zero, and the two energies coincide at $\lambda_p = \lambda^{c,1}_p$. \\
 
\begin{centering} 
{\it{Pseudogap state as a mixed state.}}\\
\end{centering}
\vspace{\baselineskip}

We now argue that the PG state  can be  described as a  mixed state,
      in which a portion of a system, with density $n_1 = n -\delta$,  remains  in a  polaron state,    while the rest of fermions,  with
         density $n-n_1 = \delta$,
        is in an FL state. 
         The full energy of such a state is $E_M (\delta) = E_{P,0} (\delta)  + E_{FL} (\delta)$ ($M$ stands for mixed),  where
         \beq
         E_{P,0} (\delta)   = -\beta \omega_0 (n-\delta)
         \eeq
          and $ E_{FL} (\delta) = E_{{\text{kin}},FL} (\delta) - \beta \omega_0 \delta^2$, where the last term is the Hartree energy of an FL. 
          The kinetic energy of an FL with density $\delta$ is  
          \beq
          E_{{\text{kin}},FL} (\delta) =   \frac{W}{\pi^2} \int_{-1}^{{\hat \mu}_{FL}} x K(1-x^2) dx
         \label{tt_4}
          \eeq
           and ${\hat \mu}_{FL} = 2 \mu_{FL}/W$ is related to density $\delta$ by
           \beq
           \delta = \frac{2}{\pi^2} \int_{-1}^{{\hat \mu}_{FL}} K(1-x^2) dx.
           \label{tt_5}
           \eeq
           To the leading order in $\delta$,
           \beq
           E_{FL} (\delta) = -\frac{W}{2} \delta.
           \label{zzz_20}
           \eeq  
            The value of $\delta$ is determined from $d E_M (\delta)/d\delta =0$.  Evaluating the derivative using 
            $d E_{{\text{kin}},FL}/d\delta = (E_{{\text{kin}},FL}/d {\hat \mu}_{FL})/(d \delta/d{\hat \mu}_{FL})$, 
            we obtain the equation
            \beq
            2 \lambda_p (2 \delta-1) = {\hat \mu}_{FL}.
            \label{tt_6}
            \eeq
             Substituting this ${\hat \mu}_{FL}$ into  (\ref{tt_5}), we 
   obtain the  closed-form equation for ${\hat \mu}_{FL}$:
        \beq
         1 + \frac{{\hat \mu}_{FL}}{2 \lambda_p} = \frac{4}{\pi^2} \int_{-1}^{{\hat \mu}_{FL}} K(1-x^2) dx.
         \label{tt_9}
       \eeq
Comparing Eqs. (\ref{tt_9}) and (\ref{d_25_4}), we see that 
\beq 
{\hat \mu}_{FL} = - {\hat E}, ~~{\text{i.e.}} \mu_{FL} = -E. 
\label{pppp}
\eeq
We also note that  ${\hat \mu}_{FL}$, and, hence, $E$   depend on $\lambda_p$, but not on $n$.  Eq. (\ref{tt_6}) then shows that $\delta$ also  does not depend on $n$. 

 At $\lambda_p \leq 1/2$, we obtain 
          \beq
    {\hat \mu}_{FL} = -1 + \pi \delta +O(\delta^2),~~ \delta = \frac{1-2\lambda_p}{\pi -2} + O\left((1-2 \lambda_p)^2\right).
          \label{d_31_3}
      \eeq    
 The energy of the mixed state $E_M (\delta)$ to order $\delta^2$ is 
             \beq
        E_M (\delta)  = \beta \omega_0 \left(-n  - \delta (1-2\lambda_p) + \delta^2 \frac{\pi-2}{2}\right).
        \label{d_31_2}
      \eeq
      With $\delta$ from (\ref{d_31_3}), it becomes    
         \beq
        E_M   = -\beta \omega_0 \left(n + \frac{(1-2\lambda_p)^2}{2(\pi-2)}\right).
        \label{d_31_4}
      \eeq
      Comparing Eq. (\ref{d_31_1}), we see that to this order,   $E_M   = E_{PG}$. 
We verified numerically that $E_M \equiv E_{PG}$  everywhere in the mixed state, i.e., for all $\lambda_p$ in  the interval $\lambda^{c,1}_p < \lambda_p < 1/2$.   

As a further evidence that the PG state can be viewed as a mixed state, we show that the Maxwell construction is satisfied, i.e., the full chemical potentials of the polaron and FL states coincide.    The full chemical potentials, which we define as $\mu^{full}_P$ and $\mu^{full}_{FL}$, are  the derivatives over the density of the  full energies $E_{P,0}$ and $E_{FL}$.  We have  $\mu^\text{full}_{P} = \partial E_{P,0} (n_1)/\partial n_1  =  - \beta \omega_0$ and 
  $\mu^\text{full}_{FL} (\delta) = \partial E_{FL}/\partial \delta = 
  =  (W/2) {\hat \mu}_{FL} -2\beta \omega_0 \delta = -\beta \omega_0 (1-2\delta) - 2\beta \omega_0 \delta  \equiv -\beta \omega_0$, i.e., is  the same as  $\mu^\text{full}_P$.   We emphasize that only the full chemical potentials  are equal, the chemical potentials $\mu_P$ and $\mu_{FL}$, obtained by differentiating the energies of the polaron and FL states without the corresponding Hartree terms , are not equal. 
  Namely, 
 \beq
~~\mu_{FL} = -\beta \omega_0 (1-2\delta),  ~~\mu_{P} =  -\beta \omega_0 (1- 2(n-\delta)).
 \label{pp_11}
 \eeq
  We discuss the relation between the full and ``kinetic" chemical potentials in more detail in Appendix \ref{app_F}.
 
   The relation  $E = - \mu_{FL}$,  Eq. (\ref{pppp})  is highly relevant  for the interpretation of the pseudogap state as a mixed state  as it shows that one fermionic  propagator in  $G^c (\omega, \epsilon_k)$ in  Eq. (\ref{qq_7_1_a_2}) has the same structure as in a Fermi liquid with density $\delta$.
       This allows  us to reexpress  $G^c (\omega, \epsilon_k)$  as
  \begin{widetext}
       \bea
  && G^c (\omega, \epsilon_k) = \frac{1- n -\delta}{(1-2\delta)} \frac{1}{\omega - E - \epsilon_k} + \frac{n -\delta}{(1-2\delta)} \frac{1}{\omega + E - \epsilon_k} =  \nonumber \\
&& \frac{1}{\omega + \mu_{FL} - \epsilon_k}  + \frac{n -\delta}{(1-2\delta)} \left( \frac{1}{\omega + E - \epsilon_k} - \frac{1}{\omega - E - \epsilon_k} \right)  =  G^{FL} (\omega, \epsilon_k) + 
    G^P (\omega, \epsilon_k) .
   \label{tt_9_1}
   \eea
    \end{widetext}
    The first term in the second line is the propagator of an FL with density $\delta$. 
   The second term is the sum of the contributions  from the polaron bands centered at $\pm E = \pm \beta \omega_0 (1-2\delta)$, both weighted with the factor $(n-\delta)/(1-2\delta)$. 
 The contributions from $G^{FL}$  and $G^P$ to the total density $n = \int_{-\infty}^0 N(\omega) d \omega$ are $\delta$ from $G^{FL}$ and  $(n-\delta)/(1-2\delta) \times  [(1-\delta) - \delta] = n-\delta$ from  $G^P$.
  precisely as we suggested.  
   We also see that Im $\int d \epsilon N_0 (\epsilon) G^P (\omega, \epsilon)$ 
   vanishes at $\omega =0$, like it does for $\lambda_p > 1/2$.
   A nonzero  spectral weight at $\omega =0$  is then entirely due to the FL component. 
 
     \begin{figure}[]
	 \includegraphics[]{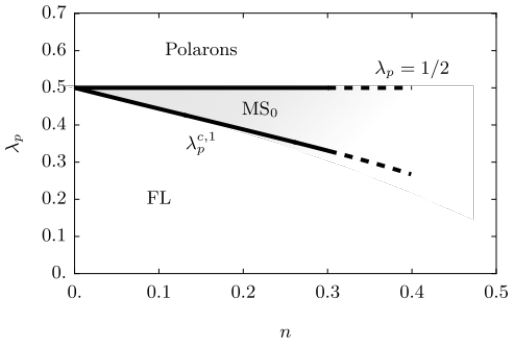}
	 \caption{Analytic phase diagram of the ancilla model for a homogeneous polaron state for small densities.
      The phase diagram contains pure Fermi-liquid (FL) and polaron states and the  the mixed FL-polaron state, labeled as  MS$_0$.}
	 \label{fig:phase_diag_ms0}
	 \end{figure}
As a summary of this Section, in  Fig.~\ref{fig:phase_diag_ms0}
we show the evolution of the system behavior with decreasing $\lambda_p$ at  $n$ far enough from half-filling. 
 At large $\lambda_p >1/2$, the  system is in the polaron state. The DOS in this state consists of narrow patches of heavy polarons at small frequencies and two continua at larger frequencies. At small $\lambda_p < \lambda^{c,1}_p$, the system is in an FL state. The DOS in this phase is a continuum
with width $W$ and a peak at $\omega  = - \mu_{FL}$.  At  intermediate $\lambda_p$ between $1/2$ and $\lambda^{c,1}_p$, the system is in the mixed state, labeled MS$_0$. Here,  a portion of a system with density $n-\delta$ is in the polaron state  and the other portion with density $\delta$ is in an FL state.   The DOS in this phase
 is a continuum with the width $W (1 +2 \lambda_p)$,  the shape of the continuum varies with $\lambda$ and a portion of the continuum disappears at $\lambda_p = \lambda^{c,1}_p$.   The DOS at $\omega =0$ is small near the upper edge of the mixed state at $\lambda_p =1/2$ and grows as $\lambda_p$ decreases toward $\lambda^{c,1}_p$.   The mixed state most likely separates into spatial regions of the polaron and FL states, but to see this we need to go beyond our mean-field description.\\

     \begin{figure}[]
	 \includegraphics[]{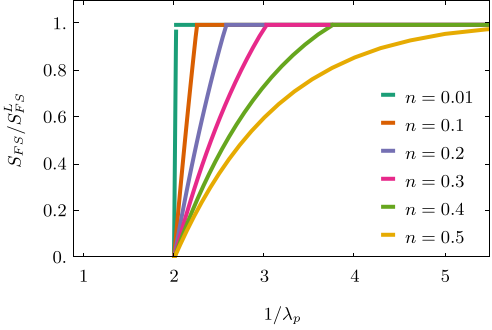}
	 \caption{Variation of the area of the Fermi surface with $\lambda_p$ in the mixed phase for several values of  the density $n$.}
	 \label{fig:Lutt}
	 \end{figure}
\begin{centering} 
 {\it{Fermi surface in the mixed state and Luttinger theorem}}\\ 
\end{centering}
\vspace{\baselineskip}

In an FL, the area of an electron Fermi surface, $S_{FS}$, is related to the fermionic density by the Luttinger theorem. In our case of spinless fermions, the relation is simply $S_{FS} = n$. In the mixed state (or equivalently, the pseudogap state), the density of the Fermi liquid component is $\delta <n$. In this situation, we expect that the  conventional relation  becomes invalid and the modified one becomes
\beq
S_{FS} = \delta,
\label{mm_1}
\eeq
where $\delta$ is given by (\ref{tt_5}).   We now  derive this expression staring from the relation between the density and 
the frequency integral of the Green's function $G^c (\epsilon, \omega)$
\beq
n = - \frac{i}{N}  \sum_{\mathbf{k}}    \lim_{t \to -0} 
\int_{-\infty}^\infty  \frac{d\omega}{2\pi}  G^c (\omega,\mathbf{k}) e^{-i\omega t}.
 \label{mm_2}
\eeq
This relation is the extension of Eq. (\ref{rr_a}) to an arbitrary Green's function.
 The Green's function of the physical fermion, $G^c (\omega, \epsilon_k)$ is given by 
 Eq. \eqref{qq_7_1_a_2}.  Substituting  it into Eq.~\eqref{mm_2}, setting $\lambda_p <1/2$, and using 
 the fact that $E = - \mu_{FL} -\beta \omega_0 (1-2\delta)$ and 
 the  relations 
 \beq
 S_{FS} = \int_{-W/2}^{\mu_{FL}} N(\omega) d \omega, 
\label{mm_3}
\eeq
we obtain
\bea
n &=& \frac{1}{2} \left(1 - \frac{\beta \omega_0 (1-2n)}{E} \right) + \frac{\beta \omega_0 (1-2n)}{E}  S_{FS}  \nonumber \\
&=&  \frac{n-\delta}{1-2\delta} + \frac{1-2n}{1-2\delta} S_{FS}.
\label{mm_4}
\eea
Solving for $S_{FS}$, we reproduce Eq. (\ref{mm_1}).    The area of  the Fermi surface varies within the mixed state from $S_{FS} =0$ at the upper edge, $\lambda_p =1/2$, where $\delta =0$   to $S_{FS} =n$ at the lower edge, $\lambda_p = \lambda^{c,1}_p$, where $\delta =n$.  We show the evolution of $S_{FS}$ with $\lambda_p$ in Fig.   \ref{fig:Lutt} for several  $n$.

 We believe  that  Luttinger theorem is broken in the mixed (pseudogap) phase because a portion of fermions with density $n_1 = n - \delta$  
  form localized  polaron  bound states.   Localization of these electrons breaks translational symmetry even if on average the system remains translationally invariant.   In this respect, 
    there is a certain similarity between our mixed phase and FL$^*$ phase proposed in spin-liquid (fractionalized FL) theory for cuprates and other strongly correlated materials~\cite{Bonetti_2025,Senthil_2003,Sachdev_2010}.  In both cases,  strong interactions transform an ordinary FL into a new state,  which  has coherent FL quasiparticles (conduction electrons) with a smaller FS along with localized excitations without a Fermi surface. These localized excitations are polarons in our case and a spin liquid of local moments in FL$^*$. 

 A word of caution.  In our analysis we did not include fluctuations of the U(1) order $\Delta_0$.  If these fluctuations somehow destroy the long-range order and produce a finite correlation length $\xi$ for the propagator of  the fluctuating field  $\Delta_0$,   Luttinger theorem would  not  be broken,  i.e., 
at a vanishing $\omega$ ,  the
  spectral function $A(k, \omega)$ would be  peaked at $k$ corresponding to the original Fermi surface whose area  is the fermionic density $n$.  Yet,  above a characteristic frequency $\omega* \propto \xi^{-1}$, the
  spectral function $A(k, \omega)$   would  develop strong peaks at momenta  $k$, which  are Fermi momenta  in an FL with density $\delta$ 
  ~\cite{Chubukov1995,*Vilk1997_a}. 

We also note that the Green's function of a physical fermion  has both poles and zero both in the mixed state and in the polaron state. Indeed, one can reexpress $G^c (\omega, \epsilon_k)$ in Eq. \eqref{qq_7_1_a_2} as
\beq
G^c (\omega, \epsilon_k) = \frac{\omega - \epsilon_k +E (1-2n)}{(\omega - \epsilon_k -E) (\omega - \epsilon_k +E)},
\label{mm_5}
\eeq
which has  poles at $\omega = \epsilon_k \pm E$ and zero at $\omega = \epsilon_k - E (1-2n)$. 
 The relation of zeros of the Green's function  to   Luttinger count attracted a substantial interest in recent years~ \cite{Altshuler1997,Georges2001,Dzyaloshinskii2003,Konik2006,*Stanescu2006,Rosch2007,Kiran2013,Sachdev_book,
 Fabrizio2022,*Skolimowski2022,*Blason_2023,*Staffieri_2025,Scheurer2018,*Peralta2023,*Bollmann2024,*Markov2026,Lehmann_2025,*Stepanov_2024,Bonetti_2025,Philips2025}).  We discuss the role of the zero of the Green's function  for our case in Appendix \ref{app:10}. \\

     \begin{figure}[]
	 \includegraphics[]{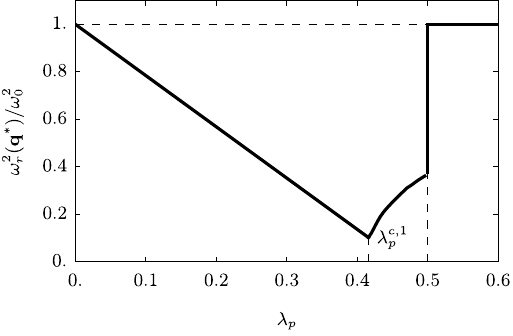}
	 \caption{Dressed phonon frequency $\omega_r ({\bf q})$ vs $\lambda_p$ at 
%AC
$n =0.15$
and ${\bf q} = {\bf q}^*$, where it has a minimum value. 
 In the polaron state,  phonon frequency is not renormalized, $\omega_r ({\bf q}) = \omega_0$.   In an FL,  $\omega_r ({\bf q}^*)$ progressively decreases 
  with increasing $\lambda_p$. In the mixed phase,    $\omega_r ({\bf q}^*)$ interpolates between the two pure phase.  It gradually increases with $\lambda_p$ inside the mixed phase, while the value of ${\bf q}^*$ decreases,  and jumps by a finite amount at the  upper edge due to discontinuity of the DOS of  2D fermions with tight-binding dispersion. }
  \label{fig:Debye}
	 \end{figure}

\begin{centering} 
{\it{Evolution of the Debye frequency in the mixed state}}\\ 
\end{centering}
\vspace{\baselineskip}

Another interesting aspect of the mixed state is  the evolution of the Debye frequency. In an FL,  fermion polarization renormalizes $\omega_0$ down to dressed 
 $\omega_r ({\bf q})$ (Eq. (\ref{or})),  which also acquires a  dispersion. The minimal value of $\omega_r ({\bf q})$ is at some density-dependent $q^* (n)$.  
 In the polaron state,  Debye frequency is not renormalized and retains its bare value $\omega_0$.  In the mixed phase, propagators of both  $\alpha$ and ${\tilde \alpha}$  fermions  have poles in either lower or upper half-plane of  frequency, depending of the value of $\epsilon_k$.   Then the static polarizations $\Pi_{st} ({\bf q}$),  made of either two $-$-fermions or two ${\tilde c}$ fermions, are non-zero, and $\omega_0$ gets renormalized into $\omega_r ({\bf q})$, with the strongest effect at $q^* (\delta)$.   The same  renormalization  is obtained  by keeping only the Fermi liquid component in the mixed phase. 
   Strictly at the  upper edge of the mixed state, $\omega_r ({\bf q}) = \omega_0$. Immediately  below the upper edge, at $\lambda_p = 1/2 -0$,  the dressed   $\omega_r (0)$ jumps down to   $\omega_0 (1-2/\pi)^{1/2}$ due to discontinuity of the DOS for tight binding dispersion $N(\omega)$ at the lower limit $\omega = -W/2$.  For other $q$,   $\omega_r ({\bf q})$ remains infinitesimally close to $\omega_0$.
   At smaller $\lambda_p$,  the momentum width, where the renormalization of $\omega_0$ is $O(1)$, increases, and ${\bf q}^*$ (the position of the minimum of 
   $\omega_r ({\bf q})$) also increases.  At the lower edge  of the mixed phase, $\omega_r ({\bf q})$  becomes the same as in a Fermi liquid.   We obtained 
   $\omega_r ({\bf q}^*)$ numerically and show the result in Fig. \ref{fig:Debye} . \\
 
\begin{centering} 
{\it{Interplay with CDW electronic order.}}\\
\end{centering}
\vspace{\baselineskip}

We next show that the phase diagram in Fig.~\ref{fig:phase_diag_ms0} 
is valid for $n< n_c$, where $n_c$ is some density smaller than $1/2$, while at larger $n$ there appear two new phases, in which an FL  state has a CDW order.  
   The   rational  for  these now phases comes from the comparison of the density dependence of  $\lambda^{c,1}_p$, above which an FL state becomes unstable toward polarons,  and the line, which we label  $\lambda^{c2}_p$, along which the dressed phonon frequency softens at some momentum $q^*$, signaling a CDW instability.
 We consider the structure and energy of the electron states in the presence of a CDW order later in Sec. \ref{CDW_el} and  here just identify 
 $\lambda^{c,2}_p$.
   
\begin{figure}[htbp]
	\includegraphics[width=0.8\linewidth]{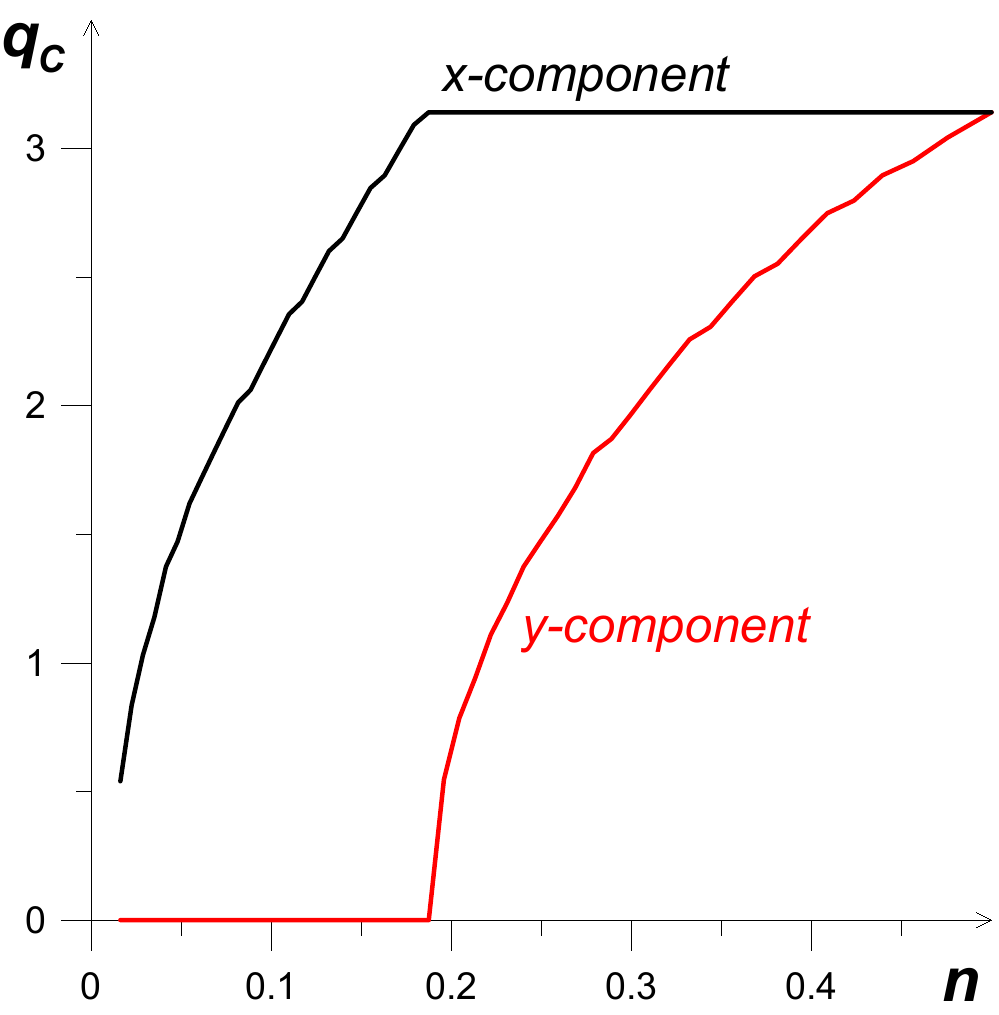}
\caption{
One of the symmetry related momentum points at which the phonon frequency softens to zero
in a 2D system with a dispersionless bare phonon spectrum $\omega_0 (q)/t =0.1$. }
\label{fig15}
\end{figure}
\begin{figure}[htbp]
	\includegraphics[width=0.8\linewidth]{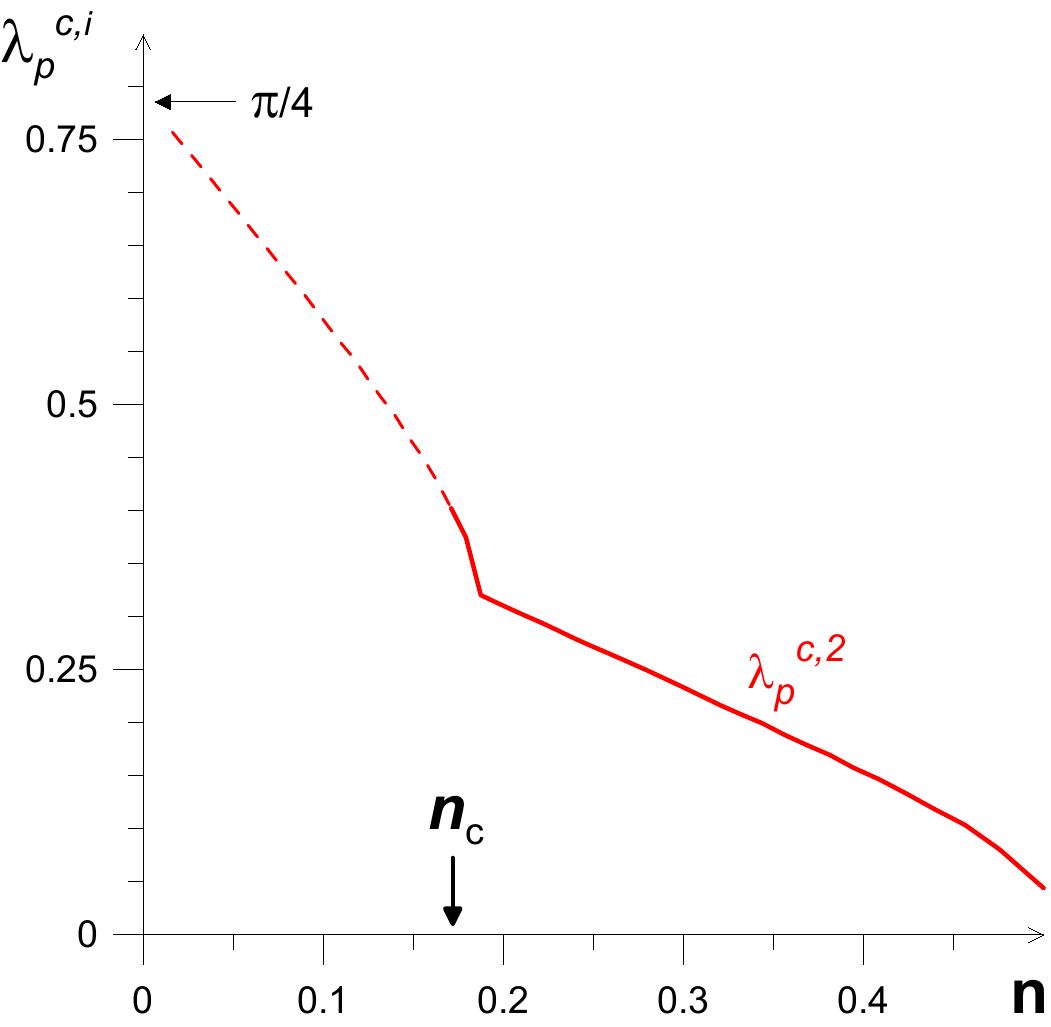}
\caption{
Critical value $\lambda^{c,2}_p$ of the electron-phonon coupling $\lambda_p$  for a CDW electronic instability 
 in a 2D system with a dispersionless bare phonon spectrum $\omega_0 (q)/t =0.1$.
The dependence of $\lambda^{c,2}_p$ on the electronic density $n$ is the same as of $\lambda_0^{c,2}$ 
  in the middle panel of the lowest row in Fig.~\ref{all2D}
   as 
  the couplings  $\lambda_0^{c,2}$ and 
  $\lambda^{c,2}_p$ differ only by a numerical factor.
}
\label{fig16}
\end{figure}

In  our analytical model with $q-$independent bare $\omega_0$ and spinless fermions, 
 the dressed $\omega_r ({\bf q}^*)$  softens  when $2 \lambda_p W \Pi_{st} (q^*) =1$,
(see Eq. (\ref{or})).
At small $n$,  we found earlier that $\lambda^{c,1}_p \leq 1/2$ is smaller than $\lambda^{c2} \approx \pi/4$,
i.e. the polaron state develops before the FL state becomes unstable toward CDW.  However,  near half-filling,  the situation is different  Here, the polarization bubble is the largest for ${\bf q}^* \approx {\mathbf Q} = (\pi,\pi)$ as $\Pi_{st} ({\mathbf Q})$ diverges as $(1/W) (\log |1/2-n|)^2$. 
 Accordingly, $\lambda^{c,2}_p$ decreases as 
 $1/(\log |1/2-n|)^2$.
The coupling  $\lambda^{c,1}_p$ also decreases  near half-filling,  but only as  $1/(\log |1/2-n|)$.
As a result, near $n =1/2$, $\lambda^{c2}_p < \lambda^{c,1}_p$, 
  and 
  an FL  becomes unstable toward CDW before the polaron state develops. The complete analysis of the density dependence of $\lambda^{c,2}_p$ is somewhat 
   involved
   because 
${\bf q}^*$
 depends on $n$.  At small $n$, the leading instability is at ${\bf q}^*= (q_c,0)$ or $(0, q_c)$, where $q_c$ increases with $n$.  At $n = n_0 \approx 0.18$, $q_c$ reaches $\pi$, and at larger $n$,
   ${\bf q}^* = (\pi, q_c)$ or $q_c, \pi)$.  At $n \to 1/2$, $q_c$ approaches $\pi$ and the CDW order develops with momentum ${\mathbf Q}$.   We show the density variation of 
   ${\bf q}^*$ in Fig.~\ref{fig15}.  In Fig.~\ref{fig16}
we plot $\lambda^{c,2}_p$ as a function of $n$, which we obtained using  Fig.~\ref{fig15} as input.
% In  Fig.~\ref{fig16}b  we plot  both $\lambda^{c,1}_p$ and $\lambda^{c,2}_p$. We see that the two critical lines 
%AC
Comparing this line with $\lambda^{c,1}_p$, we see that at small $n$, $\lambda^{c,2}_p > \lambda^{c,1}_p$, i.e., 
 the mixed state develops when the system is still stable against CDW order. In the mixed state, the dressed Debye frequency 
  renormalizes back to its bare value, which makes the system even more stable against CDW.  
 The lines $\lambda^{c,2}_p$  and $\lambda^{c,1}_p$ cross at $n_c \sim 0.17$ (see Fig. \ref{fig:phase_diag_ms1}). 
 At this $n$, $\lambda^{c,2}_p = \lambda^{c,1}_p \approx 0.40$.
At larger $n$, $\lambda^{c,2}_p < \lambda^{c,1}_p$, i.e., the CDW order develops before
the order parameter $\Delta_0$ becomes non-zero.

This observation poses a question  where is the  onset of  a CDW order  within the mixed state
 We argue that  a  CDW  electronic order holds in the lower portion of the mixed state and disappears  at the coupling, which  equals  $\lambda^{c,2}_p$, evaluated at the density $\delta (\lambda_p)$.
     Because $\delta$  does not depend on $n$,  see Eqs.
 \eqref{tt_5}  and \eqref{tt_9},  this coupling,  which we keep labeling $\lambda^{c,2}_p$,   also does not depend on $n$ and  has the same value 
  $\lambda_p \approx 0.4$ as at $n=n_c$.   For the same reason, the  electronic density along this line is $\delta_c = n_c \approx 0.17$.  
    The outcome of this consideration is the  reentrant behavior of the $\lambda^{c,2}_p$ line  at $n >n_c$. Namely, 
     CDW order emerges upon crossing from below the lower $\lambda^{c,2}_p$ line 
      in Fig. \ref{fig:phase_diag_ms1} 
      and holds up the upper $\lambda^{c,2}_p = 0.4$., above which it disappears.    For couplings in between the lower $\lambda^{c,2}_p$ and $\lambda^{c,1}_p$, the system is in a pure CDW state,  for couplings in between $\lambda^{c,1}_p$ and the upper $\lambda^{c,2}_p \approx 0.4$, the system is in the mixed    state,  in which an FL component has a CDW order. 
      At larger $0.4 \lambda_p <0.5$, the FL component in the mixed state is not ordered, and at even larger $\lambda_p$ the system is in the  pure CDW state. 
  
      We show  the full analytical phase diagram for a homogeneous polaron order  $\Delta_0$ in  the left panel of Fig. \ref{fig:phase_diag_ms1}. 
        The notations MS$_1$ and MS$_0$   refer to mixed  states with and without CDW order of an FL component.
     For $n <n_c$ only   MS$_0$   state is present.    In this figure,  we   show  $\lambda^{c,1}_p$ at $n >n_c$  as  it  would be without a CDW order. 
        In reality,  $\lambda^{c,1}_p$ gets modified.  This modification, however, is not crucial and we neglect it .
    
 We note in passing that  the emergence of a CDW electronic order in the mixed phase below $\lambda_p \approx 0.4$  can be also understood using our earlier result about the downward renormalization of the phonon frequency $\omega_0$ into $\omega_r ({\bf q})$.  
  For $n >n_c$, the dressed phonon frequency then softens at 
  $\lambda_p =0.4$ at some ${\bf q}^*$.  

\subsubsection{\bf {Finite bandwidth.  CDW polaron order}}
\label{CDW_pol}

\begin{centering} 
{\it{Qualitative reasoning.}}\\
\end{centering}
\vspace{\baselineskip}

In the analysis earlier in this Section  we assumed that the polaron condensate at $0<n<1$  is  homogeneous.   We now analyze a
  potential emergence of a 
    density-wave polaron state,  in which the order parameter $\Delta (q)$  has a finite momentum ${\bf q}$.  Such a state emerges if its energy is smaller than that of the  homogeneous polaron state. The  ground state energies of the polaron states with different ${\bf q}$ may differ for two reasons. First, there is a ${\bf q}$-dependent difference in the  kinetic energy
of polarons, which can move throughout the lattice coherently by simultaneously rearranging the
oscillator displacements. This kinetic energy is exponentially small, of order $\sim W e^{-\beta}$,
yet it may overcome an effective repulsion between fermions and at low density favors a homogeneous ${\bf q} =0$ order of  polarons.
 Second, there is an energy variation with ${\bf q}$ due to
 virtual fermion delocalization relative to the strong oscillator displacement. In leading order in $W^2$, it comes from 
the hopping of fermions to the nearest-neighbor sites \textit{without} rearranging
the phonon displacement field. At large $\lambda_p$
this energy variation can be studied perturbatively starting from a localized polaron state, i.e.,
the one  with an FLat  $\Delta_{\bf q}$.
A fermion can fluctuate to the nearest-neighbor site with a matrix element $t$ by leaving behind a large oscillator
displacement. In the virtual state, the energy increases by
 $2 \beta \omega_0$ ($+\beta \omega_0$ compared to the original $-\beta \omega_0$) 
 Assuming that all nearest-neighbor sites
on the square lattice are not occupied, the second-order correction to the energy of  a localized polaron due to these processes is
\be
\delta E_p^{(2)} = - 4\frac{t^2}{2\beta\omega_0} \equiv - \frac{W}{32\lambda_p},
\label{fluc1}
\ee
(recall that $W=8t$). This contribution to polaron energy is not exponential in $\beta$, and for large $\beta$ is the leading contribution
that determines the structure of the ground state. 
If a nearest-neighbor  site is occupied by another polaron (the case relevant for densities near half-filling),
a hopping to this site should be excluded from
$\delta E_p^{(2)}$. The corresponding change of energy is positive
\be
V =  \frac{W}{64 \lambda_p}.
\label{repulsion}
\ee
This $V$ is the repulsive nearest neighbor interaction.
Near half-filling, when there is a high probability that nearest-neighbor sites are occupied, this nearest-neighbor repulsion favors 
the checkerboard polaron state (a CDW state with momentum $Q$).  \\  

\begin{centering} 
{\it{ Mean-field analysis of a polaron order 
with  ${\mathbf q} \neq 0$.  }}\\
\end{centering}
\vspace{\baselineskip}

We follow the  same strategy  as in the earlier analysis of a  homogeneous order  and introduce a particle-hole condensate made out of $c$ and ${\tilde c}$  fermions, but this time with a finite momentum ${\mathbf  q}$.  We do not assume in this Section that ${\mathbf q} = {\bf Q} = (\pi,\pi)$ -- an order $\Delta_{\mathbf q}$ with this  momentum  will appear as the  limiting case in our consideration.   We assume that there is a single order parameter $\Delta_{\mathbf q}$  rather than two with  ${\mathbf q}$ and $-{\mathbf q}$.  In this situation, we 
can  shift the momentum of ancilla fermions by ${\mathbf q}$  (${\tilde c}_{{\bf k}+{\mathbf q}}  \to {\tilde c}_{\bf k}$),  factorize the Hamiltonian for a given $k$ and 
not worry about potential non-anticommutation between $c_{{\mathbf k}-{\mathbf q}}$  and ${\tilde c}_{\mathbf  k} $ because operators for the physical and   ancilla fermions anticommute for any momentum.  
In these notations, the order parameter is   $\Delta_{\mathbf q} = -2\beta \omega_0 (1/N) \sum_{\mathbf k} \langle c^\dagger_{\mathbf k}{\tilde c}_{k}\rangle$.   
Shifting further ${\bf k} \to {\bf k} - {\mathbf q}/2$, we obtain, as before,  the effective Hamiltonian $H^{'}_{eff} = H^{'}_{eff,2} + H^{'}_{eff,e-ph}$, where $H^{'}_{eff,e-ph}$ is still given by  Eq. (\ref{qq_9_b}), but $H^{'}_{eff,2}$ is different:
  \bea
   H^{'}_{eff,2} &=&  \sum_{\mathbf k} \left(\epsilon^{(-)}_{\mathbf k} -{\mu_n}\right) \left(c^\dagger_{\mathbf k}c_{\mathbf k}- {\tilde c}^\dagger_{\mathbf k}{\tilde c}_{\mathbf k}\right)  +  \epsilon^{(+)}_{\mathbf k}  \left(c^\dagger_{\mathbf k}c_{\mathbf k}+ {\tilde c}^\dagger_{\mathbf k}{\tilde c}_{\mathbf k}\right)   \nonumber \\
    &&+  \sum_{\mathbf k}\left[\left(\Delta_{\mathbf q} c^\dagger_{\mathbf k}{\tilde c}_{\mathbf k}+ \Delta^*_{\mathbf q} {\tilde c}^\dagger_{\mathbf k}c_{\mathbf k}\right)+ \frac{|\Delta_{\mathbf q}|^2}{2\beta\omega_0}\right]
    \label{zzz_1}
    \eea
 where  $\epsilon^{(+)}_{\mathbf k} = (\epsilon_{\mathbf k + \mathbf q/2} + \epsilon_{\mathbf k - \mathbf q/2})/2$  and $\epsilon^{(-)}_{\mathbf k} = (\epsilon_{\mathbf k + \mathbf q/2}- \epsilon_{\mathbf k - \mathbf q/2})/2$.   For tight-binding dispersion
 \bea
 \epsilon^{(+)}_{\mathbf k} &=& -\frac{W}{4} \left(\cos{k_x} \cos{q_x/2} + \cos{k_y} \cos{q_y/2} \right) \nonumber \\
\epsilon^{(-)}_{\mathbf k} &=& \frac{W}{4} \left(\sin{k_x} \sin{q_x/2} + \sin{k_y} \sin{q_y/2} \right)  
 \label{zzz_6}
 \eea
  We keep the notation ${\mu_n}$ for the chemical potential, but treat it as a variable,  determined from the condition on the fermionic density.

  The Hamiltonian $H^{'}_{eff,2}$ is diagonalized by a rotation  
 \bea
 c_{\mathbf k} &=& u_{\mathbf k} \alpha_{\mathbf k} + v_{\mathbf k} {\tilde \alpha}_{\mathbf k}  \nonumber \\
 {\tilde c}_{\mathbf k} &=& u_{\mathbf k} {\tilde \alpha}_{\mathbf k} - v_{\mathbf k} \alpha_{\mathbf k}   
 \label{zzz_14}
 \eea
 with 
 \beq
  u_{\mathbf k}  =  \sqrt{ \frac{E_{\mathbf k}-\mu_n + \epsilon^{(-)}_{\mathbf k}}{ 2 E_{\mathbf k} } },
~ v_{\mathbf k}  = -\sqrt{ \frac{E_{\mathbf k}+\mu_n - 
  \epsilon^{(-)}_{\mathbf k}}{ 2E_{\mathbf k} } },  
 \label{zzz_15}
 \eeq
 where 
 \beq
 E_{\mathbf k} = ((\mu_n - \epsilon^{(-)}_{\mathbf k} )^2 + |\Delta_{\mathbf q}|^2)^{1/2}. 
 \label{zzz_3}
 \eeq
 For simplicity we skip the $U(1)$ phase factors in  $u_{\mathbf k} $ and $v_{\mathbf k}$.   
 
 Substituting into  (\ref{zzz_1}), we obtain 
  \beq
   H^{'}_{eff,2} = \sum_{\mathbf k} \left[
   \left(\epsilon^{(+)}_{\mathbf k}  + E_k \right) c^\dagger_{\mathbf k}c_{\mathbf k} + \left(\epsilon^{(+)}_{\mathbf k}  - E_k \right)  {\tilde c}^\dagger_{\mathbf k}{\tilde c}_{\mathbf k} 
   +  \frac{|\Delta_{\mathbf q}|^2}{2\beta\omega_0} \right] .
    \label{zzz_2}
    \eeq
 We see  that now the dispersion $E_{\mathbf k}$ depends on the momentum.  The energies of $\alpha$ and ${\tilde \alpha}$ fermions are
 \beq 
 E^\alpha_{\mathbf k} = \epsilon^{(+)}_{\mathbf k}  + E_k,~ E^{\tilde \alpha}_{\mathbf k} = \epsilon^{(+)}_{\mathbf k}  - E_k.
 \label{zzz_2_1}
 \eeq
 The time-ordered Green function of the physical $c-$fermion $G^c (\omega, \epsilon_{\mathbf k})$ is, as before, a weighted combination of the Green's functions of $\alpha$ and ${\tilde \alpha}$ fermions: 
 \begin{widetext}
  \beq
  G^c (\omega, \epsilon_{\mathbf k})=
       \frac{1}{2} \left( 1 - \frac{{\mu_n} - \epsilon^{(-)}_{\mathbf k}}{E_{\mathbf k}}\right)
  \frac{1}{\omega - E_{\mathbf k}  - \epsilon^{(+)}_{\mathbf k} + i\delta_\omega} 
   + \frac{1}{2}  \left( 1 + \frac{{\mu_n} - \epsilon^{(-)}_{\mathbf k}}{E_{\mathbf k}}\right)
  \frac{1}{\omega + E_{\mathbf k} -\epsilon^{(+)}_{\mathbf k}  + i\delta_\omega}.
  \label{zzz_4}
  \eeq
The  DOS  for the $c-$fermion  is
 \beq
 N (\omega) =  \frac{1}{2}  {\int\int}_{-\pi}^{\pi} \frac{dk_x dk_y}{4\pi^2}   \left[
 \left (1 - \frac{{\mu_n} - \epsilon^{(-)}_{\mathbf k}}{E_{\mathbf k}}\right) \delta (\omega -E_{\mathbf k} -\epsilon^{(+)}_{\mathbf k} ) + 
 \left (1 + \frac{{\mu_n} - \epsilon^{(-)}_{\mathbf k}}{E_{\mathbf k}}\right)  \delta (\omega +E_{\mathbf k} -\epsilon^{(+)}_{\mathbf k} )\right]
 \label{zzz_5}
  \eeq
  \end{widetext}
 For $E_{\mathbf k} > (W/4) (\cos{q_x/2} + \cos{q_y/2})$, which holds at large enough $\lambda_p$, the DOS consists of two separate pieces, one at $\omega>0$, another at $\omega <0$. The two have 
  equal  width but not equal shape.   In this regime, the condition that the fermionic density equals $n$ is 
     \beq
  n  = \frac{1}{2}  {\int\int}_{-\pi}^{\pi} \frac{dk_x dk_y}{4\pi^2}  \left( 1 + \frac{{\mu_n} -\epsilon^{(-)}_{\mathbf k}}{E_{\mathbf k}}\right),
  \label{zzz_7}
  \eeq
  and the self-consistent condition on $|\Delta_{\mathbf q}|$ is
  \beq
  1 = \beta \omega_0  {\int\int}_{-\pi}^{\pi} \frac{dk_x dk_y}{4\pi^2}   \frac{1}{E_{\mathbf k}}. 
   \label{zzz_8}
  \eeq
 At $W=0$, we have $\mu_n = -\beta \omega_0 (1-2n)$ and $|\Delta_{\mathbf  q}| = 2 \beta \omega_0 (n(1-n))^{1/2}$  as in this limit order parameters with different ${\mathbf q}$ are all the same.   However, at  a finite $W$, the ${\mathbf q}$ dependence becomes relevant.    To the leading order in the dispersion (i.e., in $1/\lambda_p$) we find
 \bea
 \mu_n &=& -\beta \omega_0 (1-2n) \left(1 + \frac{\sin^2{q_x/2} + \sin^2{q_y/2}}{32 \lambda^2_p}\right) \label{zzz_9} \\
 |\Delta_{\mathbf  q}| &=&  2 \beta \omega_0 (n(1-n))^{1/2}  \left(1 - \frac{\sin^2{q_x/2} + \sin^2{q_y/2}}{64 \lambda^2_p}\right).  \nonumber\\
&& 
 \label{zzz_10}
 \eea
 To obtain the ground-state energy $E_{\mathbf q}$  of a polaron state  we use the same strategy  as before, i.e.,  use 
  the Green's function in (\ref{zzz_4})  to compute the grand potential for $c$ fermions $\Omega^c_{\mathbf q} (\mu_n, |\Delta_{\mathbf  q}|)$  and then 
  use the relation $E^c_{\mathbf q} (n, |\Delta_{\mathbf  q}|  ) = \Omega^c_{\mathbf q} (\mu_n, |\Delta_{\mathbf  q}|)  + \mu_n n -\beta \omega_0 n^2$.
  The grand potential is 
    \bea
  &&\Omega^c_{\mathbf q} (n, |\Delta_{\mathbf  q}| ) = \frac{|\Delta_{\mathbf q}|^2}{4\beta \omega_0} \label{zzz_11} \\
 && -\frac{1}{2} {\int\int}_{-\pi}^{\pi} \frac{dk_x dk_y}{4\pi^2} 
 \left({\mu_n} +
  \sqrt{({\mu_n} - \epsilon^{(-)})^2 + |\Delta_{\mathbf q}|^2}\right).  \nonumber  
  \eea
 We verify that conditions $\partial \Omega_{\mathbf q}/\partial {\mu_n} = -n$ and $\partial \Omega_{\mathbf q}/\partial |\Delta_{\mathbf q}| = 0$ yield the same equations as  Eqs. (\ref{zzz_7}) and (\ref{zzz_8}).
  
 To the leading order in $1/\lambda_p$, we obtain
 \beq
 E_{P, {\mathbf q}} = E^c_{\mathbf q} (n, |\Delta_{\mathbf  q}|  )  = E^c_0 - \frac{ |\Delta_{\mathbf  q}|^2}{4\beta \omega_0}    \frac{\sin^2{q_x/2} + \sin^2{q_y/2}}{32 \lambda^2_p}
 \label{zzz_12}
 \eeq
 where $E^c_0 = - \beta \omega_0 n$ is the energy of a homogeneous polaron state. 
 
 For the polaron order with momentum $Q = (\pi,\pi)$,   the second term in (\ref{zzz_12}) is  $-  |\Delta_{\mathbf  q}|^2/(64 \beta \omega_0 \lambda^2_p)
 = - 4 W (n(1-n))/(64 \lambda_p)$.      This  term has the same structure  as the  negative energy coming from a virtual  hopping  from an occupied site to a neighboring unoccupied site,  Eq.  (\ref{fluc1}):   the factor $4$ is the number of nearest neighbors and the  factor $n(1-n)$ is the product of probabilities.  For $n=1/2$, 
  the prefactor is the same as in  (\ref{fluc1}).
This virtual hopping leads to a redistribution of
the density $n_i$ between neighboring sites, which  can  be seen  directly from the expression for the Green's 
      function of a $c$-fermion,  Eq. \eqref{zzz_4}. Indeed,  for ${\mathbf q} = Q$, the coherence factors  for the two components of $G^c (\omega, \epsilon_k)$ with positive and negative energies depend on $\epsilon^{(-)}_{\mathbf k} = \epsilon_{\mathbf k}$ and their real space Fourier transforms have 
        checkerboard components on  top of $n$ and $1-n$.  Still,   keeping only $H^{'}_{eff,2}$ we find that the  DOS has a hard gap and  fermions with energies inside the
         gap are localized.  
 
 Taken at a face value,  Eq. (\ref{zzz_12}) implies that at a finite $\lambda_p$, the system prefers the order with ${\mathbf q} = Q$  (a checkerboard order) not only at half-filling, but at any  non-zero $n(1-n)$.   In Appendix~\ref{app:completeness},  we summarize the results for the ${\mathbf q} = Q$ order at arbitrary $n$.     However, we  argue
  below that  Eq.  (\ref{zzz_12})  changes  once we  go beyond mean-field and include the cubic terms for the  interaction with phonons and   conjecture
  that the checkerboard order only develops   near half-filling, while at intermediate $n$,  a polaron order  has a momentum  ${\mathbf q}$ that evolves with $n$. 
    Keeping this in mind,   we continue   with the analysis of a polaron state with some generic  ${\mathbf q}$ and study how 
   evolves upon increasing $W$ (decreasing $\lambda_p$). \\

    \begin{figure}[]
	 \includegraphics[width=0.5\textwidth]{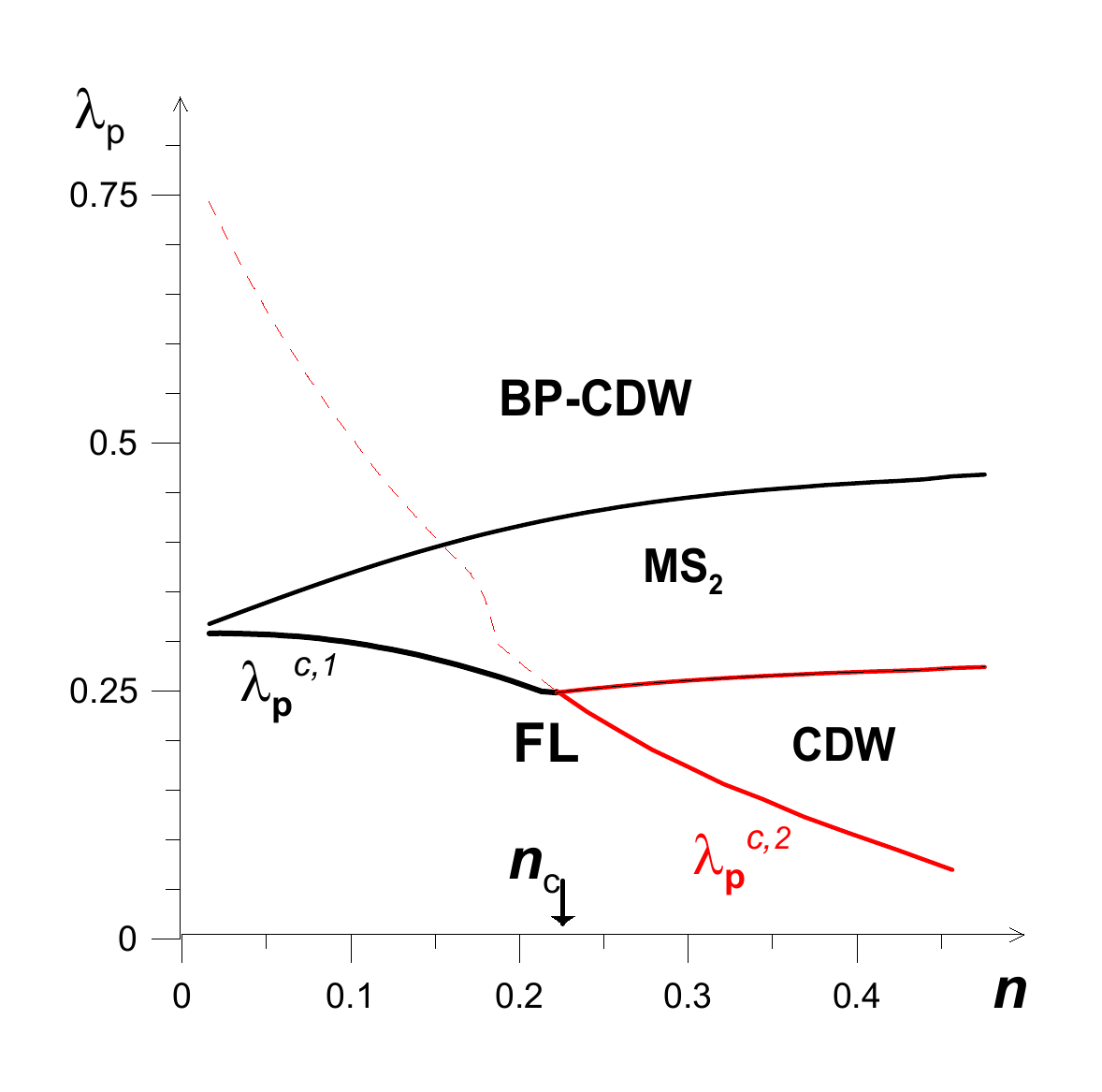}
	 \caption{Numerical (variational)  phase diagram for the case when a bare phonon dispersion has a deep minimum at $Q = (\pi,\pi)$  -- the same as in 
 the middle row of Fig. \ref{all2D}, but for spinless fermions and in terms of $\lambda_p$ rather than $\lambda_0$.  The polaron state with momentum $Q$ is labeled BP-CDW and the mixed state, consisting of these polarons and an FL, as MS$_2$. }
	 \label{fig:replot_2D_Q}
	 \end{figure} 
   
    \begin{centering} 
{\it{ Mixed state at a  non-zero ${\mathbf q}$.  }}\\
\end{centering}
\vspace{\baselineskip}

  We see from (\ref{zzz_2_1})  that the dispersion of  both $E^\alpha_{\mathbf k}$ and $E^{\tilde \alpha}_{\mathbf k}$ increases with decreasing $\lambda_p$, hence each of the two  parts of $N(\omega)$   stretches toward $\omega =0$.  At some $\lambda_p$   the two parts merge at $\omega =0$. This is  similar to the case ${\mathbf q} =0$,  but with an exception -- the merging occurs for $\lambda_p <1/2$. For small ${\mathbf q}$,  this happens at 
  \beq
  \lambda_p = \frac{1}{2} \left(1- (\sin^2{q_x/2} + \sin^2{q_y/2})\right).
  \label{zzz_19}  
  \eeq
  For ${\mathbf q} = Q$, the two continua  remain separated for all $\lambda_p >0$.
    The value $\lambda_p =1/2$ is essential here because we can use the same reasoning as for ${\mathbf q} =0$ and check   when it becomes energetically favorable 
     to create a mixed state consisting of  localized polarons  with density $n-\delta$ and an FL with density $\delta$. 
      The total energy of this state is $E_M (\delta) =  E_{P,{\mathbf q}} (\delta) +  E_{FL} (\delta)$
     The energy of an FL, linear in $\delta$, is $E_{FL}(\delta)  = - W \delta/2$  (Eq. (\ref{zzz_20})).   To obtain the energy of the polaron state $E_{P,{\mathbf q}} (\delta) = E^c_{\mathbf q} (n-\delta, |\Delta_{\mathbf  q}|  ) $ to the leading order in $\delta$ we note that  the dependence of $n$ 
      in $E^c_{\mathbf q} (n, |\Delta_{\mathbf  q}|  ) $  in (\ref{zzz_12}) is only in $E_0 = -\beta \omega_0 n$, while the rest depends on $|\Delta_{\mathbf  q}| $ but not on $n$.  We verified that it holds at arbitrary $\lambda_p$.  Because $\partial E^c/\partial |\Delta_{\mathbf  q}|  =0$,  the linear in $\delta$ term in $E_{P,{\mathbf q}} (\delta)$ comes only from $E_0$, and  $ E_{P,{\mathbf q}} (\delta)  = E^c_{\mathbf q} (n, |\Delta_{\mathbf  q}|  )  + \beta \omega_0 \delta$.  
       The prefactor for the linear in $\delta$ term in $E_M (\delta)$ is then $\beta \omega_0 - W/2 = W/2 (2\lambda_p-1)$.  It changes sign right at $\lambda_p =1/2$.
        We expanded to order $\delta^2$ and verified that it has a positive prefactor.   For example, for ${\mathbf q} =Q$,  $E_M(\delta) = E_M (0) + W/2 \left ((2 \lambda -1) \delta + 0.48 \delta^2\right)$.    As the result,  a pure nonhomogeneous polaron state becomes unstable toward a mixed state at $\lambda_p =1/2$.  We emphasize that this happens when the two continua in $N(\omega)$ in (\ref{zzz_5}) are still at some distance from each other and the states near $\omega =0$ are all gapped within mean-field.  The most extreme case here is ${\mathbf q} =Q$ as in this limit the two continua remain separated for all $\lambda_p$, yet 
         the mixed phase becomes energetically favorable for $\lambda_p <1/2$. 
         This is consistent with our variational study for the case  when  the checkerboard polaron state with ${\mathbf q} =Q$ is induced for all $n$ by choosing 
             a bare phonon dispersion  with a deep minimum at $Q$ (the middle row in Fig. \ref{all2D}).   For convenience of a reader, we plot the corresponding phase diagram  for spinless fermions and in terms of $\lambda_p$ in Fig.\ref{fig:replot_2D_Q}.   The polaron state with momentum $Q$ is labeled BP-CDW and the mixed state, consisting of these polarons and an FL, as MS$_2$.   The onset line for the mixed state deviates from $\lambda_p =1/2$  because the bare $\omega_0 $  is dispersing.  

              		  \begin{figure}[]
	 \includegraphics[]{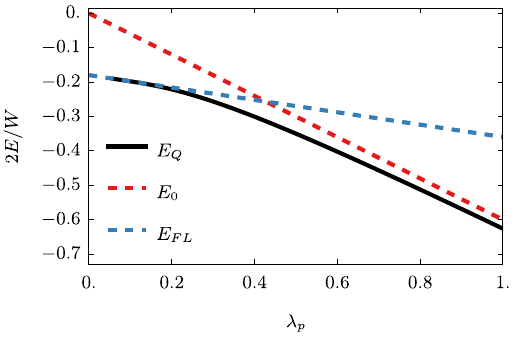}
	 \caption{Mean-field ground state energies   of the checkerboard polaron state, $E_Q$,  of the homogeneous polaron state, $E_0$,  and of a Fermi-liquid, $E_{FL}$,
 as functions of $\lambda_p$ at the representative density $n=0.3$. The mean-field checkerboard polaron state has the lowest energy over the entire range of $\lambda_p$.    }
	 \label{fig:e_polaron_cdw}
	 \end{figure}

  Like for ${\mathbf q}=0$, the  mixed state 
  extends down to $\lambda_p =\lambda^{c,1}_p$, at which $\Delta_{\mathbf q}$ vanishes. 
             To find $\lambda^{c,1}_p$, we have to solve the self-consistency equation in the regime where  the two parts of $N(\omega)$ in (\ref{zzz_5}) merge, i.e., $\langle \alpha^\dagger \alpha \rangle\neq 0$ and $\langle {\tilde \alpha}^\dagger {\tilde \alpha} \rangle \neq 1$. We find that 
              at ${\mathbf q} < Q$,   $\lambda^{c,1}_p$ is finite for $n \neq 1/2$ and at a given $n$ gradually shifts  down as the magnitude of ${\mathbf q}$ increases. 
              Also, an FL component can develop a CDW order in the  bottom part of the mixed state and below it, like it happens for ${\mathbf q} =0$.
              
              The case  ${\mathbf q} =Q$  is an exception here.  For this momentum,   $\Delta_{Q}$ remains non-zero  at arbitrary $n$,  and the two parts of $N(\omega)$ remain separated from  each other for all $\lambda_p$, no matter how small.   Comparing the energies, we  find that over the entire range of $\lambda_p$,
               $E_{P,Q}$ remains lower  than both  $E_{P,0}$ and  $E_{FL}$   
               ( see  Appendix~\ref{app:completeness} for detail) . We show the energies in   Fig. \ref{fig:e_polaron_cdw}.
                  At small $\lambda_p$  the difference  $ E_{P,Q}   = E_{FL}$ can be found analytically: 
 \beq 
     E_{P,Q}   = E_{FL}  - \frac{|\Delta_Q|^2}{4} N(\mu_n). 
    \label{zzz_25}
    \eeq
                The energy of the  mixed state is even smaller than $E_{P,Q}$,  hence an FL state does not develop even at the smallest $\lambda_p$.   
                This last result is clearly an artifact of mean-field approximation because at $n \neq 1/2$ and the smallest $\lambda_p$ we fully expect an FL state to have the lowest energy.  At $n =1/2$,  an FL state is unstable already at $\lambda_p =0+$, and  Eq. (\ref{zzz_25}) is  not  an artifact. \\
                 
\begin{centering} 
{\it{ Beyond mean-field.  The role  of the cubic term. }}\\
\end{centering}
\vspace{\baselineskip}

   We now include  the triple coupling $H^{'}_{eff,e-ph}$. In terms of $c$ and ${\tilde c}$ fermions,  it is given by  Eq. (\ref{qq_9_b}) .
  reexpressing this interaction in 
 terms of $\alpha$ and ${\tilde \alpha}$ fermions,  we find 
   \begin{widetext}
 \bea
&& H^{'}_{eff,e-ph} =  \frac{g}{\sqrt{2N\omega_0}} \sum_{{\mathbf k},{\mathbf l}}  \label{zzz_16} \\
&& 
 \left[\left( \alpha^\dagger_{{\mathbf k}+\frac{{\mathbf l}}{2}}\alpha_{{\mathbf k}-
 \frac{{\mathbf l}}{2}} + {\tilde \alpha}^\dagger_{{\mathbf k} + \frac{{\mathbf l}}{2}} {\tilde \alpha}_{{\mathbf k}-\frac{{\mathbf l}}{2}} \right) \left(u_{{\mathbf k}+\frac{{\mathbf l}}{2}} u_{{\mathbf k}-\frac{{\mathbf l}}{2}} + v_{{\mathbf k}+ \frac{{\mathbf l}}{2}} v_{{\mathbf k}-\frac{{\mathbf l}}{2}}\right) +  \left( \alpha^\dagger_{{\mathbf k}+ \frac{{\mathbf l}}{2}} {\tilde \alpha}_{{\mathbf k}-\frac{{\mathbf l}}{2}} - {\tilde \alpha}^\dagger_{{\mathbf k}+\frac{{\mathbf l}}{2}}\alpha_{{\mathbf k}-{\frac{{\mathbf l}}{2}}} \right)
 \left(u_{{\mathbf k}+ \frac{{\mathbf l}}{2}}v_{{\mathbf k}-\frac{{\mathbf l}}{2}} - v_{{\mathbf k}+ 
 \frac{{\mathbf l}}{2}}u_{{\mathbf k}-\frac{{\mathbf l}}{2}}\right) \right] \nonumber \\
 &&( b_{{\mathbf l}} + b^\dagger_{-\mathbf l} ), \nonumber
   \eea
   \end{widetext}
 where $u_{\mathbf k}$ and $v_{\mathbf k}$ are defined in (\ref{zzz_15}). 
 At $W=0$, $u_{\mathbf k}$ and $v_{\mathbf k}$ are independent of ${\mathbf k}$.
 Then $u_{\mathbf k+{\mathbf l}/2}u_{{\mathbf k}-{\mathbf l}/2} + v_{\mathbf k+{\mathbf l}/2}v_{{\mathbf k}-{\mathbf l}/2}
 = u^2+v^2 =1$ and  $u_{\mathbf k+{\mathbf l}/2}v_{{\mathbf k}-{\mathbf l}/2} - v_{\mathbf k+ {\mathbf l}/2}u_{{\mathbf k}-{\mathbf l}/2} =0$ and Eq. (\ref{zzz_16}) reduces to $ H'_{e-ph} =  \frac{g}{\sqrt{2N\omega_0}}
  \left( \alpha^\dagger \alpha + {\tilde \alpha}^\dagger {\tilde \alpha} \right)(b^\dagger + b)$,
    the same as   Eq. (\ref{10}).  
   At  a finite $W$, this formula holds for ${\mathbf q} =0$ but not at a finite ${\mathbf q}$.  There  are two new features.  First,  the  term with separate couplings  between  $c$ fermions and phonons   and between ${\tilde c}$ fermions and phonons gets dressed up by $l-$dependent combination of coherence factors.
   To the leading order in $1/\lambda_p$
\begin{widetext}
    \beq
Q_1 ({\mathbf k}, {\mathbf l})  = u_{\mathbf k+{\mathbf l}/2}u_{{\mathbf k}-{\mathbf l}/2} + v_{\mathbf k+{\mathbf l}/2}v_{{\mathbf k}-{\mathbf l}/2} 
= 
 1 - \frac{n(1-n)}{8 \lambda^2_p} \left(\sin{\frac{q_x}{2}} \sin{\frac{l_x}{2}} \cos{k_x} +
 \sin{\frac{q_y}{2}} \sin{\frac{l_y}{2}} \cos{k_y}\right) ^2.
    \label{zzz_17}
   \eeq
\end{widetext}
   
   At zero bosonic momentum ${\mathbf l} =0$, we still have $Q_1 (0) =1$, but for a generic ${\mathbf l}$ there is a correction.
   Second, there appears the term with phonon-assistant hopping between $c$ and ${\tilde c}$ fermions.    This term   generates fluctuations of the occupancy $n_i$  of a given site.  The corresponding coherence factor  is
\begin{widetext}
   \beq
 Q_2 ({\mathbf k}, {\mathbf l}) = u_{\mathbf k+{\mathbf l}/2}v_{{\mathbf k}-{\mathbf l}/2} - v_{\mathbf k+{\mathbf l}/2}u_{{\mathbf k}-{\mathbf l}/2}  =
  \frac{(n(1-n))^{1/2}}{2 \lambda_p} \left(\sin{\frac{q_x}{2}} \sin{\frac{l_x}{2}} \cos{k_x} + 
 \sin{\frac{q_y}{2}} \sin{\frac{l_y}{2}} \cos{k_y}\right). 
   \label{zzz_18}
   \eeq
   \end{widetext}
 At large $\lambda_p$, the   electron-phonon interaction in the form of Eq. (\ref{zzz_16}) still leads to the DOS, which at small $\omega$ is concentrated in 
  exponentially narrow patches centered around $|\omega| = l\omega_0$ ($l=0, 1, 2, ...$), while  the states at  $|\omega| \sim \beta$  form a  continuum 
  described by  Eq. (\ref{zzz_5}). 
   The argument here is similar to the one that we had earlier for the ${\mathbf q}=0$ polaron order.  Namely,  as long as $W \ll \beta \omega_0$,  the contributions from fermionic dispersion  do not affect the structure of $O(\beta)$ order in eikonal series and hence do not modify the exponentially large,  $e^{\beta}$, frequency renormalization.   However, a   pattern of well separated polaron patches will likely   be scrambled   due to the interaction between $\alpha$ and ${\tilde \alpha}$ fermions, which 
    mixes state with positive and negative energies.   In physics terms,  this is yet another indication of the presence  of  nearest-neighbor  (\ref{repulsion}) and longer range interactions between polarons  when the order parameter $\Delta_{\mathbf q}$ has a finite momentum.

 \begin{figure}[]
	 \includegraphics[width=0.4\textwidth]{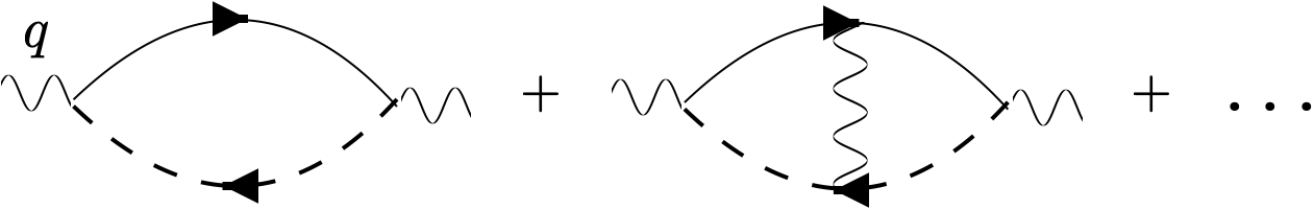}
	 \caption{Phonon self-energy diagrams that include particle-hole bubbles consisting of both $\alpha$ (solid) and $\tilde\alpha$ (dashed) fermion lines.}
	 \label{fig:ph_bubble}
	 \end{figure} 

 The same interaction between  $\alpha$ and ${\tilde \alpha}$ fermions  gives rise to the renormalization of the phonon frequency 
   $\omega_0 \to \omega_r (\mathbf l)$, which also becomes momentum-dependent.    The renormalization comes from the two mixed polarization bubbles consisting of  $\alpha$ and ${\tilde \alpha}$ fermions  (Fig. \ref{fig:ph_bubble}), each  
      with the factors of $Q_2 ({\mathbf k}, {\mathbf l})$ in the side vertices.  The calculation is straightforward and the result is 
    \bea
 && \omega_r ({\mathbf l}) = \omega_0 \nonumber \\
 &&\left (1 -\frac{n(1-n)}{8 \lambda^2_p} \sum_{i=x,y} \sin^2{\frac{q_i}{2}} (1-\cos{l_i}) \right). \nonumber \\
 &&  \label{feb_1}
  \eea
   We see that this effectively  induces  a downward  phonon  dispersion.   A dispersion of this kind  favors the checkerboard state, as we can see by comparing with the variational calculation, which  selects the checkerboard state for this type of phonon frequency (see Figs. (\ref{all2D}) and  (\ref{fig:replot_2D_Q})).

\subsubsection{\bf{Electronic  CDW order}}
\label{CDW_el}

 In the analysis above, we argued that  there is a range of $n$  around half-filling, where 
 the system first develops a CDW order upon  increasing  $\lambda_p$, and a polaron state develops at even larger $\lambda_p$ out of a CDW electron state. 
   Here we analyze the structure of the CDW-ordered  electron state.   To simplify the  calculations, we assume that  the  CDW order develops  with momentum $Q$ even at $n \neq 1/2$.  We will see that this will not lead to any artifacts within mean-field approximtion, at least to easily  recognizable ones.  One reason to analyze the CDW-ordered state  is to verify whether a CDW transition  at $n <1/2$ is continuous or first order.  Another is to compare  the CDW order with $Q$  with  the checkerboard polaron state.  

  The analysis of a CDW order is similar but not identical to that for a polaron order.  Namely, we  do not introduce an ancilla fermion and a counter-term and  instead decouple the phonon-induced  static $4-$
   fermion interaction 
\beq
U_{eff} = - \frac{ \beta \omega_0}{N} \sum_{{\mathbf k},{\mathbf p},{\mathbf q}} c^\dagger_{\mathbf k}c_{{\mathbf k}+{\mathbf q}} c^\dagger_{\mathbf p} c_{{\mathbf p}-{\mathbf q}},
\label{d_27_1}
\eeq
 by  introducing a  CDW order parameter
$\Delta^{el}_Q = -2\beta \omega_0 (1/N) \sum_{\mathbf k} \langle c^\dagger_{\mathbf k}c_{{\mathbf k}+Q} \rangle$
\footnote{This method is equivalent  to introducing a phonon condensate with momentum $Q$.}.  The order parameter is real because
$(\Delta^{el}_Q)^* = \Delta^{el}_{-Q}$, and  $Q $ and -$Q$ differ by a reciprocal lattice vector.   We reduce the summation over momenta $\sum_{\mathbf k}$ to ${\bf k}$ inside the magnetic Brillouin zone ($\sum'_{k}$) and treat $c_{{\mathbf k}+Q}$ as an independent operator. In these notations,
$\Delta^{el}_Q = -4\beta \omega_0 (1/N) \sum^{'}_{\mathbf k} \langle c^\dagger_{\mathbf k}c_{{\mathbf k}+Q} \rangle$ ,
\bea
H^{' el}_{2} &=& {\sum_{\mathbf k}}'\left(\left( \epsilon_{\mathbf k} - {\mu_n}\right) c^\dagger_{\mathbf k}c_{\mathbf k} -
(\left( \epsilon_{\mathbf k} + {\mu_n}\right) c^\dagger_{{\mathbf k}+Q} c_{{\mathbf k}+Q}\right) \nonumber \\
 & +& {\sum_{{\mathbf k}}}' \left[\Delta^{el}_Q \left(c^\dagger_{\mathbf k}c_{{\mathbf k}+Q} + c^\dagger_{{\mathbf k}+Q} c_{{\mathbf k}}\right) + \frac{(\Delta^{el}_Q)^2}{2 \beta \omega_0} \right],
\label{d_27_2}
\eea
and
      \beq
   H^{' el}_{e-ph, Q} =  \frac{g}{\sqrt{2N\omega_0}}
   \sum_{\mathbf q}  {\sum_{\mathbf k}}' \left(c_{\mathbf k}^\dagger c_{{\mathbf k}+{\mathbf q}} + c^\dagger_{{\mathbf k}+Q} c_{{\mathbf k}+Q+{\mathbf q}} \right) (b^\dagger_{\mathbf q} + b_{-{\mathbf q}}).
   \label{d_27_2_1}
   \eeq
Diagonalizing the quadratic  Hamiltonian by the Bogolyubov transformation
   \bea
   c_{\mathbf k}&=& u_{\mathbf k}\alpha_{\mathbf k}+ v_{\mathbf k}~{\tilde \alpha}_{\mathbf k}    \nonumber \\
   c_{{\mathbf k}+Q} &=& u_{\mathbf k}{\tilde \alpha}_{\mathbf k}- v_{\mathbf k}\alpha_{\mathbf k},   \eea
  we obtain
     \beq
H^{'el}_2 =  {\sum_{\mathbf k}}'
\left[\left(E_{\mathbf k}- \mu_n\right) \alpha^\dagger_{\mathbf k}\alpha_{\mathbf k}- \left(E_{\mathbf k}+ \mu_n\right)  {\tilde \alpha}^\dagger_{k} {\tilde \alpha}_{\mathbf k} + \frac{(\Delta^{el}_Q)^2}{4 \beta \omega_0} \right],
\label{d_27_8}
\eeq
where
\beq
E_{\mathbf k}= \sqrt{\epsilon^2_{\mathbf k}+ (\Delta^{el}_Q)^2},
\label{d_27_4}
\eeq
and  the coefficients $u_{\mathbf k}$ and  $v_{\mathbf k}$  are
   \beq
   u_{\mathbf k}= \sqrt{  \frac{E_{\mathbf k}+\epsilon_{\mathbf k}}{2E_{\mathbf k}} }, \;\;  
   v_{\mathbf k}= -\sqrt{ \frac{E_{\mathbf k}-\epsilon_{\mathbf k}}{2E_{\mathbf k}} }.
 \label{d_27_4_1}
\eeq
The electron-phonon interaction term becomes
  \begin{widetext}
 \bea
&& H^{' el}_{e-ph, Q} =  \label{d_27_4_2} \\
&& \frac{g}{\sqrt{2N\omega_0}} \sum_{{\mathbf q}} {\sum_{\mathbf k}}'  \left[\left( \alpha^\dagger_{\mathbf k} \alpha_{{\mathbf k}+{\mathbf q}} + {\tilde \alpha}^\dagger_{\mathbf k}{\tilde \alpha}_{{\mathbf k}+{\mathbf q}} \right) \left(u_{\mathbf k}u_{{\mathbf k}+{\mathbf  q}} + v_{\mathbf k}v_{{\mathbf k}+{\mathbf q}}\right) +  \left( \alpha^\dagger_{\mathbf k}{\tilde \alpha}_{{\mathbf k}+{\mathbf q}} - {\tilde \alpha}^\dagger_{\mathbf k}\alpha_{{\mathbf k}+{\mathbf q}} \right)
 \left(u_{\mathbf k}v_{{\mathbf k}+{\mathbf q}} - v_{\mathbf k}u_{{\mathbf k}+{\mathbf q}}\right)\right] (b^\dagger_{\mathbf q} + b_{-{\mathbf q}}). \nonumber
\eea
   \end{widetext}
We focus on $n \leq 1/2$, when $\mu_n <0$.   For a negative $\mu_n$, 
$\langle \alpha^\dagger_{\mathbf k} \alpha_{\mathbf k}\rangle =0$.  
 The self-consistency  condition on $\Delta^{el}_Q$ and  the condition on the fermionic density  $n$  are
 \bea
 &&1 = 2\beta \omega_0 \left[\frac{1}{N} {\sum_{\mathbf k}}^{'}
 \frac{1}{E_{\mathbf k}} \langle{\tilde \alpha}^\dagger_{\mathbf k}{\tilde \alpha}_{\mathbf k}\rangle\right] \label{d_27_9}  \\
  && n  = \frac{1}{N} {\sum_{\mathbf k}}^{'} \langle {\tilde \alpha}^\dagger_{\mathbf k}{\tilde \alpha}_{\mathbf k} \rangle>  
  \label{d_27_9_2} 
 \eea
The way to satisfy (\ref{d_27_9_2}) at  $n <1/2$ is to keep $|\mu_n|$ slightly above   $|\Delta^{el}_Q|$.    Then $\langle{\tilde \alpha}^\dagger_{\mathbf k}{\tilde \alpha}_{\mathbf k}\rangle$ is non-zero in a limited range  $\sqrt{{\mu_n}^2 -(\Delta^{el}_Q)^2} < |\epsilon_k| < W/2$.     Eqs. (\ref{d_27_9}  ) and (\ref{d_27_9_2}) become
\bea
   1  &=& 2 \beta \omega_0 \int_{\sqrt{{\mu_n}^2 -(\Delta^{el}_Q)^2}}^{W/2} ~~\frac{N_0 (\epsilon)}{\sqrt{\epsilon^2 + (\Delta^{el}_Q)^2}}  \nonumber \\
  n&=&  \int_{\sqrt{{\mu_n}^2 -(\Delta^{el}_Q)^2}}^{W/2} ~~N_0 (\epsilon) d \epsilon.  
  \label{d_27_9_3} 
\eea
At large $\lambda_p$, $\Delta^{el}_Q \approx 2\beta \omega_0 n$ and ${\mu_n} \approx - 2\beta \omega_0 n$ is slightly below the top of the ${\tilde \alpha}$ band. 
 This implies that away from half-filling the system is a CDW metal.  The chemical potential  
 ${\mu_n}$ jumps from $-\beta \omega_0$ to $0$ at $n = 1/2-0$ and jumps further to  
 ${\mu_n} \approx \beta \omega_0$ at $n =1/2 +0$ (one needs a small but finite $T$ to see this clearly). 

 Solving Eqs. ( \ref{d_27_9_3})  for infinitesimally small   $\Delta^{el}_Q$, we obtain 
 $\lambda^{c,2}_p$  for the onset of a CDW order.   This  $\lambda^{c,2}_p$ vanishes at $n=1/2$ as $1/(\log{(1/2-n)})^2$,  but  remains finite for all  $n$ and rapidly increases as $n$ decreases.  This is fully in line with the expectations and also consistent with the variational analysis.  
 
We next  compute the grand potential within the same computational scheme as before, and  extract from it the ground-state energy  of the CDW  electronic state.   
 For the grand potential per particle we obtain, for $n <1/2$,  
 \bea
 &&\Omega^{el}_{Q} ({\mu_n}, \Delta^{el}_Q) =  \frac{(\Delta^{el}_Q)^2}{4 \beta \omega_0}  \\
 &&
 - \frac{1}{2} \int_{\sqrt{{\mu_n}^2 - (\Delta^{el}_Q)^2}}^{W/2}
 N_0 (\epsilon) d \epsilon \left({\mu_n} + \sqrt{\epsilon^2 + (\Delta^{el}_Q)^2}\right) \nonumber 
 \eea
  Evaluating this $\Omega^{el}_Q$ for equilibrium values of $\mu_n$ and $\Delta^{el}_Q$, and adding  
 ${\mu_n} n$ and  $\beta \omega_0 n^2$, we obtain the ground-state energy per particle  of the CDW electronic state, 
 $E^{el}_Q$.  At  $W = 0$,  the calculation can be done analytically, and  the result is 
 \beq
 E^{el}_{Q}  \approx  -\beta \omega_0 n  + \frac{\beta \omega_0}{4} (1-4n^2).
  \label{d_27_10}
 \eeq
  As expected, for all $n$ away from half-filling,  $E^{el}_{Q}$ is  higher than the energy of the polaron state
   $ E_P = - \beta \omega_0 n$.   By continuity, this implies that  at large $\lambda_p$ the ground state is a polaron state 
    rather than a CDW electron state. 
 \begin{figure}[]
	 \includegraphics[]{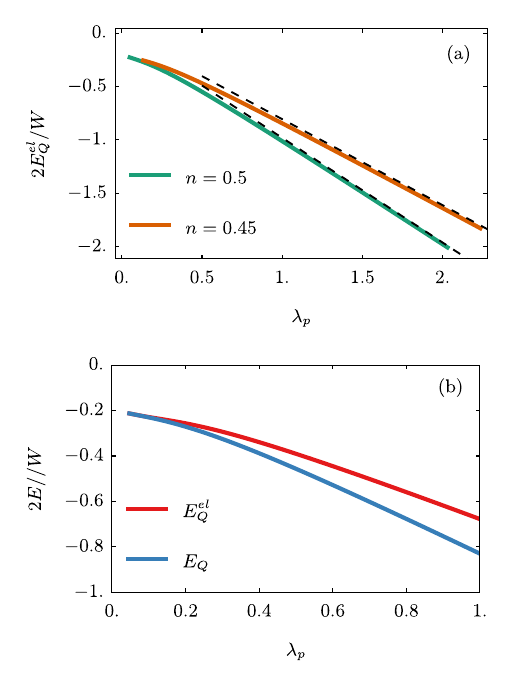}
	 \caption{(a) Electronic CDW energy $E^{el}_Q$ for $n=0.5$ (half-filling) and $n=0.45$. Dashed lines show the strong coupling result, \eqref{d_27_10}. (b) Comparison between  the energy $E_Q^{el}$ of the electronic  CDW state and $E_{P,Q}$ of the  checkerboard polaron state. We set  $n=0.4$.}
	 \label{fig:e_el_cdw}
	 \end{figure} 
	 For $W \neq 0$, we obtained $E_{el} (Q)$ numerically.  We plot the result in Fig. \ref{fig:e_el_cdw}.
We see that  for $n <1/2$,  $E_Q^{el}$, is larger than  the energy of the checkerboard polaron state $E_{P,Q}$ for all $\lambda_p$. 
 This strongly suggests that there is no first-order transition between  CDW polaron and CDW electron states.

\subsubsection{\bf{Half-filling}}
\label{degeneracy}

At  half-filling, $\mu_n =0$,  $\langle {\tilde \alpha}_{\mathbf k} \alpha_{\mathbf k}\rangle =1$ for all ${\mathbf k}$.  The  quadratic Hamiltonian  takes the form 
\beq
H^{' el}_2 =  {\sum_{\mathbf k}}' \left[E_{\mathbf k}\left(\alpha^\dagger_{\mathbf k}\alpha_{\mathbf k}- {\tilde \alpha}^\dagger_{k} {\tilde \alpha}_{\mathbf k}\right)
 + \frac{(\Delta^{el}_Q)^2}{2 \beta \omega_0} \right],
\label{d_27_3}
\eeq
and the  self-consistent equation on $\Delta^{el}_Q$ is
 \beq
 1 = \beta \omega_0  \int_{-W/2}^{W/2} \frac{N_0 (\epsilon) d \epsilon }{\sqrt{\epsilon^2 + (\Delta^{el}_Q)^2}}.
\label{d_27_5}
\eeq
For tight-binding dispersion, it reduces to
 \beq
 1 = 4\lambda_p \frac{2}{\pi^2} \int_0^1 \frac{K(1-x^2)}{\sqrt{x^2 + ({\hat \Delta}^{el}_Q)^2}},
\label{d_27_6}
\eeq
where ${\hat\Delta}^{el}_Q = (2/W) \Delta^{el}_Q$.
This equation is analogous to the  one for the SDW order parameter in the Hubbard model;  $\beta \omega_0$ plays the role of $U$ (see e.g., Refs. \cite{swz,Frenkel1992}).   The order parameter is non-zero for  all $\lambda_p$, even for the smallest one. This is consistent with the vanishing  of $\lambda^{c,2}_p$ in $n=1/2$.  At small $\lambda_p$,
 $\Delta^{el}_Q \sim W e^{-\pi/\sqrt{2\lambda_p}}$  is exponentially small  At large $\lambda_p$,
 $\Delta^{el}_Q \approx \beta \omega_0$.  The  full ground-state energy  per particle is 
 \bea
 E^{el}_Q &=& -\frac{1}{2N} \sum_{\mathbf k}E_{\mathbf k} +  \frac{(\Delta^{el}_Q)^2}{4 \beta \omega_0} - \frac{\beta \omega_0}{4} \nonumber \\
 &=& - \frac{1}{2} \int_{-W/2}^{W/2} N_0 (\epsilon) d \epsilon \sqrt{\epsilon^2 + (\Delta^{el}_Q)^2}  + \frac{(\Delta^{el}_Q)^2}{4 \beta \omega_0} - \frac{\beta \omega_0}{4}. 
  \label{d_27_6_1}
 \eea

These equations are identical to those for a checkerboard polaron order at $n=1/2$. Indeed, for the latter the  self-consistence equation for 
$|\Delta_Q|$ is 
   \beq
  1 = \beta \omega_0  \int_{-W/2}^{W/2 } \frac{N_0 (\epsilon)}{\sqrt{\epsilon^2 + |\Delta_Q|^2}} d \epsilon,  \label{d_26_2_1}
  \eeq
 and the  ground-state energy is
   \beq
 E_ {P,Q} =  = -\frac{1}{2} \int_{-W/2}^{W/2} N_0 (\epsilon) d \epsilon \sqrt{\epsilon^2 + |\Delta_Q|^2} +  \frac{|\Delta_Q|^2}{4\beta \omega_0} - \frac{\beta \omega_0}{4} 
 \label{d_26_7_a}
  \eeq
  We see that   $|\Delta_Q| = |\Delta^{el}_Q|$ and $E_{P,Q} = E^{el}_Q$.  
  The mean-field DOS  for both states  is
  \begin{widetext}
\beq
N(\omega) =     \frac{1}{2}  \int_{-W/2}^{W/2} N_0 (\epsilon) d \epsilon   \left[
 \left (1 - \frac{\epsilon{\mathbf k}}{\sqrt{\epsilon^2 + \Delta^2_Q}} \right) \delta \left (\omega -  \sqrt{\epsilon^2 + \Delta^2_Q}\right) +  
 \left (1 + \frac{\epsilon{\mathbf k}}{\sqrt{\epsilon^2 + \Delta^2_Q}} \right) \delta \left (\omega +  \sqrt{\epsilon^2 + \Delta^2_Q}\right)\right]
 \label{zzz_5_1}
  \eeq
  \end{widetext}
 consists of two parts, one at $\omega >0$, another at $\omega <0$, which remain separated for all non-zero $\lambda_p$. 
  
     We argue that this 
 equivalence is not accidental and implies   that strictly at half-filling, where the probabilities for  a given site to be occupied and to be empty are the same,  an  ancilla fermion is not  distinct  from a physical fermion.   An immediate  consequence  is that  the ordered state at half-filling breaks a discrete $Z_2$ symmetry rather than $U(1)$.  In this respect, $n=1/2$ is qualitatively different from all other $n$,  at which  a polaron order breaks $U(1)$ symmetry.     
   The type of the ordering transition at half-filling has been analyzed   numerically in  \cite{costa2018,esterlis_19} and was found to be in the Ising universality class. This is consistent  with our assertion that the order parameter at half-filling breaks $Z_2$ symmetry rather than $U(1)$.

\subsubsection{Outstanding issues and conjectures}

  There are several outstanding issues that require  going beyond our current approach.  We list them here and make several conjectures. 
    
  \begin{itemize}  
  
  \item{How to recover an FL behavior at the smallest $\lambda_p$ and $n <1/2$   when the  polaron order is with ${\mathbf q} = Q$?}\\
   
 We recall that within mean-field, the energy of the checkerboard polaron state $E_{P,Q}$ is  smaller than $E_{FL}$  for all values of $\lambda_p$ and any $n$. While  such behavior is fully expected for $n =1/2$, for smaller $n$ we expect that an FL  is a ground state at the smallest $\lambda_p$.   We  suggested  that this result is an artifact of mean-field approximation, and will be corrected once we move beyond mean-field and include the  cubic term.

        \begin{figure}[]
	 \includegraphics[width=0.9\linewidth]{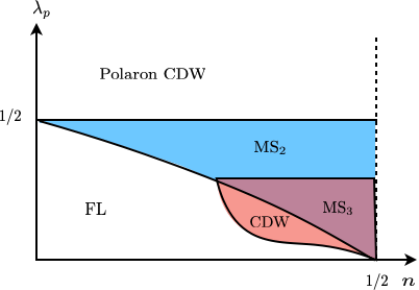}
	 \caption{Schematic  phase diagram for the case when the checkerboard  polaron order holds at all electron densities. 
      The phase diagram  includes a pure FL state, a pure checkerboard polaron state (polaron CDW), a CDW state, and two mixed states: MS$_2$ (FL and polaron CDW) and MS$_3$ (CDW  and polaron CDW). The location of the lines and the widths of  the two mixed phases are  only suggestive.   This phase diagram is qualitatively similar to the numerical one in Fig.  \ref{fig:replot_2D_Q}.  }
 \label{fig:phase_diag_ms2+ms3}
	 \end{figure} 
  
  We note in this regard that the  mixed  cubic term, taken to second order, lowers the ground-state energy.  This decrease has to be compensated by the increase of the energy from the counter-term, whose prefactor then has to be modified from Eq.  (\ref{zzz_30}).  
    Once we decouple  the counter-term  using $\Delta_{\mathbf q}$,   the 
   $|\Delta_{\mathbf  q}|^2/(4\beta \omega_0)$  term in  the energy of a physical $c-$fermion  should  acquire an additional factor 
    $1 + F_{\mathbf q} (n, \lambda_p)$.     We did not attempt to obtain this term explicitly as to get the energy gain from he mixed cubic term one has to sum up infinite diagrammatic series.       On physical grounds, we expect that  for ${\mathbf q} =Q$,  this term will render  an FL state as the ground state 
      at the smallest $\lambda_p$ and $n \neq 1/2$, i.e., 
    $ |\Delta_{\mathbf  q}|^2  F(n, \lambda_p) /(4\beta \omega_0)$   will overcome the  energy difference  between the checkerboard polaron and FL states, 
 $|\Delta_Q|^2  N(\mu_n)/4$  (see Eq.  (\ref{ww_23}) ). 
    Assuming that this is the case and combining with other results, we present in Fig. \ref{fig:phase_diag_ms2+ms3}  
    a suggestive phase diagram  for the checkerboard polaron order, assuming that it holds for all values of $n$.
    
      \item{Can a polaron state with $0< {\mathbf q} <Q$ be the ground state for some $0<n<1/2$?}\\
      
      To address this issue, we extend the analysis in the previous item to large $\lambda_p$.  
 At  such $\lambda_p$,  we expect that $F_{\mathbf q} (n, \lambda_p) = a (\sin^2{q_x/2} + \sin^2{q_y/2})/(32 \lambda^2_p)$. where $a=o(1)$, simply  because all other renormalizations have this form.  The  ground-state energy of the ${\mathbf q}$ state then gets modified  from (\ref{zzz_12}) to 
   \bea
 && E_{P, {\mathbf q}} = E^c_{\mathbf q} (n, |\Delta_{\mathbf  q}|  )  =\nonumber \\
  &&E^c_0 - \left(1-a\right)\frac{ |\Delta_{\mathbf  q}|^2}{4\beta \omega_0}    \frac{\sin^2{q_x/2} + \sin^2{q_y/2}}{32 \lambda^2_p},
 \label{zzz_12a}
 \eea
 where, we remind, $E^c_0 = - \beta \omega_0 n$ is the energy of a homogeneous polaron state.   
Accordingly,
 \bea
 \mu_n &=& -\beta \omega_0 (1-2n)  \nonumber \\
 && \left(1 + \left(1-a\right) \frac{\sin^2{q_x/2} + \sin^2{q_y/2}}{32 \lambda^2_p}\right) 
\label{zzz_9a} \\
 |\Delta_{\mathbf  q}| &=&  2 \beta \omega_0 (n(1-n))^{1/2}    \nonumber \\
 &&\left(1 -\left(1 +2a\right)  \frac{\sin^2{q_x/2} + \sin^2{q_y/2}}{64 \lambda^2_p}\right).
\label{zzz_10a}
 \eea
We did not attempt to  obtain the value of $a$ in the computations.  We just note  that for a particular $a=1$,  the polaron energy  at this order becomes independent on ${\mathbf q}$  In this situation, the selection of  the polaron order at a given $n$  has to be  done by the subleading terms   
 \footnote{To keep the ground-state energy independent on ${\mathbf q}$ to order $1/\lambda^4_p$, one has to include $1/\lambda^4_p$ correction into the 
 $F_{\mathbf q} (n, \lambda_p) |\Delta_{\mathbf q}|^2/(4\beta \omega_0)$ term and also add an additional term with the structure  ${\tilde F} _{\mathbf q} (n, \lambda_p) |\Delta_{\mathbf q}|^4/(4 \beta \omega_0)^3$.  We verified that the energies of the homogeneous and checkerboard polaron states coincide if we choose  $F_{Q} (n, \lambda_p) = 1/(16 \lambda^2_p) + 9/(1024 \lambda^4_p)$ and  ${\tilde F}_{Q} (n, \lambda_p) = - 9/(1024 \lambda^4_p)$.
  We also note  that  for this choice of $F_{Q} (n, \lambda_p)$,  the  renormalization of the chemical potential  vanishes, i.e.,  $\mu_n = -\beta \omega_0 (1-2n)$ 
   also at non-zero $W$.} 
 and the latter may  select  a gradual variation of ${\mathbf q}$  of the ground state from a homogeneous polaron state near $n=0$  (or $n=1$) 
 to a checkerboard polaron state around $n =0.5$.     We believe that a gradual evolution of  ${\mathbf q}$  of the ground state  with $n$ is the most natural scenario. 
  We also remind that the selection of the ground state at a given $n$ can be controlled by imposing a dispersion of a bare fermion, as our variational analysis demonstrated.  

\item {\it Does a nonhomogeneous polaron state  with  ${\mathbf q} \neq 0$ has a Fermi surface? }\\

           While the answer is a definitive ``No" for a spin-1/2 system where bipolarons are bosonic degrees of freedom, the situation is more subtle for 
           %AC 
            spinless fermions/polarons.
 We recall that at ${\mathbf q} =0$, the DOS at a finite $W$ consists of two non-interacting parts at $\omega >0$  (from $\alpha$-fermions) 
 and $\omega<0$ (from ${\tilde \alpha}$ fermions), and each part of $N(\omega)$ vanishes at $\omega \equiv 0$.  Then there is no Fermi surface in the pure polaron state - it  only emerges in the mixed state.    For a non-zero ${\mathbf q}$, the cubic term $H^{'}_{eff,e-ph}$, Eq. (\ref{zzz_16}), contains a piece  with phonon-assistant coupling between  
 $\alpha$ and ${\tilde \alpha}$ fermions. To rigorously analyze the effect of this term on the DOS at small $\omega$ along with the effect of 
   a momentum-dependent prefactor $Q_1 ({\mathbf k}, {\mathbf l})$  for the coupling of $\alpha$ or ${\tilde \alpha}$  fermions to phonons, one would have to again sum infinite series of self-energy and vertex correction diagrams, but with the complication that now one also has to include the loops made from one ${\alpha}$ and one ${\tilde \alpha}$ fermion.  We did not attempt to sum up these series.   We conjecture that  the resulting DOS  will be non-zero at $\omega =0$. This would mean that a state with a nonhomogeneous polaron order is a metal with some Fermi surface.  By continuity from the ${\mathbf q} =0$ case, we expect that the area of the FS in a nonhomogeneous polaron state gradually increases with increasing ${\mathbf q}$.

   \item {\it How the DOS evolves near the onset of a mixed state for ${\mathbf q} \neq 0$? }\\

 We  found above that at a nonzero ${\mathbf q}$, the mixed state, in which some fermions form localized polarons and some form an FL, becomes energetically favorable  at $\lambda_p <1/2$,   before the two pieces of the DOS touch each other.  an FL component has a non-zero spectral weight around $\omega =0$,  where it was absent at $\lambda_p >1/2$.     We conjecture that an FL DOS  emerges at $\lambda_0 =1/2+0$ with an infinitesimally small residue $Z$ , which grows as the system moves inside the mixed phase. The residue $Z$ does not explicitly show up in $E_{FL} = -(W/2) \delta$, which we used in energy consideration,  as it appears in the overall factor in  both $E_{FL} = - (Z W/2\pi) (1 + 2 \mu_{FL}/W)$ and $\delta = (Z\pi) (1 + 2 \mu_{FL}/W)$ (see Eqs. (\ref{tt_4}) and (\ref{tt_5})) and cancels out in the ratio $E_{FL}/\delta$.    At the same time,  when $\delta$ and $Z$ both increase into the mixed phase starting from infinitesimally small values,  $(1 + 2\mu/W)$ does not have to vanish at the onset of the mixed phase, i.e.,  a Fermi surface area  may already be no-zero at the onset. 
  This is consistent with the  conjecture that we made in the previous item.
 
\item{\it Is Luttinger theorem violated at a non-zero ${\mathbf q}$?} \\

We conjectured in the previous two items that a polaron state with a finite ${\mathbf q}$ is a metal  with some Fermi surface already at large $\lambda_p$ and that  the area of the Fermi surface gradually increases in the mixed state starting from a finite value at the onset of this state.   We further conjecture that by continuity from the case ${\mathbf q} =0$,   the area of the Fermi surface in the pure polaron state gradually increases with increasing ${\mathbf q}$  but remains smaller than in an FL state, i.e., Luttinger theorem is violated.  
The situation at half-filling may be different because  we argued above that the 
 checkerboard   polaron state is  not  localized beyond 
 the mean-field approximation. In this situation, there  is no physically motivated  argument for violation of the Luttinger theorem. 
We  note in  this regard that at half-filling the order parameter $\Delta_Q = \Delta^{el}_Q$ breaks $Z_2$ symmetry rather than $U(1)$.

\end{itemize}

\section{Comparison between numerical and analytic results and the full phase diagram}
 \label{sec:comp}

We begin with several observations.  First, in the analytical study we considered the 2D case and assumed that the bare $\omega_0$ is independent of $q$.  In the numerical study, we considered several forms of $\omega_0 (q)$ and both 2D and 3D systems.  For comparison between analytical and numerical results, we select the lower panel of Fig.  \ref{all2D}
  which is for the same 2D case with $q-$independent $\omega_0$.  Second, the numerical results 
   in Fig.  \ref{all2D} are presented in units of $\lambda_0 = N_F g^2/\omega^2_0$, which measures the strength of interaction near the Fermi surface in MET (see (\ref{l0})), while the analytical results are presented in terms of $\lambda_p = g^2/(2 \omega^2_0 W)$, which measures coupling to
fermions with energies comparable to the bandwidth.
Third, the numerical results have been obtained  for spin-$1/2$ fermions, 
 while the analytical results are for spinless fermions. 
To  move from $S=1/2$ to spinless fermions, one has to re-scale the density and  the electron polarization
 by a factor of two.
With this 
rescaling included, 
 the relation between $\lambda_0$ and $\lambda_p$ is
 \beq
 \lambda_0 =  \lambda_p \left(N_F  W\right).
 \label{qq_12_3}
 \eeq
For our tight-binding model,  $N_F = N (\mu_{FL})= (4/\pi^2 W) K(1-{\hat \mu}^2_{FL})$, where ${\hat \mu}_{FL} = 2 \mu_{FL}/W$ is related to the density $n$ by Eq. (\ref{zz_2}).  At small $n$,
$ {\hat \mu}_{FL} \to -1$
and $N_F \approx 2/(\pi W)$; at $n \to 1/2$, 
$ {\hat \mu}_{FL} \to 0$, and $N_F$ diverges logarithmically.

To make the comparison between analytical and numerical results more straightforward, in Fig.~\ref{fig:phase_diag_ms1}, which we presented at the beginning of this paper,   
we plot the numerical phase diagram in 2D  in units of $\lambda_p$, adjusted to spinless fermions, along with the analytical  phase diagram
 for the case of a homogeneous order of polarons with $\Delta (q) = \Delta (q=0)$.
Comparing the two phase diagrams, we see that they are essentially identical.  This is not surprising because  in the numerical study we identified a homogeneous polaron state as a localized one, with no fluctuations between the sites occupied by polarons and empty sites.  In the analytical study, we used probabilistic description and expressed Green's function at a given site (the spatial Fourier transform of $G^c (\omega, \epsilon_k)$) as the sum of the  Green's functions of the operators $\alpha$ and $\tilde \alpha$, weighted with the factors $1-n$ and $n$, respectively. Operators $\alpha$ and  ${\tilde \alpha}$ describe unoccupied states with
$n_i =0$ and occupied states with $n_i =1$. There is no phonon assisted hopping between unoccupied and occupied sites even when $W$ is finite; hence this polaron state is fully localized.

We emphasize that while the phase diagrams for spinful and spinless fermions are identical, the corresponding DOS are not identical. 
 Already in the atomic limit, the DOS for  spinful fermions contains a true gap around $\omega =0$ associated with the finite minimal energy of a bipolaron, while the DOS in the spinless case contains a set of polaron $\delta-$functional peaks down to $\omega =0$. 
  The evolution of the DOS with increasing $W$ is the same for the two cases, if we restrict with the continuum of states at $\omega \sim \pm \beta \omega_0$, but beyond this approximation, there are qualitative distinctions between spinful and spinless fermions. 
Namely,  for bipolarons,  the true gap in the DOS  still holds at at a small but  finite $W$,  i.e.
   the system of $S=1/2$ fermions remains an insulator.   The two boundaries of the mixed state are then the actual
    phase transition lines: the one at a larger $\lambda_0$ is an insulator to a metal transition, and the lower one is a transition from a metal with broken Luttinger theorem to a conventional metal in which Luttinger theorem holds.  For spinless fermions, at $W >0$, the sub-gaps between polaron patches hold within our approximation, but beyond it, there is an exponentially small damping rate for a decay of states with a finite $\omega$ 
    (see Appendix \ref{app_E} for more discussion on this).  As a  consequence, the DOS in between polaron patches  is exponentially small but still finite.  Then, strictly speaking, the system is a metal at any finite $W$.  From this perspective, the upper boundary of the mixed phase is a very sharp crossover but not a true phase transition.  The lower boundary of the mixed phase is a true phase transition for the same reason as for spinful fermions -- the Luttinger theorem gets broken. 

    The numerical and analytical phase diagrams for the checkerboard polaron state  are shown in Fig.~\ref{fig:phase_diag_ms2+ms3}. %\ref{fig17_b}.
 The analytical phase diagram is a suggested one, as we discussed in Sec. \ref{CDW_pol}.  
  We emphasize  that the  checkerboard states in the numerical and analytical studies were treated differently. For the numerical study, we selected a localized state with strict checkerboard order of polarons in a cluster
($n_i =1$ at even sites inside the cluster  and $n_i =0$ at odd sites and vise versa) and no polarons outside a cluster. 
      For Einstein phonon this state is energetically degenerate with any other configuration of polarons, and only gains extra energy if we include virtual fluctuations to the n.n. sites.
In the analytical study, we set $\Delta (q) = \Delta (q=Q)$, used probabilistic description and expressed the Green's function of a physical fermion via the Green's function of $\alpha$ and ${\tilde \alpha}$ fermions, which 
describe filled and empty polaron states at a given site.   The probability to be either  in $\alpha$ or ${\tilde \alpha}$ state depends on $\epsilon_k$;  in the real space, the probabilities vary between even and odd sites. This state has some similarities with the $(1-0)$-type  polaron state used in the numerical study; however, the variation of probabilities is weak at large $\lambda_p$. In addition, the state with $\Delta (q)  = \Delta (Q)$ is not localized as $H'_{e-ph}$ generates phonon-assisted hopping between the occupied and empty sites.  Nevertheless, a similarity between the numerical and analytic  phase diagrams shows that  both approaches capture the same physics. 

 We also emphasize that the phase diagrams that we presented, both for homogeneous and $(\pi,pi)$ polaron/bi=polaron orders,  hold for large $\beta$.   For small $\beta$, the transformation for the atomic limit to MET at large $W$ is different (see, e.g., \cite{Ciuchi_2003}).  The key reason for  the difference is that polarons-bipolarons cannot be viewed as bound states of a fermion and  a large number of phonons attracted to it.

\subsection{The full  phase diagram }
\label{phase_dia}

For the full phase diagram, we assume that the polaron order is homogeneous for $n$ away from half-filling and checkerboard near half-filling.    We assume that the phase diagrams are similar for both types of polaron order and contain the mixed phase, which in some range of densities  near half-filling is split into the regions where the electronic component is in an FL state and where it has a CDW order.  We show the candidate phase diagram in Fig.~\ref{fig:phase_diag_full}. 
In the doping range where the polaron order is homogeneous, the phase boundaries can be determined by exact calculations.  For the checkerboard polaron order, the phase boundaries are drawn by hand.  At $n=1/2$, the states with $(\pi,\pi)$ polaron and electron order are degenerate. 

 \subsection{Comparison with other works}
 \label{comp} 
 
Properties of e-ph systems at strong-coupling, and the associated tendencies toward polaron/bipolaron, have been extensively studied numerically using Monte Carlo simulation \cite{esterlis_18, esterlis_19, Nosarzewski2021, murthy_2023} and dynamical mean-field theory (DMFT) methods \cite{freericks1993, Millis_1996,Millis_1996_a, Ciuchi_1993,Ciuchi_1997,Ciuchi_1998,Ciuchi_2003,Ciuchi_2006,fratini2021,Ciuchi_2025}.
In most of these studies, the primary focus was the influence of bipolaron formation  for spinful fermions on superconductivity and 
  of polarons/bipolarons on the normal state transport at 
 nonzero temperatures.  In our study, we restricted ourselves to $T=0$, neglected the potential superconductivity in the variational analysis, and performed analytical  calculations for spinless fermions. 
  Some of the previous work also focused on CDW instabilities. 
  The authors of \cite{Ciuchi_1993,esterlis_19, Nosarzewski2021,murthy_2023} 
  found that in systems near half-filling (and, in general, away from perfect Fermi surface nesting), strong electron-phonon coupling 
   leads to commensurate $(\pi,\pi)$ charge-density wave ground states 
   At dilute concentrations,  incommensurate charge-density waves \cite{Ciuchi_1993} and phase separation \cite{murthy_2023} have been reported. 
   These findings are generally consistent with our analysis. 

A transformation from an FL to 
 polaron (or bipolaron) regime 
   was  
 analyzed in
 \cite{Millis_1996,Millis_1996_a,Ciuchi_2003, Ciuchi_2006,Carlson2004,esterlis_19,fratini2021,murthy_2023}, with connections drawn to high-temperature superconductivity \cite{Carlson2004} and the metal-insulator transition \cite{Ciuchi_2003, Ciuchi_2006,fratini2021}. In some of these works~\cite{freericks1993,Millis_1996,Millis_1996_a,fratini2021,murthy_2023} the formation of polarons / bipolarons was described in terms of 
 bimodal distribution of phonon displacements, interpreted~\cite{Millis_1996,Millis_1996_a} as the appearance of 
 frozen-in  lattice distortions. 
This scenario is qualitatively and in several aspects also quantitatively similar to our ``ancilla fermion" scenario for the homogeneous polaron state.

To be more precise here, we compare our results with those of  Ref. \cite{Millis_1996,Millis_1996_a} by Millis, Mueller, and Shraiman (MMS).  They considered FL to polaron crossover  within DMFT for several models, including the one for spinless fermions with a semicircular DOS and a dispersionless boson, which is most directly related to our model.  MMS focused on the limit where (in our notation) $\omega_0 \to 0$ while  $\beta \omega_0$ remains finite (i.e., $\beta$ tends to infinity). In this limit, a phonon propagator reduces to $\delta (\Omega)$, i.e., a phonon can be treated classically. 
Because $\omega_0 = 0+$, there are no separate polaron states at discrete $\omega = m \omega_0$ in the atomic limit, $\lambda_p = \infty$.  Yet, the minimization of energy within the DMFT yields frozen-in lattice distortions of opposite sign  on occupied and unoccupied sites, which gives rise to two $\delta-$functional peaks in the DOS at $\pm \beta \omega_0$, one with residue $n$, and another with residue $1-n$.  This fully agrees with our result at $\beta \to \infty$ (and also with the exact solution of a single-site  Holsten model at near-infinite $\beta$). 
  At large but finite $\lambda_p$, MMS found the two-patch form of the DOS $N(\omega)$,  one patch at positive and another at negative $\omega$. In between, $N(\omega) =0$.  This again agrees with our result modulo that our DOS has 
   polaron peaks at smaller $|\omega|$. Finally, MMS found an intermediate state in the range of $\lambda_p$,
   where frozen-in lattice distortions are still non-zero, yet $N(\omega)$  is non-zero for all $\omega$ and eventually becomes the same as in an FL when frozen-in lattice distortions disappear. 
   This intermediate state is very similar to our psedogap/mixed state, where our order parameter $\Delta_0$ is non-zero, yet $N(\omega)$  is non-zero for all $\omega$. 
   
     For a direct comparison with MMS, we compute the local Green's function and the DOS in our model, using a semicircular form of the DOS for free fermions instead of (\ref{wed_2}):  
\beq
     N_0 (\epsilon) = \frac{4}{\pi W} \sqrt{1 - \left(\frac{2 \epsilon}{W}\right)^2}.
 \label{wed_1}
 \eeq
The self-consistent equation on a homogeneous $\Delta_0$ and the equation on the chemical potential $\mu_n$ for this $N_0 (\epsilon)$ become
\beq
\frac{\pi}{2}  = 2 \lambda_p T({\hat E}), ~~ \frac{\pi}{2} (2n-1) = {\hat \mu}_n  
T({\hat E}),
\label{wed_3}
 \eeq  
 where, as before, ${\hat \mu}_n = 2\mu_n/W$, ${\hat E} = 2 E/W = ({\hat \mu}^2_n + |{\hat \Delta}_0|^2)^{1/2}$,
 ${\hat \Delta}_0 = 2 \Delta_0/W$, and
 \beq
 T({\hat E}) = \sqrt{1 - {\hat E}^2 } + \frac{\arcsin({\hat E})}{\hat E}. 
 \label{wed_4}
 \eeq 
 The local Green's function $G_{loc} (\omega) = \int N_0 (\epsilon) G(\omega, \epsilon)$ is
 \begin{widetext} 
 \beq
 G_{loc} (\hat \omega) = \frac{2}{W} \left(\frac{\left(1 - \frac{{\hat \mu}_n}{\hat E}\right)}{{\hat \omega} - {\hat E} +
 {\text {sign}} ({\hat \omega} - {\hat E}) \sqrt{({\hat \omega} - {\hat E})^2 -1}} + \frac{\left(1 + \frac{{\hat \mu}_n}{\hat E}\right)}{{\hat \omega} + {\hat E} +
 {\text {sign}} ({\hat \omega} + {\hat E}) \sqrt{({\hat \omega} + {\hat E})^2 -1}}\right)
 \label{wed_5}
 \eeq 
 and the DOS $N(\hat \omega) = -\frac{1}{\pi} |{\text {Im}} G_{loc} (\hat \omega)|$ is
  \beq 
  N(\hat \omega)  = \frac{2}{\pi W}  \text{{Re}} 
  \left[ \left(1 - \frac{{\hat \mu}_n}{\hat E}\right) \sqrt{1-({\hat \omega} - {\hat E})^2}  
  + \left(1 + \frac{{\hat \mu}_n}{\hat E}\right) \sqrt{1-({\hat \omega} + {\hat E})^2} \right].
   \label{wed_6}
 \eeq 
\end{widetext}
 
 The solution of (\ref{wed_3}) for $\lambda_p >1/2$ is the same as in Sec. \ref{arb_n_arb_W}:
  ${\hat \mu} = 2 \lambda (2n-1)$ and $|{\hat \Delta_0}| = 2 \lambda  (4n(1-n))$, i.e., ${\hat E} = 2 \lambda_p$. 
   The DOS $N(\hat \omega)$  consists of two patches, each with width $2$, centered at $\hat \omega = \pm 2\lambda_p$, i.e., at $\omega = \pm \beta \omega_0$.   This is not identical to the DOS obtained by MMS because they added 
   frozen-in lattice distortions to  the local Green's function, while in our  theory the order parameter $\Delta_0$ is added to the momentum-resolved Green's function $G(\omega, \epsilon)$ before we integrate over $\epsilon$ to obtain the DOS.  Still, both our DOS and the one earlier obtained by MMS describe two-patch structure of the DOS with the gap between patches.     
   
   For $\lambda_p <1/2$, the solution of  (\ref{wed_3}) is
  ${\hat \mu} = 2 \lambda_p (2n-1)$ - the same as before, but  $|{\hat \Delta_0}|$ now decreases 
   with  decreasing $\lambda_p$. Running $|{\hat \Delta_0}|$ is the solution of 
\bea
&&\frac{\pi}{4 \lambda_p} = \sqrt{1 - 4\lambda^2_p (1-2n)^2 - |{\hat \Delta}_0|^2} \nonumber \\
&&+ \frac{\arcsin{(4\lambda^2_p (1-2n)^2 + |{\hat \Delta}_0|^2)^{1/2}}}{(4\lambda^2_p (1-2n)^2 + |{\hat \Delta}_0|^2)^{1/2}}.
  \label{wed_7}
 \eea 
\begin{figure}[htbp]
	\includegraphics[]{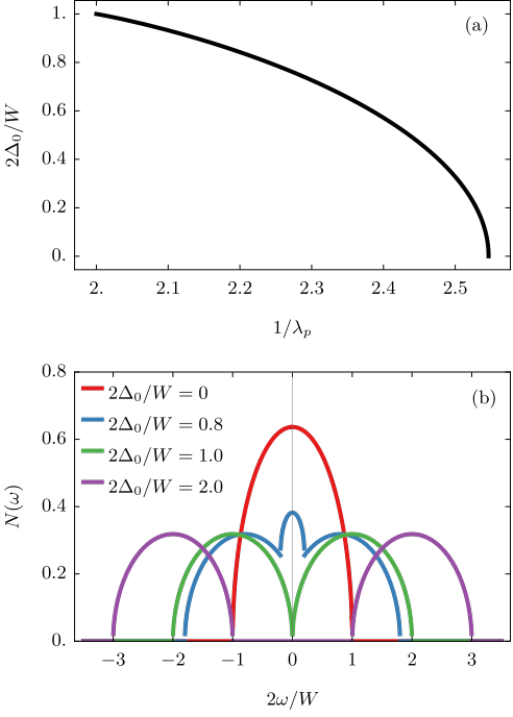}
   \caption{Results for a semicircular (non-interacting) DOS \eqref{wed_1}. 
(a)  The dependence of $2|\Delta_0|/W$ on $1/\lambda_p$ for semicircular density of states for free fermions. The order parameter vanishes for $\lambda_p = \pi/8$ ($1/\lambda_p\approx 2.55$). (b) The DOS in units of $W/2$ for different 
$2|\Delta_0|/W$. The maximum at a finite $\omega$ emerges at $2|\Delta_0|/W =1/2$.  The DOS has a gap around $\omega =0$ for $2|\Delta_0|/W >1$.}
\label{Fig_22_a1}
\end{figure}
 
 For direct comparison with Fig. 1 in ~\cite{Millis_1996}, in Fig. \ref{Fig_22_a1} we  plot, for $n=1/2$,  $|{\hat \Delta}_0|$ as a  function of $\lambda_p$ and the DOS  as a function of frequency for several values of 
 $|{\hat \Delta}_0|$. The DOS at $n=1/2$ is 
 \beq
  N(\hat \omega)  = \frac{2}{\pi W}  \text{{Re}} 
  \left(\sqrt{1-({\hat \omega} - |{\hat \Delta}_0|)^2} 
  +\sqrt{1-({\hat \omega} + |{\hat \Delta}_0|)^2} \right).
   \label{wed_8}
 \eeq 
   The intermediate phase exists for $n=1/2$ between $\lambda_p =1/2$ and $\lambda^{c,1}_p = \pi/8$.  Within this range, there is a sub-range of smaller $\lambda_p \geq \lambda^{c,1}_p$, where the DOS has a single peak at $\omega =0$ and the sub-range of larger $\lambda_p \leq 1/2$, where it has two peaks at non-zero frequencies.  MMS found the same separation into two sub-regimes  in their analysis of the DOS in the intermediate phase, although our and MMS  functional forms of the DOS  differ somewhat.  

The MMS DMFT approach has been extended to $T=0$ in a series of papers by Ciuchi et al (CEA)
\cite{Ciuchi_1997,Ciuchi_1998,Ciuchi_2003,Ciuchi_2006}. 
 They considered the same model as MMS at $n = 0$ and at $n =1/2$  and analyzed both the adiabatic limit $\omega_0 \ll W$ and the anti-adiabatic limit $\omega_0 \gg W$.  In the ``ultra-adiabatic" limit $\omega_0 \to 0$, $\beta \omega_0 \to {\text {const}}$, their  analysis is identical to that of MMS because, in this limit, the boson propagator 
  reduces to $\delta (\Omega)$. However, at finite $\omega_0$ they detected additional features.  
   Their results for spinless fermions can be directly compared to ours at $n=0$, where a CDW instability does not interfere with polaron formation.  Here, CEA found a cascade of transformations of the fermionic DOS 
   upon decreasing  $\lambda_p$, with the number of gapped regions in the DOS being reduced by one in each transformation.
    This is in full agreement with our evolution of the DOS, in which a growing continuum in the DOS
    upon decreasing $\lambda_p$ absorbs polaron patches one by one.   Their evolution of the DOS 
     within the polaron phase (e.g., Fig. 12 from \cite{Ciuchi_1997}) is also very similar to ours at $W \geq \sqrt{\beta} \omega_0$ (see Appendix~\ref{app_E}). The only difference is that CEA consider the continuum as incoherent, while we associate the continuum with a free-fermion band.  CEA also argued that the number of phonons in  the polaron 
       state at the lowest $\omega$ is of order $\beta$. This fully agrees with our analysis of the  structure of the continued fraction representation of the Green's function in Sec. \ref{n_0_W_0}. 
     
     For half-filling, their and our results 
      agree once we extend our analysis to a semicircular DOS, as described earlier in this Section,  and neglect potential development of CDW order.  One particular point of agreement here is the DMFT result~\cite{Ciuchi_2003,Ciuchi_2006} that the DOS  for spinless fermions  remains finite 
      at $\omega =0$ for any value of $\lambda_p$, as long as $\omega_0$ is finite.  This agrees 
       with our result that the zero-frequency polaron peak from the atomic limit transforms at finite $W$  into a  patch of exponentially small, but  still  finite, width of order $\sim e^{-\beta}$, and this patch eventually merges with the continua.   Their mapping of a polaron state  into a two-level system  has some similarity with the introduction of an ancilla fermion in our theory. 

In the diagrammatic approach, a fermionic Green's function at $n=0+$ and finite $W$ has also been analyzed in Ref. \cite{Berciu_2006}  in the approximation in which every diagram for the self-energy is averaged over momentum.  This yields  the spectral function with an exponentially small width ($e^{-\beta}$) of all polaron peaks, including those at $\omega \approx \beta \omega_0$.  In our view,  this form is valid for $\beta = O(1)$, when the residues of polaron peaks at $\omega =0$ and $\omega = \beta \omega_0$ are comparable. However, for large $\beta$, we argued in  this work that exponentially narrow polaron patches hold only for $\omega \sim \omega_0$, while  fermions with $\omega \approx \beta \omega_0$ form a free-particle continuum, whose width is set by $W$ for $W > \omega_0 \sqrt{\beta}$, and the evolution of the DOS at larger $W \sim \beta \omega_0$ can be viewed as the gradual absorption of polaron peaks by the continuum. 

\subsection{Additional considerations}

\subsubsection{\bf {Three-dimensional case}}

\begin{figure*}[htbp]
	\includegraphics[width=0.4\linewidth]{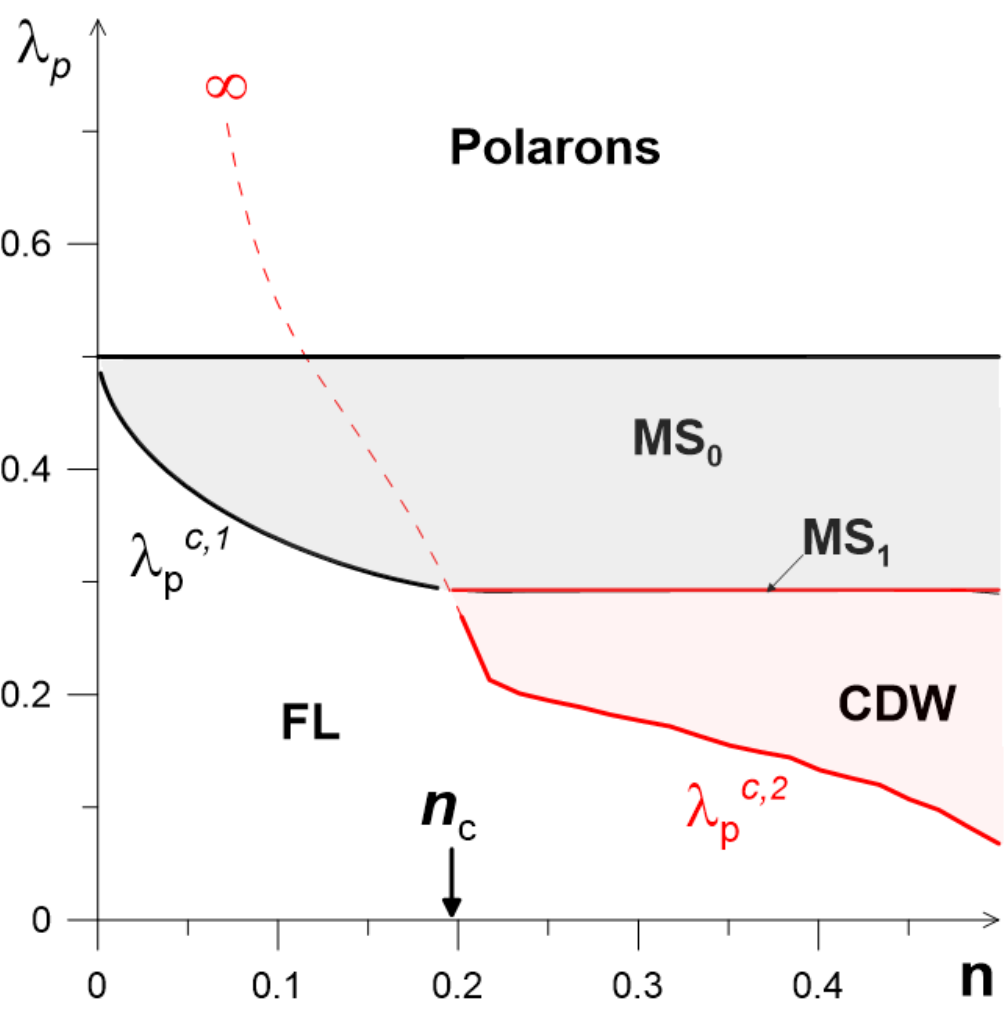}
    ~~~~~~~~~
	\includegraphics[width=0.43\linewidth]{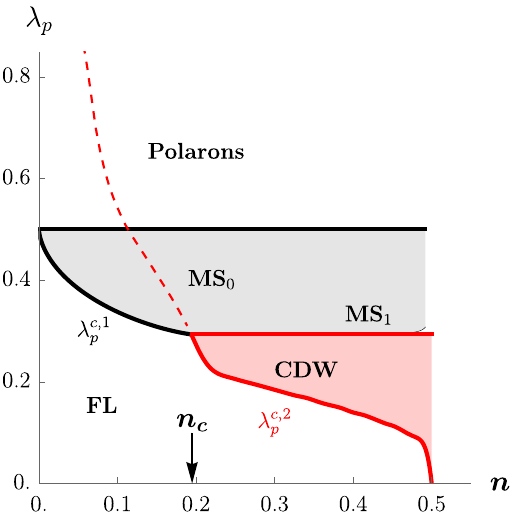}
    
\caption{
Phase diagram for a 3D system  of spinless fermions for a dispersionless bare phonon spectrum $\omega_0 (q)/t =0.1$. Left panel -- numerical (variational) phase diagram. Right panel --  analytical phase diagram. Like in 2D, they are essentially identical. 
The variational phase diagram is identical to the one shown in the middle panel of the lowest row in Fig.~\ref{Upin3D}
after expressing $\lambda_0$ in terms of $\lambda_p$ and adjusting to spinless fermions.
Like in 2D, the phase diagram contains pure polaron and FL states, a CDW state and two mixed states: $MS_0$ (polarons and FL) and  $MS_1$ (polarons and CDW).  }
\label{fig21}
\end{figure*}
 The 3D phase diagram, obtained in the variational study for dispersionless $\omega_0$ and spinful fermions, 
is presented in the  middle of the lower panel in  Fig. \ref{Upin3D}.  For convenience, in Fig.\ref{fig21}
we re-plot this phase diagram in units of $\lambda_p$ and for spinless fermions.      
We note that phase diagrams in 2D and 3D are similar, but there is one key distinction: in 3D the DOS of the FL vanishes as $n^{1/3}$ at small $n$. As a consequence,
MET remains stable against low-energy fluctuations up to $\lambda^{c,2}_p (n) \propto 1/N_F \propto 1/n^{1/3}$. In the same limit, the homogeneous polaron state  emerges  at 
  $\lambda^{c,1}_p\approx 1/2$.
This implies that at small $n$, the instability toward the polaron state occurs 
already at weak MET coupling, when the dressed phonon frequency is almost the same as 
$\omega_0$.

For larger $n$, the phase diagram is similar to that in 2D:
there again exists an intermediate mixed phase between $\lambda_p =1/2$ and $\lambda^{c,1}_p$, 
where a portion of a system is at the end point of the homogeneous polaron state and another portion in the FL state.  The line $\lambda^{c,2}_p$ again lies below  $\lambda^{c,1}_p$ at $n >n_c \approx 0.2$, and for these $n$ there exists the CDW electronic order, sandwiched  between $\lambda^{c,2}_p \equiv \lambda^{c,2}_p (n)$ and $\lambda^{c,2}_p (n_c)$.

\subsubsection{\bf{Finite T}}
At temperatures above any ordering transition, the evolution from the FL state to the polaron state is smooth. Nevertheless, numerical calculations have shown that this crossover occurs rapidly \cite{Ranninger_1993,Millis_1996,Millis_1996_a,Ciuchi_2003, Ciuchi_2006,esterlis_19,fratini2021,murthy_2023}, and can be identified by the onset of a pseudogap in the electronic density of states \cite{esterlis_19}. As detailed in Sec.\ref{arb_n_arb_W},  the pseudogap is a clear indicator of polaron formation, with the redistribution of spectral weight into high-energy peaks being associated to the presence of sites with and without large lattice distortions. Although a detailed study of finite-$T$ effects is beyond the scope of the present work, the qualitative behavior in the important regime $\omega_0 \ll T \ll W \lesssim \beta \omega_0$ can be understood by just replacing $\omega_0$ by $T$. For example, the electronic density of states has a two-peak structure with peaks at $\pm \beta \omega_0$ of width $\sim \sqrt{\beta \omega_0 T}$ \cite{murthy_2023}.

 \section{Conclusions}
\label{sec:concl}

 In this paper, we considered a system of electrons interacting with an optical phonon in the adiabatic limit
when the phonon Debye frequency is much smaller than the Fermi energy.
 For weak coupling, such  systems are traditionally described by MET.
 A conventional belief is that MET holds even at strong coupling, when the electron self-energy is large, and breaks down only near the point where the dressed phonon spectrum softens at some momentum $q$.
We argued that at small and near-maximum  electron  densities, the FL state described by MET 
 ceases to be the ground state already at  smaller couplings, when MET is  stable against low-energy fluctuations.
The ground state at these couplings is a two-component mixed state, in which an FL still exists, but with a smaller density, and the remaining portion of the system is in a 
bipolaron state  for   $S=1/2$ fermions and in a polaron state  when fermions have only one spin projection.   Such a state likely  spatially  separates into FL and bipolaron/polaron regions.  We argued that bipolaron/polaron formation is not a low-energy phenomenon and that relevant fermions have energies comparable to the bandwidth $W$.
 The mixed state emerges prior to phonon softening in both 2D and 3D systems. In 3D the effect is particularly striking as this state  appears already at weak coupling. 
  
We presented  numerical and  analytical reasoning for this behavior.  Numerical  evidence was obtained from variational
 considerations in both 3D and 2D, analytical evidence was obtained from the diagrammatic treatment of the on-site Holstein model for spinless fermions in 2D.  In this treatment, we  first  demonstrated  how to reproduce the exact results for the single-site  Holstein model  and then extended the technique to the case where fermions can hop to neighboring sites.  We argued that it is essential to keep vertex corrections in diagrammatic series on the same footing as self-energy corrections to the internal fermion lines.  We argued that eikonal treatment is sufficient at  zero and full densities, but for any other density, one has to additionally introduce ancilla fermions and form a two-particle condensate $\Delta (q)$ out of the original and ancilla fermions.    
 
 We considered two types of bipolaron/polaron states: a  homogeneous one, with  the largest possible density on the occupied sites and zero otherwise, and the state with checkerboard arrangement of bipolarons/polarons. We conjectured that the homogeneous state emerges when  the phonon dispersion has a minimum at 
  small momentum, while the checkerboard state emerges when the minimum is at momentum $Q$ near the zone boundary
  We obtained similar but equivalent phase diagrams in the two cases. In analytical diagrammatic treatment, the homogeneous and checkerboard states are those with $\Delta (q=0)$  and $\Delta (q=Q)$, respectively.  
 
 At larger densities and particularly near half-filling, 
we found that the  first instability upon increasing the coupling is into a CDW electronic state, and bipolarons/polarons develop at larger couplings out of the CDW-ordered electronic state.  
In this situation, strong coupling MET holds near the onset of a CDW order unless MET becomes unstable for some other reason.  We analyzed how the CDW electronic order evolves within the mixed state and found a reentrant behavior of the onset of this order, i.e., as the coupling increases, CDW order first appears and  then disappears within the mixed state.  We also constructed the global phase diagram for the case when the bipolaron/polaron state is homogeneous  away from half-filling and  checkerboard near half-filling. 
    
Overall, our results imply  that  the emergence of bipolarons/polarons 
imposes the most severe limitation on the applicability range of MET for all fermionic densities  except very near half-filling, and even there bipolarons/polarons appear already at weak coupling. 

%AC
A set of peaks in the DOS at low energies, which resemble the polaron peaks about to be absorbed by a continuum extending from higher energies, have been reported in Ref. \cite{Butler2021} for  1T - Ta$S_2$.   
This material is viewed as a Mott insulator.  We believe that it may well be in the 
polaron state at $\lambda_p$ slightly above $1/2$.

\section{Acknowledgements}

We acknowledge with thanks useful discussions with B. Altshuler, M. Berciu, E. Berg, S. Ciuchi, M. Fabrizio, R. Fernandes, S. Fratini, S. Ilani,  T. Hanaguri,  
T. Heikkila, W. Metzner, C. Murthy, H-Y Kee, M. Kiselev,  S. Kivelson, A. Millis, N. Nagaosa, P. Nosov, 
R. Ojajaervi, Ph. Phillips, M. Randeria,  S. Sachdev,  M.V. Sadovskii, G. Sangiovanni, D. Senechal,  J. Schmalian, A. Stern,  B. Svistunov, A-M Tremblay, Y. Wang,  M. Ye and S-S Zhang. 
 AVC was supported by the U.S.\ Department of Energy, Office of Science, Basic Energy Sciences, under Award No.~DE-SC0014402.
 ArA was supported by the DOE-Office Of Science, DE-SC0026038.
AVC and NVP acknowledge support from the Simons Foundation grant SFI-MPS-NFS-00006741-07 for the Simons Collaboration on New Frontiers in Superconductivity. 
%AC
AVC and NVP thank the Aspen Center for physics for the hospitality at the early stages of this project.  Aspen Center for Physics  is supported by National Science Foundation grant PHY-2210452.

\appendix

\section{Fermionic self-energy at $n=0$}
\label{app:Q}
In this Appendix, we will briefly discuss perturbation series for the self-energy strictly at $n=0$, when the Fermi energy vanishes. This holds deep in regime B in the notation from Sec. \ref{sec:van} in the main text 
(regime B is defined as the one where $E_F \ll \omega_0$ and regime A as the one where $\omega_0 \ll E_F \ll W$). 

The analytical analysis in Sec. \ref{sec:van}  is done in 2D, and we keep $D=2$ here (we list the results for $D=3$ at the end of this Appendix). 

The MET for a 2D electron-phonon system holds in the regime B.  In this regime, the vertex corrections are small in the Eliashberg parameter $\lambda_E$. In 2D,  
\beq
\lambda_E= \lambda_0 \frac{\omega_0}{E_F} \log{\frac{E_F}{\omega_0}}
\label{satt_4}
 \eeq 
(see \cite{Andrey_review}).  As we said in the main text,  $\lambda_E$ is small 
compared to $\lambda_0$
in $(\omega_0/E_F)$. The additional logarithm in (\ref{satt_4}) is specific to $D=2$. 

The ME parameter  $\lambda_E$ can be extracted from the ratio of the two-loop fermionic self-energy with vertex correction included and the one-loop self-energy:  $\lambda_E \sim \Sigma^{(2v)}/\Sigma^{(1)}$.  Both can be computed on the Matsubara axis. For a dispersionless  phonon, $\Sigma^{(1)}$ depends on $\omega_m/\omega_0$ and
  $\Sigma^{(2v)}$  depends  on $\omega_m/\omega_0$ and fermionic momentum $k$.  We compute 
  $\Sigma^{(2v)}$ and also the rainbow-type  two-loop diagram $\Sigma^{(2r)}$  by setting $T=0$, setting the external momentum $k$ at the bottom of the band
  and  expanding to the first order in $\omega_m$. 

The calculation of two-loop diagrams, particularly of $\Sigma^{(2r)}$, requires extra care as each self-energy 
 diagram has a dynamical and  a static part. The  latter 
 accounts for the renormalization of the 
  fermionic dispersion and also accounts for the 
 renormalization of the chemical potential
 from bare $\mu_0$ to the actual $\mu$, related to density. 
  It is customary to incorporate the static self-energy  into the bare fermionic Green's function
  and then compute only the dynamical self-energy at each order.  This procedure, however, is not rigorously justified as the static self-energy from a given order may contribute to the dynamical self-energy at a higher order. We verified that in our case the renormalization of the fermionic dispersion is not crucial and can be safely neglected, i.e., the dressed $\epsilon_k$  can be approximated by a bare tight-binding dispersion
  with bandwidth $W$. However, the  renormalization of the chemical potential cannot be just absorbed to the bare  Green's function. Namely, we find that if we do this, $\Sigma^{(2r)} (\omega_m)$ contains a contribution that depends on the fermionic bandwidth, i.e. comes from high-energy fermions. The same holds at a finite $n$ and, as such, contradicts the idea of MET that physically detectable interaction corrections come exclusively from low-energy fermions.   We will show that once we treat the renormalization of the chemical potential explicitly, we find   an  additional contribution to the {\it dynamical}  one-loop  self-energy, which 
  acts as a  counter-term to the 
  two-loop  rainbow diagram  and cancels out the contribution from this diagram that depends on $W$.
  After this,  the two-loop self-energy becomes independent of $W$,  i.e., it comes 
    exclusively from fermions with energies well below $W$ as MET anticipated.
    We note in passing that the  renormalization of the bare $\mu_0$ into $\mu$ involves fermions with energies of order $W$, however such renormalization is not  detectable as 
    only the dressed $\mu$ is related to the actual density and has a physical meaning.  

\begin{figure}[]
 \includegraphics[width=\linewidth]{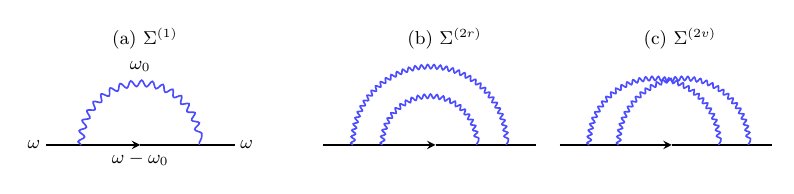}
 \caption{(a) One-loop electron self-energy. (b) Two-loop rainbow-type diagram, (c) Two-loop diagram with vertex correction included.}
 \label{fig:diag_app_A}
 \end{figure}

We now describe the calculations.
Both one-loop and two-loop self-energies are the convolutions of phonon and fermionic propagators with internal momenta and frequencies over which we have to integrate (see Fig. \ref{fig:diag_app_A}).

We begin with the calculation of the one-loop self-energy (Fig.  \ref{fig:diag_app_A}a)
 to the leading order in the e-ph coupling (i.e., to the  leading order in $\lambda_0$).   For this calculation, the renormalization $\mu_0$ into $\mu$ is irrelevant, as it by itself is  caused by e-ph coupling, and for
  $n =0+$ and momenta near the bottom of the band, $\epsilon_p - \mu_0$ 
  can be approximated by $p^2 W/8$.

The calculation of the one-loop diagram to the leading order in $\lambda_0$ is straightforward.  Integrating over phonon frequency and momentum, we  find that  $\Sigma^{(1)} (k, \omega_m)$ is
  independent on $k$ and within logarithmic accuracy is 
\beq
 \Sigma^{(1)}  (\omega_m)  = - \lambda_0 \omega_0 \log{\frac{W}{\omega_0 - i \omega_m} },
 \label{satt_1}
 \eeq 
 where $\lambda_0 = g^2 N_F/(2 \omega^2_0)$ and $N_F = (2/\pi W)$.
 This self-energy can be split into static and dynamic parts:
 \beq
 \Sigma^{(1)} (\omega_m)   = \Sigma^{(1)}  (0) + \Sigma^{(1)}_{dyn}  (\omega_m),
 \label{satt_1_1}
 \eeq 
 where
\bea
 \Sigma^{(1)} (0)   &=& - \lambda_0 \omega_0 \log{\frac{W}{\omega_0}}, \nonumber \\
 \Sigma^{(1)}_{dyn}  (\omega_m) &=& - \lambda_0 \omega_0  \log{\frac{\omega_0}{\omega_0 - i \omega_m}}. 
 \label{satt_1_2}
 \eea 
The first term, which depends on $W$,  accounts for the renormalization of the chemical potential, $\mu = \mu_0 - \Sigma^{(1)} (0)$, and the second, which does not depend on $W$,  is responsible for the physically detectable 
mass renormalization and fermionic damping.  At small $\omega_m$, $\Sigma^{(1)}_{dyn}  (\omega_m) = - i\lambda_0  \omega_m$.

We now incorporate the renormalization of the chemical potential into the new "bare" fermionic Green's function, and do this explicitly rather than formally replacing $\mu_0$ by $\mu$ and writing $\epsilon_p -\mu \approx p^2 W/8$ (doing this would keep Eq. (\ref {satt_1}) intact).  In explicit procedure, we first
reexpress  $\epsilon_p -\mu_0$ as  $\epsilon_p  -\mu- \Sigma^{(1)} (0)$ and  then use 
$\epsilon_p -\mu \approx p^2 W/8$. Substituting into the one-loop diagram and expanding in $\Sigma^{(1)} (0)$, we find the contribution to the {\it dynamical} $\Sigma^{(1)} (\omega_m)$ to order $\lambda^2_0$.
 Combining it with the $O(\lambda_0)$ contribution and expanding in $\omega_m$, we obtain:
 \beq
 \Sigma^{(1)}_{dyn}  (\omega_m) = - i \lambda_0 \omega_m + i \lambda^2_0 \omega_m \log{\frac{W}{\omega_0}}.
 \label{satt_1_3}
 \eeq 

We now compute the two two-loop self-energies, $\Sigma^{(2v)}$ and $\Sigma^{(2r)}$. 
 Both contain $\lambda^2_0$ in the prefactors, and in the calculations to this order,  the renormalization of $\mu_0$ into $\mu$ can be neglected. 
We start with the vertex correction diagram, (Fig.  \ref{fig:diag_app_A}c).
 Integrating explicitly over frequency, expanding tight-binding dispersion $\epsilon_p -\mu_0$ near the bottom of the band as
$\epsilon_p -\mu \approx p^2W/8$, introducing $x = p^2/(8W \omega_0)$ and integrating over the angles between internal fermionic momenta, we obtain for the dynamical part of the vertex correction diagram at small $\omega_m$ 
\beq
 \Sigma^{(2v)}_{dyn} = -i \lambda^2_0 \omega_m Q_v,
 \label{satt_2}
 \eeq 
 where
 \begin{widetext}
 \beq
 Q_v = \iint^\infty _0  dx dy \left[ \frac{2}{(x+1)^2} \frac{1}{y+1} \frac{1}{((x+y+2)^2 -4 xy )^{1/2}}
  + \frac{1}{(x+1)} \frac{1}{y+1} \frac{x+y+2}{((x+y+2)^2 -4 xy )^{3/2}}\right] 
  \approx 1.5.
 \label{satt_3}
 \eeq 
\end{widetext}
Taking the ratio $\Sigma^{(2v)}_{dyn}/\Sigma^{(1)}_{dyn}$, we obtain $\lambda_E \sim \lambda_0$.  We see that there is 
 no 
 smallness of $\lambda_E/\lambda_0$  at $n =0+$. At the same time, this ratio does not diverge when $E_F =0$,
 i.e.  the Eliashberg parameter saturates at $\lambda_E \sim \lambda_0$. 
 For an order of magnitude  estimate, one can describe the crossover between regimes $A$ and $B$ 
  by replacing $\omega_0/E_F$ by $\omega_0/(E_F + a \omega_0)$ with  $a = O(1)$.

 The computation of 
  the rainbow-type  diagram $\Sigma^{(2r)}_{dyn}$, Fig. \ref{fig:diag_app_A}b. 
   proceeds the same way, and the result is 
\beq
 \Sigma^{(2r)}_{dyn} = -i \lambda^2_0 \omega_m Q_r,
 \label{satt_5}
 \eeq 
 where
 \begin{widetext}
 \beq
 Q_r = \iint^{W/\omega_0}_0  dx dy \left[ \frac{2}{(x+1)^3} \frac{1}{y+2} 
  + \frac{1}{(x+1)^2} \frac{1}{(y+2)^2}\right]  =  \log{\frac{ W}{\omega_0}} + \frac{1}{2} + \log{\frac{1}{2}}.
 \label{satt_6}
 \eeq 
\end{widetext}
This diagram contains the one-loop self-energy correction as the internal part, and it is natural to combine
 $\Sigma^{(2r)}_{dyn}$ with $\Sigma^{(1)}_{dyn}$ from (\ref{satt_1_3}). Doing this, we obtain 
 \beq
 \Sigma^{(1)}_{dyn}  (\omega_m)  +   \Sigma^{(2r)}_{dyn}  (\omega_m)= - i \lambda_0 \omega_m 
 - i \lambda^2_0 \omega_m  \left(\frac{1}{2} + \log{\frac{1}{2}}\right).
 \label{satt_1_4}
 \eeq 
 The complete dependence on frequency is given by  
\be
\begin{aligned}
\Sigma^{(1)} + \Sigma^{(2r)} &= - 
\lambda_0 \omega_0 \log \frac{W}{\omega_0 - i \omega_m}   \\
& - \lambda^2_0 \omega_0^2 \frac{1}{\omega_0 - i\omega_n }
    \log  \frac{\omega_0}{2\omega_0-i\omega_m}.
\end{aligned}
 \label{satt_1_4c}
 \ee
We see that the $W-$dependent term cancels out. Combining this with $\Sigma^{(2v)}_{dyn}  (\omega_m)$, we see that 
 the self-energy is expressed as a series in $\lambda_0$. 
There is no other parameter at $n =0+$. 
 
We also emphasize that at $E_F =0$, the phonon $\omega_0$ does not soften as the polarization bubble vanishes identically.   At small but finite $E_F$, there exists a small range of momenta $q$ near $q=0$, where 
$\omega_r = \omega_0 (1-2\lambda_0)^{1/2}$, i.e., MET becomes internally unstable at $\lambda_0 =1/2$.
Still,  the instability against polarons occurs at smaller $\lambda_0 = 1/\pi$  [in terms of $\lambda_p$, used in the analytical calculations in the main text, the two instabilities occur at $\lambda_p = \pi/4$ and $1/2$, respectively ].  

For completeness, we now briefly summarize the results in 3D. To  the leading order in the e-ph coupling,  we have 
$\Sigma^{(1)}_{dyn} = -i \omega_m \lambda^{3D}_0$ 
where  $\lambda^{3D}_0 = g^2 N_F (2 \omega^2_0)$. In region A,   $N_F \sim  (E_F/W)^{1/2} /W \propto n^{1/3}$
 and in region B, $ N_F \sim (\omega_0/W)^{1/2}/W$.  
 The Eliashberg parameter $\lambda^{3D}_E$ is smaller than $\lambda^{3D}_0$ in region A by $\omega_0/E_F$ and again becomes the same as $\lambda^{3D}_0$ in regime B. In the last regime, 
 $\Sigma^{(2v)}_{dyn}$ and $\Sigma^{(2r)}_{dyn}$, corrected by the renormalization of $\mu$, scale as
$ -i \omega_m (\lambda^{3D}_0)^2$.  The canonical MET is broken because $\lambda^{3D}_E \sim \lambda^{3D}_0$, but still,   the self-energy is expressed as regular  series in $\lambda^{3D}_0$. 
There is again no other parameter at $n =0+$. 

 \section{DOS of the Holstein model at smaller $\beta$.}
\label{app_B}
 \begin{figure}[htbp]
	\includegraphics[width=0.8\linewidth]{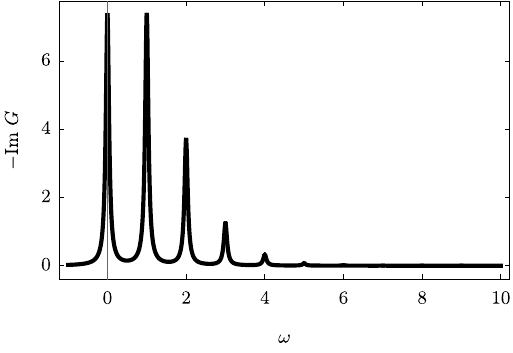}
\caption{DOS of a single site Holstein model for $\beta =1$. 
The residues of the $\delta-$functional peaks are $Z_m = e^{-\beta} \beta^m/m!$.}
\label{fig_ap_a_9}
\end{figure}

In this Appendix,   we briefly analyze the structure of the DOS of the Holstein model at $n =0+$ at small $\beta$. The expression for the Green's function $G^{H} (\omega, n=0)$,   Eq. (\ref{n_2}) in the main text, is valid for all values of $\beta$. At $\beta \ll 1$, it becomes  
 \beq
      G^{H} (\omega, 0) \approx \sum_{m=0}^\infty \frac{\beta^m}{m!} \frac{1}{\omega + i \delta -m \omega_0},
     \label{ap_b_1}
      \eeq
The DOS consists of the set of $\delta$-functions, such as at large $\beta$, but now the largest residue $Z \approx 1$ is for the polaron in $\omega =0$, while the residues of the polarons at larger $\omega = m \omega_0$ rapidly decrease as $Z_m = \beta^m/m!$. We show this DOS in Fig.~\ref{fig_ap_a_9}.
 
\section{Green's function in  the rainbow and self-consistent one-loop approximation in the atomic limit.}
\label{app_A}

In this Appendix we discuss  what would be the form of the fermionic Green's function if we neglected vertex corrections in the order-by-order perturbation theory. For definiteness, we focus on the case of vanishing density 
$n=0+$.  

We consider two approximations: the non-self-consistent rainbow approximation, in which we include only rainbow diagrams for the self-energy (Fig.~\ref{fig:sig_R})  
and the self-consistent one-loop approximation, in which we include all diagrams for the renormalization of the internal fermionic line in the one-loop diagram, but neglect vertex corrections (Fig.~\ref{fig:sig_1L}). 
 Our goal here is to  understand whether keeping vertex corrections, as we did in the eikonal calculations in the main text, is crucial for the description of polarons.  We consider the two approximations separately and in each case we first consider the limit, in which the phonon propagator can be approximated by $\chi (\Omega_m) = \delta (\Omega_m)$, and then analyze the diagrams for the actual $\chi (\Omega_m) = 2\omega_0/(\Omega^2_m + \omega^2_0)$.  As in the main text, we introduce the chemical potential $\mu$ and determine $\mu$ from the condition that the fermionic density is infinitesimally small.

\subsection{Rainbow approximation}

 \begin{figure}[]
	\includegraphics[width=0.95\linewidth]{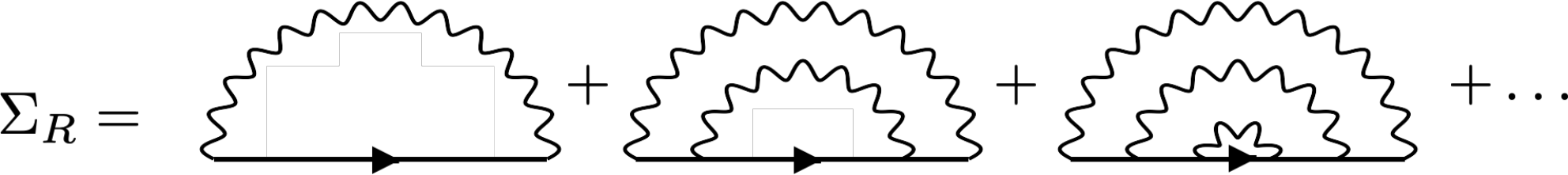}
\caption{Diagrams contributing to the rainbow approximation to the fermion self-energy, shown through third order.}
\label{fig:sig_R}
\end{figure}

  We write $G^{-1}_{R}(\omega) = \omega + \mu - \Sigma_{R} (\omega)$, where the subindex indicates that we compute the self-energy by keeping only rainbow diagrams (one diagram at each loop order), see Fig.~\ref{fig:sig_R}. 
We measure $\omega$, $\mu$ and $\Sigma_{R}$  in units of $\omega_0$ and introduce ${\bar\omega} = \omega/\omega_0$, ${\bar \mu} = \mu/\omega_0$ and ${\bar \Sigma}_{R} = \Sigma_{R}/\omega_0$. For the Green's function, we introduce ${\bar G}_{R} = G_{R}~ \omega_0$.

\subsubsection{\bf{$\delta$-functional bosonic propagator}}

For $\chi (\Omega_m) = \delta (\Omega_m)$, the analytic expression for the self-energy is
\bea
&& {\bar \Sigma}_{R} ({\bar \omega}) = \frac{\beta}{{\bar \omega} + {\bar \mu}} + \frac{\beta^2}{({\bar \omega} + {\bar \mu})^3}  \nonumber \\
&=& \frac{\beta}{{\bar \omega} + {\bar \mu}} \sum_{n=0}^\infty \left(\frac{\beta}{({\bar \omega} + {\bar \mu})^2}\right)^n    
\label{ap_5}
\eea 
 For $|{\bar \omega} + {\bar \mu}| > \sqrt{\beta}$, the series in (\ref{ap_5}) converge, the summation is elementary and yields
 \beq
{\bar \Sigma}_{R} ({\bar \omega}) = \frac{\beta}{{\bar \omega} + {\bar \mu}}  \frac{1}{1-\frac{\beta^2}{({\bar \omega} + {\bar \mu})^2}}
\label{ap_6}
\eeq
Substituting into  $({\bar G}_{R})^{-1} = {\bar \omega} + {\bar \mu} -  {\bar \Sigma}_{R} ({\bar \omega})$ and introducing 
\beq
y = \frac{{\bar \omega} + {\bar \mu}}{\sqrt{\beta}}, ~{\bar \Sigma}_{R} (y) = \sqrt{\beta}  
{\hat{\bar \Sigma}}_{R} (y),~ {\bar G}_{R} (y) =   \frac{{\hat{\bar G}}_{R} (y)}{\sqrt{\beta}},
\label{ap_6_1}
 \eeq
 we obtain
\beq
{\hat{\bar G}}_{R} (y) = \frac{y^2 -1}{y (y^2 - 2)} 
\label{ap_7}
\eeq
We see that ${\hat{\bar G}}_{R} (y)$ has  poles at $y = \pm \sqrt{2}$, i.e., at ${\bar \omega} + {\bar \mu} = \pm \sqrt{2\beta}$.  The poles are within the radius of convergence of the series in (\ref{ap_5}), which in terms of  $y$  is $|y| >1$. The residue of each pole 
 is $Z_0 =1/4$.   The condition $n = 0+$ sets the lower pole at $\omega =0$, hence ${\bar \mu} = - \sqrt{2 \beta}$. The other pole is at ${\bar \omega} = 2 \sqrt{2\beta}$. 
 
 The Green's function in (\ref{ap_7}) also has a pole at $y =0$ with residue $Z_1 =1/2$.  The corresponding retarded 
 ${\hat{\bar G}}_{R} (y + i0)$  satisfies the  requirement
 $-(1/\pi) \int_{-\infty}^{\infty} {\text {Im}} {\hat{\bar G}}_{R} (y + i0) dy = 2Z_0 + Z_1 =1$.  However, the pole at $y=0$ in (\ref{ap_7}) 
  is outside the radius of convergence of the series in (\ref{ap_5}) and from this perspective  should not be counted. 
 
 It is intuitively expected that under proper analytic continuation Eq. (\ref{ap_6}) becomes valid also for $|y| <1$. Below we explicitly show how to analytically continue the series in 
 (\ref{ap_5}) to reproduce (\ref{ap_7}).  To the best of our knowledge, this has not been demonstrated before. 
 
\begin{figure}[htbp]
	\includegraphics[width=\linewidth]{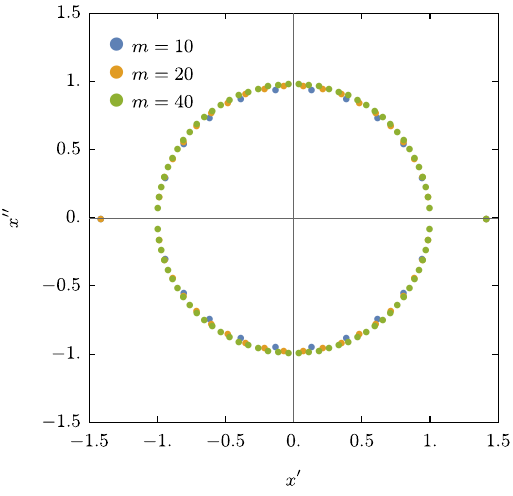}
\caption{Rainbow approximation for $\chi (\Omega_m) = \delta (\Omega_m)$. The location of the poles of the Green's function  ${\hat {\bar  G}}^{(m)}_R (z)$, $z = y' + i y''$  for 
% $m =10,20$, and $40$.
%AC
$m =10, 15$ and $25$.  
 As $m$ increases the poles become more dense and 
  their residues decrease.  At $m = \infty$, the former poles  form a continuous branch cut along a circle of radius  $r=1$.}
\label{fig_ap_a_2}
\end{figure}

 To proceed, we introduce the partial sum
 \beq
{\hat{\bar \Sigma}}^{(m)}_{R} (y) =  \frac{1}{y} \sum_{n=0}^m \left(\frac{1}{y^2}\right)^n =  y \frac{1-\left(\frac{1}{y^2}\right)^{m+1}}{y^2-1}   
\label{ap_8}
\eeq
 and find the location of the poles of ${\hat{\bar G}}^{(m)}_{R} (y) = 1/(y  - {\hat{\bar \Sigma}}^{(m)}_{R} (y))$.  A straightforward analysis shows that at a given $m$,  these are two poles at real $y_{1,2} = \pm a$, with $a>1$, inside the region of convergence of the summation in (\ref{ap_5}), and $2(m-2)$ poles in the complex plane of $z = y' + i y''$,  located inside the circle $|z| =1$, i.e., outside the radius of convergence of the infinite sum.   For even $m$, the poles appear in "quadrants",  at $z= \pm c \pm id$,  for odd $m$, there are two additional poles at $z = \pm ie$. (see Fig. \ref{fig_ap_a_2})
 The presence of the poles in the upper half-plane of $y$ (i.e., the upper half-plane of $\omega$)  indicates that ${\hat{\bar G}}^{(m)}_{R} (y +i0)$ is not a retarded function. The actual retarded function has to be obtained by taking the limit $m \to \infty$ {\it combined} with the analytic continuation from larger $|z|$, where the series converge.  
  
  As $m$ increases, the residue of each pole decreases as $1/m$ and the number of poles increases as $m$. 
   At $m = \infty$, there appears a contour  in the complex plane,  at which ${\hat{\bar G}}^{(m=\infty)}_R (z) =  {\hat {\bar G}}_R (z) $ has a branch cut (Fig. \ref{fig_ap_a_2}). To analyze how to analytically continue through it, we consider as an example an even $m$, when there is a pole along the imaginary axis, and vary $z = i {\tilde z}$ along this axis.  Using (\ref{ap_8}), we obtain
    \beq
   {\hat{\bar G}}^{(m)}_{R} (i{\tilde z}) = -i \frac{{\tilde z}^2 +1}{{\tilde z} ({\tilde z}^2 +2)} J ({\tilde z}), 
   \label{ap_9}
\eeq
 where
 \beq
 J ({\tilde z})  = \frac{{\tilde z}^{2 m}}{{\tilde z}^{2m} - \frac{1}{{\tilde z}^2+2}}.   
    \label{ap_10}
\eeq 
 We focus on positive ${\tilde z}$ in the upper half-plane of the complex frequency, where the true retarded function must be analytic. 
   
At large $m$,  $J ({\tilde z})$ has a pole at ${\tilde z} \leq 1$. 
%(see Fig. \ref{fig_ap_a_2}). 
Introducing ${\tilde z} = 1-\epsilon$ and approximating ${\tilde z}^{2m}   = e^{2 m \log{\tilde z}}$ as 
 $e^{-2 m \epsilon}$, we reexpress (\ref{ap_10}) as
\beq
 J (\epsilon)  = \frac{e^{-2 m \epsilon}}{e^{-2 m \epsilon} - \frac{1}{3}}.
    \label{ap_11}
\eeq                  
The function $ J (\epsilon)$ has a pole at $\epsilon = \epsilon_0 = \log{3}/(2 m)$. Near the pole, 
\beq
 J (\epsilon)  \approx \frac{1}{2m} \frac{1}{\epsilon_0 - \epsilon}.
    \label{ap_12}
\eeq  
  We now keep $\epsilon$ finite and take the limit $m \to \infty$. 
  The residue of the pole  vanishes at $m = \infty$ and transforms into a branch cut.  
A simple analysis of (\ref{ap_11}) shows that 
$J(\epsilon)$ jumps by a finite number between $\epsilon <0$ and $\epsilon >0$: 
\beq
J (\epsilon) \approx  1 + \frac{1}{3} e^{-2m |\epsilon|} ~~{\text {at}}~~ \epsilon <0 ({\tilde z} >1)
\eeq
and 
\beq
J (\epsilon) \approx -3 e^{-2 m \epsilon} ~~ {\text {at}}~~ \epsilon >0 ({\tilde z} <1)
\eeq
Because all derivatives of $J (\epsilon <0)$ vanish at $m = \infty$,  the natural analytical continuation of $J$ from  $\epsilon <0$ (${\tilde z} >1$), where the series in (\ref{ap_5})) converge to $\epsilon >0$ (${\tilde z} <1$) is to just set $J (\epsilon) \equiv  1$ for ${\epsilon} >0$. Substituting into (\ref{ap_9}) we find that the corresponding ${\hat{\bar G}}_{R} (z)$ has  the same form as the
analytic ${\hat{\bar G}}_{R} (y)$ from (\ref{ap_7}),  rotated to the imaginary axis.
We verify that Eq. (\ref{ap_7}) is  reproduced in the  analytical continuation through the branch cut along any direction in the plane of the complex $z$.  

The outcome of this analysis is that  the summation of the infinite series for the self-energy outside the 
 range of convergence  should be accompanied by a proper analytical continuation. The latter can be done  by considering a finite number of terms in the sum, performing analytical continuation,  and then taking the limit in which the number of terms in the sum  approaches infinity. 

The analytic Green's function ${\hat{\bar G}}_{R} (\omega+i0)$ has three $\delta-$functional peaks: a central peak at $\omega = \sqrt{2\beta} \omega_0$, with the residue $Z_1=1/2$ and two satellite peaks at $\omega =0$ and $\omega = 2\sqrt{2\beta} \omega_0$, each with the residue $Z_0 =1/4$.  The central peak is at the same location as for free fermions (with the same $\mu$). Half of the weight of this peak moves to satellite peaks.   This is quite different from the result of eikonal summation for the same $\chi (\Omega_m) = \delta(\Omega_m)$. There, the free-fermion peak at $\omega = -\mu = \beta \omega_0$ broadens into a Gaussian. 

\subsubsection{\bf{Einstein boson}}
 \begin{figure}
  \centering
    \noindent
    \includegraphics[]{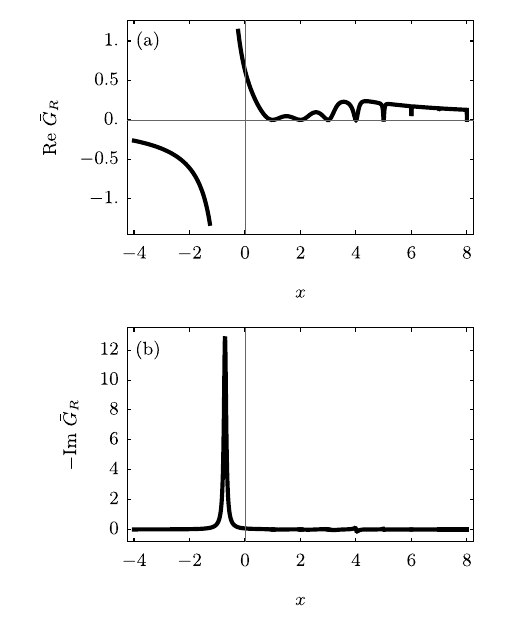}
\caption{Real (a) and imaginary (b)  parts of the  Green's function ${\bar G}_{R} (x+i0)$ on the real axis, obtained by the numerical summation of infinite diagrammatic series in (\ref{ap_2}).   The function ${\bar G}_{R} (x+i0)$ has a single pole at a negative $x$ and a number of zeros at integer positive $x=m$. Around each zero the function is quadratic and the slope rapidly increases with increasing $m$.}
\label{fig_ap_a_10}
\end{figure} 

For $\chi (\Omega_m) = 2\omega_0/(\Omega^2_m + \omega^2_0)$ the rainbow series yields
\beq
{\bar \Sigma}_{R} ({\bar \omega}) = \frac{\beta}{{\bar \omega} + {\bar \mu} - 1} + \frac{\beta^2}{({\bar \omega} + {\bar \mu} - 1)^2  ({\bar \omega} + {\bar \mu} - 2) }  + ....
\label{ap_1_a}
\eeq
 where, we recall, ${\bar \omega} = \omega/\omega_0$ and ${\bar \mu} = \mu/\omega_0$.  
 Eq. (\ref{ap_1_a})  be reexpressed in terms of $x = {\bar \omega} + {\bar \mu}$ as
\bea
{\bar \Sigma}_{R} (x, \beta) &=&  \sum_{n=1}^\infty \frac{\beta^n}{x-n} \left(\prod_{m=1}^{n-1} \frac{1}{(x-m)}\right)^2 \nonumber \\
&& = \sum_{n=1}^\infty \frac{\beta^n}{x-n}  \frac{1}{((1-x)_{n-1})^2}
 \label{ap_2}
\eea 
 where $(a)_b$ is a Pochhammer function.
   The sum converges for all non-integer $x$ and diverges quadratically at positive integer $x$. Around each $x =m$, 
  the Green's function ${\bar G}_{R} (x)$ scales as $(x-m)^2$.  
   The summation in (\ref{ap_2}) can be easily done numerically. We show the result in Fig. \ref{fig_ap_a_10}.  
\begin{figure}
   	\includegraphics[width=\linewidth]{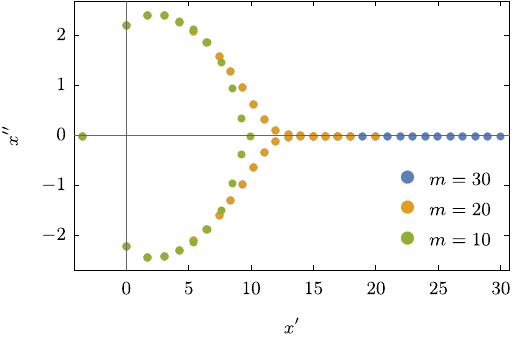}
\caption{Location of the poles in ${\bar G}^{(m)}_{R} (z)$ in the plane of complex $z=x' + i x''$ for $m =10,20,30$.  
The structure of the poles at a finite distance from the real axis does not  change with $m$, the new poles as $m$ increases appear at larger $z$ very near the real axis. }
\label{fig_ap_a_4}
\end{figure}

We see that ${\bar G}_{R} (x,\beta)$ has a single pole at negative $x = -x_0$ and a number of zeros at integer positive $x=m$.
The condition that the density  $n = 0+$ confines this pole to $\omega =0$, i.e., it sets  ${\bar \mu} = -x_0$.
The residue of the pole $Z$ can be obtained by expanding each term of the series for $\Sigma_{R} (\omega,\beta)$ to the linear order in $\omega$.   Collecting  the prefactors, we obtain
 \beq
 Z^{-1} = 1 + \sum_{n=1}^\infty \beta^n D_n
 \label{ap_3}
 \eeq
  where  
  \beq
  D_n  =2 \frac{\Psi (n-{\bar \mu}) - \Psi (1-{\bar \mu})}{((1-{\bar \mu})_{n-1})^2 (n-{\bar \mu})} + 
  \frac{1}{((1-{\bar \mu})_{n})^2},     
  \label{ap_4}
 \eeq
 and $\Psi (x)$ is the di-$\Gamma$ function. 
  For $\beta \gg 1$, $Z\approx 0.25$ (one needs a truly large $\beta \sim 10^3$ to reach this value). For smaller $\beta$, $Z$ is larger but still smaller than one.  Because there is only one pole with residue $Z <1$, the condition $\int_{-\infty}^\infty dx {\text {Im}} {\bar G}_R (x + i0,\beta) =1$ is not satisfied, suggesting that
 ${\bar G}_R (z,\beta)$ has poles in the upper half-plane of complex  $z = x' + i x''$, i.e., it is not a 
  retarded function.  There is no requirement that the full Green's function in the rainbow approximation has to be retarded because this approximation is not a self-consistent one (the reducible diagrams for $G_R (x,\beta)$ are not included into the dressing of the propagator of an internal fermion in the diagrammatic series for $\Sigma_R (x,\beta)$). 
      
To analyze the structure of ${\bar G}_R (z,\beta)$  
 we again introduce the partial sum
 \beq
{\bar \Sigma}^{(m)}_{R} (x,\beta) =- \sum_{n=1}^m \frac{\beta^n}{n-x}  \frac{1}{((1-x)_{n-1})^2}
 \label{ap_2_a}
\eeq  
 and analyze the positions of the poles of ${\bar G}^{(m)}_{R} (z,\beta) = 1/(z - {\bar \Sigma}^{(m)}_{R} (z,\beta))$  at a given $m$. We show the results for $m= 10, 20, 30$ in Fig. \ref{fig_ap_a_4}.  
   For each $m$, there are two poles strictly along the real axis, one at negative and one at positive $x$,  and $2(m-1)$ poles in the complex plane,  with $m-1$ poles in the upper half-plane of $z$.  This is similar to the case of $\delta-$functional $\chi (\Omega_m)$, but there is one crucial difference:  the positions and residues of a 
   finite number $m_0 \sim \sqrt{\beta}$  of these poles  do not vary with $m$ at $m \gg m_0$ and remain separated 
     with finite residues $Z_i \sim 1/\beta^{1/4}$ in the limit $m \to \infty$.  The residues of all other poles vanish exponentially with $m$ as   $Z_i \sim (m_0/m)^{m_0}$.  Of the two poles on the real axis, the one at negative $x =-x_0$
   ($x_0 \approx - 1.35 \sqrt{\beta}$ at large $\beta$) has a $m-$independent residue $Z$, which approaches $1/4$ at large $\beta$, as we just said,  while the residue of the other pole vanishes exponentially at $m \to \infty$.
  Because a finite number of poles in the upper half-plane remain at $m = \infty$, the function 
  ${\bar G}^{(m =\infty)}_{R} (z,\beta)  = {\bar G}_{R} (z,\beta)$ is obviously non-analytic. 

 \begin{figure}
  \centering
    \noindent
    \includegraphics[width=\linewidth]{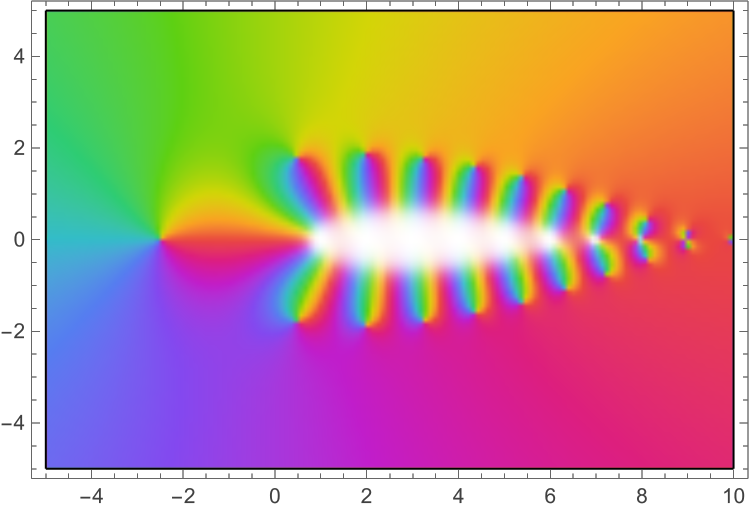}
\caption{Color plot of the argument of $1/{\bar G}_{R} (z,\beta)$ in the complex plane.  The points,  around which  the argument varies by $2\pi$ under anticlockwise rotation, are the dynamical vorices -- the zeros of $1/{\bar G}_{R} (z,\beta)$, i.e., the poles of ${\bar G}_{R} (z,\beta)$. }
\label{fig_ap_a_30}
\end{figure} 

After some extensive analysis, we found the closed-form analytical expression for the sum in 
\eqref{ap_2}:
\beq
{\bar \Sigma}_{R} (x,\beta)  \equiv \frac{\beta}{x-1}  _{1}F_{2} \left[1; 1 - x, 2 - x; \beta\right]
 \label{ap_2_a_1}
\eeq
where  $_{1} F_{2} \left[a; b;c\right]$ is a generalized Hypergeometric function.  This function perfectly matches the result of the numerical summation in \eqref{ap_2}.   The exact solution is  advantageous in that it can be easily extended to the upper complex plane by replacing $x$ by $z$.  In Fig. \ref{fig_ap_a_30} we plot the argument of 
$1/{\bar G}_{R} (z,\beta)$ in the complex plane. The points around which  the argument varies by $2\pi$ under an anticlockwise rotation are the zeros of $1/{\bar G}_{R} (z,\beta)$, i.e., the poles of ${\bar G}_{R} (z,\beta)$.
We see that there are indeed poles in the upper half-plane at $m \to \infty$. This confirms that ${\bar G}_{R} (z,\beta)$ is not an analytic function.   The conclusion here is that the rainbow approximation for the interaction with an Einstein phonon  should not be used, as it leads to the unphysical result for the Green's  function. 

 \begin{figure}[htbp]
	\includegraphics[width=\linewidth]{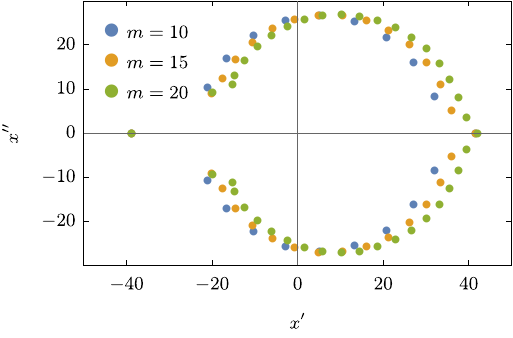}
\caption{The location of the poles of ${\bar G}^{(m)}_R (z)$, $z = x' + i x''$, for $\beta = 800$ and 
 $m =10,15$, and $20$. 
 The structure of the poles is essentially the same as in Fig. \ref{fig_ap_a_2}.}
\label{fig_ap_a_5}
\end{figure}

  We note in passing  that the results that we obtained earlier for the $\delta-$functional propagator are reproduced  if we take the double limit $m \to \infty$ and $\beta \to \infty$ and keep 
 $\beta^2 >> m$. In  this limit,   the density of the poles in the upper half-plane  increases with $m$, and the poles eventually 
 form a branch cut along a circle with a radius equal to $\sqrt{\beta}$, see Fig. \ref{fig_ap_a_5}. This is 
   the same structure as we found above.   The two poles on the real axis move to  $x = \pm \sqrt{2\beta}$ outside of the circle, and their residues become equal and approach $1/4$. The  analytic Green's function  in this limit 
    is obtained via analytical continuation through the branch cut and has an additional pole at  $x=0$ with the residue $Z=1/2$. 

\subsection{Self-consistent one-loop approximation}

 \begin{figure}[]
	\includegraphics[width=0.95\linewidth]{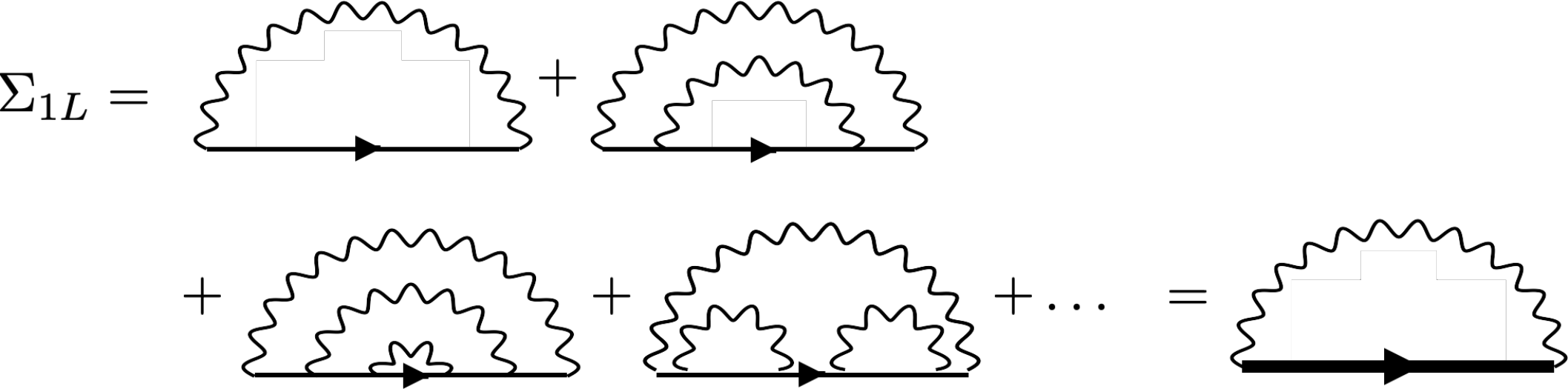}
\caption{Diagrams contributing to the self-consistent one-loop self-energy, shown through third order. Bold arrowed line denotes the fully dressed (self-consistent) Green's function.}
\label{fig:sig_1L}
\end{figure}

 We now extend the perturbation series to include all renormalizations of the internal fermionic propagator in the self-energy diagram , e.g., two subsequent rainbow renormalizations.  We show the corresponding diagrams in Fig.~\ref{fig:sig_1L}. 
  The sum of the diagrammatic series can be formally represented in a compact form by replacing the internal fermionic line by the fully renormalized Green's function.  However,  this substitution is fully justified  only when the series converge. When they do not, the series have to be analytically continued, as we did in the previous section.
  
  As before, we first consider a boson with a $\delta$-functional propagator and then show the results for an Einstein boson.   
  
\subsubsection{\bf {$\delta$-functional bosonic propagator}}
We use the same notations, Eq. (\ref{ap_6_1}),  as before.
The analytic expression for the self-energy is 
\bea
&&{\hat{\bar \Sigma}}_{1L} (y) = \frac{1}{y}\left(1 + \frac{1}{y^2} + \frac{2}{y^4} + \frac{5}{y^6} +  \frac{14}{y^8} + ...\right)\nonumber \\
&&= \frac{1}{y} \left(\sum_{n=0}^\infty  \left(\frac{1}{y^2}\right)^n \frac{(2n)!}{n! (n+1)!}\right).
\label{ap_15_1}
\eea 
Using the Stirling formula for factorials, one can easily verify that the series converges for $y >2$. 
 For such $y$, the summation of the infinite series yields
\beq
{\hat {\bar \Sigma}}_{1L} (y) = \frac{1}{y}  \frac{2}{1 + \sqrt{1-\frac{4}{y^2}}}.
\label{ap_15_2}
\eeq 
Substituting into $({\hat{\bar G}}_{1L}(y))^{-1} = y -{\hat {\bar \Sigma}}_{1L} (y)$, we obtain
\beq
{\hat{\bar G}}_{1L}(y)  \frac{2}{y + \sqrt{y^2-4} {\text{sign}} y}.
\label{ap_15}
\eeq
The same result is obtained if we replace the series for ${\hat{\bar \Sigma}}_{1L}$ with an effective one-loop 
 expression
 \beq
 {\hat{\bar \Sigma}}_{1L} (y) = {\hat{\bar G}}_{1L} (y)
 \eeq
  and solve the self-consistent equation
  \beq
  ({\hat{\bar G}}_{1L} (y))^{-1} = y - {\hat{\bar G}}_{1L} (y)
 \label{ap_16}
\eeq    
subject to ${\hat{\bar G}}_{1L} (y) =1/y$ at $|y| \to \infty$. 

  Taken on its own, the function ${\hat {\bar G}}_{1L}(y)$ in (\ref{ap_15}) is an analytic function of $y$
  in the upper half-plane of $y$.   
 The imaginary part of this Green's function is  
  \beq
 {\text {Im}} {\hat{\bar G}} (y + i0) = - \sqrt{1 -\frac{y^2}{4}} \theta (|y|-2)  
 \label{ap_17}
\eeq 
where $\theta (x) =1$ when $x <0$. 
 The corresponding DOS in units of the actual $\omega$ is
 \beq
 N(\omega) = \frac{2}{\pi \omega_0 \sqrt{\beta}} \sqrt{\frac{\omega}{4\omega_0 \sqrt{\beta}} \left(1 - \frac{\omega}{4\omega_0 \sqrt{\beta}}\right)},~~ x < 4 \omega_0 \sqrt{\beta} 
 \eeq    
   The normalization condition $-(1/\pi) \int_{-\infty}^\infty {\text {Im}}{\hat{\bar G}} (y + i0)=1$ is satisfied. 

It is tempting to associate ${\hat{\bar G}}_{1L} (y + i0)$ from (\ref{ap_15}) with the true retarded Green's function, and we argue below that this is indeed the case.  However, we recall that the perturbation series converge only for $|y| >2$, while $ {\text {Im}}{\hat{\bar G}} (y + i0)$ is non-zero at $y <2$, when there is no convergence.  This again implies that an analytical continuation is required.

 \begin{figure}[htbp]
	\includegraphics[width=\linewidth]{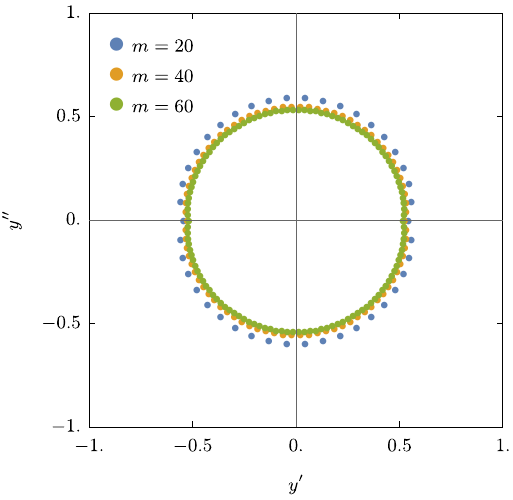}
\caption{Poles of ${\hat {\bar G}}^{(m)}_{1L} (1/z)$  for complex $z = y' + i y''$ in the self-consistent one-loop approximation for phonon $\chi (\Omega_m) = \delta (\Omega_m)$.  The results are for $m= 20, 40$ and $60$.
  As $m$ increases, the poles get denser and their residues decrease. At $m =\infty$ the former poles form a branch cut along a circle with radius $|z|=2$ ($1/|z| =1/2$). }
\label{fig_ap_a_7}
\end{figure}
 
To proceed,  we again introduce the partial sum for the self-energy
\beq
  {\hat {\bar \Sigma}}^{(m)}_{1L} (y) =  \frac{1}{y} \sum_{n=0}^m \left(\frac{1}{y^2}\right)^n \frac{(2n)!}{n! (n+1)!}
\label{ap_18}
\eeq 
The sum can be evaluated exactly. To shorten the presentation, we present the result right for ${\hat{\bar G}}^{(m)}_{1L} (y) = 1/(y - {\hat {\bar \Sigma}}^{(m)}_{1L} (y))$. It is   
\begin{widetext}
\beq
  \left({\hat{\bar G}}^{(m)}_{1L} (y)\right)^{-1} = \frac{y +  \sqrt{y^2-4} {\text{sign}} y}{2}  + \frac{1}{y^{2m+1}} \frac{(2m)!}{m! (m+1)!} \left(-1+ {_2}F_{1} \left(1, \frac{1}{2} +m, 2+m, \frac{4}{y^2}\right)\right)
  \label{ap_19}
\eeq 
\end{widetext}
where $_2F_1$ is a Hypergeometric function. 
Analyzing the location of the poles of ${\hat{\bar G}}^{(m)}_{1L}$ (zeros of  $({\hat{\bar G}}^{(m)}_{1L})^{-1}$ from (\ref{ap_19})) at a given $m$,  we again find a set of poles in the complex plane of $z = y' + i y''$ (see Fig. \ref{fig_ap_a_7}). 
  When $m$ increases, the poles become more dense and their residues decrease. At $m \to \infty$, there appears a circle in the complex plane of $z$ with a radius equal to $2$, where  $ {\hat {\bar \Sigma}}^{(m)}_{1L} (z)$ has a cut branch. Then one needs to continue analytically through the branch cut to obtain an analytic retarded function 
 ${\hat{\bar \Sigma}}^{ret}_{1L} (z)$.  To see how to do this, let us move through the branch cut along the real $z=y$.  We set $y = 2(1 -\epsilon)$ and treat $\epsilon$ as small.   Substituting into 
 (\ref{ap_19}), we find 
 \beq
  \left({\hat{\bar G}}^{(m)}_{1L} (y)\right)^{-1} = \frac{y +  \sqrt{y^2-4} {\text{sign}} y}{2}  + 
  \frac{e^{(2m+1) \epsilon}}{\sqrt{\pi m}}.
  \label{ap_20}
\eeq 
For $\epsilon <0$ the last term exponentially approaches zero at $m \to \infty$, indicating that the series 
of perturbations converge. At $\epsilon <0$, it diverges, signaling the divergence of the perturbation series.  Because 
 all  derivatives of the second term in (\ref{ap_20}) vanish at $m \to \infty$ taken at $\epsilon <0$, the natural analytic continuation through the branch cut is to set this term equal to zero at $\epsilon >0$.  Doing this and then setting $m \to \infty$, we obtain 
 \beq
  \left({\hat{\bar G}}_{1L} (y)\right)^{-1} = \frac{y +  \sqrt{y^2-4} {\text{sign}} y}{2},
  \label{ap_21}
  \eeq
   which coincides with (\ref{ap_15}).   We see therefore that the solution of the self-consistent one-loop equation 
   for ${\hat{\bar G}}_{1L} (y)$ is valid for all $y$ and yields a semicircular DOS centered at $y=0$, yet to justify this result,  one has to identify the branch but and analytically continue through it.
   
\subsubsection{\bf {Einstein boson}}
       \begin{figure}[htbp]
	\includegraphics[width=\linewidth]{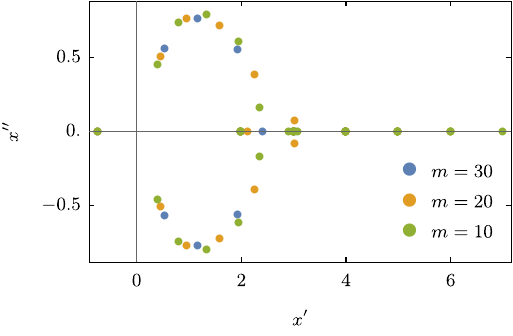}
\caption{Self-consistent one-loop approximation for an Einstein phonon. Poles of ${\bar \Sigma}^{(m)}_{1L} (z)$ for complex $z = x'+ix''$  for $m=5,6$ and $7$. Contrary to the case of rainbow approximation, Fig. \ref{fig_ap_a_4},  the poles away from the real axis get more dense with 
 increasing $m$. }
\label{fig_ap_a_11}
\end{figure}
For $\chi (\Omega_m) = 2\omega_0/(\Omega^2_m + \omega^2_0)$ we again use $x = {\tilde \omega} + {\tilde \mu}$. The perturbation  series for the self-energy  yield
\begin{widetext}
\bea
&&{\bar \Sigma}_{1L} (x,\beta) = \frac{\beta}{x-1} + \frac{\beta^2}{(x-1)^2 (x-2)} + \frac{\beta^3}{(x-1)^2 (x-2)^2} \left(\frac{1}{(x-3)} + \frac{1}{x-1}\right) \nonumber \\
&& + \frac{\beta^4}{(x-1)^2 (x-2)^2} \left(\frac{1}{(x-3)^2 (x-4)} + \frac{1}{(x-2) (x-3)^2} + \frac{2}{(x-1)(x-2) (x-3)} + \frac{1}{(x-1)^2 (x-2)} \right)  + ... 
 \label{ap_22}
\eea    
\end{widetext}
As before, we introduce a partial sum of $m$ terms, ${\bar \Sigma}^{(m)}_{1L} (x, \beta)$ and  analyze the poles of ${\bar G}^{(m)}_{1L} (z, \beta)$ in the complex plane of $z = x' + i x''$.  We show the results for $m=5,6$ and $7$ in Fig. \ref{fig_ap_a_11}.  We see that there are poles very close to the integer $x'=n$ on the real axis and poles in the complex plane, half of which are in the upper half-plane.  This implies that ${\bar \Sigma}^{(m)}_{1L} (z, \beta)$ is a non-analytic function at a finite $m$,  like we found in the rainbow approximation for an Einstein phonon.  However, contrary to the rainbow approximation, the number of poles away from the real axis increases with $m$, i.e., the poles become more dense (this is best seen in Fig. \ref{fig_ap_a_11} for $0<x'<2$).   We conjecture that at $m = \infty$ these poles  form a branch cut along some closed contour, in which case the analytic ${\bar \Sigma}_{1L} (x, \beta)$ is obtained by 
 an analytical continuation through the branch cut, as for the case of a $\delta-$functional $\chi (\Omega_m)$. 
 
        \begin{figure}[htbp]
        \centering
    \noindent
    \includegraphics[]{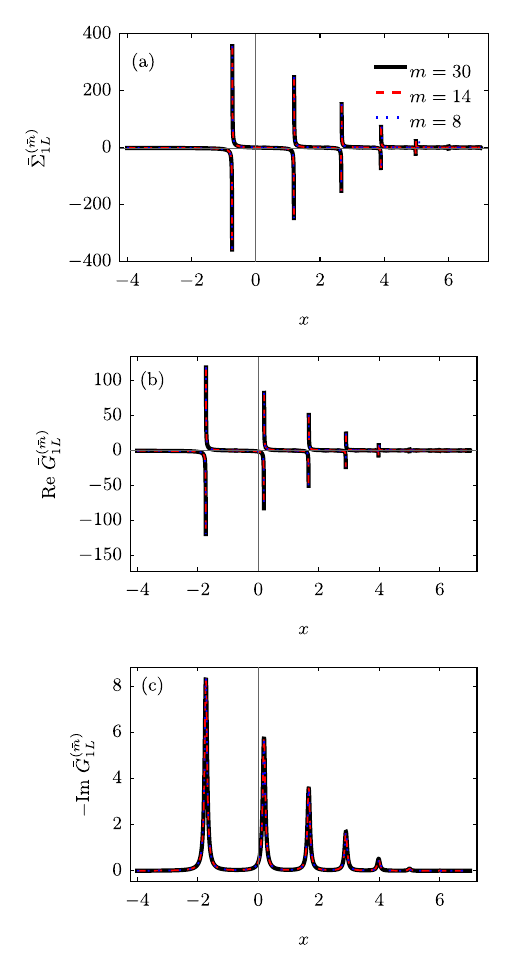} 
\caption{Self-consistent one-loop approximation for an Einstein phonon. 
 (a) Self-energy ${\bar \Sigma}^{({\bar m})}_{1L} (x, \beta)$ recast as continued fractions, with ${\bar m}$ terms retained in the fractions.  The results are for $\beta =3$ and  ${\bar m} =8, 14, 30$. There is essentially no difference between the expressions for these ${\bar m}$, which implies that this is the correct result for the actual ${\bar \Sigma}_{1L} (x, \beta)$.  (b) The real part of the Green's function  $\text{Re}~{\bar G}^{({\bar m})}_{1L} (x, \beta)$. The results for all three values of ${\bar m}$ are essentially identical implying that this is the true ${\bar G}_{1L} (x, \beta)$. (c) The imaginary part of the Green's function  $-\text{Im}~{\bar G}^{({\bar m})}_{1L} (x, \beta)$, consisting of a set of $\delta$-functions.  For better presentation we  added the damping $\gamma =0.05$.   }
\label{fig_ap_a_12}
\end{figure}
 
To proceed, we note that the diagrammatic series for ${\bar \Sigma}_{1L} (x, \beta)$ can be cast into the form of continued fractions, like in the eikonal calculation in the main text.  Namely, Eq. (\ref{ap_22}) can be reexpressed as 
  \beq
     {\bar \Sigma}_{1L} (x, \beta) = \frac{\beta}{x-1 - \beta G_2 (x, \beta)}, 
     \eeq
      \bea
      G_2 (x, \beta) &=&  \frac{1}{x-2 - \beta G_3 (x, \beta)} \nonumber \\
      && .... \nonumber \\
      G_n (x, \beta) &=&  \frac{1}{x-n -\beta G_{n+1} (x, \beta)}
      \label{ap_25}
      \eea
 We argue that, as in the eikonal case, the high-order terms in the series with ${\bar m} > O({\sqrt{\beta}})$ are small and irrelevant. To see this, we introduce partial sums of ${\bar m}$ terms and label then as
  ${\bar \Sigma}^{{\bar m}}_{1L} (x, \beta)$.  We plot ${\bar \Sigma}^{{\bar m}}_{1L} (x, \beta)$ in Fig. \ref{fig_ap_a_12}a  for $\beta =3$ and $m =10,14$ and $30$. We see that the results are virtually indistinguishable between these ${\bar m}$. We verified that the self-energy stays the same for larger ${\bar m}$. In panels (b) and (c) of this figure we plot the real and imaginary parts of the Green's function 
  for the same ${\bar m}$.  reexpressing the last result in terms of $\omega$, we see  that the DOS consists of series of $\delta-$functional peaks, like the eikonal DOS in the main text. The verified that the first peak is at $x =-x_0$, and $x_0  \propto \sqrt{\beta}$. This sets the value of the chemical potential at $\mu =- \omega_0 x_0$. We do a more detailed comparison with the eikonal DOS in the following. 
 
        \begin{figure}[htbp]
	\includegraphics[width=\linewidth]{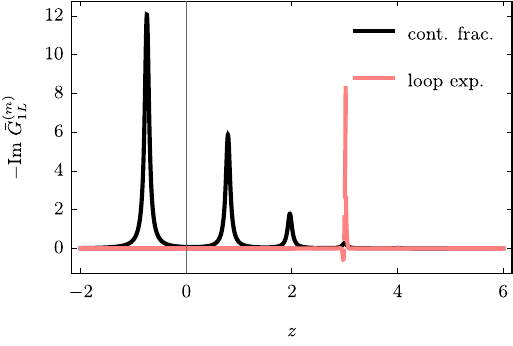}
\caption{Self-consistent one-loop approximation for an Einstein phonon.  The 
The DOS obtained (i) using ${\bar G}^{(m)}_{1L} (z, \beta)$  from the loop expansion for $m=8$ (an orange line) and (ii) using the continued fractions series with the same number of terms (a blue line). The DOS from continued fractions is very close to the exact DOS, while the one from the loop expansion is very different.  }
\label{fig_ap_a_16}
\end{figure}
It is instructive to compare the direct perturbative expansion and continued fraction series taken to the same order.
  In Fig. \ref{fig_ap_a_16} we compare the DOS in the two approaches for $m = {\bar m}=8$.   We see that the DOS obtained using the continued fractions form for the self-energy (the blue line) essentially reproduces the  true DOS, whereas the perturbative result is very far from it.   This is another indication that  the direct loop expansion does not converge and requires an analytical continuation, while the continued fraction series for the self-energy nicely converge.  
  
  Direct loop expansion still yields reliable results  at  small $\beta$. Here, one can expand near each integer $x=n$ as $x = n + \delta_n$ and obtain the set of poles at small $\delta_n$ and find their residues.  As an example, consider $n=1$.  A straightforward analysis of the diagrammatic series leading to  (\ref{ap_22}) shows that the self-energy ${\bar \Sigma}_{1L} (1+ \delta, \beta)$  is
  \beq
  {\bar \Sigma}_{1L} (1+ \delta, \beta) =1 -\frac{\beta}{\delta} + \left(\frac{\beta}{\delta}\right)^2 - \left(\frac{\beta}{\delta}\right)^3 + \left(\frac{\beta}{\delta}\right)^4 + ...
  \label{ap_24}
  \eeq
    The series are conditionally convergent and yield  ${\bar \Sigma}_{1L} (1+ \delta, \beta) = 1/(1 + \beta/\delta)$. The Green's function for vanishing $\delta$ is then
  \beq
  {\bar G}_{1L} (1 + \delta, \beta) \approx \frac{\beta}{\delta}
  \eeq
  The Green's function has a pole at $x=1$, with residue $\beta$. 
  Equivalently, one can obtain  $ {\bar G}_{1L} (\delta, \beta) \approx 1/(\delta + \beta)$,
 $ {\bar G}_{1L} (2 + \delta, \beta) \approx (\beta^2/4)/(\delta - \beta^2/2)$, etc.  The resulting DOS consists of the set of poles  near integer $x$ with  residues $Z_n =1/(n!)^2$.    This is qualitatively similar to the DOS at larger $\beta =3$ in Fig. \ref{fig_ap_a_12}. 
    
       \begin{figure}[htbp]
        \centering
    \noindent
    \includegraphics[]{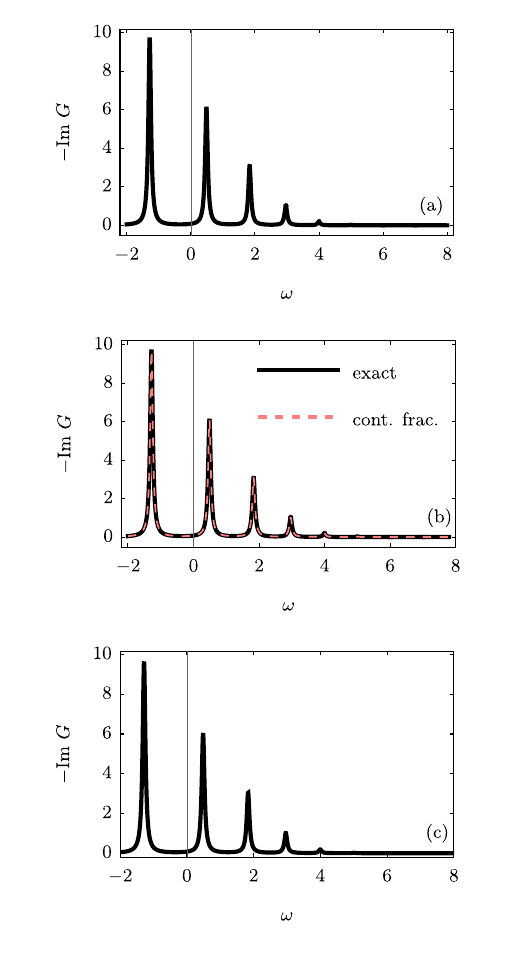} 
\caption{(a) Exact Im ${\bar G}_{1L} (x + 0.01i,\beta)$ within one-loop approximation for an Einstein phonon (see Eq. \ref{ap_25}); (b) The comparison between the exact Im ${\bar G}_{1L} (x + 0.01i,\beta )$  and Im ${\bar G}^{({\bar m})}_{1L} (x + 0.01i,\beta)$ obtained from the continued fractions series  for $m=30$; (c) Im ${\bar G}_{1L} (x + 0.01I,\beta)$ , 
  obtained from the numerical solution of Eq. \ref{ap_23}. Here $\beta=2$. }
\label{fig_ap_a_20}
\end{figure}

 \begin{figure}
  \centering
    \noindent
   \includegraphics[width=\linewidth]{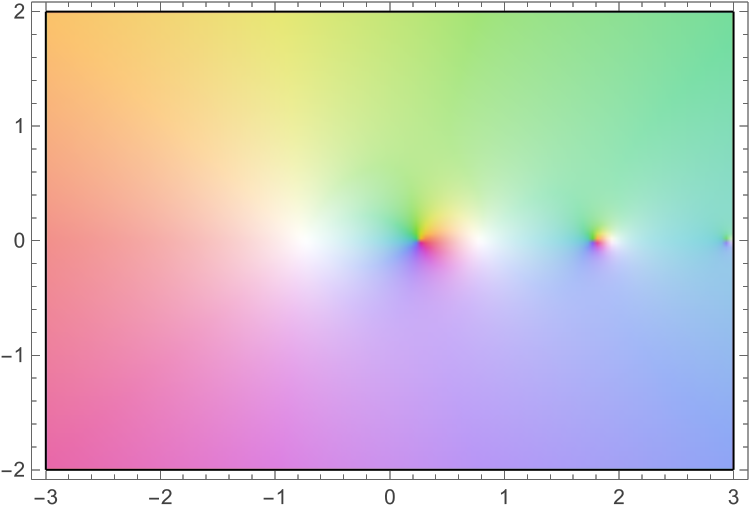}
\caption{Color plot of the argument of $1/{\bar G}_{1L} (z,\beta)$ in the complex plane,  where ${\bar G}_{1L} (z,\beta)$ is the solution of \eqref{ap_25_a}. The points around which the phase varies by $2\pi$ under anticloclwise rotation are the zeros of 
$1/{\bar G}_{1L} (z,\beta)$, i.e., the poles of ${\bar G}_{1L} (z,\beta)$, We see that the poles are confined to the real axis, and there are no  poles in the upper half-plane.  This implies that ${\bar G}_{1L} (z,\beta)$  is an analytic function.
 }
\label{fig_ap_a_31}
\end{figure} 
 
 Finally, after  extensive analysis, we found the exact analytical expression for 
$ {\bar G}_{1L} (x, \beta)$.  For this we reexpressed \eqref{ap_22} as  the non-linear equation 
\beq
  \left({\bar G}_{1L} (x)\right)^{-1} = x - \beta {\bar G}_{1L} (x-1)
  \label{ap_23}
  \eeq 
   and analyzed it.  
 We found that this equation allows for an infinite set of solutions in the form 
 \bea
 &&{\bar G}_{1L} (x, \beta) = \label{ap_24_a} \\
 &&- \frac{1}{\sqrt{\beta}} \frac{J_x (2 \sqrt{\beta}) \left(\cos{\pi x} + C(x) \sin{\pi x} \right) - J_{-x} (2\sqrt{\beta})}{J_{x+1} (2 \sqrt{\beta}) \left(\cos{\pi x} + C(x) \sin{\pi x} \right) + J_{-(x+1)} (2\sqrt{\beta})},
 \nonumber 
  \eea
where $C(x)$ is an arbitrary function, which can even be  complex.   The functional form of $C(x)$ is fixed by the requirement that a physically meaningful solution must satisfy the boundary condition  ${\bar G}_{1L} (x, \beta) \approx 1/x$ at the largest $|x|$. A simple experimentation shows  that this holds when $C(x) = - \cot{\pi x}$.  With this choice,
   \beq
 {\bar G}_{1L} (x, \beta) =\frac{1}{\sqrt{\beta}} \frac{J_{-x} (2\sqrt{\beta})}{ J_{-(x+1)} (2\sqrt{\beta})}.
 \label{ap_25_a}
  \eeq   
A very similar result has been obtained in the analysis of the Green's function of a single hole in an Ising antiferromagnet, also within a self-consistent one-loop approximation~\cite{Starykh_1996}. 
  
 We plot Im ${\bar G}_{1L} (x, \beta)$ in  Fig. \ref{fig_ap_a_20} a. We see that it consists of the set of $\delta-$functions at some $\beta-$dependent $x$.  In Fig. \ref{fig_ap_a_20}b we compare Im ${\bar G}_{1L} (x, \beta)$ with Im ${\bar G}^{\bar m}_{1L} (x, \beta)$ obtained from the continued fraction series for $m=30$.  We see that the two functions are essentially identical.  For larger $\beta \sim 10^2$, more terms in the continuation fractions are needed to reproduce the exact result. For completeness, in Fig. \ref{fig_ap_a_20}c we present the numerical solution of Eq. \eqref{ap_23}.  We see that it coincides with the exact result.   
   In Fig. \ref{fig_ap_a_31} 
      we plot the argument of the complex $1/{\bar G}_{1L} (z, \beta)$ from \eqref{ap_25_a}. We see that there are no  dynamical vorices in the upper half-plane, which means that ${\bar G}_{1L} (z, \beta)$  is an analytic function. We verify that condition $(1/\pi) \int_{-\infty}^\infty {\text {Im}} {\bar G}_{1L} (x, \beta) dx =1$ is satisfied. 
    
 \subsection{Comparison with the eikonal calculation}  
  
  We now compare the results of the rainbow and self-consistent one-loop approximations, which both neglect vertex corrections, with the Green's function that we obtained in the main text using the eikonal computational technique which includes vertex corrections on equal footings with the renormalization of the propagator of an intermediate fermion. 
  We recall that eikonal calculation reproduces the exact Green's function of the Holstein model at zero density. 
  
  The Green's function, obtained within the rainbow approximation for the interaction with an Einstein phonon is non-analytic and qualitatively  different from the ones 
   obtained in the eikonal and self-consitent one-loop approaches.  This  shows that neglecting the renormalization of the Green's function of an intermediate fermion is a very crude approximation, which  does not reproduce the very basic features of the true Green's function. 
   
\begin{figure}[htbp]
	\includegraphics[width=\linewidth]{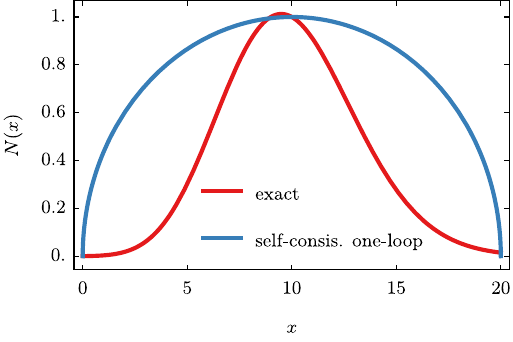}
\caption{Exact DOS  for the $\delta$-functional phonon propagator,  $N(x) \propto  e^{-\beta} \beta^x/x!$, plotted as a function of a continuous $x$  vs  the DOS in the self-consistent one-loop approximation for the same phonon propagator,  $N(x) \propto \sqrt{1 -(x-x_{max})^2/x^2_{max}}$.  Both DOS are normalized to their maximal values, and we adjusted $x_{max}$ to match the positions of the maxima.  The exact DOS is substantially more narrow. }
\label{fig_ap_a8}
\end{figure}

   The Green's function obtained in the self-consistent one-loop approximation is closer to the exact result, but still  substantially differs from it, particularly at large $\beta$. 
   For a $\delta-$functional phonon propagator,  the DOS in this approximation   is a  broad continuum, whose width is the same as the frequency at which Im $G (\omega)$ has a maximum -- both are $\sqrt{2\beta} \omega_0$ at large $\beta$. 
    The exact DOS is also continuous and is peaked at a finite frequency  $\omega = \beta \omega_0$, however  the peak is narrow at large $\beta$ -- its width,
     $\sqrt{2\beta} \omega_0$, is  smaller by $\sqrt {2/\beta}$ than its position (see Fig. \ref{fig_ap_a8}). 
      
      For an Einstein phonon, both the exact DOS and the one obtained within the self-consistent one-loop approximation consists of $\delta$-functional peaks at discrete $\omega$.
      However, the positions of the poles   and the structure of their residues are different. 
      The peaks in the exact DOS, reproduced in the eikonal calculation, are exactly at $\omega = m \omega_0$ and their 
       residues evolve with $m$ as $e^{-\beta} \beta^m/m!$. The poles in the DOS obtained in the self-consistent one-loop approximation are away from $\omega =m$ and their residues decrease with increasing $m$.   
      The difference is not that substantial at small $\beta$, where  
      the poles in  $G_{1L} (\omega +i0)$ are near  
       integer $\omega = m \omega_0$ and  the residues $Z_m$ decrease with increasing $m$ even in the  exact DOS, see Appendix \ref{app_B}.  For large $\beta$, the difference gets larger  as in the exact DOS, the residues of the peaks are exponentially small at small $m$  and increase to be of order $1/\beta$ at $m \approx \beta$, while
         in the self-consistent one-loop approximation  the residues of the peaks gradually decrease with increasing $m$.
 As a consequence, the description in terms of a narrow coherent peak at $\omega = \beta \omega_0$ and a set of polarons with exponentially small residues, which we used in the main text as the point of departure for the evolution of the DOS with increasing bandwidth, is not applicable to the DOS in the self-consistent one-loop approximation. Because the two computational schemes differ in their treatment of vertex corrections, the results in this Appendix imply that  keeping vertex corrections is crucial for  proper description of the evolution of the DOS with increasing $W$. 

        \begin{figure}[htbp]
        \centering
    \noindent
	\includegraphics[]{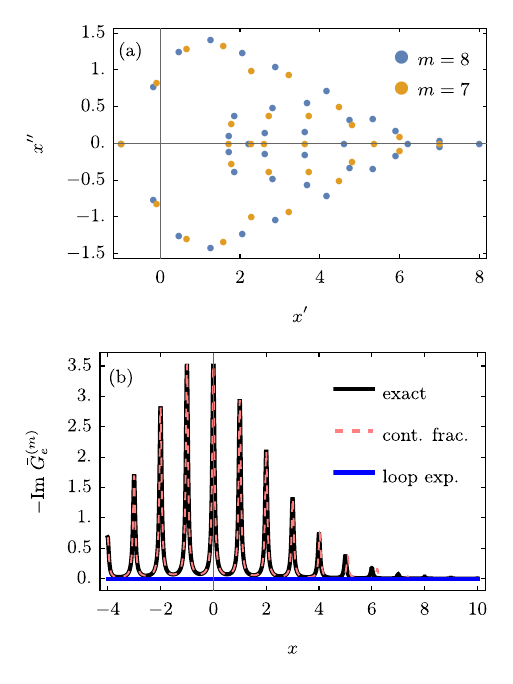} 
    \caption{(a) Poles of the partial ${\bar G}^{(m)}_{e} (z)$ in the complex plane of $z = x' + i x''$  (Green's function in the eikonal approach up to $m$-th order in the loop expansion) for $m =7$ and $8$ (blue and orange dots). (b) The DOS obtained using ${\bar G}^{(m)}_{e} (z)$ for $m=8$ and continued fractions series with 8 terms. The DOS from the continued fraction approximation is much closer to the exact DOS than that from the loop expansion. }
    \label{fig_ap_a_18}
\end{figure}
        
We note in passing that within the eikonal approach, one can also introduce a  partial sum of $m$ terms in the order-by-order perturbation theory ${\bar \Sigma}^{(m)}_{e} (x)$ and  analyze the poles of ${\bar G}^{(m)}_{e} (z)$ in the complex plane of $z = x' + i x''$.  We show the results in Fig. \ref{fig_ap_a_18}a.  We again see poles in the upper half-plane of $z$, whose number increases with $m$ indicating that the correct ${\bar \Sigma}_{e} (x)$ is obtained by  taking the limit $m \to \infty$ together with the analytical continuation.  In Fig. \ref{fig_ap_a_18}b we compare the DOS obtained by restricting the loop expansion for the self-energy at order $m=8$ and by taking equal number of terms in the continued fractions. We see that the continued fractions yield a DOS close to the exact one, while the DOS obtained from the perturbative expansion is very  different.   

\section{ Eikonal Green's function $G^c (\omega)$ in terms of $\Gamma$-functions}
\label{app_D}

\begin{figure}[]
	\includegraphics[]{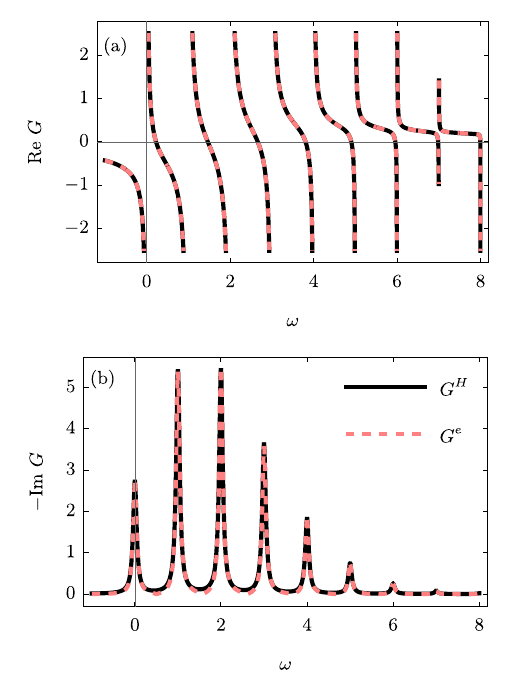}
\caption{Comparison of the (a) real and (b) imaginary parts of $G^e$ with $G^H$. Here $n \to 0^+$ and $\beta =2$.}
\label{fig:ge_vs_gh}
\end{figure}

The Kummer confluent hypergeometric function is related to the lower
 incomplete $\Gamma$-function $\Gamma_L (s,x)$ by
\beq
\Gamma_L (s,x) = \frac{1}{s} e^{-x} x^s _{1}F_{1} (1,1 +s, x).
\label{ii_2_b}
\eeq
The function $\Gamma_L (x,x)$ is given by 
\beq
\Gamma_L (s,x)=\int_{0}^{x } t^{s-1} e^{-t} dt.
\label{ii_2_a}
\eeq
It can be reexpressed  as $\Gamma_L (s,x) = \Gamma (s) - \Gamma_U (s,x)$, where
 $\Gamma (s)$ is the ordinary $\Gamma$-function, 
 \beq
\Gamma_L (s,x)=\int_{0}^{\infty} t^{s-1} e^{-t} dt,
\label{ii_2_c}
\eeq
and  $\Gamma_U (s,x)$ is the  upper  incomplete $\Gamma$-function
\beq
\Gamma_U (s,x)=\int_{x}^{\infty} t^{s-1} e^{-t} dt.
\label{ii_2_a_1}
\eeq
The relation between the Kummer confluent hypergeometric function and $\Gamma$ and $\Gamma_U$  is
\bea
&& \frac{1}{\omega} _{1}F_{1} (1,1-\frac{\omega}{\omega_0} -\beta) = \nonumber \\
 &&- \frac{e^{-\beta}}{\omega_0} (-\beta)^{\omega/\omega_0} \left[\Gamma(-\frac{\omega}{\omega_0}) -\Gamma_U(-\frac{\omega}{\omega_0},-\beta)\right].  
\label{ii_1}
\eea
The r.h.s. of (\ref{ii_1}) can be reexpressed as
\beq
 - \frac{e^{-\beta}}{\omega_0} \frac{(\beta)^{\omega/\omega_0}}{\cos{(\pi \omega/\omega_0)}} \left[\Gamma(-\frac{\omega}{\omega_0}) -{\text {Re}} \Gamma_U(-\frac{\omega}{\omega_0},-\beta)\right].   
\label{ii_3}
\eeq
The Green's function $G^e (\omega)$ from Eq. (\ref{k_4}) is expressed via these $\Gamma$-functions as
\beq
G^e (\omega)= - \frac{e^{-\beta}}{\omega_0} \frac{(\beta)^{\omega/\omega_0}}{\cos{(\pi \omega/\omega_0)}} \left[\Gamma(-\frac{\omega}{\omega_0}) -{\text {Re}} \Gamma_U(-\frac{\omega}{\omega_0},-\beta)\right].  
\label{ii_4}
\eeq
For the retarded Green's function, we should shift $\omega$ to $\omega + i0$. This should be done only in the ordinary $\Gamma$-function.  In other places, the shift is irrelevant. We note that there is no divergence of $G^e$ at $\omega = -\omega_0 (1/2 +n)$, where $n$ is an integer, as the vanishing of $\cos{(\pi \omega/\omega_0)}$ is compensated for by the vanishing of the numerator because
${\text {Re}} \Gamma_U(-(1/2+n),-\beta) = \Gamma(-(1/2+n))$ for any $\beta$.  We also note that 
${\text {Re}} \Gamma_U(-\frac{\omega}{\omega_0},-\beta)$ is non-singular for any $\omega/\omega_0$, including integer values.  The poles of $G^e (\omega)$ then come from the poles of the ordinary $\Gamma$-function $\Gamma (- \omega/\omega_0 - i0)$.  The poles are at $\omega = - m \omega_0$, where $m \geq 0$ is an integer. 
 In Fig.~\ref{fig:ge_vs_gh}
 we plot real and imaginary parts of $G^e (\omega)$ from (\ref{ii_4}) along with real and imaginary parts of 
  the exact Green's function of the single-site Holstein model  at vanishing density $n =0+$
  \beq
      G^{H} (\omega, 0) = e^{-\beta} \sum_{m=0}^\infty \frac{\beta^m}{m!} \frac{1}{\omega + i \delta -m \omega_0},
     \label{ii_5}
      \eeq
(Eq. (\ref{n_2}) in the main text).   We clearly see that $G^e (\omega)$ coincides with $G^{H} (\omega,0)$.
      
Next, we verified that the exact Green's function is  the Fourier transform of $G^e (\tau)$ from Eq. (\ref{Gtau}). 
Indeed, taking the Fourier transform, we obtain
 \beq
      G^e (\omega) = - \frac{e^{-\beta}}{\omega_0} \int_0^\infty dt e^{i t\omega/\omega_0} e^{\beta e^{t}}
 \label{ii_6}
      \eeq      
Evaluating the integral, we reproduce Eq. (\ref{ii_4}).   For completeness, we also verified, by expanding in $\omega_0/\omega$, that Eq. (\ref{ii_6}) reproduces the eikonal series order by order.  From this perspective, the most straightforward proof that eikonal series reproduce the exact solution of the single-site Holstein model is 
 (i) the observation that eikonal series are 
 reproduced by (\ref{ii_6}) in order-by-order expansion, (ii) the equivalence between (\ref{ii_6}) and (\ref{ii_4}) 
  and (iii) the equivalence between (\ref{ii_4}) and (\ref{ii_5}).        

\section{ Melting of polaron states near $\omega = \beta \omega_0$}       
\label{app_E}

In this Appendix, we discuss in more detail how individual polaron peaks at energies $\omega \approx \beta \omega_0$ 
 melt into a continuum when the bandwidth $W$ becomes of order $\sqrt{\beta} \omega_0$. 
 The argument, which we present in the main text, is that at $W=0$, the residues of the peaks,  specified by $m \approx \beta$,    
  are  $Z_m \propto 1/\sqrt{\beta}$,  At a finite $W$, each individual peak then transforms into a patch of width $W_m \sim W Z_m \sim W/\sqrt{\beta}$.  The distance between individual peaks is $\omega_0$, so the patches start to overlap at $W \sim \sqrt{\beta} \omega_0$.   Here we discuss in more detail how this process happens and how the DOS near $\omega = \beta \omega_0$ evolves with $W$  in this range of  $W$.  In particular, we obtain the imaginary part of the self-energy for these fermions.
  
  We find it more advantageous here to use the Lang-Firsov canonical transformation~\cite{lang_firsov,Ranninger_1993} rather than the diagrammatic theory. For convenience of a reader, we first list the known results and then describe our calculations. 

  \subsection{Known results}
  
  Lang-Firsov transformation is a convenient way to diagonalize the single-site Holstein Hamiltonian $H$:  
\begin{equation}
H =   \omega_0 \sum_i a_i^\dagger a_i + \sqrt{\beta} \omega_0  \sum_i c_i^\dagger c_i (a_i + a_i^\dagger).
\label{eq:e1}
\end{equation}
We consider the Hamiltonian here and do not add the chemical potential $\mu$.   
Lang and Firsov demonstrated that this Hamiltonian can be exactly diagonalized via the 
unitary transformation
\begin{equation}
S = \exp\left[ -\sqrt{\beta}  \sum_i n_i (a_i - a_i^\dagger) \right].
\label{eq:e2}
\end{equation}
This transformation physically represents a shift in the equilibrium positions of the local oscillators in the presence of the electron. The transformed Hamiltonian $\tilde{H} = S^\dagger H S$ becomes:
\begin{equation}
\tilde{H} = -\sum_i E_p n_i + \omega_0 \sum_i a_i^\dagger a_i,
\label{eq:e3}
\end{equation}
where  $E_p = \beta \omega_0$. The retarded electron Green's function $G^H(\omega)$ can be evaluated exactly by evaluating  the overlap between 
 the  original and displaced oscillator states. The result is the same as Eq. (\ref{n_2}): 
\begin{equation}
G^H(\omega) = \sum_{m=0}^{\infty}  \frac{Z_m}{\omega - E_m + i\eta},
\label{eq:e4}
\end{equation}
where $E_m = (m - \beta)\omega_0$ and   $Z_m$ has the Poisson distribution:
\begin{equation}
Z_m = e^{-\beta} \frac{\beta^m}{m!}.
\label{eq:e5}
\end{equation}
When the bandwidth is finite, $H$ has the extra hopping term  $V = -t  \sum_{i,j}  c_i^\dagger c_j$, which acts between different sites. Under the Lang-Firsov transformation, this term term transforms into 
\begin{equation}
\tilde{V} = S^\dagger V S = \sum_{i,j} t_{ij} c_i^\dagger c_j X_i^\dagger X_j
\label{eq:e6}
\end{equation}
where $t_{ij} = -t$ for nearest neighbors, and $X_i = \exp[\sqrt{\beta}(a_i - a_i^\dagger)]$ is the phonon shift operator at site $i$.

To construct the Green's function for a finite $W$, it is convenient to 
transform to a momentum representation, i.e., transform from a localized state $|j, m_j\rangle$ with 
    $m_j$ phonons at site $j$ to the corresponding Bloch state with momentum $k$: 
\begin{equation}
|k, m\rangle = \frac{1}{\sqrt{N}} \sum_j e^{i k R_j} |j, m_j\rangle
\label{eq:e7}
\end{equation}
The first-order correction to $E_m$ in Eq. \eqref{eq:e4} is given by \cite{Mahan00}
\begin{equation}
\Delta E_m(k) = \frac{1}{N} \sum_{i,j} t_{ij} e^{i k (R_i - R_j)} \langle 0_i, m_j | X_i^\dagger X_j | m_i, 0_j \rangle
\label{eq:e8}
\end{equation}
The lattice matrix element factors into a product of local overlaps, $\langle 0_i | X_i^\dagger | m_i \rangle \langle m_j | X_j | 0_j \rangle$. Using 
\begin{equation}
|\langle m | X | 0 \rangle|^2 = e^{-\beta} \frac{\beta^m}{m!} = Z_m
\label{eq:e9}
\end{equation}
 one obtains the dressed $E_m$ in the form
 \begin{equation}
E_m(k) = (m - \beta)\omega_0 + Z_m \epsilon_k
\label{eq:e10}
\end{equation}
This shows that, at least to the first order in $W$, the correction to the level $m$ from fermionic hopping comes with the factor $Z_m$.  This is the same result that we obtained by analyzing the diagrammatic series at a finite $W$. 

\subsection{Our results}

We compute  the damping rate of  fermions in patch $m$  due to transitions to other patches. 
To do this,  we  evaluate the off-diagonal elements of the dressed hopping operator $\tilde{V} = \sum_{i,j} t_{ij} c_i^\dagger c_j X_i^\dagger X_j$. From physics perspective, these terms describe incoherent scattering processes in which a polaron changes its momentum and  patch number, leaving behind a localized lattice excitation.  

We evaluate the probability scattering rate $\Gamma_m(k)$ using standard Fermi's golden 
rule. This calculation is exact at the leading order in $W$. We assume that the result for $\Gamma_m(k)$ holds, at  least qualitatively, at $W \sim \sqrt{\beta} \omega$. 
We consider an initial state $|k, m\rangle$  and a final state $|q, m'; p_j\rangle$ describing a scattered polaron  with momentum $q$ in patch $m'$ and  $p$ phonons left at a lattice site $j$. The scattering rate $\Gamma_m(k)$ is 
\begin{eqnarray}
&&\Gamma_m(k) = 
2\pi \sum_{m', p} \sum_j \sum_q 
\label{eq:e11}
\\
&&
|\langle q, m'; p_j | \tilde{V} | k, m \rangle|^2 \delta(E_m(k) - E_{m'}(q) - p\omega_0),
\nonumber
\end{eqnarray}
 where 
 \begin{equation}
\label{eq:e12}
\langle q, m'; p_j | \tilde{V} | k, m \rangle = \frac{1}{N} e^{i(k-q)R_j} \epsilon_q \langle m' | X^\dagger | 0 \rangle \langle p | X | m \rangle .
\end{equation}
Using (\ref{eq:e9}), we obtain 
\begin{equation}
\label{eq:e14}
\sum_j |\langle q, m'; p_j | \tilde{V} | k, m \rangle|^2 = \frac{1}{N} \epsilon_q^2 Z_{m'} F_{m,p}(\beta),
\end{equation}
 where $F_{m,p}(\beta) = |\langle p | X | m \rangle|^2$ is the Franck-Condon factor for the overlap between the initial $m$-phonon state and the remaining $p$-phonon state.  For patches with $m \approx \beta$ and $p = O(1)$, 
$F_{m,p} \sim 1/\sqrt{\beta}$.
Combing the components, we obtain
\begin{eqnarray}
&&
\Gamma_m (k) = 2\pi \sum_{m', p} Z_{m'} F_{m,p}(\beta) \times 
\label{eq:e15}\\
&&
\int d\epsilon N_0 (\epsilon) \epsilon^2 \delta((m - m' - p)\omega_0 + Z_m \epsilon_k - Z_{m'} \epsilon),
\nonumber
\end{eqnarray}
where $N_0 (\epsilon)$ is the DOS of the free fermions.  Using $N_0 (\epsilon \epsilon^2 \sim 1/W$, we find that the scattering rate of the patches with $m \approx \beta$ is  $\Gamma_m \sim W/\sqrt{\beta}$.  We see that at $W \sim \sqrt{\beta} \omega_0$, when patches with $m \approx \beta$ start to overlap, the damping rate turns out to be of the same order $\omega_0$ as the dressed bandwidth $W Z_m \sim W/\sqrt{\beta}$.  

\begin{figure}[t]
    \centering
    \includegraphics[width=\linewidth]{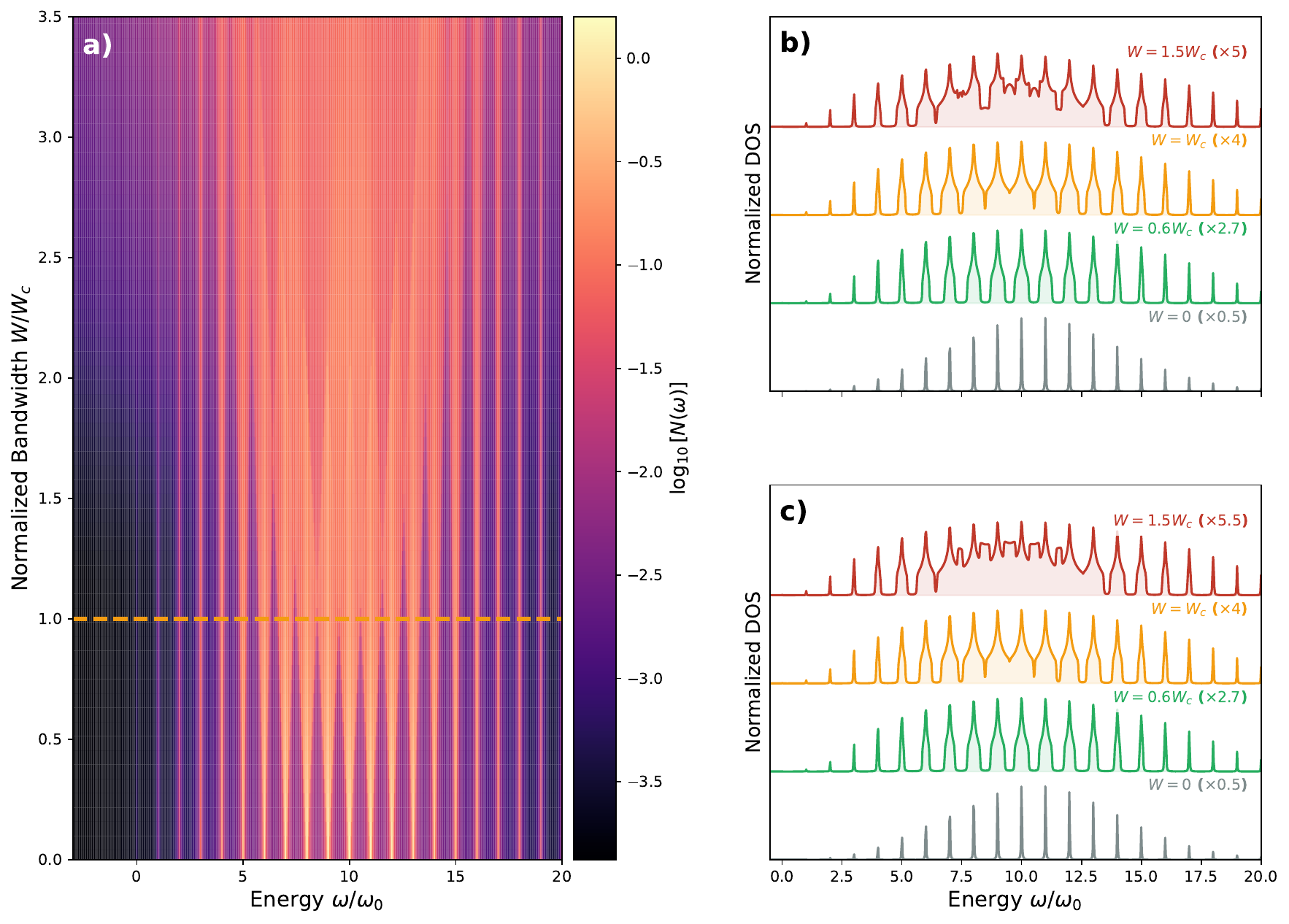}
    \caption{Evolution of the polaron density of states in a 2D  system with tight-binding dispersion at vanishing fermionic density $n = 0+$ 
    We  set $\beta=10$ and vary  the bandwidth $W$. Panel a:  Heatmap of the DOS as a function of $W$, showing the
     melting of the polaron patches for $\omega \sim \beta \omega_0$, while the patches at smaller (and also larger) $\omega$  remain well separated.  Panel b:  The DOS as a function of frequency for several $W$ measured in units of $W_c = \omega_0 \sqrt{2\pi \beta}$.  Panel c: the same as in b, but assuming a constant $\Gamma_m (k) = 0.02$. }
    \label{fig:2dmelting}
\end{figure}

In Fig.  \ref{fig:2dmelting}  we show the results of the calculation of the DOS for different $W$ for our model of 
 fermions with tight-binding dispersion $N_0  (\epsilon)$, given by (\ref{tt_3}) in the main text.
We   set  
 the Green's function for fermions with $\omega \approx \omega_0$ to be 
\beq
G^H (\omega, k) = \sum_m \frac{Z_m}{\omega - (m-\beta) \omega_0 + i \Gamma_m (k)/2},
\eeq
explicitly compute $\Gamma_m (k)$, 
and use the same relation between the DOS $N(\omega)$  and the Green's function as in the main text:
\beq
N(\omega) = -\frac{1}{\pi} \int d\epsilon N_0 (\epsilon) {\text {Im}} G (\omega, k).
\eeq
We  set $\beta =10$ and vary $W$, which we measure in these plots in units of $W_c = \omega_0 \sqrt{2\pi \beta}$.  We see that at $W \ll W_c$, the spectrum consists of separate narrow patches, centered at $\omega = m \omega_0$.   As $W$ increases, the patches at $m << \beta$ remain narrow and separated, while the patches in the middle begin to overlap.  
 Note that this process does not immediately lead to one continuum DOS with the  same structure as for free fermions.  Rather, the DOS has a smooth continuum background on top of which there are mini free-Fermion DOS's with van Hove singularities in the middle. 
  For comparison, in Panel c of this Figure we show the DOS obtained by approximating the actual $\Gamma_m (k)$ by  a small constant.
   We see that qualitatively the behavior is the same as  the one that we obtained with the actual $\Gamma_m (k)$, but there are some differences. 

We caution that this approach works only as long as $W$ remains of order $\omega_0 \sqrt{\beta}$. In larger $W$, and in particular 
in $W \sim \beta \omega_0$, it  does not reproduce the DOS of the free fermion continuum with width $W$ and also does not 
   describe how the continuum absorbs polaron peaks at the smallest energies.   To describe this physics, the diagrammatic approach is advantageous.  

\section{Full and "kinetic" chemical potentials}
\label{app_F}

In this appendix, we compare in more detail the full and kinetic chemical potentials
for a system with density-dependent Hartree potential energy $E^{pot} =-\beta \omega_0 n^2$.
 To shorten the presentation, we consider an FL state with  the same kinetic energy $E^{kin}$  as for free fermions.
  The analysis of the polaron state is very similar.
  
We definite the full chemical potential $\mu^{full}$ as the derivative of the full ground-state energy
 $E^{full} (n) = E^{kin} + E^{pot}$ over density, $\mu^{full} = d E^{full}/dn$, and define the 
 kinetic chemical potential $\mu^{kin} = \mu$ as  the 
  chemical potential of a Fermi gas,  dressed by the Hartree term. 
 The Hamiltonian of the system is
 \beq
 H = \sum_k \epsilon_k c^\dagger_k c_k - \beta \omega_0 n^2.
 \label{mmm_1}
  \eeq
  The grand potential $H' = H - \mu_0 n$, where $\mu_0$ is a bare chemical potential, is 
\beq
 H' = \sum_k (\epsilon_k - \mu_0) c^\dagger_k c_k - \beta \omega_0 n^2.
 \label{mmm_2}
  \eeq
 The Hartree term  gives rise to static self-energy $\Sigma = -2 \beta \omega_0 n$.
  Absorbing  this self-energy into the $c^\dagger_k c_k$ term, we reexpress $H'$ as
\beq
 H' = \sum_k (\epsilon_k - \mu) c^\dagger_k c_k  +  \beta \omega_0 n^2,
 \label{mmm_3}
  \eeq  
  where $\mu = \mu_0 + 2 \beta \omega_0 n$ is the kinetic chemical potential.  The latter is directly related 
   to density by 
   \beq
   n = \int_{-W/2}^\mu N_0 (\epsilon) d \epsilon .
  \label{mmm_4}
  \eeq   
 Note that we have to change the sign of the leftover Hartree term in (\ref{mmm_3}) compared to (\ref{mmm_2}) to avoid triple counting.  We discussed this necessity in the main text. 
 
To simplify the presentation,  we neglect the frequency dependence of the DOS of free fermions, i.e., approximate $N(\omega)$ by $N_0$.  This allows us  to simplify the relation between  $\mu$ and $n$ to
  \beq
  n = N_0 \left(\mu + \frac{W}{2}\right), {\text{i. e.}}~~ \mu = \frac{n}{N_0} - \frac{W}{2}
  \label{mmm_5}
  \eeq
  The ground-state energy $E^{full} = <H>$ is given by 
\bea
E^{full} &=& \frac{N_0}{2} \left(\mu^2 + \frac{W^2}{4}\right) -\beta \omega_0 n^2 \nonumber \\
&=& \frac{n^2}{2N_0} - n \frac{W}{2} - \beta \omega_0 n^2
 \label{mmm_6}
  \eea
Differentiating over $n$, we obtain
\beq
\mu^{full} = \frac {d E^{full}}{d n} = - 2 \beta \omega_0 n  + \frac{n}{N_0} - \frac{W}{2}
\label{mmm_7}
  \eeq
  Comparing with (\ref{mmm_5}), we find
  \beq
  \mu^{full} = \mu - 2 \beta \omega_0 n
  \label{mmm_8}
  \eeq
We see that the two  chemical potentials are not equal in the case where the potential energy (the Hartree term) 
 depends on $n$.

 For completeness, we show that the thermodynamic relation 
 \beq
 E^{full} = \Omega + \mu^{full} n, 
 \label{mmm_11}
 \eeq
 where the grand potential $\Omega = <H'>$,  is satisfied,
    The grand potential of free fermions with dressed $\mu$ is a function of the chemical potential $\mu$.
   Averaging Eq. (\ref{mmm_3}) and using (\ref{mmm_5}), we obtain
   \beq
   \Omega (\mu, n) = - \frac{N_0}{2} \left(\mu + \frac{W}{2}\right)^2  + \beta \omega_0 n^2.
   \label{mmm_9}
  \eeq
   The relation $\partial \Omega/\partial \mu = -n$ is satisfied. 
 Converting $\Omega (\mu,n)$ into $\Omega (n)$ using (\ref{mmm_5}), we find
 \beq
 \Omega (n)  = -\frac{n^2}{2N_0} + \beta \omega_0 n^2.
 \label{mmm_10}
  \eeq
  Substituting into (\ref{mmm_11}) and using \eqref{mmm_8} and \eqref{mmm_5}, we obtain
  \bea
  E^{full} &=& -\frac{n^2}{2N_0} + \beta \omega_0 n + n \left(\frac{n}{N_0} - \frac{W}{2} - 2 \beta \omega_0 n\right)  \nonumber \\
  &=& \frac{n^2}{2N_0}  - n \frac{W}{2}  - \beta \omega_0 n^2, 
  \label{mmm_12}
  \eea
  which coincides with (\ref{mmm_6}).  One can also verify that $d \Omega (n)/dn + n d \mu^{full} (n) /dn =0$, which explains why one obtains $d E^{full}/dn = \mu^{full}$ when one uses (\ref{mmm_11}) for $E^{full}$.

  \section{Role of the nodes of the  Green's function  for the Luttinger's theorem}
  \label{app:10}
  
\subsection{Generic consideration} 

In Luttinger-Ward approach to an FL~\cite{Luttinger1960}, the  relation between fermionic density $n$  per spin projection  and the frequency integral of the Green's function: 
\beq
n = -\frac{i}{N}  \sum_k \lim_{t \to -0}  \int_{-\infty}^\infty  \frac{d\omega}{2\pi}  G (\omega, \epsilon_k) e^{-i\omega t}
 \label{mm_2_1}
\eeq
(Eq.( \ref{mm_2}) in the main text),    is reexpressed using the 
 Dyson equation $  G^{-1} (\omega, \epsilon_k)  =  G^{-1}_0 (\omega, \epsilon_k)  - \Sigma (\omega, \epsilon_k) $
 and the  relation
$G (\omega, \epsilon_k) = \partial \log[G^{-1} (\omega, \epsilon_k)]/\partial \omega  + G(\omega, \epsilon_k) \partial \Sigma (\omega, \epsilon_k)/\partial \omega$. The bare Green's function here is   $ G^{-1}_0 (\omega, \epsilon_k)  = \omega  + i\delta_\omega 
 - (\epsilon_k - \mu_{FL})$, where $\delta_\omega = \delta  {\text{sign}} \omega$ such that   $\partial G^{-1}_0  (\omega, \epsilon_k)/\partial \omega =1$. 
Substituting into (\ref{mm_2_1}),  one obtains 
\beq
n = I_1 + I_2,
 \label{mm_2_2}
\eeq
 where 
\beq
I_1  =  -\frac{i}{N} \sum_k  \lim_{t \to -0}    \int_{-\infty}^\infty \frac{d\omega}{2\pi} \frac{\partial}{\partial \omega} 
\log{ \left[G^{-1} (\omega, \epsilon_k)\right]} e^{-i\omega t}
 \label{mm_2_3}
 \eeq
  and 
  \beq
I_2  = -  -\frac{i}{N} \sum_k   \int_{-\infty}^\infty \frac{d\omega}{2\pi}  G(\omega, \epsilon_k)   \frac{\partial}{\partial \omega} 
\Sigma  (\omega, \epsilon_k).
 \label{mm_2_4}
 \eeq
  In the last term, the  condition on how to take the limit $t \to -0$ is not relevant as at large $\omega$, the integrand decays  faster than $1/\omega$. 
Luttinger  and Ward ~\cite{Luttinger1960} and Luttinger~\cite{Luttinger1960a} argued that in an arbitrary FL,  $I_2 =0$ to an infinite order in perturbation theory, as long as $\Sigma (\omega,\epsilon_k)$ does not develop a pole~\cite{Altshuler1997}.  Then $n = I_1$.  A more generic formulation  of   Luttinger theorem has been given by Oshikawa in ~\cite{Oshikawa2000},  see also~\cite{Sachdev_book}.    
For a Fermi liquid with a single pole in the Geeen's function, a careful consideration~\cite{Luttinger1960} yields  
   \beq
I_1  =  \frac{1}{2\pi i N} \sum_k   \log{\left[\frac{i \delta + \epsilon_k -\mu}{-i \delta + \epsilon_k -\mu}\right]}.
 \label{mm_2_5}
 \eeq 
   Using 
    \beq
 \log{\left[\frac{i \delta + C}{-i \delta + C}\right]} = 2\pi i \Theta(C),
 \label{mm_2_6}
 \eeq 
   where $\Theta (C) =1$ for $C <0$ and zero otherwise,  one obtains   
   \beq
   n =   \frac{1}{N} \sum_k \Theta(\epsilon_k - \mu),
   \label{mm_2_7}
 \eeq  
  which is known as  Luttinger theorem.
  
  For further convenience we combine Eqs. (\ref{mm_2_3}), (\ref{mm_2_5}) and (\ref{mm_2_6}) into 
  \bea
  && \lim_{t \to -0}    \int_{-\infty}^\infty \frac{d\omega}{2\pi i } \frac{\partial}{\partial \omega}  
  \log{ (\omega + i \delta_\omega -C_k )} e^{-i\omega t} \nonumber \\
  && =    \Theta  (C_k). 
  \label{mm_2_17}
  \eea
 Also, for a generic Green's function,  a straightforward  analysis based on Cauchi principle  
  yields 
  \beq 
  I_1 =  \frac{1}{N} \sum_k   \left(P_k -Z_k\right),
    \label{mm_2_7_9}
  \eeq
where $P_k$ and $Z_k$ are the total numbers of poles and 
zeros in the upper half-plane of frequency at a given $k$. 
    
\subsection{Our case: A homogeneous polaron order} 

In our case, the Green's function of a physical fermion, $G^c (\omega, \epsilon_k)$,  has two poles:
\beq
G^c (\omega, \epsilon_k) = \frac{1}{2}  \left[\frac{1 + \frac{\mu_n}{E }}{\omega + i\delta_\omega +E-\epsilon_k} + \frac{1 - \frac{\mu_n}{E}}{\omega + i\delta_\omega -E - \epsilon_k}\right],
   \label{mm_2_8}
 \eeq 
where $E = \sqrt{\mu^2_n + |\Delta_0|^2}$.  
Like we said in the main text, upon substituting (\ref{mm_2_8}) into (\ref{mm_2_1}), we obtain  different results for $\lambda_p >1/2$ and $\lambda_p <1/2$.
In the first case, $E = \beta \omega_0 > W/2$, hence $E  > |\epsilon_k|$ for all momenta.   In this situation,  one of the two poles  at $\omega = -E + \epsilon_k + i\delta$ is in the upper half-plane of frequency, while the other is in the lower half-plane.  Taking the contribution from this pole, we obtain 
    \beq
    n = \frac{1}{2} \left(1 + \frac{\mu_n}{E }\right).  
       \label{mm_2_9}
 \eeq  
   With $E = \beta \omega_0$, this relation  yields $\mu_n  = -\beta \omega_0 (1-2n)$, which we used in the main text.    There is no pole at $\omega =0$ for any ${\bf k}$, hence no Fermi surface.

    For $\lambda_p <1/2$,   $E = \beta \omega_0 < W/2$. In this situation,  each of the two poles in (\ref{mm_2_8}) is in the upper half-plane of frequency in some range of ${\bf k}$,  and each 
     pole there is at $\omega >0$  for some ${\bf k}$ and at $\omega <0$ for  other ${\bf k}$, which is the condition  for the existence of a Fermi surface. 
        We define the Fermi surface area in a conventional way, as 
         \beq
 S_{FS} = \int_{-W/2}^{-E} N_0 (\epsilon) d \epsilon =  \int_{E}^{W/2} N_0 (\epsilon) d \epsilon
\label{mm_2_10}
\eeq
 (Eq. (\ref{mm_3})  in the main text). The relation between $n$ and $G^c$, Eq. (\ref{mm_2_1}), then yields
 \bea
n &=& \frac{1}{2} \left(1 - \frac{\mu_n}{E} \right) + \frac{\mu_n}{E}  S_{FS}.  \nonumber \\
\label{mm_2_11}
\eea
Simultaneously,  the self-consistent equation on $\Delta_0$, which  in our case sets the value of $E$,  becomes
\bea 
E &=& \beta \omega_0  \left(\int_{-W/2}^E N_0 (\epsilon_k) d \epsilon_k  - \int_{E}^{W/2} N_0 (\epsilon_k) d \epsilon_k\right)  \nonumber \\
&=& \beta \omega_0 (1 - 2S_{FS})
\label{mm_2_12}
\eea
(Eq. (\ref{d_25_1}) in the main text).   Solving (\ref{mm_2_11}) and (\ref{mm_2_12}), we obtain $\mu_n = -\beta \omega_0 (1-2n)$, the same as for larger $\lambda$, and $S_{FS} = \delta$, where  $\delta  = \delta (\lambda_p) \approx (1-2 \lambda_p)/(\pi -2)$   at  $\lambda_p \leq 1/2$ and $\delta (\lambda) = n$ at $\lambda_p = \lambda^{c,1}_p$ where $\Delta_0$  vanishes and the system becomes a conventional Fermi liquid.   In the range $1/2 > \lambda_p > \lambda^{c,1}_p$,   $S_{FL} <n$, hence  Luttinger theorem is violated.   In the main text we showed that in this range of $\lambda_p$, the system with density $n$ can be viewed as being split into an FL with density $\delta$ (hence $S_{FL} = \delta$) and  localized polarons with density $n-\delta$. 

We now show how this result is reproduced if we use Luttinger-Ward  reasoning and express the density $n$ as $n = I_1 + I_2$, like in (\ref{mm_2_2}).  
 We can exactly reexpress our $G^c$ from (\ref{mm_2_8}) as
 \beq
 G^c (\omega, \epsilon_k) = \frac{\partial}{\partial \omega} \log{\left[(G^c (\omega,\epsilon_k))^{-1}\right]}
   +   G^c (\omega, \epsilon_k)  \frac{\partial}{\partial \omega}  \Sigma^c (\omega, \epsilon_k),
\label{mm_2_14}
\eeq
 where we introduced 
 \beq
 \Sigma^c (\omega, \epsilon_k)  =  \frac{|\Delta_0|^2}{\omega - \epsilon_k - \mu_n}.
 \label{mm_2_15}
\eeq
 The function $G^c$  in (\ref{mm_2_8}) can in turn be reexpressed as 
 \beq
 G^c (\omega, \epsilon_k) = \frac{\omega + i\delta_\omega - \epsilon_k - \mu_n}{(\omega + i\delta_\omega +E-\epsilon_k)(\omega + i\delta_\omega -E-\epsilon_k)}
 \label{mm_2_16}
\eeq
 We see that $ G^c (\omega, \epsilon_k) $ has both poles and zeros.   Both are equally relevant for $\partial \log{[(G^c (\omega, \epsilon_k))^{-1}]}/\partial \omega$. 
  This point has been emphasized in \cite{Altshuler1997} and in series of recent publications  \cite{Blason_2023,*Staffieri_2025,Lehmann_2025,*Stepanov_2024,Bonetti_2025,Philips2025}. 
  Moreover,  because $\mu_n = - \beta \omega_0 (1-2n)$,   there exists a line 
 $\lambda_p = 1/(2 (1-2n))$ inside the polaron phase,  below which (i.e., at smaller $\lambda_p$) there appears  a range of ${\bf k}$, for which the zero of $G^c$ is at $\omega <0$ and the range where  the zero is at $\omega >0$.    For a  pole, this would set the condition for a Fermi surface.
  By analogy, the locus of points  for which $\epsilon_k = \beta \omega_0 (1-2n)$ is often termed  Luttinger Fermi surface or Luttinger zero~\cite{Dzyaloshinskii2003,Sachdev_book,Fabrizio2022,*Skolimowski2022,*Blason_2023}. 
    
  Luttinger Fermi surface exists in other ordered systems, like itinerant antiferromagnets and  BCS superconductors.  In these systems, however, Luttinger Fermi surface is present in the whole ordered phase.   Our case is somewhat unique  because   Luttinger Fermi surface emerges below a certain line inside the ordered  phase, which in our case is the polaron phase.   This offers an opportunity to check explicitly   how the emergence of such  Fermi surface affects Luttinger reasoning.  
   The issue we address  is whether the appearance of  Luttinger Fermi surface  has any observable effect. We argue that in our case it doesn't. 
   Specifically, we   argue that  
   \begin{itemize}
   \item
       A zero of the Green's function (\ref{mm_2_16}) originates from a  pole in the  self-energy (\ref{mm_2_15}),  and both have to be kept  in the Luttinger-type analysis;  
    \item 
    In our case, the  self-energy has a pole at $\omega = \epsilon_k + \mu_n$ for all values of $\lambda_p$, even the largest ones, for which $|\mu_n| \gg W/2$.  The derivative of the self-energy then has a double pole. The double pole is regularized in the product 
     $ G^c (\omega, \epsilon_k)  \frac{\partial}{\partial \omega}  \Sigma^c (\omega, \epsilon_k)$ because $ G^c (\omega, \epsilon_k)  $ vanishes at the double pole frequency. The remainder  contains three poles, 
      which  are not in the same half-plane of frequency.  As a result, $I_2 \neq 0$ even before Luttinger Fermi surface develops; 
      \item
       In the presence of the Luttinger Fermi surface, $I_1$ acquires an additional contribution, but also 
        $I_2$ changes. In the 
        polaron phase at $\lambda_p >1/2$, the extra contributions to 
        $I_1$  and $I_2$  exactly cancel each other. As a result,  the formula for the density   $n$ is  the same  Eq. (\ref{mm_2_9}) at $\lambda_p > 1/(2(1-2n))$ and at $1/(2(1-2n)) > \lambda_p >1/2$;
     \item 
     For  $\lambda_p <1/2$,   one again can identify the part of $I_2$ that cancels out with the part of $I_1$ coming from the Luttinger Fermi surface. The net result for $n = I_1 + I_2$ is Eq.   (\ref{mm_2_11}).
     \end{itemize}

  We now show explicitly how the cancellation occurs. 
  The expressions for $I_1$ and $I_2$ in our case are 
\begin{widetext}
\beq
  I_1  =  \frac{1}{N} \sum_k   \lim_{t \to -0}  \int_{-\infty}^\infty \frac{d\omega}{2\pi i} \frac{\partial}{\partial \omega} 
\log{\left[\frac{\omega + i \delta_\omega  -\epsilon_k  -\mu_n}{(\omega + i \delta_\omega -E -\epsilon_k)(\omega + i \delta_\omega  + E -\epsilon_k)}\right]}
 e^{-i\omega t}
 \label{mm_2_20}
 \eeq
and 
        \beq
        I_2 = - |\Delta_0|^2    \frac{1}{N} \sum_k  \int_{-\infty}^\infty \frac{d\omega}{2\pi i } \frac{1}{(\omega + i \delta_\omega -E -\epsilon_k)(\omega + i \delta_\omega  + E -\epsilon_k)(\omega + i \delta_\omega  -\epsilon_k  -\mu_n)}.
     \label{mm_2_18}
     \eeq 
  \end{widetext}
  Consider first  large $\lambda_p > 1/(2(1-2n))$, when there is no Luttinger Fermi surface.  
      Using   Eq. (\ref{mm_2_17}), we find that $I_1$  vanishes due to cancellation between 
       the  contributions from 
        the numerator in $G^c$, which has the form of (\ref{mm_2_17}) with $C_k = \epsilon_k  + \mu_n = \epsilon_k - \beta \omega_0 (1-2n)$, which is negative for all ${\bf k}$ and one of the terms in the denominator, for which the corresponding $C_k = \epsilon_k - E =  \epsilon_k -\beta \omega_0$ is also negative for all ${\bf k}$.   
In $I_2$  given by (\ref{mm_2_18}),   two poles are in the upper half-plane of frequency and one is in the lower half-plane, at $\omega = E + \epsilon - i \delta$.   Closing the integration contour in the lower half-plane, we  obtain
 \beq
     I_{2} =  \frac{1}{2N } \sum_k \left(1 + \frac{\mu_n}{E}\right) =  \frac{1}{2} \left(1 + \frac{\mu_n}{E}\right). 
    \label{mm_2_22}
     \eeq  
Substituting into (\ref{mm_2_2}) and using $I_1 =0$,  we reproduce  Eq.  (\ref{mm_2_9}).  We emphasize that  the result comes solely  from $I_2$.
     
     We next   reduce $\lambda_p$ to $1/(2(1-2n)) > \lambda_p >1/2$, when the Luttinger Fermi surface appears.   Now $I_1$ becomes non-zero because for the term in the numerator in (\ref{mm_2_16}),  $C_k = \epsilon_k  + \mu_n = \epsilon_k - \beta \omega_0 (1-2n)$ is negative  for some ${\bf k}$ and positive for the others. An elementary computation yields  
     \beq
     I_1 = \int_{-W/2}^{W/2} N_0 (\epsilon) d \epsilon -  \int_{-W/2}^{|\mu_n|} N_0 (\epsilon) d \epsilon =  \int_{|\mu_n|}^{W/2} N_0 (\epsilon) d \epsilon 
     \label{mm_2_19}
     \eeq 
     Simultaneously, in $I_2$,  there is either one pole in the lower half-plane of frequency  and two in the upper one, or two in the lower one and one in the upper one, depending on whether $\epsilon_k < |\mu| = \beta \omega_0 (1-2n)$  or $\epsilon_k> |\mu|$.    Evaluating  the integrals from these two ranges of $\epsilon_k$ separately, we obtain $I_2 = I_{2a} + I_{2b}$, where
     \beq
     I_{2a} =  \frac{1}{2}\int_{-W/2}^{W/2} N_0 (\epsilon) d \epsilon \left(1 + \frac{\mu_n}{E}\right) =  \frac{1}{2} \left(1 + \frac{\mu_n}{E}\right), 
    \label{mm_2_21}
     \eeq  
    is the same as (\ref{mm_2_22}), and
     \beq
     I_{2b} =  -\int_{|\mu_n|}^{W/2} N_0 (\epsilon) d \epsilon
    \label{mm_2_23}
     \eeq    
     Comparing (\ref{mm_2_19}) and (\ref{mm_2_23}), we see that $I_1 + I_{2b} =0$, i.e., a non-zero contribution to $I_1$  due to the emergence of the  Luttinger Fermi surface cancels out with the corresponding additional contribution to $I_2$. The net result is $n = I_{2a}$, which is the same as for larger $\lambda_p$ 
  
The calculations for $\lambda_p <1/2$ are a bit more involved but still straightforward.  Now  there appear ranges of  ${\bf k}$ for which  $E  = \beta \omega_0 (1-2\delta) < \epsilon_k$.  There is still a cancellation between the part of $I_1$ associated with the Luttinger Fermi surface and the extension  of $I_{2b}$. The net result is     Eq. (\ref{mm_2_11}).

   \section{Details of  mean-field analysis of a checkerboard polaron order at  a generic $n$}
  \label{app:completeness}
  
   For completeness, in this Appendix we collect the results for a checkerboard ${\mathbf q} =Q$ polaron order at a generic density $n$.
  For such momentum, $\epsilon^{(+)}_k =0,  ~\epsilon^{(-)}_k = \epsilon_k = -(W/4) (\cos{k_x} + \cos{k_y})$.  
 The effective quadratic Hamiltonian is 
      \bea
   H^{'}_{eff,2} &=&  \sum_{\mathbf k} \left(\epsilon_{\mathbf k} -{\mu_n}\right) \left(c^\dagger_{\mathbf k}c_{\mathbf k}- {\tilde c}^\dagger_{\mathbf k}{\tilde c}_{\mathbf k}\right)  \nonumber \\
    &&+  \sum_{\mathbf k}\left[\left(\Delta_Q c^\dagger_{\mathbf k}{\tilde c}_{\mathbf k}+ \Delta^*_Q {\tilde c}^\dagger_{\mathbf k}c_{\mathbf k}\right)+ \frac{|\Delta_Q|^2}{2\beta\omega_0}\right].
    \label{qq_9_a_1}
    \eea
  Diagonalizing this Hamiltonian, we obtain 
     \beq
     H^{'}_{eff,2}  = \sum_{\mathbf k}\left[E_{\mathbf k}\left( \alpha^\dagger_{\mathbf k}\alpha_{\mathbf k}-  {\tilde \alpha}^\dagger_{\mathbf k}{\tilde \alpha}_{\mathbf k}\right) + \frac{|\Delta_Q|^2}{2\beta\omega_0}\right],
 \label{qq_10_a_1}
 \eeq
  where
  \beq
  E_{\mathbf k}=  \sqrt{\left({\mu_n} -\epsilon_{\mathbf k}\right)^2 + |\Delta_Q|^2}.
 \label{qq_10_b_a}
  \eeq
The Green function of the physical $c-$fermion $G^c (\omega, \epsilon_{\mathbf k})$ is
  \bea
  G^c (\omega, \epsilon_{\mathbf k})&=&
       \left( \frac{E_{\mathbf k} - {\mu_n} + \epsilon_{\mathbf k}}{2E_{\mathbf k}}\right)
  \frac{1}{\omega - E_{\mathbf k} + i\delta_\omega} \nonumber \\
   &&+ \left( \frac{E_{\mathbf k} + {\mu_n} - \epsilon_{\mathbf k}}{2E_{\mathbf k}}\right)
  \frac{1}{\omega + E_{\mathbf k}+ i\delta_\omega}.
  \label{qq_11}
  \eea
The corresponding continuum DOS is
\begin{widetext}
 \beq
 N (\omega) =  \int_{-W/2}^{W/2} N_0 (\epsilon) d \epsilon  \left[\frac{1}{2}
 \left (1 - \frac{{\mu_n} - \epsilon}{E_\epsilon}\right) \delta (\omega -E_\epsilon) + \frac{1}{2}
 \left (1 + \frac{{\mu_n} - \epsilon}{E_\epsilon}\right) \delta (\omega +E_\epsilon)\right],
 \label{qq_11_a}
  \eeq
  \end{widetext}
   where $E_\epsilon \equiv E_{\mathbf k}=  \sqrt{\left({\mu_n} -\epsilon\right)^2 + |\Delta_Q|^2}$.
 As long as $|\Delta_Q|$ is non-zero, the DOS consists of two  continua, one at $\omega>0$, another at $\omega <0$, which do not overlap with each other.  
  The  width  of each  continuum is controlled by the  fermionic dispersion and is the same for positive and negative $\omega$, but the shape of $ N (\omega)$ is not symmetric.  
  
 The values of $|\Delta_Q|$ and $\mu_n$ at a given  density $n$ and the coupling $\lambda_p$ are obtained from the 
  self-consistency condition on the order parameter and the requirement that the fermionic density be $n$.   
  The self-consistency equation on  $|\Delta_Q|$ is
  \beq
  1 = \beta \omega_0 \int_{-W/2}^{W/2} \frac{N_0 (\epsilon)}{E_\epsilon} d \epsilon.
  \label{d_26_1}
  \eeq
    For tight-binding dispersion,  it  can be reexpressed as 
   \beq
  1 =  \frac{4 \lambda_p}{\pi^2} \int_{-1}^{1} \frac{K(1-x^2)}{\sqrt{(x- {\hat \mu})^2 + |{\hat \Delta}_Q|^2}}  dx,
  \label{d_26_2}
  \eeq
  where ${\hat \Delta}_Q = (2/W) \Delta_Q$ and we recall,
 ${\hat \mu} = (2/W) {\mu_n}$.
  The condition that the fermionic density equals  $n$ is  
     \beq
  n  = \frac{1}{2}  \int_{-W/2}^{W/2} N_0 (\epsilon) d \epsilon \left( 1 + \frac{{\mu_n} -\epsilon}{E_\epsilon}\right).
  \label{d_26_3}
  \eeq
    For tight-binding dispersion  we have
   \beq
   2n-1 = \frac{2}{\pi^2} \int_{-1}^{1} K(1-x^2)  \frac{{\hat \mu} -x}{\sqrt{(x- {\hat \mu})^2 + |{\hat \Delta}_Q|^2}}  dx.
  \label{d_26_4}
  \eeq
  Solving Eqs. (\ref{d_26_1}) and (\ref{d_26_2}), we find that both ${\mu_n}$ and $|\Delta_Q|$ vary  with  both $n$ 
   and $\lambda_p$. Expanding to the second order in $1/\lambda_p$, we find for the tight-binding model 
   \bea
   {\mu_n} &=& \beta \omega_0 (2n-1) \left(1 + \frac{1}{16\lambda^2_p} \right) \label{d_26_5} \\
   |\Delta_Q| &=& 2 \beta \omega_0 \sqrt{n(1-n)} \left(1 - \frac{1}{32 \lambda^2_p} \right). \label{d_26_5_a}
  \eea
 Analyzing Eqs. (\ref{d_26_3}) and (\ref{d_26_4}) at arbitrary $\lambda_p$,   we find that
  $|\Delta_Q|$ remains non-zero even when $\lambda_p$ is small. In this limit, $|\hat \mu| <1$, the integral in 
  (\ref{d_26_2}) diverges  logarithmically at $\Delta_Q =0$,  and the solution of (\ref{d_26_2}) yields  exponentially small but still non-zero $|\Delta_Q| \sim  W \exp[-\pi^2/(8 \lambda_p K(1 - {\hat \mu}^2_{FL}))]$  even for the smallest $\lambda_p$. 
  
 	\begin{figure*}[]
	 \includegraphics[]{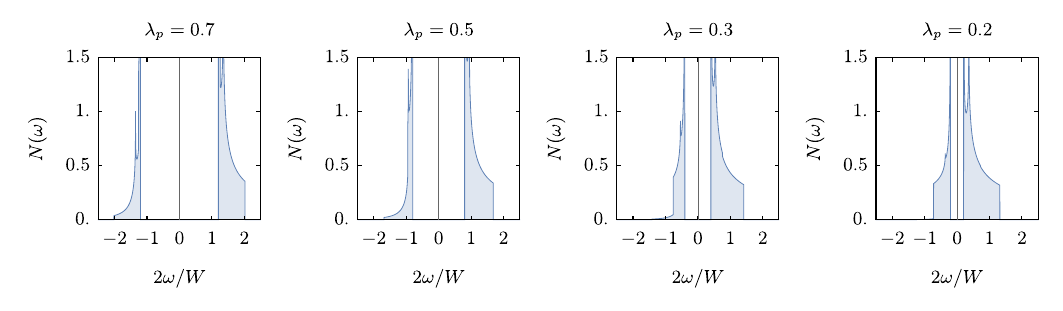}
	 \caption{Evolution of the mean-field  DOS of the checkerboard polaron state  with decreasing $\lambda_p$.  We set the density $n =0.3$.  The coherence peaks remain separated for all $\lambda_p$ as the order parameter $\Delta_Q$  remains non-zero down to $\lambda_p =0$. We argue that this is  an artifact of the mean-field approximation. }
	 \label{fig:pcdw_dos}
	 \end{figure*}
     
 In Fig. \ref{fig:pcdw_dos} we show the  evolution of the  mean-field DOS of the checkerboard polaron state with decreasing $\lambda_p$. We see that the  the two continua of the DOS do not overlap at any non-zero $\lambda_p$.  

  We next evaluate the ground-state energy 
   We compute the grand potential $\Omega_Q$ using the free-fermion Green's function as input and use the relation  $E_Q (n, |\Delta_Q|) = \Omega^{eq}_Q + \mu n - \beta 
     \omega_0 n^2$, where $ \Omega^{eq}_Q$ is the grand potential for  equilibrium values of $\mu_n$ and $|\Delta_Q|$. 
 We use Eq. (\ref{qq_11}) relating  the grand potential $\Omega_Q ({\mu_n}, |\Delta_Q|)$   to the Green's function and use Eq. (\ref{qq_11}) for the Green's function of the  physical $c-$fermion $ G^c (\omega, \epsilon_{\mathbf k})$. A straightforward calculation yields the grand potential per unit volume
  \bea
  &&\Omega_{Q} ({\mu_n}, |\Delta_Q|) = \frac{|\Delta_Q|^2}{4\beta \omega_0} \label{d_26_7} \\
 && -\frac{1}{2} \int_{-W/2}^{W/2} N_0 (\epsilon) d \epsilon \left({\mu_n} +
  \sqrt{({\mu_n} - \epsilon)^2 + |\Delta_Q|^2}\right).  \nonumber  
  \eea
 We verify that conditions $\partial \Omega_{Q}/\partial {\mu_n} = -n$ and $\partial \Omega_{Q}/\partial |\Delta_Q| = 0$ yield the same equations as  Eqs. (\ref{d_26_1}) and (\ref{d_26_2}).
    Evaluating $\Omega^{eq}_{Q} ({\mu_n}, |\Delta_Q|) + {\mu_n} n - \beta \omega_0 n^2$ for equilibrium ${\mu_n}$ and $|\Delta_Q|$, we obtain
    $  E_{Q} (n, |\Delta_Q|) \equiv E_Q (n, \lambda_p)$. At large $\lambda_p$, we  reproduce Eq. (\ref{zzz_12}): 
       \be
     E_{Q} (n,  \lambda_p) =  E_{0} (n)-\beta \omega_0 \frac{n(1-n)}{16 \lambda^2_p},
        \label{d_26_8}
  \ee
  where, we recall, $E_0 (n) = -\beta \omega_0 n$.   In the opposite limit $\lambda_p \ll 1$ and $\mu_n \approx \mu_{FL}$,  it is instructive to compare  $E_{Q} (n, \lambda_p)$ with 
  $E_{FL}$. Expanding in equilibrium $|\Delta_Q|^2$, which is exponentially small at small $\lambda_p$, we obtain
            \beq
              E_{P,Q} - E_{FL} = \frac{|\Delta_Q|^2}{4\beta \omega_0}  L,
              \label{zzz_24}
              \eeq
                   where 
      \beq
      L = 1 - \frac{8 \lambda_p}{\pi^2}  \int_{-1}^{1} 
      \frac{K(1-x^2)d x}{|{\hat \mu}_n -x| + \sqrt{({\hat \mu}_n -x)^2 + |{\hat \Delta}_Q|^2}} 
     \label{zzz_21}
        \eeq
      and ${\hat \Delta}_Q = 2\Delta_{Q}/W$,  ${\hat \nu}_n = 2\mu_n/W$.  Combining (\ref{zzz_21})  with the self-consistency condition on $|\Delta_Q|$:
      \beq
      1= \frac{4 \lambda_p}{\pi^2}  \int_{-1}^{1} 
      \frac{K(1-x^2)d x}{ \sqrt{({\hat \mu}_n -x)^2 + |{\hat \Delta}_Q|^2}}, 
      \label{zzz_22}
       \eeq
        we find after a simple algebra      
   \beq
   L = - \beta \omega_0 N(\mu_n). 
  \label{zzz_23}
  \eeq
  We see that $L$ is negative and independent of $|\Delta_Q|$.  Substituting into (\ref{zzz_24}), we obtain
      \beq 
     E_{Q}  (n, \lambda_p) = E_{FL}  - \frac{1}{4} |\Delta_Q|^2 N(\mu_n). 
        \label{ww_23}
     \eeq
      We note in passing  that 
  this computation is very similar to the one for the condensation energy of a BCS superconductor—in both cases, the dependence of the order parameter is quadratic rather than fourth order; see e.g. Ref. \cite{Mozyrsky_2019}.
 
    We argue in the main text that Eq. (\ref{ww_23})  is clearly an artifact of the mean-field approximation, as for $n  \neq 1/2$ we fully expect that  at the lowest $\lambda_p$, the ground state is  an FL. 
    
For arbitrary $n$ and $\lambda_p$ we evaluate $E_{Q} (n )$ numerically. We show the result in   Fig. \ref{fig:e_polaron_cdw}
 in the main text.

 \bibliography{ref_phonons}

\begin{thebibliography}{121}%
\makeatletter
\providecommand \@ifxundefined [1]{%
 \@ifx{#1\undefined}
}%
\providecommand \@ifnum [1]{%
 \ifnum #1\expandafter \@firstoftwo
 \else \expandafter \@secondoftwo
 \fi
}%
\providecommand \@ifx [1]{%
 \ifx #1\expandafter \@firstoftwo
 \else \expandafter \@secondoftwo
 \fi
}%
\providecommand \natexlab [1]{#1}%
\providecommand \enquote  [1]{``#1''}%
\providecommand \bibnamefont  [1]{#1}%
\providecommand \bibfnamefont [1]{#1}%
\providecommand \citenamefont [1]{#1}%
\providecommand \href@noop [0]{\@secondoftwo}%
\providecommand \href [0]{\begingroup \@sanitize@url \@href}%
\providecommand \@href[1]{\@@startlink{#1}\@@href}%
\providecommand \@@href[1]{\endgroup#1\@@endlink}%
\providecommand \@sanitize@url [0]{\catcode `\\12\catcode `\$12\catcode
  `\&12\catcode `\#12\catcode `\^12\catcode `\_12\catcode `\%12\relax}%
\providecommand \@@startlink[1]{}%
\providecommand \@@endlink[0]{}%
\providecommand \url  [0]{\begingroup\@sanitize@url \@url }%
\providecommand \@url [1]{\endgroup\@href {#1}{\urlprefix }}%
\providecommand \urlprefix  [0]{URL }%
\providecommand \Eprint [0]{\href }%
\providecommand \doibase [0]{https://doi.org/}%
\providecommand \selectlanguage [0]{\@gobble}%
\providecommand \bibinfo  [0]{\@secondoftwo}%
\providecommand \bibfield  [0]{\@secondoftwo}%
\providecommand \translation [1]{[#1]}%
\providecommand \BibitemOpen [0]{}%
\providecommand \bibitemStop [0]{}%
\providecommand \bibitemNoStop [0]{.\EOS\space}%
\providecommand \EOS [0]{\spacefactor3000\relax}%
\providecommand \BibitemShut  [1]{\csname bibitem#1\endcsname}%
\let\auto@bib@innerbib\@empty
%</preamble>
\bibitem [{\citenamefont {Devreese}(2005)}]{Devreese_2005}%
  \BibitemOpen
  \bibfield  {author} {\bibinfo {author} {\bibfnamefont {J.}~\bibnamefont
  {Devreese}},\ }\bibfield  {title} {\bibinfo {title} {Electron–phonon
  interactions and the response of polarons},\ }in\ \href
  {https://doi.org/https://doi.org/10.1016/B0-12-369401-9/00664-1} {\emph
  {\bibinfo {booktitle} {Encyclopedia of Condensed Matter Physics}}},\ \bibinfo
  {editor} {edited by\ \bibinfo {editor} {\bibfnamefont {F.}~\bibnamefont
  {Bassani}}, \bibinfo {editor} {\bibfnamefont {G.~L.}\ \bibnamefont {Liedl}},\
  and\ \bibinfo {editor} {\bibfnamefont {P.}~\bibnamefont {Wyder}}}\ (\bibinfo
  {publisher} {Elsevier},\ \bibinfo {address} {Oxford},\ \bibinfo {year}
  {2005})\ pp.\ \bibinfo {pages} {99--109}\BibitemShut {NoStop}%
\bibitem [{\citenamefont {Aleksandrov}\ and\ \citenamefont
  {Derreese}(2010)}]{Aleksandrov_2010}%
  \BibitemOpen
  \bibfield  {author} {\bibinfo {author} {\bibfnamefont {A.}~\bibnamefont
  {Aleksandrov}}\ and\ \bibinfo {author} {\bibfnamefont {J.}~\bibnamefont
  {Derreese}},\ }\href {https://doi.org/10.1007/978-3-642-01896-1} {\emph
  {\bibinfo {title} {Advances in Polaron Physics}}}\ (\bibinfo  {publisher}
  {Springer Berlin Heidelberg},\ \bibinfo {address} {Springer-Verlag, Berlin
  Heidelberg},\ \bibinfo {year} {2010})\BibitemShut {NoStop}%
\bibitem [{\citenamefont {Franchini}\ \emph {et~al.}(2021)\citenamefont
  {Franchini}, \citenamefont {Reticcioli}, \citenamefont {Setvin},\ and\
  \citenamefont {Diebold}}]{Franchini_2021}%
  \BibitemOpen
  \bibfield  {author} {\bibinfo {author} {\bibfnamefont {C.}~\bibnamefont
  {Franchini}}, \bibinfo {author} {\bibfnamefont {M.}~\bibnamefont
  {Reticcioli}}, \bibinfo {author} {\bibfnamefont {M.}~\bibnamefont {Setvin}},\
  and\ \bibinfo {author} {\bibfnamefont {U.}~\bibnamefont {Diebold}},\
  }\bibfield  {title} {\bibinfo {title} {Polarons in materials},\ }\href@noop
  {} {\bibfield  {journal} {\bibinfo  {journal} {Nature Reviews Materials}\
  }\textbf {\bibinfo {volume} {6}},\ \bibinfo {pages} {560} (\bibinfo {year}
  {2021})}\BibitemShut {NoStop}%
\bibitem [{\citenamefont {Dai}\ \emph {et~al.}(2025)\citenamefont {Dai},
  \citenamefont {Lafuente-Bartolome},\ and\ \citenamefont
  {Giustino}}]{Dai_2025}%
  \BibitemOpen
  \bibfield  {author} {\bibinfo {author} {\bibfnamefont {Z.}~\bibnamefont
  {Dai}}, \bibinfo {author} {\bibfnamefont {J.}~\bibnamefont
  {Lafuente-Bartolome}},\ and\ \bibinfo {author} {\bibfnamefont
  {F.}~\bibnamefont {Giustino}},\ }\href {https://arxiv.org/abs/2512.06176}
  {\bibinfo {title} {Polarons from first principles}} (\bibinfo {year}
  {2025}),\ \Eprint {https://arxiv.org/abs/2512.06176} {arXiv:2512.06176
  [cond-mat.mtrl-sci]} \BibitemShut {NoStop}%
\bibitem [{\citenamefont {Migdal}(1958)}]{Migdal}%
  \BibitemOpen
  \bibfield  {author} {\bibinfo {author} {\bibfnamefont {A.~B.}\ \bibnamefont
  {Migdal}},\ }\bibfield  {title} {\bibinfo {title} {Interactions between
  electrons and lattice vibrations in a superconductor},\ }\href
  {http://www.jetp.ac.ru/cgi-bin/e/index/e/7/6/p996?a=list} {\bibfield
  {journal} {\bibinfo  {journal} {Sov. Phys. JETP}\ }\textbf {\bibinfo {volume}
  {7}},\ \bibinfo {pages} {996} (\bibinfo {year} {1958})}\BibitemShut {NoStop}%
\bibitem [{\citenamefont {Eliashberg}(1960)}]{Eliashberg}%
  \BibitemOpen
  \bibfield  {author} {\bibinfo {author} {\bibfnamefont {G.~M.}\ \bibnamefont
  {Eliashberg}},\ }\bibfield  {title} {\bibinfo {title} {Interactions between
  electrons and lattice vibrations in a superconductor},\ }\href
  {http://www.jetp.ac.ru/cgi-bin/e/index/e/11/3/p696?a=list} {\bibfield
  {journal} {\bibinfo  {journal} {JETP}\ }\textbf {\bibinfo {volume} {11}},\
  \bibinfo {pages} {696} (\bibinfo {year} {1960})}\BibitemShut {NoStop}%
\bibitem [{\citenamefont {Abrikosov}\ \emph {et~al.}(1965)\citenamefont
  {Abrikosov}, \citenamefont {Gorkov},\ and\ \citenamefont
  {Dzyaloshinski}}]{AGD}%
  \BibitemOpen
  \bibfield  {author} {\bibinfo {author} {\bibfnamefont {A.~A.}\ \bibnamefont
  {Abrikosov}}, \bibinfo {author} {\bibfnamefont {L.~P.}\ \bibnamefont
  {Gorkov}},\ and\ \bibinfo {author} {\bibfnamefont {I.~E.}\ \bibnamefont
  {Dzyaloshinski}},\ }\href@noop {} {\emph {\bibinfo {title} {Methods of
  Quantum Feld Theory in Statistical Physics}}}\ (\bibinfo  {publisher}
  {Pergamon Oxford},\ \bibinfo {year} {1965})\BibitemShut {NoStop}%
\bibitem [{\citenamefont {Dolgov}\ \emph {et~al.}(2005)\citenamefont {Dolgov},
  \citenamefont {Mazin}, \citenamefont {Golubov}, \citenamefont {Savrasov},\
  and\ \citenamefont {Maksimov}}]{Mazin_2005}%
  \BibitemOpen
  \bibfield  {author} {\bibinfo {author} {\bibfnamefont {O.~V.}\ \bibnamefont
  {Dolgov}}, \bibinfo {author} {\bibfnamefont {I.~I.}\ \bibnamefont {Mazin}},
  \bibinfo {author} {\bibfnamefont {A.~A.}\ \bibnamefont {Golubov}}, \bibinfo
  {author} {\bibfnamefont {S.~Y.}\ \bibnamefont {Savrasov}},\ and\ \bibinfo
  {author} {\bibfnamefont {E.~G.}\ \bibnamefont {Maksimov}},\ }\bibfield
  {title} {\bibinfo {title} {Critical temperature and enhanced isotope effect
  in the presence of paramagnons in phonon-mediated superconductors},\ }\href
  {https://doi.org/10.1103/PhysRevLett.95.257003} {\bibfield  {journal}
  {\bibinfo  {journal} {Phys. Rev. Lett.}\ }\textbf {\bibinfo {volume} {95}},\
  \bibinfo {pages} {257003} (\bibinfo {year} {2005})}\BibitemShut {NoStop}%
\bibitem [{\citenamefont {Wang}\ and\ \citenamefont
  {Chubukov}(2013)}]{Wang_Ch_2013}%
  \BibitemOpen
  \bibfield  {author} {\bibinfo {author} {\bibfnamefont {Y.}~\bibnamefont
  {Wang}}\ and\ \bibinfo {author} {\bibfnamefont {A.}~\bibnamefont
  {Chubukov}},\ }\bibfield  {title} {\bibinfo {title} {Quantum-critical pairing
  in electron-doped cuprates},\ }\href
  {https://doi.org/10.1103/PhysRevB.88.024516} {\bibfield  {journal} {\bibinfo
  {journal} {Phys. Rev. B}\ }\textbf {\bibinfo {volume} {88}},\ \bibinfo
  {pages} {024516} (\bibinfo {year} {2013})}\BibitemShut {NoStop}%
\bibitem [{\citenamefont {Chubukov}\ \emph {et~al.}(2020)\citenamefont
  {Chubukov}, \citenamefont {Abanov}, \citenamefont {Esterlis},\ and\
  \citenamefont {Kivelson}}]{Andrey_review}%
  \BibitemOpen
  \bibfield  {author} {\bibinfo {author} {\bibfnamefont {A.~V.}\ \bibnamefont
  {Chubukov}}, \bibinfo {author} {\bibfnamefont {A.}~\bibnamefont {Abanov}},
  \bibinfo {author} {\bibfnamefont {I.}~\bibnamefont {Esterlis}},\ and\
  \bibinfo {author} {\bibfnamefont {S.~A.}\ \bibnamefont {Kivelson}},\
  }\bibfield  {title} {\bibinfo {title} {Eliashberg theory of phonon-mediated
  superconductivity -- when it is valid and how it breaks down},\ }\href
  {https://doi.org/https://doi.org/10.1016/j.aop.2020.168190} {\bibfield
  {journal} {\bibinfo  {journal} {Annals of Physics}\ }\textbf {\bibinfo
  {volume} {417}},\ \bibinfo {pages} {168190} (\bibinfo {year}
  {2020})}\BibitemShut {NoStop}%
\bibitem [{\citenamefont {Mirabi}\ \emph {et~al.}(2020)\citenamefont {Mirabi},
  \citenamefont {Boyack},\ and\ \citenamefont {Marsiglio}}]{Mirabi_2020}%
  \BibitemOpen
  \bibfield  {author} {\bibinfo {author} {\bibfnamefont {S.}~\bibnamefont
  {Mirabi}}, \bibinfo {author} {\bibfnamefont {R.}~\bibnamefont {Boyack}},\
  and\ \bibinfo {author} {\bibfnamefont {F.}~\bibnamefont {Marsiglio}},\
  }\bibfield  {title} {\bibinfo {title} {Thermodynamics of eliashberg theory in
  the weak-coupling limit},\ }\href
  {https://doi.org/10.1103/PhysRevB.102.214505} {\bibfield  {journal} {\bibinfo
   {journal} {Phys. Rev. B}\ }\textbf {\bibinfo {volume} {102}},\ \bibinfo
  {pages} {214505} (\bibinfo {year} {2020})}\BibitemShut {NoStop}%
\bibitem [{\citenamefont {Kiessling}\ \emph
  {et~al.}(2025{\natexlab{a}})\citenamefont {Kiessling}, \citenamefont
  {Altshuler},\ and\ \citenamefont {Yuzbashyan}}]{Kiessling2025}%
  \BibitemOpen
  \bibfield  {author} {\bibinfo {author} {\bibfnamefont {M.~K.-H.}\
  \bibnamefont {Kiessling}}, \bibinfo {author} {\bibfnamefont {B.~L.}\
  \bibnamefont {Altshuler}},\ and\ \bibinfo {author} {\bibfnamefont {E.~A.}\
  \bibnamefont {Yuzbashyan}},\ }\bibfield  {title} {\bibinfo {title} {Bounds on
  tc in the eliashberg theory of superconductivity. ii: Dispersive phonons},\
  }\href {https://doi.org/10.1007/s10955-025-03468-z} {\bibfield  {journal}
  {\bibinfo  {journal} {Journal of Statistical Physics}\ }\textbf {\bibinfo
  {volume} {192}},\ \bibinfo {pages} {94} (\bibinfo {year}
  {2025}{\natexlab{a}})}\BibitemShut {NoStop}%
\bibitem [{\citenamefont {Kiessling}\ \emph
  {et~al.}(2025{\natexlab{b}})\citenamefont {Kiessling}, \citenamefont
  {Altshuler},\ and\ \citenamefont {Yuzbashyan}}]{Kiessling2025_1}%
  \BibitemOpen
  \bibfield  {author} {\bibinfo {author} {\bibfnamefont {M.~K.-H.}\
  \bibnamefont {Kiessling}}, \bibinfo {author} {\bibfnamefont {B.~L.}\
  \bibnamefont {Altshuler}},\ and\ \bibinfo {author} {\bibfnamefont {E.~A.}\
  \bibnamefont {Yuzbashyan}},\ }\bibfield  {title} {\bibinfo {title} {Bounds on
  tc in the eliashberg theory of superconductivity. iii: Einstein phonons},\
  }\href {https://doi.org/10.1007/s10955-025-03469-y} {\bibfield  {journal}
  {\bibinfo  {journal} {Journal of Statistical Physics}\ }\textbf {\bibinfo
  {volume} {192}},\ \bibinfo {pages} {93} (\bibinfo {year}
  {2025}{\natexlab{b}})}\BibitemShut {NoStop}%
\bibitem [{\citenamefont {Gnezdilov}\ and\ \citenamefont
  {Boyack}(2025)}]{Gnezdilov_2025}%
  \BibitemOpen
  \bibfield  {author} {\bibinfo {author} {\bibfnamefont {N.~V.}\ \bibnamefont
  {Gnezdilov}}\ and\ \bibinfo {author} {\bibfnamefont {R.}~\bibnamefont
  {Boyack}},\ }\href {https://arxiv.org/abs/2505.02894} {\bibinfo {title}
  {Upper bound on $t_c$ in a strongly coupled electron-boson superconductor}}
  (\bibinfo {year} {2025}),\ \Eprint {https://arxiv.org/abs/2505.02894}
  {arXiv:2505.02894} \BibitemShut {NoStop}%
\bibitem [{Note1()}]{Note1}%
  \BibitemOpen
  \bibinfo {note} {At weak coupling, $\lambda \ll 1$, Eliashberg $T_c = 1.13
  \protect \sqrt {e} \omega _0 e^{-1/\lambda } = 0.69 \omega e^{-1/\lambda }$,
  where $\omega _0$ is the Debye frequency (see \cite
  {Mazin_2005,*Wang_Ch_2013} and references therein), differs by $\protect
  \sqrt {e}$ from the often cited value 1.13, which holds for the case when the
  phonon susceptibility is set to be constant at frequencies smaller than
  $\omega _0$ and zero otherwise.}\BibitemShut {Stop}%
\bibitem [{\citenamefont {Yuzbashyan}\ \emph {et~al.}(2025)\citenamefont
  {Yuzbashyan}, \citenamefont {Altshuler},\ and\ \citenamefont
  {Patra}}]{Yuzbashyan}%
  \BibitemOpen
  \bibfield  {author} {\bibinfo {author} {\bibfnamefont {E.~A.}\ \bibnamefont
  {Yuzbashyan}}, \bibinfo {author} {\bibfnamefont {B.~L.}\ \bibnamefont
  {Altshuler}},\ and\ \bibinfo {author} {\bibfnamefont {A.}~\bibnamefont
  {Patra}},\ }\bibfield  {title} {\bibinfo {title} {Instability of metals with
  respect to strong electron-phonon interaction},\ }\href
  {https://doi.org/10.1103/sf5p-2g5l} {\bibfield  {journal} {\bibinfo
  {journal} {Phys. Rev. Lett.}\ }\textbf {\bibinfo {volume} {135}},\ \bibinfo
  {pages} {026503} (\bibinfo {year} {2025})}\BibitemShut {NoStop}%
\bibitem [{\citenamefont {Yuzbashyan}\ and\ \citenamefont
  {Altshuler}(2022{\natexlab{a}})}]{Yuzbashyan_1}%
  \BibitemOpen
  \bibfield  {author} {\bibinfo {author} {\bibfnamefont {E.~A.}\ \bibnamefont
  {Yuzbashyan}}\ and\ \bibinfo {author} {\bibfnamefont {B.~L.}\ \bibnamefont
  {Altshuler}},\ }\bibfield  {title} {\bibinfo {title} {Breakdown of the
  migdal-eliashberg theory and a theory of lattice-fermionic superfluidity},\
  }\href {https://doi.org/10.1103/PhysRevB.106.054518} {\bibfield  {journal}
  {\bibinfo  {journal} {Phys. Rev. B}\ }\textbf {\bibinfo {volume} {106}},\
  \bibinfo {pages} {054518} (\bibinfo {year} {2022}{\natexlab{a}})}\BibitemShut
  {NoStop}%
\bibitem [{\citenamefont {Yuzbashyan}\ and\ \citenamefont
  {Altshuler}(2022{\natexlab{b}})}]{Yuzbashyan_2}%
  \BibitemOpen
  \bibfield  {author} {\bibinfo {author} {\bibfnamefont {E.~A.}\ \bibnamefont
  {Yuzbashyan}}\ and\ \bibinfo {author} {\bibfnamefont {B.~L.}\ \bibnamefont
  {Altshuler}},\ }\bibfield  {title} {\bibinfo {title} {Migdal-eliashberg
  theory as a classical spin chain},\ }\href
  {https://doi.org/10.1103/PhysRevB.106.014512} {\bibfield  {journal} {\bibinfo
   {journal} {Phys. Rev. B}\ }\textbf {\bibinfo {volume} {106}},\ \bibinfo
  {pages} {014512} (\bibinfo {year} {2022}{\natexlab{b}})}\BibitemShut
  {NoStop}%
\bibitem [{\citenamefont {Semenok}\ \emph {et~al.}(2025)\citenamefont
  {Semenok}, \citenamefont {Altshuler},\ and\ \citenamefont
  {Yuzbashyan}}]{Yuzbashyan_3}%
  \BibitemOpen
  \bibfield  {author} {\bibinfo {author} {\bibfnamefont {D.~V.}\ \bibnamefont
  {Semenok}}, \bibinfo {author} {\bibfnamefont {B.~L.}\ \bibnamefont
  {Altshuler}},\ and\ \bibinfo {author} {\bibfnamefont {E.~A.}\ \bibnamefont
  {Yuzbashyan}},\ }\bibfield  {title} {\bibinfo {title} {Fundamental limits on
  the electron-phonon coupling and superconducting tc},\ }\href
  {https://doi.org/https://doi.org/10.1002/adma.202507013} {\bibfield
  {journal} {\bibinfo  {journal} {Advanced Materials}\ }\textbf {\bibinfo
  {volume} {37}},\ \bibinfo {pages} {2507013} (\bibinfo {year}
  {2025})}\BibitemShut {NoStop}%
\bibitem [{\citenamefont {Alexandrov}\ \emph {et~al.}(1994)\citenamefont
  {Alexandrov}, \citenamefont {Kabanov},\ and\ \citenamefont
  {Ray}}]{Alexandrov}%
  \BibitemOpen
  \bibfield  {author} {\bibinfo {author} {\bibfnamefont {A.~S.}\ \bibnamefont
  {Alexandrov}}, \bibinfo {author} {\bibfnamefont {V.~V.}\ \bibnamefont
  {Kabanov}},\ and\ \bibinfo {author} {\bibfnamefont {D.~K.}\ \bibnamefont
  {Ray}},\ }\bibfield  {title} {\bibinfo {title} {From electron to small
  polaron: An exact cluster solution},\ }\href
  {https://doi.org/10.1103/PhysRevB.49.9915} {\bibfield  {journal} {\bibinfo
  {journal} {Phys. Rev. B}\ }\textbf {\bibinfo {volume} {49}},\ \bibinfo
  {pages} {9915} (\bibinfo {year} {1994})}\BibitemShut {NoStop}%
\bibitem [{\citenamefont {Alexandrov}\ and\ \citenamefont {Mott}(1994)}]{Mott}%
  \BibitemOpen
  \bibfield  {author} {\bibinfo {author} {\bibfnamefont {A.~S.}\ \bibnamefont
  {Alexandrov}}\ and\ \bibinfo {author} {\bibfnamefont {N.~F.}\ \bibnamefont
  {Mott}},\ }\bibfield  {title} {\bibinfo {title} {Bipolarons},\ }\href
  {https://doi.org/10.1088/0034-4885/57/12/001} {\bibfield  {journal} {\bibinfo
   {journal} {Reports on Progress in Physics}\ }\textbf {\bibinfo {volume}
  {57}},\ \bibinfo {pages} {1197} (\bibinfo {year} {1994})}\BibitemShut
  {NoStop}%
\bibitem [{\citenamefont {Alexandrov}\ and\ \citenamefont
  {Mott}(1996)}]{Mott_1}%
  \BibitemOpen
  \bibfield  {author} {\bibinfo {author} {\bibfnamefont {A.~S.}\ \bibnamefont
  {Alexandrov}}\ and\ \bibinfo {author} {\bibfnamefont {N.~F.}\ \bibnamefont
  {Mott}},\ }\href {https://doi.org/10.1142/2784} {\emph {\bibinfo {title}
  {Polarons and Bipolarons}}}\ (\bibinfo  {publisher} {WORLD SCIENTIFIC},\
  \bibinfo {year} {1996})\ \Eprint
  {https://arxiv.org/abs/https://www.worldscientific.com/doi/pdf/10.1142/2784}
  {https://www.worldscientific.com/doi/pdf/10.1142/2784} \BibitemShut {NoStop}%
\bibitem [{\citenamefont {Mahan}(2000)}]{Mahan00}%
  \BibitemOpen
  \bibfield  {author} {\bibinfo {author} {\bibfnamefont {G.~D.}\ \bibnamefont
  {Mahan}},\ }\href@noop {} {\emph {\bibinfo {title} {Many-Particle
  Physics}}},\ \bibinfo {edition} {3rd}\ ed.,\ Physics of Solids and Liquids\
  (\bibinfo  {publisher} {Kluwer Academic/Plenum Publishers},\ \bibinfo
  {address} {New York},\ \bibinfo {year} {2000})\BibitemShut {NoStop}%
\bibitem [{\citenamefont {Prokof'ev}\ \emph {et~al.}(2026)\citenamefont
  {Prokof'ev}, \citenamefont {Esterlis}, \citenamefont {Abanov},\ and\
  \citenamefont {Chubukov}}]{chubukov_prl_2026}%
  \BibitemOpen
  \bibfield  {author} {\bibinfo {author} {\bibfnamefont {N.}~\bibnamefont
  {Prokof'ev}}, \bibinfo {author} {\bibfnamefont {I.}~\bibnamefont {Esterlis}},
  \bibinfo {author} {\bibfnamefont {A.}~\bibnamefont {Abanov}},\ and\ \bibinfo
  {author} {\bibfnamefont {A.}~\bibnamefont {Chubukov}},\ }\bibfield  {title}
  {\bibinfo {title} {Limits of validity for migdal-eliashberg theory: Role of
  polarons and bipolarons},\ }\href {https://doi.org/10.1103/zd4s-cgnj}
  {\bibfield  {journal} {\bibinfo  {journal} {Phys. Rev. Lett.}\ }\textbf
  {\bibinfo {volume} {137}},\ \bibinfo {pages} {096501} (\bibinfo {year}
  {2026})}\BibitemShut {NoStop}%
\bibitem [{Note2()}]{Note2}%
  \BibitemOpen
  \bibinfo {note} {We emphasize that vertex corrections are still small in this
  regime, as $\omega _r (q)$ is smaller than $E_F$ for all $q$. In this respect
  strong coupling MET for electron-phonon interaction is different from an
  effective MET for electrons interacting with soft fluctuations in spin or
  charge channel (see Refs. \cite {Zhang_1,*Zhang_2} for more
  details).}\BibitemShut {Stop}%
\bibitem [{\citenamefont {Combescot}(1995)}]{combescot}%
  \BibitemOpen
  \bibfield  {author} {\bibinfo {author} {\bibfnamefont {R.}~\bibnamefont
  {Combescot}},\ }\bibfield  {title} {\bibinfo {title} {Strong-coupling limit
  of eliashberg theory},\ }\href {https://doi.org/10.1103/PhysRevB.51.11625}
  {\bibfield  {journal} {\bibinfo  {journal} {Phys. Rev. B}\ }\textbf {\bibinfo
  {volume} {51}},\ \bibinfo {pages} {11625} (\bibinfo {year}
  {1995})}\BibitemShut {NoStop}%
\bibitem [{\citenamefont {Wu}\ \emph {et~al.}(2021)\citenamefont {Wu},
  \citenamefont {Zhang}, \citenamefont {Abanov},\ and\ \citenamefont
  {Chubukov}}]{paper_5}%
  \BibitemOpen
  \bibfield  {author} {\bibinfo {author} {\bibfnamefont {Y.-M.}\ \bibnamefont
  {Wu}}, \bibinfo {author} {\bibfnamefont {S.-S.}\ \bibnamefont {Zhang}},
  \bibinfo {author} {\bibfnamefont {A.}~\bibnamefont {Abanov}},\ and\ \bibinfo
  {author} {\bibfnamefont {A.~V.}\ \bibnamefont {Chubukov}},\ }\bibfield
  {title} {\bibinfo {title} {Interplay between superconductivity and non-fermi
  liquid at a quantum critical point in a metal. v. the $\gamma$ model and its
  phase diagram. the case $\gamma = 2$},\ }\href
  {https://doi.org/10.1103/PhysRevB.103.024522} {\bibfield  {journal} {\bibinfo
   {journal} {Phys. Rev. B}\ }\textbf {\bibinfo {volume} {103}},\ \bibinfo
  {pages} {024522} (\bibinfo {year} {2021})}\BibitemShut {NoStop}%
\bibitem [{\citenamefont {Chakraverty}\ \emph {et~al.}(1998)\citenamefont
  {Chakraverty}, \citenamefont {Ranninger},\ and\ \citenamefont
  {Feinberg}}]{Chakraverty1}%
  \BibitemOpen
  \bibfield  {author} {\bibinfo {author} {\bibfnamefont {B.~K.}\ \bibnamefont
  {Chakraverty}}, \bibinfo {author} {\bibfnamefont {J.}~\bibnamefont
  {Ranninger}},\ and\ \bibinfo {author} {\bibfnamefont {D.}~\bibnamefont
  {Feinberg}},\ }\bibfield  {title} {\bibinfo {title} {Experimental and
  theoretical constraints of bipolaronic superconductivity in high ${T}_{c}$
  materials: An impossibility},\ }\href
  {https://doi.org/10.1103/PhysRevLett.81.433} {\bibfield  {journal} {\bibinfo
  {journal} {Phys. Rev. Lett.}\ }\textbf {\bibinfo {volume} {81}},\ \bibinfo
  {pages} {433} (\bibinfo {year} {1998})}\BibitemShut {NoStop}%
\bibitem [{\citenamefont {Chakraverty}\ \emph {et~al.}(1999)\citenamefont
  {Chakraverty}, \citenamefont {Ranninger},\ and\ \citenamefont
  {Feinberg}}]{Chakraverty2}%
  \BibitemOpen
  \bibfield  {author} {\bibinfo {author} {\bibfnamefont {B.~K.}\ \bibnamefont
  {Chakraverty}}, \bibinfo {author} {\bibfnamefont {J.}~\bibnamefont
  {Ranninger}},\ and\ \bibinfo {author} {\bibfnamefont {D.}~\bibnamefont
  {Feinberg}},\ }\bibfield  {title} {\bibinfo {title} {Chakraverty et al.
  reply:},\ }\href {https://doi.org/10.1103/PhysRevLett.82.2621} {\bibfield
  {journal} {\bibinfo  {journal} {Phys. Rev. Lett.}\ }\textbf {\bibinfo
  {volume} {82}},\ \bibinfo {pages} {2621} (\bibinfo {year}
  {1999})}\BibitemShut {NoStop}%
\bibitem [{\citenamefont {Zhang}\ \emph {et~al.}(2023)\citenamefont {Zhang},
  \citenamefont {Berg},\ and\ \citenamefont {Chubukov}}]{Zhang_1}%
  \BibitemOpen
  \bibfield  {author} {\bibinfo {author} {\bibfnamefont {S.-S.}\ \bibnamefont
  {Zhang}}, \bibinfo {author} {\bibfnamefont {E.}~\bibnamefont {Berg}},\ and\
  \bibinfo {author} {\bibfnamefont {A.~V.}\ \bibnamefont {Chubukov}},\
  }\bibfield  {title} {\bibinfo {title} {Free energy and specific heat near a
  quantum critical point of a metal},\ }\href
  {https://doi.org/10.1103/PhysRevB.107.144507} {\bibfield  {journal} {\bibinfo
   {journal} {Phys. Rev. B}\ }\textbf {\bibinfo {volume} {107}},\ \bibinfo
  {pages} {144507} (\bibinfo {year} {2023})}\BibitemShut {NoStop}%
\bibitem [{\citenamefont {Zhang}\ \emph {et~al.}(2024)\citenamefont {Zhang},
  \citenamefont {Raines},\ and\ \citenamefont {Chubukov}}]{Zhang_2}%
  \BibitemOpen
  \bibfield  {author} {\bibinfo {author} {\bibfnamefont {S.-S.}\ \bibnamefont
  {Zhang}}, \bibinfo {author} {\bibfnamefont {Z.~M.}\ \bibnamefont {Raines}},\
  and\ \bibinfo {author} {\bibfnamefont {A.~V.}\ \bibnamefont {Chubukov}},\
  }\bibfield  {title} {\bibinfo {title} {Applicability of eliashberg theory for
  systems with electron-phonon and electron-electron interaction: A comparative
  analysis},\ }\href {https://doi.org/10.1103/PhysRevB.109.245132} {\bibfield
  {journal} {\bibinfo  {journal} {Phys. Rev. B}\ }\textbf {\bibinfo {volume}
  {109}},\ \bibinfo {pages} {245132} (\bibinfo {year} {2024})}\BibitemShut
  {NoStop}%
\bibitem [{\citenamefont {Abanov}\ and\ \citenamefont
  {Chubukov}(2020)}]{paper_1}%
  \BibitemOpen
  \bibfield  {author} {\bibinfo {author} {\bibfnamefont {A.}~\bibnamefont
  {Abanov}}\ and\ \bibinfo {author} {\bibfnamefont {A.~V.}\ \bibnamefont
  {Chubukov}},\ }\bibfield  {title} {\bibinfo {title} {Interplay between
  superconductivity and non-{Fermi} liquid at a quantum critical point in a
  metal. i. the $\gamma$ model and its phase diagram at $\ensuremath{T=0}$: The
  case $0<\gamma <1$},\ }\href {https://doi.org/10.1103/PhysRevB.102.024524}
  {\bibfield  {journal} {\bibinfo  {journal} {Phys. Rev. B}\ }\textbf {\bibinfo
  {volume} {102}},\ \bibinfo {pages} {024524} (\bibinfo {year}
  {2020})}\BibitemShut {NoStop}%
\bibitem [{\citenamefont {Marsiglio}(2020)}]{marsiglio2020eliashberg}%
  \BibitemOpen
  \bibfield  {author} {\bibinfo {author} {\bibfnamefont {F.}~\bibnamefont
  {Marsiglio}},\ }\bibfield  {title} {\bibinfo {title} {Eliashberg theory: A
  short review},\ }\href@noop {} {\bibfield  {journal} {\bibinfo  {journal}
  {Annals of Physics}\ }\textbf {\bibinfo {volume} {417}},\ \bibinfo {pages}
  {168102} (\bibinfo {year} {2020})}\BibitemShut {NoStop}%
\bibitem [{\citenamefont {Lang}\ and\ \citenamefont
  {Firsov}(1963)}]{lang_firsov}%
  \BibitemOpen
  \bibfield  {author} {\bibinfo {author} {\bibfnamefont {I.}~\bibnamefont
  {Lang}}\ and\ \bibinfo {author} {\bibfnamefont {Y.~A.}\ \bibnamefont
  {Firsov}},\ }\bibfield  {title} {\bibinfo {title} {Kinetic theory of
  semiconductors with low mobility},\ }\href@noop {} {\bibfield  {journal}
  {\bibinfo  {journal} {JETP}\ }\textbf {\bibinfo {volume} {16}},\ \bibinfo
  {pages} {1301} (\bibinfo {year} {1963})}\BibitemShut {NoStop}%
\bibitem [{\citenamefont {Holstein}(1959)}]{Holstein1959}%
  \BibitemOpen
  \bibfield  {author} {\bibinfo {author} {\bibfnamefont {T.}~\bibnamefont
  {Holstein}},\ }\bibfield  {title} {\bibinfo {title} {Studies of polaron
  motion: Part i. the molecular-crystal model},\ }\href
  {https://doi.org/https://doi.org/10.1016/0003-4916(59)90002-8} {\bibfield
  {journal} {\bibinfo  {journal} {Annals of Physics}\ }\textbf {\bibinfo
  {volume} {8}},\ \bibinfo {pages} {325} (\bibinfo {year} {1959})}\BibitemShut
  {NoStop}%
\bibitem [{\citenamefont {Ranninger}(1993)}]{Ranninger_1993}%
  \BibitemOpen
  \bibfield  {author} {\bibinfo {author} {\bibfnamefont {J.}~\bibnamefont
  {Ranninger}},\ }\bibfield  {title} {\bibinfo {title} {Spectral properties of
  small-polaron systems},\ }\href {https://doi.org/10.1103/PhysRevB.48.13166}
  {\bibfield  {journal} {\bibinfo  {journal} {Phys. Rev. B}\ }\textbf {\bibinfo
  {volume} {48}},\ \bibinfo {pages} {13166} (\bibinfo {year}
  {1993})}\BibitemShut {NoStop}%
\bibitem [{\citenamefont {Prokof'ev}\ and\ \citenamefont
  {Svistunov}(1998)}]{polarons1998}%
  \BibitemOpen
  \bibfield  {author} {\bibinfo {author} {\bibfnamefont {N.~V.}\ \bibnamefont
  {Prokof'ev}}\ and\ \bibinfo {author} {\bibfnamefont {B.~V.}\ \bibnamefont
  {Svistunov}},\ }\bibfield  {title} {\bibinfo {title} {Polaron problem by
  diagrammatic quantum monte carlo},\ }\href
  {https://doi.org/10.1103/PhysRevLett.81.2514} {\bibfield  {journal} {\bibinfo
   {journal} {Phys. Rev. Lett.}\ }\textbf {\bibinfo {volume} {81}},\ \bibinfo
  {pages} {2514} (\bibinfo {year} {1998})}\BibitemShut {NoStop}%
\bibitem [{\citenamefont {Mishchenko}\ \emph {et~al.}(2000)\citenamefont
  {Mishchenko}, \citenamefont {Prokof'ev}, \citenamefont {Sakamoto},\ and\
  \citenamefont {Svistunov}}]{polarons2000}%
  \BibitemOpen
  \bibfield  {author} {\bibinfo {author} {\bibfnamefont {A.~S.}\ \bibnamefont
  {Mishchenko}}, \bibinfo {author} {\bibfnamefont {N.~V.}\ \bibnamefont
  {Prokof'ev}}, \bibinfo {author} {\bibfnamefont {A.}~\bibnamefont
  {Sakamoto}},\ and\ \bibinfo {author} {\bibfnamefont {B.~V.}\ \bibnamefont
  {Svistunov}},\ }\bibfield  {title} {\bibinfo {title} {Diagrammatic quantum
  monte carlo study of the fr\"ohlich polaron},\ }\href
  {https://doi.org/10.1103/PhysRevB.62.6317} {\bibfield  {journal} {\bibinfo
  {journal} {Phys. Rev. B}\ }\textbf {\bibinfo {volume} {62}},\ \bibinfo
  {pages} {6317} (\bibinfo {year} {2000})}\BibitemShut {NoStop}%
\bibitem [{\citenamefont {{{Sadovskii}, M.~V.}}(2006)}]{SadovskiiBook}%
  \BibitemOpen
  \bibfield  {author} {\bibinfo {author} {\bibnamefont {{{Sadovskii},
  M.~V.}}},\ }\href@noop {} {\emph {\bibinfo {title} {Diagrammatics}}}\
  (\bibinfo  {publisher} {{World Scientific Publishing Co (2006)}},\ \bibinfo
  {year} {2006})\BibitemShut {NoStop}%
\bibitem [{\citenamefont {Kuchinskiy}\ and\ \citenamefont
  {Sadovskiy}(2024)}]{Kuchinskiy_2024}%
  \BibitemOpen
  \bibfield  {author} {\bibinfo {author} {\bibfnamefont {E.~Z.}\ \bibnamefont
  {Kuchinskiy}}\ and\ \bibinfo {author} {\bibfnamefont {M.~V.}\ \bibnamefont
  {Sadovskiy}},\ }\bibfield  {title} {\bibinfo {title} {Generalized dynamical
  keldysh model},\ }\href {https://doi.org/10.31857/s004445102407006x}
  {\bibfield  {journal} {\bibinfo  {journal} {Journal of Experimental and
  Theoretical Physics}\ }\textbf {\bibinfo {volume} {166}},\ \bibinfo {pages}
  {45–62} (\bibinfo {year} {2024})}\BibitemShut {NoStop}%
\bibitem [{\citenamefont {L\'evy}\ and\ \citenamefont
  {Sucher}(1969)}]{Levy1969}%
  \BibitemOpen
  \bibfield  {author} {\bibinfo {author} {\bibfnamefont {M.}~\bibnamefont
  {L\'evy}}\ and\ \bibinfo {author} {\bibfnamefont {J.}~\bibnamefont
  {Sucher}},\ }\bibfield  {title} {\bibinfo {title} {Eikonal approximation in
  quantum field theory},\ }\href {https://doi.org/10.1103/PhysRev.186.1656}
  {\bibfield  {journal} {\bibinfo  {journal} {Phys. Rev.}\ }\textbf {\bibinfo
  {volume} {186}},\ \bibinfo {pages} {1656} (\bibinfo {year}
  {1969})}\BibitemShut {NoStop}%
\bibitem [{\citenamefont {Lee}\ \emph {et~al.}(1973)\citenamefont {Lee},
  \citenamefont {Rice},\ and\ \citenamefont {Anderson}}]{Lee1973}%
  \BibitemOpen
  \bibfield  {author} {\bibinfo {author} {\bibfnamefont {P.~A.}\ \bibnamefont
  {Lee}}, \bibinfo {author} {\bibfnamefont {T.~M.}\ \bibnamefont {Rice}},\ and\
  \bibinfo {author} {\bibfnamefont {P.~W.}\ \bibnamefont {Anderson}},\
  }\bibfield  {title} {\bibinfo {title} {Fluctuation effects at a peierls
  transition},\ }\href {https://doi.org/10.1103/PhysRevLett.31.462} {\bibfield
  {journal} {\bibinfo  {journal} {Phys. Rev. Lett.}\ }\textbf {\bibinfo
  {volume} {31}},\ \bibinfo {pages} {462} (\bibinfo {year} {1973})}\BibitemShut
  {NoStop}%
\bibitem [{\citenamefont {Efros}(1971)}]{Efros1971}%
  \BibitemOpen
  \bibfield  {author} {\bibinfo {author} {\bibfnamefont {A.~L.}\ \bibnamefont
  {Efros}},\ }\bibfield  {title} {\bibinfo {title} {Theory of electron states
  in heavily doped semiconductors},\ }\href@noop {} {\bibfield  {journal}
  {\bibinfo  {journal} {Sov. Phys. JETP}\ }\textbf {\bibinfo {volume} {32}},\
  \bibinfo {pages} {479} (\bibinfo {year} {1971})}\BibitemShut {NoStop}%
\bibitem [{\citenamefont {Sadovskii}(1974{\natexlab{a}})}]{Sadovskii1974}%
  \BibitemOpen
  \bibfield  {author} {\bibinfo {author} {\bibfnamefont {M.~V.}\ \bibnamefont
  {Sadovskii}},\ }\bibfield  {title} {\bibinfo {title} {A model of a disordered
  system (a contribution to the theory of "liquid semiconductors")},\
  }\href@noop {} {\bibfield  {journal} {\bibinfo  {journal} {Sov. Phys. JETP}\
  }\textbf {\bibinfo {volume} {39}},\ \bibinfo {pages} {845} (\bibinfo {year}
  {1974}{\natexlab{a}})}\BibitemShut {NoStop}%
\bibitem [{\citenamefont {Posazhennikova}\ and\ \citenamefont
  {Coleman}(2003)}]{Posazhennikova2003}%
  \BibitemOpen
  \bibfield  {author} {\bibinfo {author} {\bibfnamefont {A.}~\bibnamefont
  {Posazhennikova}}\ and\ \bibinfo {author} {\bibfnamefont {P.}~\bibnamefont
  {Coleman}},\ }\bibfield  {title} {\bibinfo {title} {Quenched disorder
  formulation of the pseudogap problem},\ }\href
  {https://doi.org/10.1103/PhysRevB.67.165109} {\bibfield  {journal} {\bibinfo
  {journal} {Phys. Rev. B}\ }\textbf {\bibinfo {volume} {67}},\ \bibinfo
  {pages} {165109} (\bibinfo {year} {2003})}\BibitemShut {NoStop}%
\bibitem [{\citenamefont {Kiselev}\ and\ \citenamefont
  {Kikoin}(2009)}]{Kiselev2009}%
  \BibitemOpen
  \bibfield  {author} {\bibinfo {author} {\bibfnamefont {M.~N.}\ \bibnamefont
  {Kiselev}}\ and\ \bibinfo {author} {\bibfnamefont {K.~A.}\ \bibnamefont
  {Kikoin}},\ }\bibfield  {title} {\bibinfo {title} {Scalar and vector keldysh
  models in the time domain},\ }\href
  {https://doi.org/10.1134/S0021364009030047} {\bibfield  {journal} {\bibinfo
  {journal} {JETP Letters}\ }\textbf {\bibinfo {volume} {89}},\ \bibinfo
  {pages} {114} (\bibinfo {year} {2009})}\BibitemShut {NoStop}%
\bibitem [{\citenamefont {Efremov}\ and\ \citenamefont
  {Kiselev}(2022)}]{Efremov2022}%
  \BibitemOpen
  \bibfield  {author} {\bibinfo {author} {\bibfnamefont {D.~V.}\ \bibnamefont
  {Efremov}}\ and\ \bibinfo {author} {\bibfnamefont {M.~N.}\ \bibnamefont
  {Kiselev}},\ }\bibfield  {title} {\bibinfo {title} {{Seven Études on
  dynamical Keldysh model}},\ }\href
  {https://doi.org/10.21468/SciPostPhysLectNotes.65} {\bibfield  {journal}
  {\bibinfo  {journal} {SciPost Phys. Lect. Notes}\ ,\ \bibinfo {pages} {65}}
  (\bibinfo {year} {2022})}\BibitemShut {NoStop}%
\bibitem [{\citenamefont {Sadovskii}(1974{\natexlab{b}})}]{Sadovskii1974SS}%
  \BibitemOpen
  \bibfield  {author} {\bibinfo {author} {\bibfnamefont {M.~V.}\ \bibnamefont
  {Sadovskii}},\ }\bibfield  {title} {\bibinfo {title} {Theory of
  quasi-one-dimensional systems undergoing a peierls transition},\ }\href@noop
  {} {\bibfield  {journal} {\bibinfo  {journal} {Sov. Phys. -- Solid State}\
  }\textbf {\bibinfo {volume} {16}},\ \bibinfo {pages} {1632} (\bibinfo {year}
  {1974}{\natexlab{b}})}\BibitemShut {NoStop}%
\bibitem [{\citenamefont {Tchernyshyov}(1999)}]{Tchernyshyov1999}%
  \BibitemOpen
  \bibfield  {author} {\bibinfo {author} {\bibfnamefont {O.}~\bibnamefont
  {Tchernyshyov}},\ }\bibfield  {title} {\bibinfo {title} {Pseudogap in one
  dimension},\ }\href {https://doi.org/10.1103/PhysRevB.59.1358} {\bibfield
  {journal} {\bibinfo  {journal} {Phys. Rev. B}\ }\textbf {\bibinfo {volume}
  {59}},\ \bibinfo {pages} {1358} (\bibinfo {year} {1999})}\BibitemShut
  {NoStop}%
\bibitem [{\citenamefont {Yamase}\ and\ \citenamefont
  {Metzner}(2012)}]{Yamase_2012}%
  \BibitemOpen
  \bibfield  {author} {\bibinfo {author} {\bibfnamefont {H.}~\bibnamefont
  {Yamase}}\ and\ \bibinfo {author} {\bibfnamefont {W.}~\bibnamefont
  {Metzner}},\ }\bibfield  {title} {\bibinfo {title} {Fermi-surface truncation
  from thermal nematic fluctuations},\ }\href
  {https://doi.org/10.1103/PhysRevLett.108.186405} {\bibfield  {journal}
  {\bibinfo  {journal} {Phys. Rev. Lett.}\ }\textbf {\bibinfo {volume} {108}},\
  \bibinfo {pages} {186405} (\bibinfo {year} {2012})}\BibitemShut {NoStop}%
\bibitem [{\citenamefont {{Y.M. Vilk}}\ and\ \citenamefont {{A.-M.S.
  Tremblay}}(1997)}]{Vilk1997}%
  \BibitemOpen
  \bibfield  {author} {\bibinfo {author} {\bibnamefont {{Y.M. Vilk}}}\ and\
  \bibinfo {author} {\bibnamefont {{A.-M.S. Tremblay}}},\ }\bibfield  {title}
  {\bibinfo {title} {Non-perturbative many-body approach to the hubbard model
  and single-particle pseudogap},\ }\href {https://doi.org/10.1051/jp1:1997135}
  {\bibfield  {journal} {\bibinfo  {journal} {J. Phys. I France}\ }\textbf
  {\bibinfo {volume} {7}},\ \bibinfo {pages} {1309} (\bibinfo {year}
  {1997})}\BibitemShut {NoStop}%
\bibitem [{\citenamefont {Schmalian}\ \emph {et~al.}(1998)\citenamefont
  {Schmalian}, \citenamefont {Pines},\ and\ \citenamefont
  {Stojkovi\ifmmode~\acute{c}\else \'{c}\fi{}}}]{Schmalian1998}%
  \BibitemOpen
  \bibfield  {author} {\bibinfo {author} {\bibfnamefont {J.}~\bibnamefont
  {Schmalian}}, \bibinfo {author} {\bibfnamefont {D.}~\bibnamefont {Pines}},\
  and\ \bibinfo {author} {\bibfnamefont {B.}~\bibnamefont
  {Stojkovi\ifmmode~\acute{c}\else \'{c}\fi{}}},\ }\bibfield  {title} {\bibinfo
  {title} {Weak pseudogap behavior in the underdoped cuprate superconductors},\
  }\href {https://doi.org/10.1103/PhysRevLett.80.3839} {\bibfield  {journal}
  {\bibinfo  {journal} {Phys. Rev. Lett.}\ }\textbf {\bibinfo {volume} {80}},\
  \bibinfo {pages} {3839} (\bibinfo {year} {1998})}\BibitemShut {NoStop}%
\bibitem [{\citenamefont {Schmalian}\ \emph {et~al.}(1999)\citenamefont
  {Schmalian}, \citenamefont {Pines},\ and\ \citenamefont
  {Stojkovi\ifmmode~\acute{c}\else \'{c}\fi{}}}]{Schmalian1999}%
  \BibitemOpen
  \bibfield  {author} {\bibinfo {author} {\bibfnamefont {J.}~\bibnamefont
  {Schmalian}}, \bibinfo {author} {\bibfnamefont {D.}~\bibnamefont {Pines}},\
  and\ \bibinfo {author} {\bibfnamefont {B.}~\bibnamefont
  {Stojkovi\ifmmode~\acute{c}\else \'{c}\fi{}}},\ }\bibfield  {title} {\bibinfo
  {title} {Microscopic theory of weak pseudogap behavior in the underdoped
  cuprate superconductors: General theory and quasiparticle properties},\
  }\href {https://doi.org/10.1103/PhysRevB.60.667} {\bibfield  {journal}
  {\bibinfo  {journal} {Phys. Rev. B}\ }\textbf {\bibinfo {volume} {60}},\
  \bibinfo {pages} {667} (\bibinfo {year} {1999})}\BibitemShut {NoStop}%
\bibitem [{\citenamefont {Sadovskii}(2001)}]{Sadovskii_review}%
  \BibitemOpen
  \bibfield  {author} {\bibinfo {author} {\bibfnamefont {M.~V.}\ \bibnamefont
  {Sadovskii}},\ }\bibfield  {title} {\bibinfo {title} {Pseudogap in
  high-temperature superconductors},\ }\href
  {https://doi.org/10.1070/PU2001v044n05ABEH000902} {\bibfield  {journal}
  {\bibinfo  {journal} {Phys. Usp.}\ }\textbf {\bibinfo {volume} {44}},\
  \bibinfo {pages} {515} (\bibinfo {year} {2001})}\BibitemShut {NoStop}%
\bibitem [{\citenamefont {Sadovskii}\ \emph {et~al.}(2005)\citenamefont
  {Sadovskii}, \citenamefont {Nekrasov}, \citenamefont {Kuchinskii},
  \citenamefont {Pruschke},\ and\ \citenamefont {Anisimov}}]{Sadovskii_extra}%
  \BibitemOpen
  \bibfield  {author} {\bibinfo {author} {\bibfnamefont {M.~V.}\ \bibnamefont
  {Sadovskii}}, \bibinfo {author} {\bibfnamefont {I.~A.}\ \bibnamefont
  {Nekrasov}}, \bibinfo {author} {\bibfnamefont {E.~Z.}\ \bibnamefont
  {Kuchinskii}}, \bibinfo {author} {\bibfnamefont {T.}~\bibnamefont
  {Pruschke}},\ and\ \bibinfo {author} {\bibfnamefont {V.~I.}\ \bibnamefont
  {Anisimov}},\ }\bibfield  {title} {\bibinfo {title} {Pseudogaps in strongly
  correlated metals: A generalized dynamical mean-field theory approach},\
  }\href {https://doi.org/10.1103/PhysRevB.72.155105} {\bibfield  {journal}
  {\bibinfo  {journal} {Phys. Rev. B}\ }\textbf {\bibinfo {volume} {72}},\
  \bibinfo {pages} {155105} (\bibinfo {year} {2005})}\BibitemShut {NoStop}%
\bibitem [{\citenamefont {Kuchinskii}\ \emph {et~al.}(2006)\citenamefont
  {Kuchinskii}, \citenamefont {Nekrasov},\ and\ \citenamefont
  {Sadovskii}}]{Sadovskii_extra_1}%
  \BibitemOpen
  \bibfield  {author} {\bibinfo {author} {\bibfnamefont {E.~Z.}\ \bibnamefont
  {Kuchinskii}}, \bibinfo {author} {\bibfnamefont {I.~A.}\ \bibnamefont
  {Nekrasov}},\ and\ \bibinfo {author} {\bibfnamefont {M.~V.}\ \bibnamefont
  {Sadovskii}},\ }\bibfield  {title} {\bibinfo {title} {{Pseudogaps:
  introducing the length scale into dynamical mean-field theory}},\ }\href
  {https://doi.org/10.1063/1.2199442} {\bibfield  {journal} {\bibinfo
  {journal} {Low Temperature Physics}\ }\textbf {\bibinfo {volume} {32}},\
  \bibinfo {pages} {398} (\bibinfo {year} {2006})},\ \Eprint
  {https://arxiv.org/abs/https://pubs.aip.org/aip/ltp/article-pdf/32/4/398/13901426/398\_1\_online.pdf}
  {https://pubs.aip.org/aip/ltp/article-pdf/32/4/398/13901426/398\_1\_online.pdf}
  \BibitemShut {NoStop}%
\bibitem [{\citenamefont {Rohe}\ and\ \citenamefont
  {Metzner}(2005)}]{Rohe_2005}%
  \BibitemOpen
  \bibfield  {author} {\bibinfo {author} {\bibfnamefont {D.}~\bibnamefont
  {Rohe}}\ and\ \bibinfo {author} {\bibfnamefont {W.}~\bibnamefont {Metzner}},\
  }\bibfield  {title} {\bibinfo {title} {Pseudogap at hot spots in the
  two-dimensional hubbard model at weak coupling},\ }\href
  {https://doi.org/10.1103/PhysRevB.71.115116} {\bibfield  {journal} {\bibinfo
  {journal} {Phys. Rev. B}\ }\textbf {\bibinfo {volume} {71}},\ \bibinfo
  {pages} {115116} (\bibinfo {year} {2005})}\BibitemShut {NoStop}%
\bibitem [{\citenamefont {Sedrakyan}\ and\ \citenamefont
  {Chubukov}(2010)}]{Sedrakyan2010}%
  \BibitemOpen
  \bibfield  {author} {\bibinfo {author} {\bibfnamefont {T.~A.}\ \bibnamefont
  {Sedrakyan}}\ and\ \bibinfo {author} {\bibfnamefont {A.~V.}\ \bibnamefont
  {Chubukov}},\ }\bibfield  {title} {\bibinfo {title} {Pseudogap in underdoped
  cuprates and spin-density-wave fluctuations},\ }\href
  {https://doi.org/10.1103/PhysRevB.81.174536} {\bibfield  {journal} {\bibinfo
  {journal} {Phys. Rev. B}\ }\textbf {\bibinfo {volume} {81}},\ \bibinfo
  {pages} {174536} (\bibinfo {year} {2010})}\BibitemShut {NoStop}%
\bibitem [{\citenamefont {Yamase}\ \emph {et~al.}(2016)\citenamefont {Yamase},
  \citenamefont {Eberlein},\ and\ \citenamefont {Metzner}}]{Yamase_2016}%
  \BibitemOpen
  \bibfield  {author} {\bibinfo {author} {\bibfnamefont {H.}~\bibnamefont
  {Yamase}}, \bibinfo {author} {\bibfnamefont {A.}~\bibnamefont {Eberlein}},\
  and\ \bibinfo {author} {\bibfnamefont {W.}~\bibnamefont {Metzner}},\
  }\bibfield  {title} {\bibinfo {title} {Coexistence of incommensurate
  magnetism and superconductivity in the two-dimensional hubbard model},\
  }\href {https://doi.org/10.1103/PhysRevLett.116.096402} {\bibfield  {journal}
  {\bibinfo  {journal} {Phys. Rev. Lett.}\ }\textbf {\bibinfo {volume} {116}},\
  \bibinfo {pages} {096402} (\bibinfo {year} {2016})}\BibitemShut {NoStop}%
\bibitem [{\citenamefont {Ye}\ \emph {et~al.}(2023)\citenamefont {Ye},
  \citenamefont {Wang}, \citenamefont {Fernandes},\ and\ \citenamefont
  {Chubukov}}]{Ye2023}%
  \BibitemOpen
  \bibfield  {author} {\bibinfo {author} {\bibfnamefont {M.}~\bibnamefont
  {Ye}}, \bibinfo {author} {\bibfnamefont {Z.}~\bibnamefont {Wang}}, \bibinfo
  {author} {\bibfnamefont {R.~M.}\ \bibnamefont {Fernandes}},\ and\ \bibinfo
  {author} {\bibfnamefont {A.~V.}\ \bibnamefont {Chubukov}},\ }\bibfield
  {title} {\bibinfo {title} {Location and thermal evolution of the pseudogap
  due to spin fluctuations},\ }\href
  {https://doi.org/10.1103/PhysRevB.108.115156} {\bibfield  {journal} {\bibinfo
   {journal} {Phys. Rev. B}\ }\textbf {\bibinfo {volume} {108}},\ \bibinfo
  {pages} {115156} (\bibinfo {year} {2023})}\BibitemShut {NoStop}%
\bibitem [{\citenamefont {Ye}\ and\ \citenamefont {Chubukov}(2023)}]{Ye2023_1}%
  \BibitemOpen
  \bibfield  {author} {\bibinfo {author} {\bibfnamefont {M.}~\bibnamefont
  {Ye}}\ and\ \bibinfo {author} {\bibfnamefont {A.~V.}\ \bibnamefont
  {Chubukov}},\ }\bibfield  {title} {\bibinfo {title} {Crucial role of thermal
  fluctuations and vertex corrections for the magnetic pseudogap},\ }\href
  {https://doi.org/10.1103/PhysRevB.108.L081118} {\bibfield  {journal}
  {\bibinfo  {journal} {Phys. Rev. B}\ }\textbf {\bibinfo {volume} {108}},\
  \bibinfo {pages} {L081118} (\bibinfo {year} {2023})}\BibitemShut {NoStop}%
\bibitem [{Note3()}]{Note3}%
  \BibitemOpen
  \bibinfo {note} {The series diverge for {\protect \it any} $\omega $,
  including the largest $\omega \gg \beta \omega _0$.}\BibitemShut {Stop}%
\bibitem [{\citenamefont {Ciuchi}\ \emph {et~al.}(1997)\citenamefont {Ciuchi},
  \citenamefont {de~Pasquale}, \citenamefont {Fratini},\ and\ \citenamefont
  {Feinberg}}]{Ciuchi_1997}%
  \BibitemOpen
  \bibfield  {author} {\bibinfo {author} {\bibfnamefont {S.}~\bibnamefont
  {Ciuchi}}, \bibinfo {author} {\bibfnamefont {F.}~\bibnamefont {de~Pasquale}},
  \bibinfo {author} {\bibfnamefont {S.}~\bibnamefont {Fratini}},\ and\ \bibinfo
  {author} {\bibfnamefont {D.}~\bibnamefont {Feinberg}},\ }\bibfield  {title}
  {\bibinfo {title} {Dynamical mean-field theory of the small polaron},\ }\href
  {https://doi.org/10.1103/PhysRevB.56.4494} {\bibfield  {journal} {\bibinfo
  {journal} {Phys. Rev. B}\ }\textbf {\bibinfo {volume} {56}},\ \bibinfo
  {pages} {4494} (\bibinfo {year} {1997})}\BibitemShut {NoStop}%
\bibitem [{\citenamefont {Pairault}\ \emph {et~al.}(1998)\citenamefont
  {Pairault}, \citenamefont {S\'en\'echal},\ and\ \citenamefont
  {Tremblay}}]{Pairault1998}%
  \BibitemOpen
  \bibfield  {author} {\bibinfo {author} {\bibfnamefont {S.}~\bibnamefont
  {Pairault}}, \bibinfo {author} {\bibfnamefont {D.}~\bibnamefont
  {S\'en\'echal}},\ and\ \bibinfo {author} {\bibfnamefont {A.-M.~S.}\
  \bibnamefont {Tremblay}},\ }\bibfield  {title} {\bibinfo {title}
  {Strong-coupling expansion for the hubbard model},\ }\href
  {https://doi.org/10.1103/PhysRevLett.80.5389} {\bibfield  {journal} {\bibinfo
   {journal} {Phys. Rev. Lett.}\ }\textbf {\bibinfo {volume} {80}},\ \bibinfo
  {pages} {5389} (\bibinfo {year} {1998})}\BibitemShut {NoStop}%
\bibitem [{\citenamefont {Pairault}\ \emph {et~al.}(2000)\citenamefont
  {Pairault}, \citenamefont {S{\'e}n{\'e}chal},\ and\ \citenamefont
  {Tremblay}}]{Pairault2000}%
  \BibitemOpen
  \bibfield  {author} {\bibinfo {author} {\bibfnamefont {S.}~\bibnamefont
  {Pairault}}, \bibinfo {author} {\bibfnamefont {D.}~\bibnamefont
  {S{\'e}n{\'e}chal}},\ and\ \bibinfo {author} {\bibfnamefont {A.-M.~S.}\
  \bibnamefont {Tremblay}},\ }\bibfield  {title} {\bibinfo {title}
  {Strong-coupling perturbation theory of the hubbard model},\ }\href
  {https://doi.org/10.1007/s100510070253} {\bibfield  {journal} {\bibinfo
  {journal} {The European Physical Journal B - Condensed Matter and Complex
  Systems}\ }\textbf {\bibinfo {volume} {16}},\ \bibinfo {pages} {85} (\bibinfo
  {year} {2000})}\BibitemShut {NoStop}%
\bibitem [{\citenamefont {Goodvin}\ \emph {et~al.}(2006)\citenamefont
  {Goodvin}, \citenamefont {Berciu},\ and\ \citenamefont
  {Sawatzky}}]{Goodvin2006}%
  \BibitemOpen
  \bibfield  {author} {\bibinfo {author} {\bibfnamefont {G.~L.}\ \bibnamefont
  {Goodvin}}, \bibinfo {author} {\bibfnamefont {M.}~\bibnamefont {Berciu}},\
  and\ \bibinfo {author} {\bibfnamefont {G.~A.}\ \bibnamefont {Sawatzky}},\
  }\bibfield  {title} {\bibinfo {title} {Green's function of the holstein
  polaron},\ }\href {https://doi.org/10.1103/PhysRevB.74.245104} {\bibfield
  {journal} {\bibinfo  {journal} {Phys. Rev. B}\ }\textbf {\bibinfo {volume}
  {74}},\ \bibinfo {pages} {245104} (\bibinfo {year} {2006})}\BibitemShut
  {NoStop}%
\bibitem [{\citenamefont {Berciu}(2006)}]{Berciu_2006}%
  \BibitemOpen
  \bibfield  {author} {\bibinfo {author} {\bibfnamefont {M.}~\bibnamefont
  {Berciu}},\ }\bibfield  {title} {\bibinfo {title} {Green's function of a
  dressed particle},\ }\href {https://doi.org/10.1103/PhysRevLett.97.036402}
  {\bibfield  {journal} {\bibinfo  {journal} {Phys. Rev. Lett.}\ }\textbf
  {\bibinfo {volume} {97}},\ \bibinfo {pages} {036402} (\bibinfo {year}
  {2006})}\BibitemShut {NoStop}%
\bibitem [{Note4()}]{Note4}%
  \BibitemOpen
  \bibinfo {note} {Numerical calculations at a finite ${\protect \tilde \omega
  }$ require some care as one has to select the contribution from a particular
  pole at $\omega = k \omega _0$, which can be done by either taking a very
  small ${\protect \tilde \omega }$ or keeping it small but finite and
  comparing $G^e (k \omega _0 + {\protect \tilde \omega })$ with the full form
  of $G^H (k \omega _0 + {\protect \tilde \omega }, 0)$ from (\ref {n_2}). For
  negative $k$, we verified that $(G^e (k \omega _0))^{-1}$ does not vanish,
  i.e., there are no poles at negative $\omega $. This is what we expect at the
  density $n =0+$.}\BibitemShut {Stop}%
\bibitem [{\citenamefont {Alexandrov}\ and\ \citenamefont
  {Ranninger}(1992)}]{Ranninger_1992}%
  \BibitemOpen
  \bibfield  {author} {\bibinfo {author} {\bibfnamefont {A.~S.}\ \bibnamefont
  {Alexandrov}}\ and\ \bibinfo {author} {\bibfnamefont {J.}~\bibnamefont
  {Ranninger}},\ }\bibfield  {title} {\bibinfo {title} {Polaronic effects in
  the photoemission spectra of strongly coupled electron-phonon systems},\
  }\href {https://doi.org/10.1103/PhysRevB.45.13109} {\bibfield  {journal}
  {\bibinfo  {journal} {Phys. Rev. B}\ }\textbf {\bibinfo {volume} {45}},\
  \bibinfo {pages} {13109} (\bibinfo {year} {1992})}\BibitemShut {NoStop}%
\bibitem [{\citenamefont {Sadovskii}(2022)}]{Sadovskii:2022_ufn}%
  \BibitemOpen
  \bibfield  {author} {\bibinfo {author} {\bibfnamefont {M.~V.}\ \bibnamefont
  {Sadovskii}},\ }\bibfield  {title} {\bibinfo {title} {Limits of eliashberg
  theory and bounds for superconducting transition temperature},\ }\href
  {https://doi.org/10.3367/UFNe.2021.05.039007} {\bibfield  {journal} {\bibinfo
   {journal} {Phys. Usp.}\ }\textbf {\bibinfo {volume} {65}},\ \bibinfo {pages}
  {724} (\bibinfo {year} {2022})}\BibitemShut {NoStop}%
\bibitem [{\citenamefont {Bonetti}\ \emph {et~al.}(2025)\citenamefont
  {Bonetti}, \citenamefont {Christos}, \citenamefont {Nikolaenko},
  \citenamefont {Patel},\ and\ \citenamefont {Sachdev}}]{Bonetti_2025}%
  \BibitemOpen
  \bibfield  {author} {\bibinfo {author} {\bibfnamefont {P.~M.}\ \bibnamefont
  {Bonetti}}, \bibinfo {author} {\bibfnamefont {M.}~\bibnamefont {Christos}},
  \bibinfo {author} {\bibfnamefont {A.}~\bibnamefont {Nikolaenko}}, \bibinfo
  {author} {\bibfnamefont {A.~A.}\ \bibnamefont {Patel}},\ and\ \bibinfo
  {author} {\bibfnamefont {S.}~\bibnamefont {Sachdev}},\ }\href
  {https://arxiv.org/abs/2508.20164} {\bibinfo {title} {Critical quantum
  liquids and the cuprate high temperature superconductors}} (\bibinfo {year}
  {2025}),\ \Eprint {https://arxiv.org/abs/2508.20164} {arXiv:2508.20164
  [cond-mat.str-el]} \BibitemShut {NoStop}%
\bibitem [{Note5()}]{Note5}%
  \BibitemOpen
  \bibinfo {note} {We follow~\cite {Lifshitz2006} and call $H'$ the
  Hamiltonian.}\BibitemShut {Stop}%
\bibitem [{\citenamefont {Lifshitz}\ and\ \citenamefont
  {Pitaevskii}(2006)}]{Lifshitz2006}%
  \BibitemOpen
  \bibfield  {author} {\bibinfo {author} {\bibfnamefont {E.~M.}\ \bibnamefont
  {Lifshitz}}\ and\ \bibinfo {author} {\bibfnamefont {L.~P.}\ \bibnamefont
  {Pitaevskii}},\ }\href@noop {} {\emph {\bibinfo {title} {Course of
  Theoretical Physics Vol. 9}}},\ \bibinfo {edition} {reprinted}\ ed.\
  (\bibinfo  {publisher} {Elsevier},\ \bibinfo {address} {Oxford},\ \bibinfo
  {year} {2006})\ p.\ \bibinfo {pages} {387}\BibitemShut {NoStop}%
\bibitem [{\citenamefont {Senthil}\ \emph {et~al.}(2003)\citenamefont
  {Senthil}, \citenamefont {Sachdev},\ and\ \citenamefont
  {Vojta}}]{Senthil_2003}%
  \BibitemOpen
  \bibfield  {author} {\bibinfo {author} {\bibfnamefont {T.}~\bibnamefont
  {Senthil}}, \bibinfo {author} {\bibfnamefont {S.}~\bibnamefont {Sachdev}},\
  and\ \bibinfo {author} {\bibfnamefont {M.}~\bibnamefont {Vojta}},\ }\bibfield
   {title} {\bibinfo {title} {Fractionalized fermi liquids},\ }\href
  {https://doi.org/10.1103/PhysRevLett.90.216403} {\bibfield  {journal}
  {\bibinfo  {journal} {Phys. Rev. Lett.}\ }\textbf {\bibinfo {volume} {90}},\
  \bibinfo {pages} {216403} (\bibinfo {year} {2003})}\BibitemShut {NoStop}%
\bibitem [{\citenamefont {Sachdev}(2010)}]{Sachdev_2010}%
  \BibitemOpen
  \bibfield  {author} {\bibinfo {author} {\bibfnamefont {S.}~\bibnamefont
  {Sachdev}},\ }\bibfield  {title} {\bibinfo {title} {Holographic metals and
  the fractionalized fermi liquid},\ }\href
  {https://doi.org/10.1103/PhysRevLett.105.151602} {\bibfield  {journal}
  {\bibinfo  {journal} {Phys. Rev. Lett.}\ }\textbf {\bibinfo {volume} {105}},\
  \bibinfo {pages} {151602} (\bibinfo {year} {2010})}\BibitemShut {NoStop}%
\bibitem [{\citenamefont {Chubukov}(1995)}]{Chubukov1995}%
  \BibitemOpen
  \bibfield  {author} {\bibinfo {author} {\bibfnamefont {A.~V.}\ \bibnamefont
  {Chubukov}},\ }\bibfield  {title} {\bibinfo {title} {Quasiparticle spectrum
  in a nearly antiferromagnetic fermi liquid: Shadow and flat bands},\ }\href
  {https://doi.org/10.1103/PhysRevB.52.R3840} {\bibfield  {journal} {\bibinfo
  {journal} {Phys. Rev. B}\ }\textbf {\bibinfo {volume} {52}},\ \bibinfo
  {pages} {R3840(R)} (\bibinfo {year} {1995})}\BibitemShut {NoStop}%
\bibitem [{\citenamefont {Vilk}(1997)}]{Vilk1997_a}%
  \BibitemOpen
  \bibfield  {author} {\bibinfo {author} {\bibfnamefont {Y.~M.}\ \bibnamefont
  {Vilk}},\ }\bibfield  {title} {\bibinfo {title} {Shadow features and shadow
  bands in the paramagnetic state of cuprate superconductors},\ }\href
  {https://doi.org/10.1103/PhysRevB.55.3870} {\bibfield  {journal} {\bibinfo
  {journal} {Phys. Rev. B}\ }\textbf {\bibinfo {volume} {55}},\ \bibinfo
  {pages} {3870} (\bibinfo {year} {1997})}\BibitemShut {NoStop}%
\bibitem [{\citenamefont {Altshuler}\ \emph {et~al.}(1998)\citenamefont
  {Altshuler}, \citenamefont {Chubukov}, \citenamefont {Dashevskii},
  \citenamefont {Finkel'stein},\ and\ \citenamefont {Morr}}]{Altshuler1997}%
  \BibitemOpen
  \bibfield  {author} {\bibinfo {author} {\bibfnamefont {B.~L.}\ \bibnamefont
  {Altshuler}}, \bibinfo {author} {\bibfnamefont {A.~V.}\ \bibnamefont
  {Chubukov}}, \bibinfo {author} {\bibfnamefont {A.}~\bibnamefont
  {Dashevskii}}, \bibinfo {author} {\bibfnamefont {A.~M.}\ \bibnamefont
  {Finkel'stein}},\ and\ \bibinfo {author} {\bibfnamefont {D.~K.}\ \bibnamefont
  {Morr}},\ }\bibfield  {title} {\bibinfo {title} {Luttinger theorem for a
  spin-density-wave state},\ }\href {https://doi.org/10.1209/epl/i1998-00164-y}
  {\bibfield  {journal} {\bibinfo  {journal} {Europhysics Letters}\ }\textbf
  {\bibinfo {volume} {41}},\ \bibinfo {pages} {401} (\bibinfo {year}
  {1998})}\BibitemShut {NoStop}%
\bibitem [{\citenamefont {Georges}\ \emph {et~al.}(2001)\citenamefont
  {Georges}, \citenamefont {Parcollet},\ and\ \citenamefont
  {Sachdev}}]{Georges2001}%
  \BibitemOpen
  \bibfield  {author} {\bibinfo {author} {\bibfnamefont {A.}~\bibnamefont
  {Georges}}, \bibinfo {author} {\bibfnamefont {O.}~\bibnamefont {Parcollet}},\
  and\ \bibinfo {author} {\bibfnamefont {S.}~\bibnamefont {Sachdev}},\
  }\bibfield  {title} {\bibinfo {title} {Quantum fluctuations of a nearly
  critical heisenberg spin glass},\ }\href
  {https://doi.org/10.1103/PhysRevB.63.134406} {\bibfield  {journal} {\bibinfo
  {journal} {Phys. Rev. B}\ }\textbf {\bibinfo {volume} {63}},\ \bibinfo
  {pages} {134406} (\bibinfo {year} {2001})}\BibitemShut {NoStop}%
\bibitem [{\citenamefont {Dzyaloshinskii}(2003)}]{Dzyaloshinskii2003}%
  \BibitemOpen
  \bibfield  {author} {\bibinfo {author} {\bibfnamefont {I.}~\bibnamefont
  {Dzyaloshinskii}},\ }\bibfield  {title} {\bibinfo {title} {Some consequences
  of the luttinger theorem: The luttinger surfaces in non-fermi liquids and
  mott insulators},\ }\href {https://doi.org/10.1103/PhysRevB.68.085113}
  {\bibfield  {journal} {\bibinfo  {journal} {Phys. Rev. B}\ }\textbf {\bibinfo
  {volume} {68}},\ \bibinfo {pages} {085113} (\bibinfo {year}
  {2003})}\BibitemShut {NoStop}%
\bibitem [{\citenamefont {Konik}\ \emph {et~al.}(2006)\citenamefont {Konik},
  \citenamefont {Rice},\ and\ \citenamefont {Tsvelik}}]{Konik2006}%
  \BibitemOpen
  \bibfield  {author} {\bibinfo {author} {\bibfnamefont {R.~M.}\ \bibnamefont
  {Konik}}, \bibinfo {author} {\bibfnamefont {T.~M.}\ \bibnamefont {Rice}},\
  and\ \bibinfo {author} {\bibfnamefont {A.~M.}\ \bibnamefont {Tsvelik}},\
  }\bibfield  {title} {\bibinfo {title} {Doped spin liquid: Luttinger sum rule
  and low temperature order},\ }\href
  {https://doi.org/10.1103/PhysRevLett.96.086407} {\bibfield  {journal}
  {\bibinfo  {journal} {Phys. Rev. Lett.}\ }\textbf {\bibinfo {volume} {96}},\
  \bibinfo {pages} {086407} (\bibinfo {year} {2006})}\BibitemShut {NoStop}%
\bibitem [{\citenamefont {Stanescu}\ and\ \citenamefont
  {Kotliar}(2006)}]{Stanescu2006}%
  \BibitemOpen
  \bibfield  {author} {\bibinfo {author} {\bibfnamefont {T.~D.}\ \bibnamefont
  {Stanescu}}\ and\ \bibinfo {author} {\bibfnamefont {G.}~\bibnamefont
  {Kotliar}},\ }\bibfield  {title} {\bibinfo {title} {Fermi arcs and hidden
  zeros of the green function in the pseudogap state},\ }\href
  {https://doi.org/10.1103/PhysRevB.74.125110} {\bibfield  {journal} {\bibinfo
  {journal} {Phys. Rev. B}\ }\textbf {\bibinfo {volume} {74}},\ \bibinfo
  {pages} {125110} (\bibinfo {year} {2006})}\BibitemShut {NoStop}%
\bibitem [{\citenamefont {Rosch}(2007)}]{Rosch2007}%
  \BibitemOpen
  \bibfield  {author} {\bibinfo {author} {\bibfnamefont {A.}~\bibnamefont
  {Rosch}},\ }\bibfield  {title} {\bibinfo {title} {Breakdown of luttinger's
  theorem in two-orbital mott insulators},\ }\href
  {https://doi.org/10.1140/epjb/e2007-00312-3} {\bibfield  {journal} {\bibinfo
  {journal} {The European Physical Journal B: Condensed Matter and Complex
  Systems}\ }\textbf {\bibinfo {volume} {59}},\ \bibinfo {pages} {495}
  (\bibinfo {year} {2007})}\BibitemShut {NoStop}%
\bibitem [{\citenamefont {Dave}\ \emph {et~al.}(2013)\citenamefont {Dave},
  \citenamefont {Phillips},\ and\ \citenamefont {Kane}}]{Kiran2013}%
  \BibitemOpen
  \bibfield  {author} {\bibinfo {author} {\bibfnamefont {K.~B.}\ \bibnamefont
  {Dave}}, \bibinfo {author} {\bibfnamefont {P.~W.}\ \bibnamefont {Phillips}},\
  and\ \bibinfo {author} {\bibfnamefont {C.~L.}\ \bibnamefont {Kane}},\
  }\bibfield  {title} {\bibinfo {title} {Absence of luttinger's theorem due to
  zeros in the single-particle green function},\ }\href
  {https://doi.org/10.1103/PhysRevLett.110.090403} {\bibfield  {journal}
  {\bibinfo  {journal} {Phys. Rev. Lett.}\ }\textbf {\bibinfo {volume} {110}},\
  \bibinfo {pages} {090403} (\bibinfo {year} {2013})}\BibitemShut {NoStop}%
\bibitem [{\citenamefont {Sachdev}(2023)}]{Sachdev_book}%
  \BibitemOpen
  \bibfield  {author} {\bibinfo {author} {\bibfnamefont {S.}~\bibnamefont
  {Sachdev}},\ }\href {https://doi.org/10.1017/9781009212687} {\emph {\bibinfo
  {title} {Quantum Phases of Matter}}}\ (\bibinfo  {publisher} {Cambridge
  University Press},\ \bibinfo {address} {Cambridge, United Kingdom},\ \bibinfo
  {year} {2023})\BibitemShut {NoStop}%
\bibitem [{\citenamefont {Fabrizio}(2022)}]{Fabrizio2022}%
  \BibitemOpen
  \bibfield  {author} {\bibinfo {author} {\bibfnamefont {M.}~\bibnamefont
  {Fabrizio}},\ }\bibfield  {title} {\bibinfo {title} {Emergent quasiparticles
  at luttinger surfaces},\ }\href {https://doi.org/10.1038/s41467-022-29190-y}
  {\bibfield  {journal} {\bibinfo  {journal} {Nature Communications}\ }\textbf
  {\bibinfo {volume} {13}},\ \bibinfo {pages} {1561} (\bibinfo {year}
  {2022})}\BibitemShut {NoStop}%
\bibitem [{\citenamefont {Skolimowski}\ and\ \citenamefont
  {Fabrizio}(2022)}]{Skolimowski2022}%
  \BibitemOpen
  \bibfield  {author} {\bibinfo {author} {\bibfnamefont {J.}~\bibnamefont
  {Skolimowski}}\ and\ \bibinfo {author} {\bibfnamefont {M.}~\bibnamefont
  {Fabrizio}},\ }\bibfield  {title} {\bibinfo {title} {Luttinger's theorem in
  the presence of luttinger surfaces},\ }\href
  {https://doi.org/10.1103/PhysRevB.106.045109} {\bibfield  {journal} {\bibinfo
   {journal} {Phys. Rev. B}\ }\textbf {\bibinfo {volume} {106}},\ \bibinfo
  {pages} {045109} (\bibinfo {year} {2022})}\BibitemShut {NoStop}%
\bibitem [{\citenamefont {Blason}\ and\ \citenamefont
  {Fabrizio}(2023)}]{Blason_2023}%
  \BibitemOpen
  \bibfield  {author} {\bibinfo {author} {\bibfnamefont {A.}~\bibnamefont
  {Blason}}\ and\ \bibinfo {author} {\bibfnamefont {M.}~\bibnamefont
  {Fabrizio}},\ }\bibfield  {title} {\bibinfo {title} {Unified role of green's
  function poles and zeros in correlated topological insulators},\ }\href
  {https://doi.org/10.1103/PhysRevB.108.125115} {\bibfield  {journal} {\bibinfo
   {journal} {Phys. Rev. B}\ }\textbf {\bibinfo {volume} {108}},\ \bibinfo
  {pages} {125115} (\bibinfo {year} {2023})}\BibitemShut {NoStop}%
\bibitem [{\citenamefont {Staffieri}\ and\ \citenamefont
  {Fabrizio}(2025)}]{Staffieri_2025}%
  \BibitemOpen
  \bibfield  {author} {\bibinfo {author} {\bibfnamefont {G.}~\bibnamefont
  {Staffieri}}\ and\ \bibinfo {author} {\bibfnamefont {M.}~\bibnamefont
  {Fabrizio}},\ }\bibfield  {title} {\bibinfo {title} {Signatures of the fermi
  surface reconstruction of a doped mott insulator in a slab geometry},\ }\href
  {https://doi.org/10.1103/hyth-626t} {\bibfield  {journal} {\bibinfo
  {journal} {Phys. Rev. B}\ }\textbf {\bibinfo {volume} {112}},\ \bibinfo
  {pages} {155155} (\bibinfo {year} {2025})}\BibitemShut {NoStop}%
\bibitem [{\citenamefont {Scheurer}\ \emph {et~al.}(2018)\citenamefont
  {Scheurer}, \citenamefont {Chatterjee}, \citenamefont {Wu}, \citenamefont
  {Ferrero}, \citenamefont {Georges},\ and\ \citenamefont
  {Sachdev}}]{Scheurer2018}%
  \BibitemOpen
  \bibfield  {author} {\bibinfo {author} {\bibfnamefont {M.~S.}\ \bibnamefont
  {Scheurer}}, \bibinfo {author} {\bibfnamefont {S.}~\bibnamefont
  {Chatterjee}}, \bibinfo {author} {\bibfnamefont {W.}~\bibnamefont {Wu}},
  \bibinfo {author} {\bibfnamefont {M.}~\bibnamefont {Ferrero}}, \bibinfo
  {author} {\bibfnamefont {A.}~\bibnamefont {Georges}},\ and\ \bibinfo {author}
  {\bibfnamefont {S.}~\bibnamefont {Sachdev}},\ }\bibfield  {title} {\bibinfo
  {title} {Topological order in the pseudogap metal},\ }\href
  {https://doi.org/10.1073/pnas.1720580115} {\bibfield  {journal} {\bibinfo
  {journal} {PNAS}\ }\textbf {\bibinfo {volume} {115}},\ \bibinfo {pages}
  {E3665} (\bibinfo {year} {2018})}\BibitemShut {NoStop}%
\bibitem [{\citenamefont {Peralta~Gavensky}\ \emph {et~al.}(2023)\citenamefont
  {Peralta~Gavensky}, \citenamefont {Sachdev},\ and\ \citenamefont
  {Goldman}}]{Peralta2023}%
  \BibitemOpen
  \bibfield  {author} {\bibinfo {author} {\bibfnamefont {L.}~\bibnamefont
  {Peralta~Gavensky}}, \bibinfo {author} {\bibfnamefont {S.}~\bibnamefont
  {Sachdev}},\ and\ \bibinfo {author} {\bibfnamefont {N.}~\bibnamefont
  {Goldman}},\ }\bibfield  {title} {\bibinfo {title} {Connecting the many-body
  chern number to luttinger's theorem through st\ifmmode \check{r}\else
  \v{r}\fi{}eda's formula},\ }\href
  {https://doi.org/10.1103/PhysRevLett.131.236601} {\bibfield  {journal}
  {\bibinfo  {journal} {Phys. Rev. Lett.}\ }\textbf {\bibinfo {volume} {131}},\
  \bibinfo {pages} {236601} (\bibinfo {year} {2023})}\BibitemShut {NoStop}%
\bibitem [{\citenamefont {Bollmann}\ \emph {et~al.}(2024)\citenamefont
  {Bollmann}, \citenamefont {Setty}, \citenamefont {Seifert},\ and\
  \citenamefont {K\"onig}}]{Bollmann2024}%
  \BibitemOpen
  \bibfield  {author} {\bibinfo {author} {\bibfnamefont {S.}~\bibnamefont
  {Bollmann}}, \bibinfo {author} {\bibfnamefont {C.}~\bibnamefont {Setty}},
  \bibinfo {author} {\bibfnamefont {U.~F.~P.}\ \bibnamefont {Seifert}},\ and\
  \bibinfo {author} {\bibfnamefont {E.~J.}\ \bibnamefont {K\"onig}},\
  }\bibfield  {title} {\bibinfo {title} {Topological green's function zeros in
  an exactly solved model and beyond},\ }\href
  {https://doi.org/10.1103/PhysRevLett.133.136504} {\bibfield  {journal}
  {\bibinfo  {journal} {Phys. Rev. Lett.}\ }\textbf {\bibinfo {volume} {133}},\
  \bibinfo {pages} {136504} (\bibinfo {year} {2024})}\BibitemShut {NoStop}%
\bibitem [{\citenamefont {Markov}\ \emph {et~al.}(2026)\citenamefont {Markov},
  \citenamefont {Nikishin}, \citenamefont {Cooper}, \citenamefont {Goldman},\
  and\ \citenamefont {Gavensky}}]{Markov2026}%
  \BibitemOpen
  \bibfield  {author} {\bibinfo {author} {\bibfnamefont {A.~A.}\ \bibnamefont
  {Markov}}, \bibinfo {author} {\bibfnamefont {A.~M.}\ \bibnamefont
  {Nikishin}}, \bibinfo {author} {\bibfnamefont {N.~R.}\ \bibnamefont
  {Cooper}}, \bibinfo {author} {\bibfnamefont {N.}~\bibnamefont {Goldman}},\
  and\ \bibinfo {author} {\bibfnamefont {L.~P.}\ \bibnamefont {Gavensky}},\
  }\href {https://arxiv.org/abs/2603.17006} {\bibinfo {title} {Luttinger's
  theorem violation and green's function topological invariants in a fractional
  chern insulator}} (\bibinfo {year} {2026}),\ \Eprint
  {https://arxiv.org/abs/2603.17006} {arXiv:2603.17006 [cond-mat.str-el]}
  \BibitemShut {NoStop}%
\bibitem [{\citenamefont {Lehmann}\ \emph {et~al.}(2025)\citenamefont
  {Lehmann}, \citenamefont {Crippa}, \citenamefont {Sangiovanni},\ and\
  \citenamefont {Budich}}]{Lehmann_2025}%
  \BibitemOpen
  \bibfield  {author} {\bibinfo {author} {\bibfnamefont {C.}~\bibnamefont
  {Lehmann}}, \bibinfo {author} {\bibfnamefont {L.}~\bibnamefont {Crippa}},
  \bibinfo {author} {\bibfnamefont {G.}~\bibnamefont {Sangiovanni}},\ and\
  \bibinfo {author} {\bibfnamefont {J.~C.}\ \bibnamefont {Budich}},\ }\bibfield
   {title} {\bibinfo {title} {Probing green's function zeros by cotunneling
  through mott insulators},\ }\href {https://doi.org/10.1103/jnq4-sykq}
  {\bibfield  {journal} {\bibinfo  {journal} {Phys. Rev. Lett.}\ }\textbf
  {\bibinfo {volume} {135}},\ \bibinfo {pages} {106303} (\bibinfo {year}
  {2025})}\BibitemShut {NoStop}%
\bibitem [{\citenamefont {Stepanov}\ \emph {et~al.}(2024)\citenamefont
  {Stepanov}, \citenamefont {Chatzieleftheriou}, \citenamefont {Wagner},\ and\
  \citenamefont {Sangiovanni}}]{Stepanov_2024}%
  \BibitemOpen
  \bibfield  {author} {\bibinfo {author} {\bibfnamefont {E.~A.}\ \bibnamefont
  {Stepanov}}, \bibinfo {author} {\bibfnamefont {M.}~\bibnamefont
  {Chatzieleftheriou}}, \bibinfo {author} {\bibfnamefont {N.}~\bibnamefont
  {Wagner}},\ and\ \bibinfo {author} {\bibfnamefont {G.}~\bibnamefont
  {Sangiovanni}},\ }\bibfield  {title} {\bibinfo {title} {Interconnected
  renormalization of hubbard bands and green's function zeros in mott
  insulators induced by strong magnetic fluctuations},\ }\href
  {https://doi.org/10.1103/PhysRevB.110.L161106} {\bibfield  {journal}
  {\bibinfo  {journal} {Phys. Rev. B}\ }\textbf {\bibinfo {volume} {110}},\
  \bibinfo {pages} {L161106} (\bibinfo {year} {2024})}\BibitemShut {NoStop}%
\bibitem [{\citenamefont {La~Nave}\ \emph {et~al.}(2025)\citenamefont
  {La~Nave}, \citenamefont {Zhao},\ and\ \citenamefont
  {Phillips}}]{Philips2025}%
  \BibitemOpen
  \bibfield  {author} {\bibinfo {author} {\bibfnamefont {G.}~\bibnamefont
  {La~Nave}}, \bibinfo {author} {\bibfnamefont {J.}~\bibnamefont {Zhao}},\ and\
  \bibinfo {author} {\bibfnamefont {P.~W.}\ \bibnamefont {Phillips}},\
  }\bibfield  {title} {\bibinfo {title} {Luttinger count is the homotopy not
  the physical charge: Generalized anomalies characterize non-fermi liquids},\
  }\href {https://doi.org/10.1103/lfxt-fspg} {\bibfield  {journal} {\bibinfo
  {journal} {Phys. Rev. Lett.}\ }\textbf {\bibinfo {volume} {135}},\ \bibinfo
  {pages} {246501} (\bibinfo {year} {2025})}\BibitemShut {NoStop}%
\bibitem [{Note6()}]{Note6}%
  \BibitemOpen
  \bibinfo {note} {This method is equivalent to introducing a phonon condensate
  with momentum $Q$.}\BibitemShut {Stop}%
\bibitem [{\citenamefont {Schrieffer}\ \emph {et~al.}(1989)\citenamefont
  {Schrieffer}, \citenamefont {Wen},\ and\ \citenamefont {Zhang}}]{swz}%
  \BibitemOpen
  \bibfield  {author} {\bibinfo {author} {\bibfnamefont {J.~R.}\ \bibnamefont
  {Schrieffer}}, \bibinfo {author} {\bibfnamefont {X.~G.}\ \bibnamefont
  {Wen}},\ and\ \bibinfo {author} {\bibfnamefont {S.~C.}\ \bibnamefont
  {Zhang}},\ }\bibfield  {title} {\bibinfo {title} {Dynamic spin fluctuations
  and the bag mechanism of high-${T}_{c}$ superconductivity},\ }\href
  {https://doi.org/10.1103/PhysRevB.39.11663} {\bibfield  {journal} {\bibinfo
  {journal} {Phys. Rev. B}\ }\textbf {\bibinfo {volume} {39}},\ \bibinfo
  {pages} {11663} (\bibinfo {year} {1989})}\BibitemShut {NoStop}%
\bibitem [{\citenamefont {Chubukov}\ and\ \citenamefont
  {Frenkel}(1992)}]{Frenkel1992}%
  \BibitemOpen
  \bibfield  {author} {\bibinfo {author} {\bibfnamefont {A.~V.}\ \bibnamefont
  {Chubukov}}\ and\ \bibinfo {author} {\bibfnamefont {D.~M.}\ \bibnamefont
  {Frenkel}},\ }\bibfield  {title} {\bibinfo {title} {Renormalized perturbation
  theory of magnetic instabilities in the two-dimensional hubbard model at
  small doping},\ }\href {https://doi.org/10.1103/PhysRevB.46.11884} {\bibfield
   {journal} {\bibinfo  {journal} {Phys. Rev. B}\ }\textbf {\bibinfo {volume}
  {46}},\ \bibinfo {pages} {11884} (\bibinfo {year} {1992})}\BibitemShut
  {NoStop}%
\bibitem [{\citenamefont {Costa}\ \emph {et~al.}(2018)\citenamefont {Costa},
  \citenamefont {Blommel}, \citenamefont {Chiu}, \citenamefont {Batrouni},\
  and\ \citenamefont {Scalettar}}]{costa2018}%
  \BibitemOpen
  \bibfield  {author} {\bibinfo {author} {\bibfnamefont {N.~C.}\ \bibnamefont
  {Costa}}, \bibinfo {author} {\bibfnamefont {T.}~\bibnamefont {Blommel}},
  \bibinfo {author} {\bibfnamefont {W.-T.}\ \bibnamefont {Chiu}}, \bibinfo
  {author} {\bibfnamefont {G.}~\bibnamefont {Batrouni}},\ and\ \bibinfo
  {author} {\bibfnamefont {R.~T.}\ \bibnamefont {Scalettar}},\ }\bibfield
  {title} {\bibinfo {title} {Phonon dispersion and the competition between
  pairing and charge order},\ }\href
  {https://doi.org/10.1103/PhysRevLett.120.187003} {\bibfield  {journal}
  {\bibinfo  {journal} {Phys. Rev. Lett.}\ }\textbf {\bibinfo {volume} {120}},\
  \bibinfo {pages} {187003} (\bibinfo {year} {2018})}\BibitemShut {NoStop}%
\bibitem [{\citenamefont {Esterlis}\ \emph {et~al.}(2019)\citenamefont
  {Esterlis}, \citenamefont {Kivelson},\ and\ \citenamefont
  {Scalapino}}]{esterlis_19}%
  \BibitemOpen
  \bibfield  {author} {\bibinfo {author} {\bibfnamefont {I.}~\bibnamefont
  {Esterlis}}, \bibinfo {author} {\bibfnamefont {S.~A.}\ \bibnamefont
  {Kivelson}},\ and\ \bibinfo {author} {\bibfnamefont {D.~J.}\ \bibnamefont
  {Scalapino}},\ }\bibfield  {title} {\bibinfo {title} {Pseudogap crossover in
  the electron-phonon system},\ }\href
  {https://doi.org/10.1103/PhysRevB.99.174516} {\bibfield  {journal} {\bibinfo
  {journal} {Phys. Rev. B}\ }\textbf {\bibinfo {volume} {99}},\ \bibinfo
  {pages} {174516} (\bibinfo {year} {2019})}\BibitemShut {NoStop}%
\bibitem [{Note7()}]{Note7}%
  \BibitemOpen
  \bibinfo {note} {To keep the ground-state energy independent on ${\protect
  \mathbf q}$ to order $1/\lambda ^4_p$, one has to include $1/\lambda ^4_p$
  correction into the $F_{\protect \mathbf q} (n, \lambda _p) |\Delta
  _{\protect \mathbf q}|^2/(4\beta \omega _0)$ term and also add an additional
  term with the structure ${\protect \tilde F} _{\protect \mathbf q} (n,
  \lambda _p) |\Delta _{\protect \mathbf q}|^4/(4 \beta \omega _0)^3$. We
  verified that the energies of the homogeneous and checkerboard polaron states
  coincide if we choose $F_{Q} (n, \lambda _p) = 1/(16 \lambda ^2_p) + 9/(1024
  \lambda ^4_p)$ and ${\protect \tilde F}_{Q} (n, \lambda _p) = - 9/(1024
  \lambda ^4_p)$. We also note that for this choice of $F_{Q} (n, \lambda _p)$,
  the renormalization of the chemical potential vanishes, i.e., $\mu _n =
  -\beta \omega _0 (1-2n)$ also at non-zero $W$.}\BibitemShut {Stop}%
\bibitem [{\citenamefont {Capone}\ and\ \citenamefont
  {Ciuchi}(2003)}]{Ciuchi_2003}%
  \BibitemOpen
  \bibfield  {author} {\bibinfo {author} {\bibfnamefont {M.}~\bibnamefont
  {Capone}}\ and\ \bibinfo {author} {\bibfnamefont {S.}~\bibnamefont
  {Ciuchi}},\ }\bibfield  {title} {\bibinfo {title} {Polaron crossover and
  bipolaronic metal-insulator transition in the half-filled holstein model},\
  }\href {https://doi.org/10.1103/PhysRevLett.91.186405} {\bibfield  {journal}
  {\bibinfo  {journal} {Phys. Rev. Lett.}\ }\textbf {\bibinfo {volume} {91}},\
  \bibinfo {pages} {186405} (\bibinfo {year} {2003})}\BibitemShut {NoStop}%
\bibitem [{\citenamefont {Esterlis}\ \emph {et~al.}(2018)\citenamefont
  {Esterlis}, \citenamefont {Nosarzewski}, \citenamefont {Huang}, \citenamefont
  {Moritz}, \citenamefont {Devereaux}, \citenamefont {Scalapino},\ and\
  \citenamefont {Kivelson}}]{esterlis_18}%
  \BibitemOpen
  \bibfield  {author} {\bibinfo {author} {\bibfnamefont {I.}~\bibnamefont
  {Esterlis}}, \bibinfo {author} {\bibfnamefont {B.}~\bibnamefont
  {Nosarzewski}}, \bibinfo {author} {\bibfnamefont {E.~W.}\ \bibnamefont
  {Huang}}, \bibinfo {author} {\bibfnamefont {B.}~\bibnamefont {Moritz}},
  \bibinfo {author} {\bibfnamefont {T.~P.}\ \bibnamefont {Devereaux}}, \bibinfo
  {author} {\bibfnamefont {D.~J.}\ \bibnamefont {Scalapino}},\ and\ \bibinfo
  {author} {\bibfnamefont {S.~A.}\ \bibnamefont {Kivelson}},\ }\bibfield
  {title} {\bibinfo {title} {Breakdown of the migdal-eliashberg theory: A
  determinant quantum monte carlo study},\ }\href
  {https://doi.org/10.1103/PhysRevB.97.140501} {\bibfield  {journal} {\bibinfo
  {journal} {Phys. Rev. B}\ }\textbf {\bibinfo {volume} {97}},\ \bibinfo
  {pages} {140501} (\bibinfo {year} {2018})}\BibitemShut {NoStop}%
\bibitem [{\citenamefont {Nosarzewski}\ \emph {et~al.}(2021)\citenamefont
  {Nosarzewski}, \citenamefont {Huang}, \citenamefont {Dee}, \citenamefont
  {Esterlis}, \citenamefont {Moritz}, \citenamefont {Kivelson}, \citenamefont
  {Johnston},\ and\ \citenamefont {Devereaux}}]{Nosarzewski2021}%
  \BibitemOpen
  \bibfield  {author} {\bibinfo {author} {\bibfnamefont {B.}~\bibnamefont
  {Nosarzewski}}, \bibinfo {author} {\bibfnamefont {E.~W.}\ \bibnamefont
  {Huang}}, \bibinfo {author} {\bibfnamefont {P.~M.}\ \bibnamefont {Dee}},
  \bibinfo {author} {\bibfnamefont {I.}~\bibnamefont {Esterlis}}, \bibinfo
  {author} {\bibfnamefont {B.}~\bibnamefont {Moritz}}, \bibinfo {author}
  {\bibfnamefont {S.~A.}\ \bibnamefont {Kivelson}}, \bibinfo {author}
  {\bibfnamefont {S.}~\bibnamefont {Johnston}},\ and\ \bibinfo {author}
  {\bibfnamefont {T.~P.}\ \bibnamefont {Devereaux}},\ }\bibfield  {title}
  {\bibinfo {title} {Superconductivity, charge density waves, and bipolarons in
  the holstein model},\ }\href {https://doi.org/10.1103/PhysRevB.103.235156}
  {\bibfield  {journal} {\bibinfo  {journal} {Phys. Rev. B}\ }\textbf {\bibinfo
  {volume} {103}},\ \bibinfo {pages} {235156} (\bibinfo {year}
  {2021})}\BibitemShut {NoStop}%
\bibitem [{\citenamefont {Murthy}\ \emph {et~al.}(2023)\citenamefont {Murthy},
  \citenamefont {Pandey}, \citenamefont {Esterlis},\ and\ \citenamefont
  {Kivelson}}]{murthy_2023}%
  \BibitemOpen
  \bibfield  {author} {\bibinfo {author} {\bibfnamefont {C.}~\bibnamefont
  {Murthy}}, \bibinfo {author} {\bibfnamefont {A.}~\bibnamefont {Pandey}},
  \bibinfo {author} {\bibfnamefont {I.}~\bibnamefont {Esterlis}},\ and\
  \bibinfo {author} {\bibfnamefont {S.~A.}\ \bibnamefont {Kivelson}},\
  }\bibfield  {title} {\bibinfo {title} {A stability bound on the t-linear
  resistivity of conventional metals},\ }\href
  {https://doi.org/10.1073/pnas.2216241120} {\bibfield  {journal} {\bibinfo
  {journal} {Proceedings of the National Academy of Sciences}\ }\textbf
  {\bibinfo {volume} {120}},\ \bibinfo {pages} {e2216241120} (\bibinfo {year}
  {2023})},\ \Eprint
  {https://arxiv.org/abs/https://www.pnas.org/doi/pdf/10.1073/pnas.2216241120}
  {https://www.pnas.org/doi/pdf/10.1073/pnas.2216241120} \BibitemShut {NoStop}%
\bibitem [{\citenamefont {Freericks}\ \emph {et~al.}(1993)\citenamefont
  {Freericks}, \citenamefont {Jarrell},\ and\ \citenamefont
  {Scalapino}}]{freericks1993}%
  \BibitemOpen
  \bibfield  {author} {\bibinfo {author} {\bibfnamefont {J.~K.}\ \bibnamefont
  {Freericks}}, \bibinfo {author} {\bibfnamefont {M.}~\bibnamefont {Jarrell}},\
  and\ \bibinfo {author} {\bibfnamefont {D.~J.}\ \bibnamefont {Scalapino}},\
  }\bibfield  {title} {\bibinfo {title} {Holstein model in infinite
  dimensions},\ }\href {https://doi.org/10.1103/PhysRevB.48.6302} {\bibfield
  {journal} {\bibinfo  {journal} {Phys. Rev. B}\ }\textbf {\bibinfo {volume}
  {48}},\ \bibinfo {pages} {6302} (\bibinfo {year} {1993})}\BibitemShut
  {NoStop}%
\bibitem [{\citenamefont {Millis}\ \emph
  {et~al.}(1996{\natexlab{a}})\citenamefont {Millis}, \citenamefont {Mueller},\
  and\ \citenamefont {Shraiman}}]{Millis_1996}%
  \BibitemOpen
  \bibfield  {author} {\bibinfo {author} {\bibfnamefont {A.~J.}\ \bibnamefont
  {Millis}}, \bibinfo {author} {\bibfnamefont {R.}~\bibnamefont {Mueller}},\
  and\ \bibinfo {author} {\bibfnamefont {B.~I.}\ \bibnamefont {Shraiman}},\
  }\bibfield  {title} {\bibinfo {title} {Fermi-liquid-to-polaron crossover. i.
  general results},\ }\href {https://doi.org/10.1103/PhysRevB.54.5389}
  {\bibfield  {journal} {\bibinfo  {journal} {Phys. Rev. B}\ }\textbf {\bibinfo
  {volume} {54}},\ \bibinfo {pages} {5389} (\bibinfo {year}
  {1996}{\natexlab{a}})}\BibitemShut {NoStop}%
\bibitem [{\citenamefont {Millis}\ \emph
  {et~al.}(1996{\natexlab{b}})\citenamefont {Millis}, \citenamefont {Mueller},\
  and\ \citenamefont {Shraiman}}]{Millis_1996_a}%
  \BibitemOpen
  \bibfield  {author} {\bibinfo {author} {\bibfnamefont {A.~J.}\ \bibnamefont
  {Millis}}, \bibinfo {author} {\bibfnamefont {R.}~\bibnamefont {Mueller}},\
  and\ \bibinfo {author} {\bibfnamefont {B.~I.}\ \bibnamefont {Shraiman}},\
  }\bibfield  {title} {\bibinfo {title} {Fermi-liquid-to-polaron crossover. ii.
  double exchange and the physics of colossal magnetoresistance},\ }\href
  {https://doi.org/10.1103/PhysRevB.54.5405} {\bibfield  {journal} {\bibinfo
  {journal} {Phys. Rev. B}\ }\textbf {\bibinfo {volume} {54}},\ \bibinfo
  {pages} {5405} (\bibinfo {year} {1996}{\natexlab{b}})}\BibitemShut {NoStop}%
\bibitem [{\citenamefont {Ciuchi}\ \emph {et~al.}(1993)\citenamefont {Ciuchi},
  \citenamefont {de~Pasquale}, \citenamefont {Masciovecchio},\ and\
  \citenamefont {Feinberg}}]{Ciuchi_1993}%
  \BibitemOpen
  \bibfield  {author} {\bibinfo {author} {\bibfnamefont {S.}~\bibnamefont
  {Ciuchi}}, \bibinfo {author} {\bibfnamefont {F.}~\bibnamefont {de~Pasquale}},
  \bibinfo {author} {\bibfnamefont {C.}~\bibnamefont {Masciovecchio}},\ and\
  \bibinfo {author} {\bibfnamefont {D.}~\bibnamefont {Feinberg}},\ }\bibfield
  {title} {\bibinfo {title} {Superconductivity and density waves in high
  dimensions},\ }\href {https://doi.org/10.1209/0295-5075/24/7/012} {\bibfield
  {journal} {\bibinfo  {journal} {Europhysics Letters}\ }\textbf {\bibinfo
  {volume} {24}},\ \bibinfo {pages} {575} (\bibinfo {year} {1993})}\BibitemShut
  {NoStop}%
\bibitem [{\citenamefont {{Capone, M.}}\ \emph {et~al.}(1998)\citenamefont
  {{Capone, M.}}, \citenamefont {{Ciuchi, S.}},\ and\ \citenamefont {{Grimaldi,
  C.}}}]{Ciuchi_1998}%
  \BibitemOpen
  \bibfield  {author} {\bibinfo {author} {\bibnamefont {{Capone, M.}}},
  \bibinfo {author} {\bibnamefont {{Ciuchi, S.}}},\ and\ \bibinfo {author}
  {\bibnamefont {{Grimaldi, C.}}},\ }\bibfield  {title} {\bibinfo {title} {The
  small-polaron crossover: Comparison between exact results and vertex
  correction approximation},\ }\href
  {https://doi.org/10.1209/epl/i1998-00283-5} {\bibfield  {journal} {\bibinfo
  {journal} {Europhys. Lett.}\ }\textbf {\bibinfo {volume} {42}},\ \bibinfo
  {pages} {523} (\bibinfo {year} {1998})}\BibitemShut {NoStop}%
\bibitem [{\citenamefont {Capone}\ \emph {et~al.}(2006)\citenamefont {Capone},
  \citenamefont {Carta},\ and\ \citenamefont {Ciuchi}}]{Ciuchi_2006}%
  \BibitemOpen
  \bibfield  {author} {\bibinfo {author} {\bibfnamefont {M.}~\bibnamefont
  {Capone}}, \bibinfo {author} {\bibfnamefont {P.}~\bibnamefont {Carta}},\ and\
  \bibinfo {author} {\bibfnamefont {S.}~\bibnamefont {Ciuchi}},\ }\bibfield
  {title} {\bibinfo {title} {Dynamical mean field theory of polarons and
  bipolarons in the half-filled holstein model},\ }\href
  {https://doi.org/10.1103/PhysRevB.74.045106} {\bibfield  {journal} {\bibinfo
  {journal} {Phys. Rev. B}\ }\textbf {\bibinfo {volume} {74}},\ \bibinfo
  {pages} {045106} (\bibinfo {year} {2006})}\BibitemShut {NoStop}%
\bibitem [{\citenamefont {Fratini}\ and\ \citenamefont
  {Ciuchi}(2021)}]{fratini2021}%
  \BibitemOpen
  \bibfield  {author} {\bibinfo {author} {\bibfnamefont {S.}~\bibnamefont
  {Fratini}}\ and\ \bibinfo {author} {\bibfnamefont {S.}~\bibnamefont
  {Ciuchi}},\ }\bibfield  {title} {\bibinfo {title} {{Displaced Drude peak and
  bad metal from the interaction with slow fluctuations.}},\ }\href
  {https://doi.org/10.21468/SciPostPhys.11.2.039} {\bibfield  {journal}
  {\bibinfo  {journal} {SciPost Phys.}\ }\textbf {\bibinfo {volume} {11}},\
  \bibinfo {pages} {039} (\bibinfo {year} {2021})}\BibitemShut {NoStop}%
\bibitem [{\citenamefont {Moghadas}\ \emph {et~al.}(2025)\citenamefont
  {Moghadas}, \citenamefont {Reitner}, \citenamefont {Wehling}, \citenamefont
  {Sangiovanni}, \citenamefont {Ciuchi},\ and\ \citenamefont
  {Toschi}}]{Ciuchi_2025}%
  \BibitemOpen
  \bibfield  {author} {\bibinfo {author} {\bibfnamefont {E.}~\bibnamefont
  {Moghadas}}, \bibinfo {author} {\bibfnamefont {M.}~\bibnamefont {Reitner}},
  \bibinfo {author} {\bibfnamefont {T.}~\bibnamefont {Wehling}}, \bibinfo
  {author} {\bibfnamefont {G.}~\bibnamefont {Sangiovanni}}, \bibinfo {author}
  {\bibfnamefont {S.}~\bibnamefont {Ciuchi}},\ and\ \bibinfo {author}
  {\bibfnamefont {A.}~\bibnamefont {Toschi}},\ }\href
  {https://arxiv.org/abs/2503.12113} {\bibinfo {title} {Effective enhancement
  of the electron-phonon coupling driven by nonperturbative electronic density
  fluctuations}} (\bibinfo {year} {2025}),\ \Eprint
  {https://arxiv.org/abs/2503.12113} {arXiv:2503.12113} \BibitemShut {NoStop}%
\bibitem [{\citenamefont {Carlson}\ \emph {et~al.}(2004)\citenamefont
  {Carlson}, \citenamefont {Kivelson}, \citenamefont {Orgad},\ and\
  \citenamefont {Emery}}]{Carlson2004}%
  \BibitemOpen
  \bibfield  {author} {\bibinfo {author} {\bibfnamefont {E.~W.}\ \bibnamefont
  {Carlson}}, \bibinfo {author} {\bibfnamefont {S.~A.}\ \bibnamefont
  {Kivelson}}, \bibinfo {author} {\bibfnamefont {D.}~\bibnamefont {Orgad}},\
  and\ \bibinfo {author} {\bibfnamefont {V.~J.}\ \bibnamefont {Emery}},\
  }\bibinfo {title} {Concepts in high temperature superconductivity},\ in\
  \href {https://doi.org/10.1007/978-3-642-18914-2_6} {\emph {\bibinfo
  {booktitle} {The Physics of Superconductors: Vol. II}}},\ \bibinfo {editor}
  {edited by\ \bibinfo {editor} {\bibfnamefont {K.~H.}\ \bibnamefont
  {Bennemann}}\ and\ \bibinfo {editor} {\bibfnamefont {J.~B.}\ \bibnamefont
  {Ketterson}}}\ (\bibinfo  {publisher} {Springer Berlin Heidelberg},\ \bibinfo
  {address} {Berlin, Heidelberg},\ \bibinfo {year} {2004})\ pp.\ \bibinfo
  {pages} {275--451}\BibitemShut {NoStop}%
\bibitem [{\citenamefont {Butler}\ \emph {et~al.}(2021)\citenamefont {Butler},
  \citenamefont {Yoshida}, \citenamefont {Hanaguri},\ and\ \citenamefont
  {Iwasa}}]{Butler2021}%
  \BibitemOpen
  \bibfield  {author} {\bibinfo {author} {\bibfnamefont {C.~J.}\ \bibnamefont
  {Butler}}, \bibinfo {author} {\bibfnamefont {M.}~\bibnamefont {Yoshida}},
  \bibinfo {author} {\bibfnamefont {T.}~\bibnamefont {Hanaguri}},\ and\
  \bibinfo {author} {\bibfnamefont {Y.}~\bibnamefont {Iwasa}},\ }\bibfield
  {title} {\bibinfo {title} {Doublonlike excitations and their phononic
  coupling in a mott charge-density-wave system},\ }\href
  {https://doi.org/10.1103/PhysRevX.11.011059} {\bibfield  {journal} {\bibinfo
  {journal} {Phys. Rev. X}\ }\textbf {\bibinfo {volume} {11}},\ \bibinfo
  {pages} {011059} (\bibinfo {year} {2021})}\BibitemShut {NoStop}%
\bibitem [{\citenamefont {Starykh}\ and\ \citenamefont
  {Reiter}(1996)}]{Starykh_1996}%
  \BibitemOpen
  \bibfield  {author} {\bibinfo {author} {\bibfnamefont {O.~A.}\ \bibnamefont
  {Starykh}}\ and\ \bibinfo {author} {\bibfnamefont {G.~F.}\ \bibnamefont
  {Reiter}},\ }\bibfield  {title} {\bibinfo {title} {Hole motion in the ising
  antiferromagnet: An application of the recursion method},\ }\href
  {https://doi.org/10.1103/PhysRevB.53.2517} {\bibfield  {journal} {\bibinfo
  {journal} {Phys. Rev. B}\ }\textbf {\bibinfo {volume} {53}},\ \bibinfo
  {pages} {2517} (\bibinfo {year} {1996})}\BibitemShut {NoStop}%
\bibitem [{\citenamefont {Luttinger}\ and\ \citenamefont
  {Ward}(1960)}]{Luttinger1960}%
  \BibitemOpen
  \bibfield  {author} {\bibinfo {author} {\bibfnamefont {J.~M.}\ \bibnamefont
  {Luttinger}}\ and\ \bibinfo {author} {\bibfnamefont {J.~C.}\ \bibnamefont
  {Ward}},\ }\bibfield  {title} {\bibinfo {title} {Ground-state energy of a
  many-fermion system. ii},\ }\href {https://doi.org/10.1103/PhysRev.118.1417}
  {\bibfield  {journal} {\bibinfo  {journal} {Phys. Rev.}\ }\textbf {\bibinfo
  {volume} {118}},\ \bibinfo {pages} {1417} (\bibinfo {year}
  {1960})}\BibitemShut {NoStop}%
\bibitem [{\citenamefont {Luttinger}(1960)}]{Luttinger1960a}%
  \BibitemOpen
  \bibfield  {author} {\bibinfo {author} {\bibfnamefont {J.~M.}\ \bibnamefont
  {Luttinger}},\ }\bibfield  {title} {\bibinfo {title} {Fermi surface and some
  simple equilibrium properties of a system of interacting fermions},\ }\href
  {https://doi.org/10.1103/PhysRev.119.1153} {\bibfield  {journal} {\bibinfo
  {journal} {Phys. Rev.}\ }\textbf {\bibinfo {volume} {119}},\ \bibinfo {pages}
  {1153} (\bibinfo {year} {1960})}\BibitemShut {NoStop}%
\bibitem [{\citenamefont {Oshikawa}(2000)}]{Oshikawa2000}%
  \BibitemOpen
  \bibfield  {author} {\bibinfo {author} {\bibfnamefont {M.}~\bibnamefont
  {Oshikawa}},\ }\bibfield  {title} {\bibinfo {title} {Topological approach to
  luttinger's theorem and the fermi surface of a kondo lattice},\ }\href
  {https://doi.org/10.1103/PhysRevLett.84.3370} {\bibfield  {journal} {\bibinfo
   {journal} {Phys. Rev. Lett.}\ }\textbf {\bibinfo {volume} {84}},\ \bibinfo
  {pages} {3370} (\bibinfo {year} {2000})}\BibitemShut {NoStop}%
\bibitem [{\citenamefont {Mozyrsky}\ and\ \citenamefont
  {Chubukov}(2019)}]{Mozyrsky_2019}%
  \BibitemOpen
  \bibfield  {author} {\bibinfo {author} {\bibfnamefont {D.}~\bibnamefont
  {Mozyrsky}}\ and\ \bibinfo {author} {\bibfnamefont {A.~V.}\ \bibnamefont
  {Chubukov}},\ }\bibfield  {title} {\bibinfo {title} {Dynamic properties of
  superconductors: Anderson-bogoliubov mode and berry phase in the bcs and bec
  regimes},\ }\href {https://doi.org/10.1103/PhysRevB.99.174510} {\bibfield
  {journal} {\bibinfo  {journal} {Phys. Rev. B}\ }\textbf {\bibinfo {volume}
  {99}},\ \bibinfo {pages} {174510} (\bibinfo {year} {2019})}\BibitemShut
  {NoStop}%
\end{thebibliography}%
\end{document}